\documentclass{irmaart}

\usepackage{amsmath}
\usepackage{clpsref}
\usepackage{color}
\usepackage{graphicx}
\definecolor{purple}{rgb}{0.5,0,0.5}

\numberwithin{equation}{section}
\numberwithin{figure}{section}
\numberwithin{table}{section}

\newcommand{\biggL}{\mbox{$\large \left[\rule{0ex}{2ex} \right.$}}

\newcommand{\biggR}{\mbox{$\large \left.\rule{0ex}{2ex} \right]$}}

\newcommand{\biggerL}{\mbox{$\large \left[\rule{0ex}{3ex} \right.$}}
\newcommand{\biggerR}{\mbox{$\large \left.\rule{0ex}{3ex} \right]$}}

\begin{document}
\title{Strong QCD and Dyson-Schwinger Equations}
\markboth{Craig D.~Roberts}{Strong QCD and Dyson-Schwinger Equations}
\author{Craig D.\ Roberts}
\address{Physics Division, Argonne National Laboratory, \\ Argonne, Illinois 60439, USA;\\
and
Department of Physics, Illinois Institute of Technology, \\ Chicago, Illinois 60616-3793, USA.}

\maketitle

\begin{abstract}
The real-world properties of quantum chromodynamics (QCD) -- the strongly-interacting piece of the Standard Model -- are dominated by two emergent phenomena: confinement; namely, the theory's elementary degrees-of-freedom -- quarks and gluons -- have never been detected in isolation; and dynamical chiral symmetry breaking (DCSB), which is a remarkably effective mass generating mechanism, responsible for the mass of more than 98\% of visible matter in the Universe.  These phenomena are not apparent in the formulae that define QCD, yet they play a principal role in determining Nature's observable characteristics.  Much remains to be learnt before confinement can properly be understood.  On the other hand,the last decade has seen important progress in the use of relativistic quantum field theory, so that we can now explain the origin of DCSB and are beginning to demonstrate its far-reaching consequences.  Dyson-Schwinger equations have played a critical role in these advances.  These lecture notes provide an introduction to Dyson-Schwinger equations (DSEs), QCD and hadron physics, and illustrate the use of DSEs to predict phenomena that are truly observable.\bigskip

\hspace*{-\parindent}2010 Mathematics Subject Classification:
81Q40, 
81T16, 
81T18 
\end{abstract}



\medskip

\centerline{\Large\bf Contents}
\contentsline {section}{\numberline {1}Introduction}{2}
\contentsline {section}{\numberline {2}Emergent Phenomena}{14}
\contentsline {section}{\numberline {3}Gap and Bethe-Salpeter Equations}{27}
\contentsline {section}{\numberline {4}Exact Results in Hadron Physics}{30}
\contentsline {section}{\numberline {5}Condensates are Confined within Hadrons}{38}
\contentsline {section}{\numberline {6}Many Facets of DCSB}{42}
\contentsline {section}{\numberline {7}Probing the Hadronic Interior}{55}
\contentsline {section}{\numberline {8}Charting the Interaction between Light-Quarks}{63}
\contentsline {section}{\numberline {9}Describing Baryons and Mesons Simultaneously}{66}
\contentsline {section}{\numberline {10}Epilogue}{76}
\contentsline {section}{\numberline {\rule{1em}{0ex}}References}{77}

\section{Introduction}
\label{sec:introduction}
The context for this contribution is provided by recent progress in mathematics, which followed a realisation that the process of renormalisation is naturally expressed via a Hopf algebra structure \cite{Connes:1999yr}.  Indeed, the Hopf algebra approach allows for a comprehensive description of the algebraic and combinatorial structures that underpin renormalisation, which is critical in four-dimensional quantum field theory.  This enables a mathematically sound approach to the problem of computing the $\beta$-function, which is based on two elements: the existence of quantum equations of motion, namely, the Dyson-Schwinger equations (DSEs); and the consequences of the renormalisation group for local field theories.  One needs the DSE umbrella in order to guarantee sufficient recursive structure in the theory such that a nonperturbative approach becomes feasible.  The Hopf algebraic foundations of these phenomena make the approach possible.  Finally, the Hopf algebra description permits one to comprehend and explore basic ideas of renormalisation, driving new applications of those ideas in the context of pure and applied mathematics.  The progressive mathematical reformulation of the established physical procedure of renormalisation has motivated people to imagine that this framework has the power to provide deeper insights into fundamental problems in quantum field theory.  However, practical successes are rare \cite{vanBaalen:2009hu}.

This contribution is primarily an overview of the use of DSEs in hadron physics; i.e., in nonperturbative quantum chromodynamics (QCD).  As such, the aims and problems are quite distinct from those discussed in connection with perturbation theory.  Hadron physics practitioners are primarily concerned with finding a truncation of the infinite tower of DSEs that provides a realistic description of bound-states, rather than with bringing order to the recursive tower itself.  A goal of these lecture notes is therefore to highlight the practical problems faced in the real-world use of DSEs and the physical connections between formalism and phenomena.

A springboard to the widespread application of DSEs in hadron physics was provided by Ref.\,\cite{Roberts:1994dr}.  The approach has since been successful.  Hence numerous other reviews have subsequently been written; e.g., Refs.\,\cite{Roberts:2000aa,Maris:2003vk,Pennington:2005be,Holl:2006ni,Fischer:2006ub,%
Roberts:2007jh,Roberts:2007ji,Holt:2010vj,Chang:2010jq,Swanson:2010pw,Chang:2011vu,%
Boucaud:2011ug}.  Whilst those articles are a resource, it should not be necessary to read them before delving into this one.

QCD is the strong-interaction part of the so-called Standard Model of Particle Physics.  It concerns length-scales on the order of $1\,$fm$\, = 10^{-15}$m, or smaller; i.e., scales less than one femtometre (one ``fermi'').  This corresponds to an energy of $\Lambda_{\rm QCD} \simeq 0.2\,$GeV or greater.\footnote{``eV'' is the abbreviation for electron-volt, which is used as a standard unit of measure in physics because of its usefulness in particle accelerator science: a particle with charge $q$ has an energy $E=qV$ after passing through a potential V; if q is quoted in units of the positron's charge and the potential in volts, one obtains an energy in eV.  $1\,$GeV$\,=10^9$eV.  Hadron physicists typically work with a system of ``natural units'' in which $\hbar=1=c$, so that length- and mass-energy-scales are related as follows: 1\,fm$\,\approx 1/\Lambda_{\rm QCD}$.}  Perturbation theory is a valid tool within QCD at energies$\,\gtrsim 10\,\Lambda_{\rm QCD}$.  It is widely hoped that phenomena outside the realm of the Standard Model will be encountered at energies above $1\,$TeV$\,=1000\,$GeV.

In order to understand the origin of the Standard Model, one should recall that in the early twentieth century the only matter particles known to exist were the proton, neutron and electron.  However, with the advent of cosmic ray science and particle accelerators, numerous additional particles were discovered: muon (1937), pion (1947), kaon (1947), Roper resonance (1963), etc., so that by the mid-1960s it was apparent that not all the particles could be fundamental.  A new paradigm was necessary.  The constituent-quark theory of
Gell-Mann \cite{GellMann:1964nj} and Zweig \cite{Zweig:2010jf} was a critical step forward.  This was recognised with the Nobel Prize to Gell-Mann in 1969: ``for his contributions and discoveries concerning the classification of elementary particles and their interactions''.  Over the more than forty intervening years, the theory now called the Standard Model of Particle Physics has passed almost all tests.

There are three pieces to the Standard Model.  The first is the theory of electromagnetism; i.e., quantum electrodynamics (QED), which was developed in the period 1946-1950 by Feynman, Schwinger and Tomanaga, who shared the 1965 Nobel Prize in physics: ``for their fundamental work in quantum electrodynamics, with deep-ploughing consequences for the physics of elementary particles'' \cite{Nobel65}.  The second, developed in the period 1963-1973, consumed QED through the unification of a description of electromagnetism and weak interactions, such as radioactive decays, parity-violating decays and electron-neutrino scattering.  This body of work, which produced the electroweak theory, was recognised with award of the 1979 Nobel Prize in physics to Glashow, Salam and Weinberg: ``for their contributions to the theory of the unified weak and electromagnetic interaction between elementary particles, including, \emph{inter alia}, the prediction of the weak neutral current'' \cite{Nobel79}.  The final piece is QCD, which is \emph{supposed} to describe: the existence and composition of the vast bulk of visible matter in the Universe -- the proton, neutron, pion, etc.; the forces that form them; and the structure of nuclei.  The development of QCD is far from complete but the theory is understood within the domain for which perturbation theory is a valid tool, following work by Politzer \cite{Politzer:1973fx,Politzer:1974fr}, Gross and Wilczek \cite{Gross:1973id} in 1973-1974 that brought the 2004 Nobel Prize in physics: ``for the discovery of asymptotic freedom in the theory of the strong interaction''.  As we shall see, however, hadron physics requires an understanding of nonperturbative QCD.

Constituted from these elements, the Standard Model is a local quantum gauge field theory, which can be expressed in a very compact form.  The Lagrangian possesses the following gauge symmetry:
\begin{equation}
\begin{array}{ccc}
SU_c(3) & \times & SU_L(2) \times U_Y(1)\\
\mbox{QCD} & & \mbox{electroweak}
\end{array}
\end{equation}
and contains $19$ parameters, which must be determined through comparison with experiment.  Physics is, after all, an experimental science.  The electroweak portion of the Standard Model involves $17$ of the parameters.  Most of these are tied to the Higgs boson, the model's only fundamental scalar, which has never been seen.  The electroweak sector is essentially perturbative, so the parameters are readily determined.  Only two of the Standard Model's parameters are intrinsic to QCD.  However, as QCD is Nature's only example of a truly and essentially nonperturbative fundamental theory, the impact of these parameters is not yet fully known.

\begin{figure}[t]
\vspace*{-0ex}

\includegraphics[clip,width=0.6\textwidth]{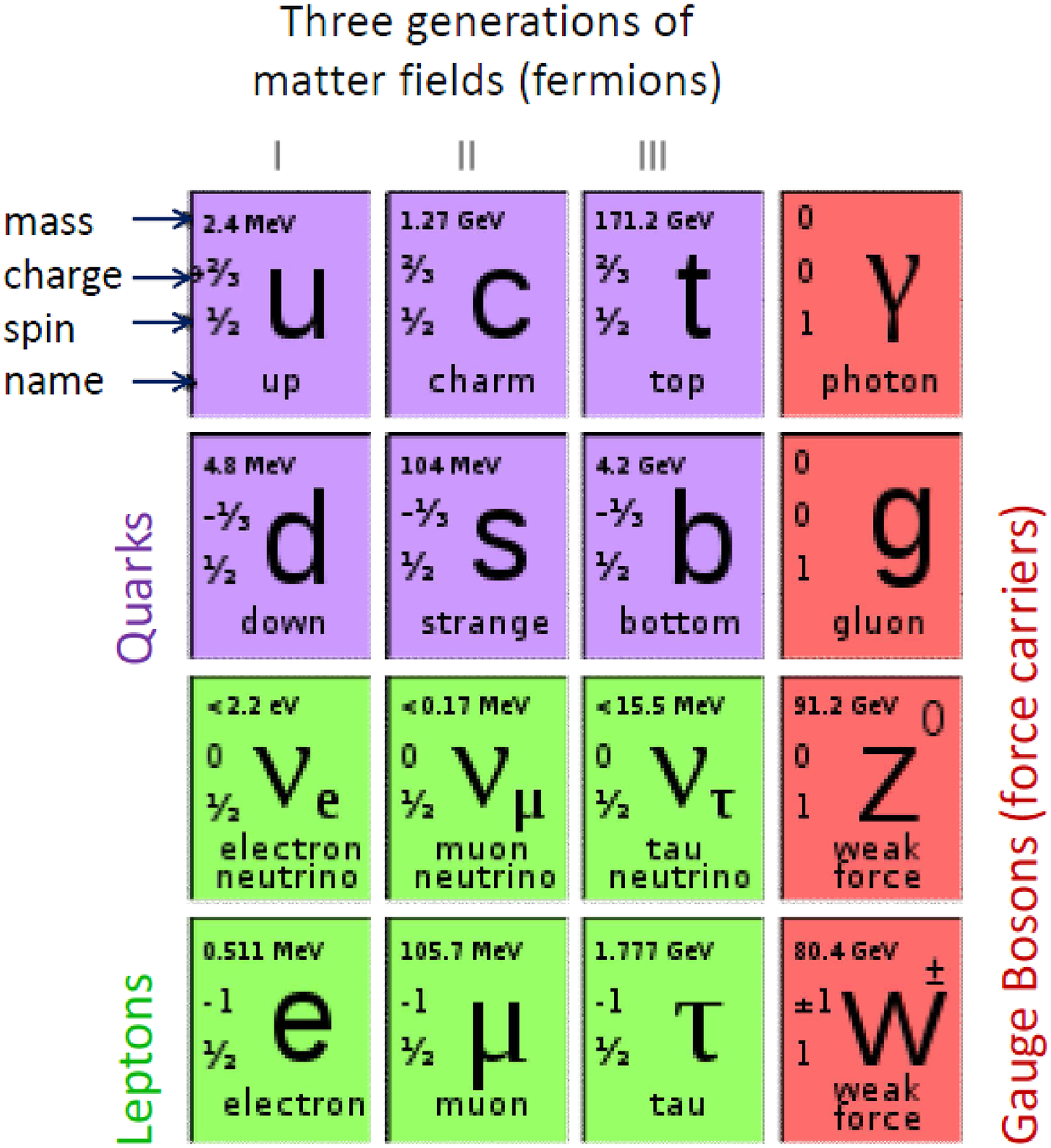}

\caption{\label{F1} Known particle content of the Standard Model.  The Higgs boson is conspicuously absent.  It is possible that within nonperturbative QCD the distinction between matter- and force-fields disappears because bound-states formed solely from the force carrying gluons (gauge-bosons) are conceivable.}

\end{figure}

The known particle content of the Standard Model is depicted in Fig.\,\ref{F1}.  The Higgs boson is obvious by its omission.  Discovery of the Higgs is a primary mission for the Large Hadron Collider (LHC).  Built at a cost of \$\,7-billion, this machine accelerates two antiparallel beams of particles to almost the speed of light in a 27km tunnel that is 175m underground, before colliding them at four interaction points.  During a ten hour experiment, each beam will travel 10-billion\,km; i.e., almost 100-times the earth-sun distance.  The energy of each collision will reach 14\,TeV.  It is the hope of many that this energy is high enough to expose the new particles or structures that are predicted by existing extensions of the Standard Model.  There is also a fear that it is not.

It is noteworthy that, at the time of writing, there is a growing acceptance that the Higgs-boson might not exist: the ATLAS and CMS detectors have excluded the existence of a Higgs over most of the mass region $145 - 466\,$GeV with 95\% certainty.
Optimists view this as an opportunity, arguing that if the Higgs does not exist, its absence will point the way to new physics.\footnote{This was the case until Tues., 13 Dec. 2011.  On that day, amidst great fanfare, CERN announced that they believe they have seen a hint of the Higgs boson.  The results are inconclusive.  In that announcement, the following observations was made: ``The Higgs field is often said to give mass to everything. That is wrong. The Higgs field only gives mass to some very simple particles. The field accounts for only one or two percent of the mass of more complex things like atoms, molecules and everyday objects, from your mobile phone to your pet llama. The vast majority of mass comes from the energy needed to hold quarks together inside atoms.'' This recognises the fundamental role of confinement and dynamical chiral symmetry breaking, key emergent phenomena within the strongly interacting part of the Standard Model, upon which much of this discourse will focus.}

There is certainly physics \emph{beyond the Standard Model}; i.e., observational facts that are inexplicable within the model.  For example, neutrinos have mass, which is not the case in the Standard Model.  This was first detected in the Homestake mine experiment of the late 1960s \cite{Davis:1968cp,Cleveland:1998nv}, which observed a deficit in the flux of solar neutrinos $\nu_e$; i.e., neutrinos associated with the electron.  Controversial from the first moment, this observation was verified conclusively at the Sudbury neutrino observatory \cite{Ahmad:2001an}.  It is explained by neutrino oscillations \cite{Pontecorvo:1957cp,Maki:1962mu,Pontecorvo:1967fh}; viz., so long as neutrinos have mass, a neutrino created in association with a specific lepton flavour (electron, muon or tau) can transform into that associated with another flavour: this is mixing between neutrino flavour- and mass-eigenstates, a phenomenon also seen in the weak interactions of quarks.

Neutrino masses are one concrete example of physics beyond the Standard Model.  Our existence is another because we represent a macroscopic excess of matter over antimatter.  Sakharov first observed \cite{Sakharov:1967dj} that in order to produce an excess of baryons, there must be processes that change baryon number.  No such processes have yet been observed.  Moreover, those processes must occur out of thermal equilibrium, otherwise they would merely balance matter and antimatter.  (Actually, the electroweak component of the Standard Model is capable of satisfying Sakharov's conditions in principle but it is too weak to explain the observed matter-antimatter asymmetry.)  The very likely existence of dark matter and dark energy are two more indications that the Standard Model is incomplete \cite{Murayama:2007ek}.  With this admission of incompleteness, one has opened Pandora's box: the known problems are few but the number of people working on physics beyond the Standard Model is practically innumerable.

One means by which to gauge the foci of contemporary research in physics is to consider the \emph{top-problems} lists that various communities have compiled: one can readily find a top-10 or top-24, etc.  (Such lists have a long history, for who can forget Hilbert's 1900 list \cite{Hilbert:1900} of the most important problems in mathematics for the twentieth century?)  Here I choose to identify just five.
\begin{description}
\item[What is dark matter?] There appears to be a halo of invisible material engulfing galaxies, which is commonly referred to as dark matter.  Existence of dark (=invisible) matter is inferred from the observation of its gravitational pull, which causes the stars in the outer regions of a galaxy to orbit faster than they would if only visible matter were present.  Another indication is that we see galaxies in our own local cluster moving toward each other.
\item[What is dark energy?] The discovery of dark energy goes back to 1998.  A group of scientists had recorded several dozen supernovae, including some so distant that their light had started to travel toward Earth when the universe was only a fraction of its present age.  Contrary to expectations, the scientists found that the expansion of the universe is not slowing, but accelerating \cite{1538-3881-116-3-1009,0004-637X-517-2-565}.  This year, the leaders of those teams shared the Nobel Prize in Physics.
\item[What is the lifetime of the proton and how do we understand it?]�It was once supposed that protons, unlike, neutrons, are stable, never decaying into smaller pieces.  Then, in the 1970's, theorists realised that their candidates for a grand unified theory, merging all the forces except gravity, implied that protons must be unstable.  Wait long enough and, very occasionally, one should break apart.  Is this truly a necessary consequence of Grand Unification?
\item[What underlies the enormous disparity between the gravitational] \textbf{scale and} \\ \mbox{\bf the typical} \textbf{mass scale of the elementary particles?}~~Expressed differently, why is gravity so much weaker than the other forces?  A small magnet can raise a paper clip even though the gravity of the whole earth is resisting.
\item[Can we quantitatively understand quark and gluon confinement in quantum] \textbf{chromodynamics and the existence of a mass gap?}~~QCD is the theory describing the strong nuclear force. Carried by gluons, it binds quarks into particles like protons and neutrons.  Apparently, the quarks and gluons are permanently confined: one cannot remove a quark or a gluon from a proton because the restraining force grows enormously with increasing distance and prevents them from escaping.
\end{description}
This last brings me to the point of these lectures.

\subsection{Quantum chromodynamics}
\label{sec:QCD}
It is straightforward to express the local Lagrangian density that defines QCD:
\begin{eqnarray}
\label{LQCD}
{\cal L}_{\rm QCD} &=& \bar q_i [i \gamma^\mu [D_\mu]_{ij} - M \delta_{ij} ] q_j - \frac{1}{4} G_{\mu\nu}^a G_a^{\mu\nu}\\
&=& \bar q_i [i \gamma^\mu \partial_\mu - m] q_i - g G_\mu^a\bar q_i \gamma^\mu T^a_{ij} q_j - \frac{1}{4} G_{\mu\nu}^a G_a^{\mu\nu}
\label{LQCD2}
\end{eqnarray}
where $G$ is the gluon gauge-field, $q$ is the quark matter-field, $g$ is the coupling constant, and the gluon field-strength tensor is
\begin{equation}
\label{gluonFST}
G_{\mu\nu}^a = \partial_\mu G_\nu^a - \partial_\nu G_\mu^a + \underline{g f^{abc} G_\mu^b G_\nu^c}.
\end{equation}

\begin{figure}[t]
\vspace*{-1ex}

\includegraphics[clip,width=0.2\textwidth]{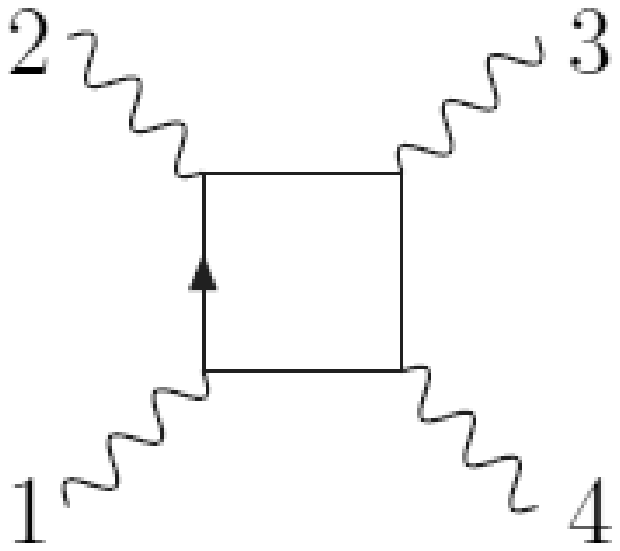}
\hspace*{0.15\textwidth}{\includegraphics[clip,width=0.4\textwidth]{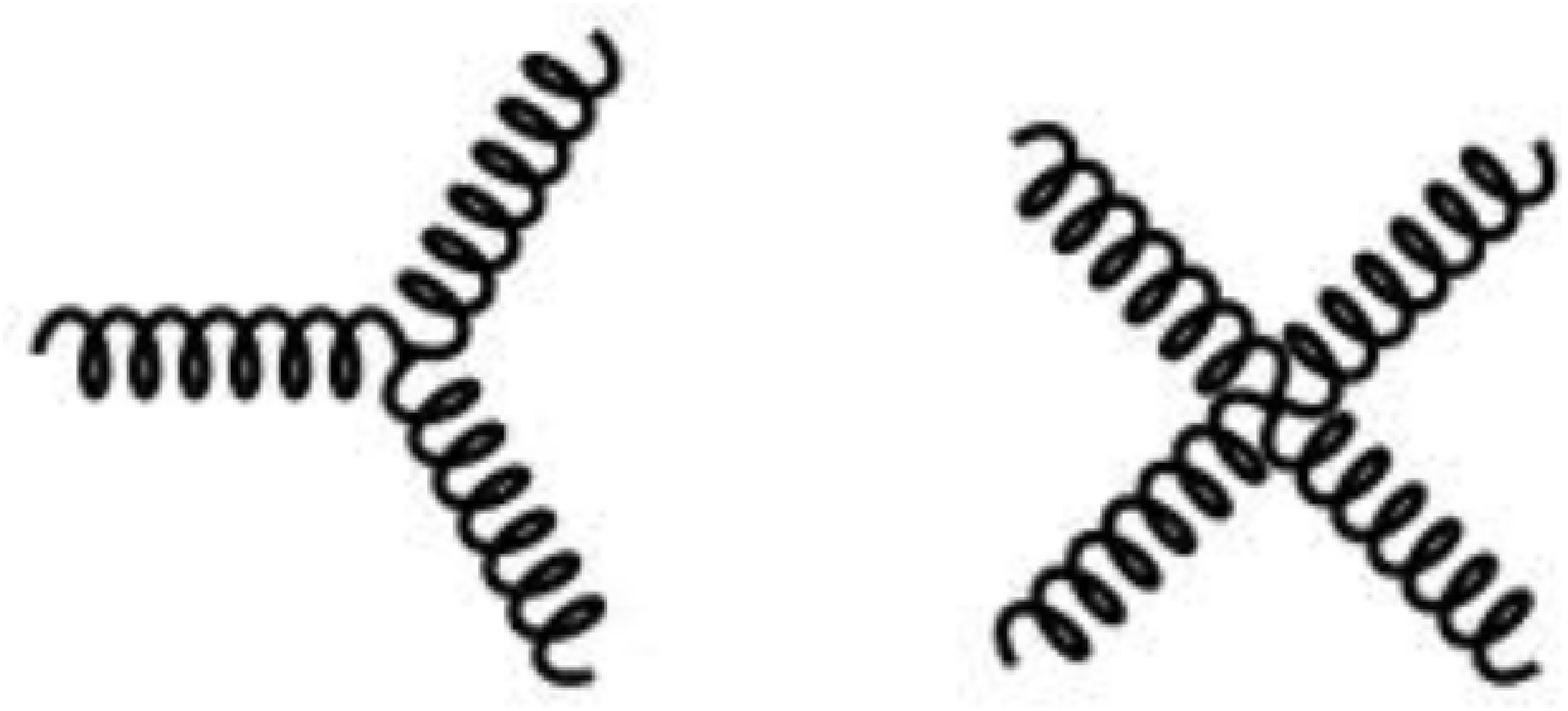}}

\caption{\label{F2} \emph{Left diagram} -- The leading-order contribution to light-by-light scattering is mediated by fermion-antifermion production and annihilation, which occurs at $O(\alpha^4)$; i.e., with probability $\sim 10^{-9}$.
\emph{Right diagrams} -- In QCD, gauge-boson self-interactions occur at tree-level, with these three-gluon [$O(g)$] and four-gluon vertices [$O(g^2)$].}

\end{figure}

The underlined term in Eq.\,\eqref{gluonFST} generates gauge-field self-interactions, which have extraordinary consequences.  This is readily elucidated through a comparison with QED, an Abelian gauge theory, which is perturbatively simple but, alone, undefined nonperturbatively.\footnote{Four-fermion operators become relevant in strong-coupling QED and must be included in order to obtain a well-defined continuum limit \cite{Rakow:1990jv,Reenders:1999bg}.}  A characteristic feature of QED is that gauge-boson self interactions do not occur at tree level. The leading contribution occurs at order $\alpha^4$, Fig.\,\ref{F2}, and since $\alpha \sim 1/137$, such interactions take place with extremely small probability.  In contrast, the non-Abelian character of QCD produces tree-level interactions between gauge bosons -- the three- and four-gluon vertices generated by $G_{\mu\nu}^a G_a^{\mu\nu}$ in Eq.\,\eqref{LQCD}; and one might guess this will have a big impact.  Doing better than guessing led to the 2004 Nobel Prize in Physics \cite{Politzer:1973fx,Politzer:1974fr,Gross:1973id}.

\begin{figure}[t]
\vspace*{-1ex}

\includegraphics[clip,width=0.5\textwidth,height=0.222\textheight]{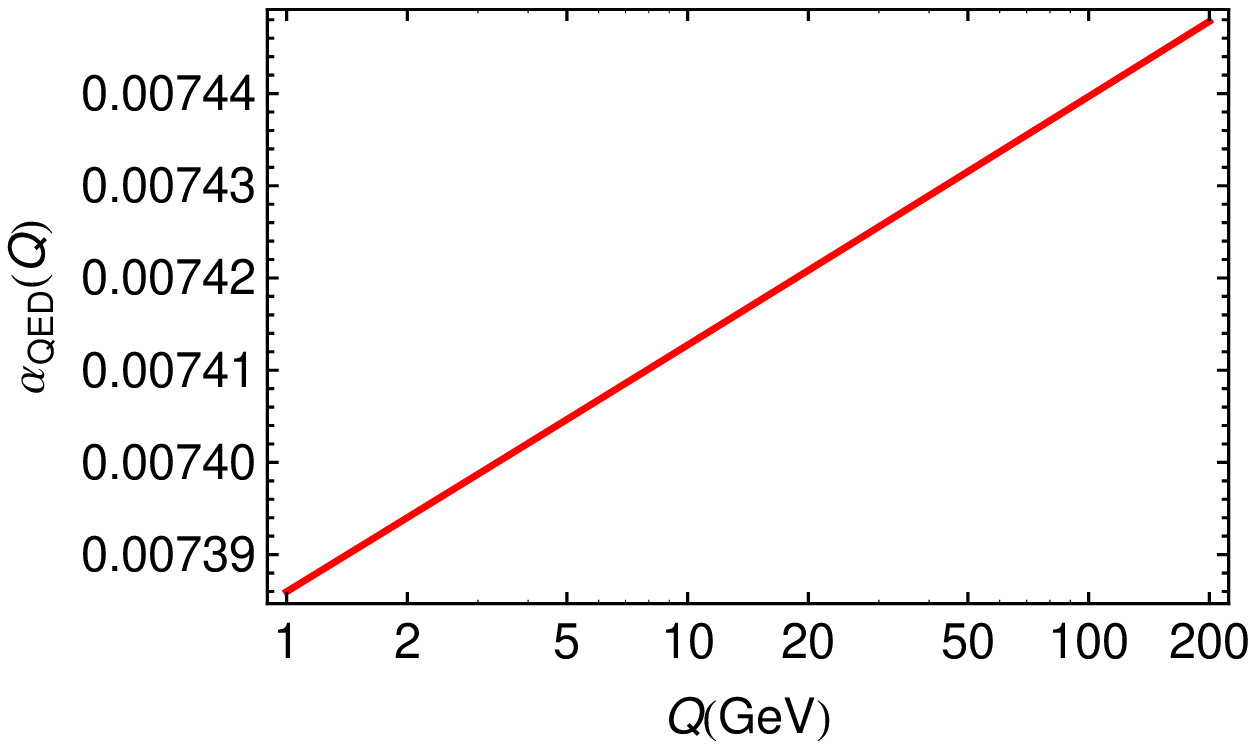}
\includegraphics[clip,width=0.5\textwidth]{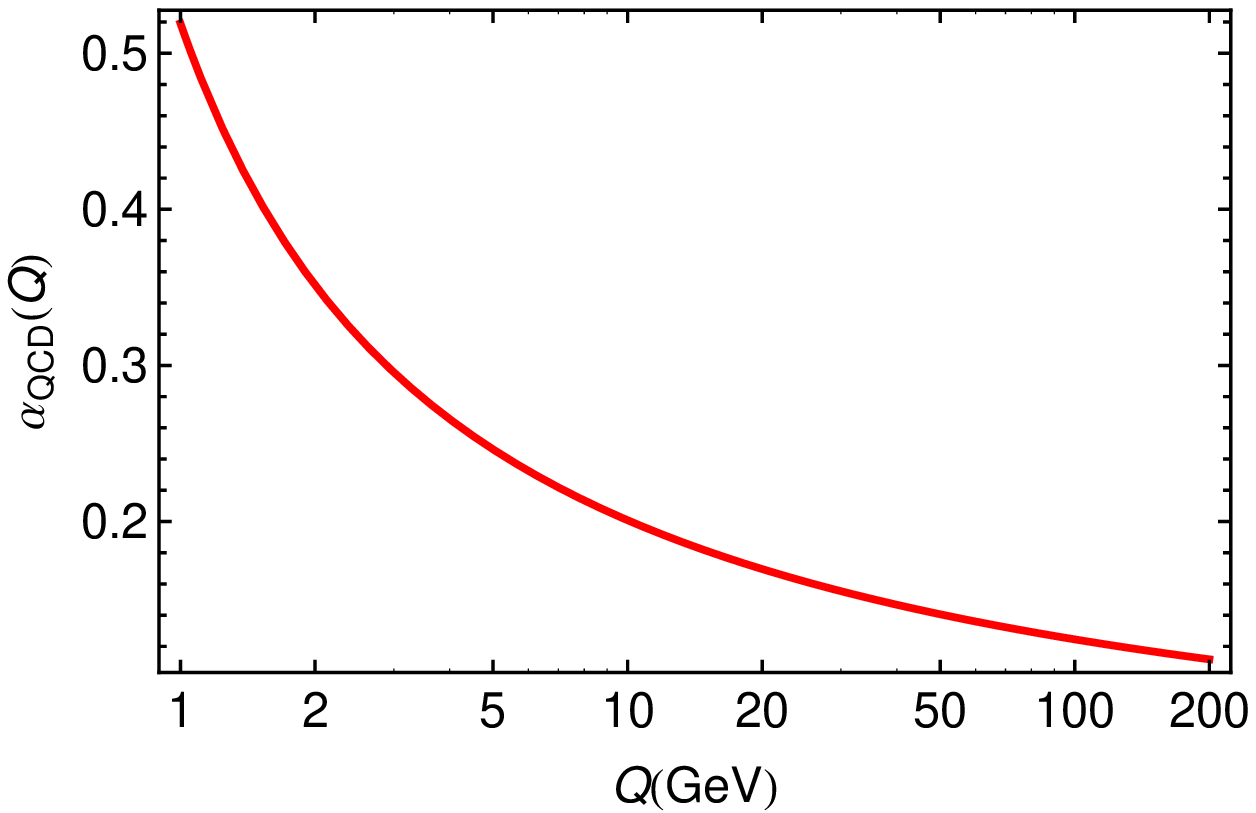}

\caption{\label{F3} \emph{Left panel} -- Running coupling in QED; and \emph{right panel} -- running coupling in QCD.  N.B.\ On the same momentum domain, the QCD coupling changes 500\,000-times more than the QED coupling and runs in the opposite direction.}
\end{figure}

Quantum gauge-field theories are typified by the feature that nothing is constant.  The distribution of charge and mass, the number of particles, etc.; indeed, all the things that quantum mechanics holds fixed, depend upon the wavelength of the tool being used to measure them. Couplings and masses are renormalised via processes involving virtual-particles.  Such effects make these quantities depend on the energy scale at which one observes them, and produce the running couplings illustrated in Fig.\,\ref{F3}.  At one-loop order, the QED coupling is
\begin{equation}
\alpha_{\rm QED}(Q) = \frac{\alpha}{\displaystyle 1-\frac{2 \alpha}{3\pi}\ln \frac{Q}{m_e}},
\end{equation}
where $\alpha$ and $m_e$ are renormalised at the physical electron mass.  The negative sign in the denominator signals that fermions screen electric charge.  In QCD, the analogous quantity is
\begin{equation}
\alpha_{\rm QCD}(Q) = \frac{6 \pi}{\displaystyle (33 - 2 N_f) \ln \frac{Q}{\Lambda_{\rm QCD}}},
\end{equation}
where $N_f$ is the number of quark flavours that are active in the process under consideration and $\Lambda_{\rm QCD}$ is a dynamically generated mass-scale, whose value is experimentally determined; viz., $\Lambda_{\rm QCD} \sim 200\,$MeV.  Plainly, quarks screen the colour charge.  However, owing to their self-interactions,  gluons antiscreen this charge.  The enormous impact of that effect is evident in Fig.\,\ref{F3}.

The running coupling depicted in the right panel of Fig.\,\ref{F3} can be translated into an interaction between quarks and gluons that depends strongly on separation; i.e., an interaction which grows rapidly with separation between the sources.  Taking the experimental value of $\Lambda_{\rm QCD}$, the coupling is found to be huge at a separation $r \simeq 0.2\,{\rm fm} \simeq \frac{1}{4}r_p$, where $r_p$ is the proton's charge radius.  This is a peculiar circumstance; viz., an interaction that becomes stronger as the participants try to separate.  It leads one to explore some fascinating possibilities: If the coupling grows so strongly with separation, then perhaps it is unbounded; and perhaps it would require an infinite amount of energy in order to extract a quark or gluon from the interior of a hadron.  Such thinking has led to the \\
\hspace*{2em}\parbox[t]{0.9\textwidth}{\textit{Confinement Hypothesis}:\\ Colour-charged particles cannot be isolated and therefore cannot be directly observed.  They clump together in colour-neutral bound-states.}\\

\hspace*{-\parindent}This is hitherto an empirical fact.  In the hope that it will become rather more, in 2000 the Clay Mathematics offered a prize of \$1-million for a proof that $SU_c(3)$ gauge theory is mathematically well-defined \cite{Jaffe:Clay}, one necessary consequence of which will be an answer to the question of whether or not the confinement conjecture is correct.

I would like to note here that pure-glue QCD is far simpler than real-world QCD, and that this is crucial in connection with the nature of confinement.  Since bosons possess a classical analogue, it is relatively straightforward to define the pure-glue theory via a discrete probability measure that is amenable to direct numerical simulation using Monte-Carlo methods.  In the absence of light-quarks, there is no perniciously nonlocal fermion determinant to interfere with the simulation.  The quenched theory has long been subject to analysis and the ``confinement of quarks'' addressed in that context \cite{Wilson:1974sk}, with the result that a linearly-rising potential is known to exist between static sources.  It is unfortunate, therefore, that this potential has no known connection with the physics of real-world QCD, wherein light-quarks are omnipresent and between which, owing to their low mass relative to the natural scale of QCD, it is impossible to define a potential.

QCD is supposedly the theory underlying the strong interaction.  It is asymptotically free, so that a chiral limit is defined and perturbation theory is a valid tool for $Q^2\gg \Lambda_{\rm QCD}^2$.  On the other hand, it is associated with an interaction that is strong over 98\% of the interior of a typical hadron, such as the proton, and therefore impossible to treat using perturbation theory.  QCD is Nature's only example of a truly nonperturbative, fundamental theory and, \emph{a priori}, no one has any idea as to what such a theory is capable of producing.  The study of nonperturbative QCD is the purview of Hadron Physics.

\subsection{Hadron physics}
\label{sec:HadronPhysics}
The term ``hadron,'' first used in 1962 \cite{1962hep..conf..845O}, refers to a class of subatomic particles that are composed of quarks and/or gluons and take part in the strong interaction.�  Well known examples are the proton, neutron, and pion.  The word is a derivative of the Greek word \emph{hadros} ($\grave{\alpha}\delta\rho\acute{o}\varsigma$), which variously means ``large'' or ``massive,'' and is intended to contrast with \emph{leptos} ($\lambda\epsilon\pi\tau\acute{o}\varsigma$), which means ``small'' or ``light.''  The world of hadrons is subdivided into baryons, which are hadrons with half-integer spin; and mesons, hadrons with integer spin.

Hadron physics is unique at the cutting edge of modern science because Nature has provided us with just one instance of a fundamental strongly-interacting theory; i.e., QCD.  The community of science has never before confronted such a challenge as solving this theory.  In 2007, the USA's Nuclear Science Advisory Council delivered a long range plan to the Department of Energy's Office of Nuclear Physics, which included the statement: ``A central goal of (the DOE Office of) Nuclear Physics is to understand the structure and properties of protons and neutrons, and ultimately atomic nuclei, in terms of the quarks and gluons of QCD.'' Internationally, this is an approximately \$1-billion/year effort in experiment and theory, with roughly \$375-million/year spent in the USA.  On the order of 90\% of these funds are spent on experiment.  To provide a context, \$1-billion/year is the operating budget of CERN.

The hadron physics community now has a range of major facilities that are accumulating data, of unprecedented accuracy and precision, which pose important challenges for theory.  The opportunities for researchers in hadron physics promise to expand with the use of extant accelerators, and upgraded and new machines and detectors that will appear on a five-to-ten-year time-scale, in China, Germany, Japan, Switzerland and the USA.
A short list of facilities may readily be compiled: Beijing's electron-positron collider; in Germany --
COSY (J\"ulich Cooler Synchrotron),
ELSA (Bonn Electron Stretcher and Accelerator),
MAMI (Mainz Microtron), and
FAIR (Facility for Antiproton and Ion Research) under construction near Darmstadt, with new generation experiments in 2015 (or later, as now seems more probable);
in Japan -- J-PARC (Japan Proton Accelerator Research Complex) under construction in Tokai-Mura, 150km NE of Tokyo, expected to begin operation toward the end of 2012, and
KEK, Tsukuba;
in Switzerland, the ALICE and COMPASS detectors at CERN;
and in the USA, both RHIC (Relativistic Heavy Ion Collider) at Brookhaven National Laboratory, which focuses on the strong phase transition, physics just 10$\mu$s after the Big Bang,
and the Thomas Jefferson National Accelerator Facility (JLab), exploring the nature of cold hadronic matter, currently being upgraded at a cost of \$310-million, with new generation experiments expected in 2016.  The interpretation and prediction of phenomena at JLab provide the themes for much of the material presented herein.

A theoretical understanding of the phenomena of hadron physics requires use of the full machinery of relativistic quantum field theory, which is the \emph{only} way to reconcile quantum mechanics with special relativity.  Relativistic quantum field theory is based on the relativistic quantum mechanics of Dirac and its unification with quantum mechanics took some time.  Questions still remain as to a practical implementation of a Hamiltonian formulation of the relativistic quantum mechanics of interacting systems.

The Poincar\'e group has ten generators: six associated with the Lorentz transformations (rotations and boosts); and four associated with translations.  Quantum mechanics describes the time evolution of a system with interactions and that evolution is generated by the Hamiltonian, or some generalisation thereof.  However, the Hamiltonian is one of the generators of the Poincar\'e group, and it is apparent from the Poincar\'e algebra that boosts do not commute with the Hamiltonian.
This is partly because the existence of antiparticles is concomitant with relativistic quantum mechanics; i.e., the equations of relativistic quantum mechanics admit \emph{negative energy} solutions.  However, once one allows for negative energy, then particle number conservation is lost:
\begin{equation}
E_{\rm system} = E_{\rm system} + (E_{p_1} + E_{\bar{p}_1}) + \ldots\ \mbox{\emph{ad~infinitum}},
\end{equation}
where $E_{\bar k}= - E_{k}$.  This poses a fundamental problem for relativistic quantum mechanics: few particle systems can be studied, but the study of (infinitely) many bodies is difficult and no general theory currently exists.
Following from these observations, the state vector calculated in one momentum frame will not be kinematically related to the state in another frame, a fact that makes a new calculation necessary in every momentum frame.  The discussion of scattering, which takes a state of momentum $p$ to a different state with momentum $p^\prime$ is therefore problematic \cite{Keister:1991sb,Coester:1992cg}.

Relativistic quantum field theory provides a way forward.  In this framework the fundamental entities are fields, which can simultaneously represent infinitely many particles.  The neutral scalar field, $\phi(x)$, provides an example.  One may write
\begin{equation}
\phi(x) = \int\frac{d^3 k}{(2\pi)^3 }\frac{1}{2 \omega_k} \left[ a(k) {\rm e}^{-i k\cdot x} + a^\dagger(k) {\rm e}^{i k\cdot x} \right],
\end{equation}
where: $\omega_k=\sqrt{|\vec{k}|^2+m^2}$ is the relativistic dispersion relation for a massive particle; the four-vector $(k^\mu)= (\omega_k,\vec{k})$; $a(k)$ is an annihilation (creation) operator for a particle (antiparticle) with four-momentum $k$ ($-k$); and $a^\dagger(k)$ is a creation (annihilation) operator for a particle (antiparticle) with four-momentum $k$ ($-k$).  With this plane-wave expansion of the field one may proceed to develop a framework in which the nonconservation of particle number is not a problem.  That is crucial because key observable phenomena in hadron physics are essentially connected with the existence of \emph{virtual} particles.

Relativistic quantum field theory has its own problems, however.  For example, the question of whether a given quantum field theory is rigorously well defined is an \emph{unsolved} mathematical problem.  All relativistic quantum field theories admit analysis via perturbation theory, and perturbative renormalisation is a well-defined procedure that has long been used in quantum electrodynamics (QED) and quantum chromodynamics (QCD), as emphasised by the 2004 Nobel Prize in Physics \cite{Politzer:1973fx,Politzer:1974fr,Gross:1973id}.  However, a rigorous definition of a theory means proving that the theory makes sense \emph{nonperturbatively}.  This is equivalent to proving that all the theory's renormalisation constants are nonperturbatively well-behaved.  I have written that understanding the properties of hadrons requires solving QCD.  However, QCD is not known to be a rigorously well-defined theory and hence it might not have a solution.  Thus, more properly, the task of hadron physics is to determine whether QCD is truly the theory of the strong interaction.

Physics is an experimental science and experiment advances quickly.  Hence, development of an understanding of observable phenomena cannot wait on mathematical rigour.  Assumptions must be made and their consequences explored.  Hadron physics practitioners therefore assume that QCD is (somehow) well-defined and follow where that idea may lead.  That means exploring and mapping the hadron physics landscape with well-understood probes, such as the electron at JLab; and employing established mathematical tools, refining and inventing others, in order to use the Lagrangian density of QCD to predict what should be observable real-world phenomena.

A primary aim of the world's current hadron physics programmes in experiment and theory is to determine whether there are any contradictions with what can actually be \emph{proved} in QCD.  Today, there are none which are uncontroversial.  Indeed, in my view there are none and those who claim otherwise are misguided; see, e.g., Refs.\,\cite{Hecht:2000xa,Aicher:2010cb,Nguyen:2011jy,Roberts:2010rn}.
Another is to identify what QCD will probably predict.  In this field the interplay between experiment and theory is the engine of discovery and progress, and the \emph{discovery potential} of both is high.  Much has been learnt in the last five years -- these notes will provide a perspective on the meaning of some of those discoveries -- and one can safely expect that many surprises remain in hadron physics.

\subsection{Euclidean metric}
In the first formulation of (quenched-)QCD as a numerical problem, a Euclidean metric was employed \cite{Wilson:1974sk}.  The Euclidean-QCD action defines a probability measure, for which many numerical simulation algorithms are available.  Working in Euclidean space is, however, more than simply pragmatic: Euclidean lattice field theory is currently a primary candidate for the rigorous definition of an interacting quantum field theory.  This relies on the plausibility of it being possible to define the generating functional via a proper limiting procedure.  The moments of the Euclidean measure; i.e., ``vacuum expectation values'' of the Euclidean fields, are the $n$-point Schwinger functions; and the quantum field theory is completely determined once all its Schwinger functions are known.  The time-ordered Green functions of the associated Minkowski space theory can be obtained in a formally well-defined fashion from the Schwinger functions \cite{Symanzik69,SW80,GJ81,SE82}.  This is always at the back of the minds of some  theorists who are interested in essentially nonperturbative QCD.

However, there is another very important reason to work in Euclidean space; viz., owing to asymptotic freedom, all results of perturbation theory are strictly valid only at spacelike-momenta.  The set of spacelike momenta correspond to a Euclidean vector space; and the continuation to Minkowski space rests on many assumptions about Schwinger functions that are demonstrably correct only in perturbation theory.  It is assumed, e.g., that a Wick rotation is valid.  Namely, that QCD dynamics do not nonperturbatively generate any singularity structure that is not encountered in perturbative diagrams.  This is a brave assumption, which appears to be completely false in the case of coloured states.  If the appearance is reality, then QCD must be defined in Euclidean space; and the properties of the real-world are then determined only from a continuation of colour-singlet quantities.  This will pose problems because nothing elementary in QCD is a colour singlet.  One must somehow solve real-world problems, such as determining the spectrum and interactions of complicated two- and three-body bound-states, before returning to the real world.  That will require imagination and a good toolbox.  The DSEs provide the latter.

The Euclidean conventions used herein are easily made plain.  For $4$-vectors $a$, $b$:
\begin{equation}
a\cdot b := a_\mu\,b_\nu\,\delta_{\mu\nu} := \sum_{i=1}^4\,a_i\,b_i\,,
\end{equation}
where $\delta_{\mu\nu}$ is the Kronecker delta and the metric tensor.  Hence, a spacelike vector, $Q_\mu$, has $Q^2>0$.  The Dirac matrices are Hermitian and defined by the algebra
\begin{equation}
\{\gamma_\mu,\gamma_\nu\} = 2\,\delta_{\mu\nu}\,.
\end{equation}
I use
\begin{equation}
\sigma_{\mu\nu} = \frac{i}{2} [ \gamma_\mu , \gamma_\nu ] \;\; \mbox{and} \;\;
\gamma_5 := -\,\gamma_1\gamma_2\gamma_3\gamma_4\,,
\end{equation}
so that
\begin{equation}
{\rm tr}\left[ \gamma_5 \gamma_\mu\gamma_\nu\gamma_\rho\gamma_\sigma \right] =
- 4 \,\varepsilon_{\mu\nu\rho\sigma}\,,\; \varepsilon_{1234}= 1\,.
\end{equation}
One may obtain the Euclidean version of any Minkowski space expression by using the following \textit{transcription rules}:
\begin{center}
\parbox{30em}{
\parbox{16em}{Configuration Space
\begin{enumerate}
\item $\displaystyle \int^M \!d^4x^M \, \rightarrow \,-i \int^E \!d^4x^E$
\item $\slash\!\!\! \partial \,\rightarrow \, i\gamma^E\cdot \partial^E $
\item $\slash \!\!\!\! A \, \rightarrow\, -i\gamma^E\cdot A^E$
\item $A_\mu B^\mu\,\rightarrow\,-A^E\cdot B^E$
\item $x^\mu\partial_\mu \to x^E\cdot \partial^E$
\end{enumerate}}\hspace*{0.5em}
\parbox{16em}{Momentum Space
\begin{enumerate}
\item $\displaystyle \int^M\! d^4k^M \, \rightarrow \,i \int^E\! d^4k^E$
\item $\slash\!\!\! k \,\rightarrow \, -i\gamma^E\cdot k^E $
\item $\slash \!\!\!\! A \, \rightarrow\, -i\gamma^E\cdot A^E$
\item $k_\mu q^\mu \, \rightarrow\, - k^E\cdot q^E$
\item $k_\mu x^\mu\,\rightarrow\,-k^E\cdot x^E$
\end{enumerate}}}
\end{center}
These rules are legitimate in perturbation theory; i.e., the correct Euclidean space integral for a given diagram will be obtained by applying them to the Minkowski integral.  The rules take account of the change of variables and Wick rotation of the contour.  When one begins with Euclidean space, as I do, the reverse is also true.  However, for diagrams that represent DSEs which involve dressed $n$-point functions, whose analytic structure is not known \textit{a priori}, the Minkowski space equation obtained using this prescription will have the right appearance but it's solutions may bear no relation to the analytic continuation of the solution of the Euclidean equation.  Any such differences will be nonperturbative in origin.  As mentioned above, it is this fact which makes a choice of metric crucial at the outset.

\subsection{Quarks and contemporary nuclear physics}
\label{sec:QCNP}
The Standard Model contains six quark flavours but for the bulk of observable material in the universe it is only the $u$-, $d-$ and $s$-quarks which are relevant.  Whilst there are nonetheless good reasons to study systems involving $c$- and $b$-quarks, e.g., exploring just which states the Standard Model can support \cite{Brambilla:2010cs} and searching for physics beyond the Standard Model, I will focus herein on the lighter quarks.

There is arguably no hadron more interesting than the pion.  It was discovered by Cecil Powell \cite{Lattes:1947mw}, who was awarded the 1950 Nobel Prize in Physics: ``for his development of the photographic method of studying nuclear processes and his discoveries regarding mesons made with this method.''  The Abstract of Ref.\,\cite{Lattes:1947mw} remarks that: ``Several features of these processes remain to be elucidated, but we present the following account of the experiments because the results appear to bear closely on the important problem of developing a satisfactory meson theory of nuclear forces.''  Indeed they did.

\begin{figure}[t]
\vspace*{-1ex}

\includegraphics[clip,width=0.4\textwidth]{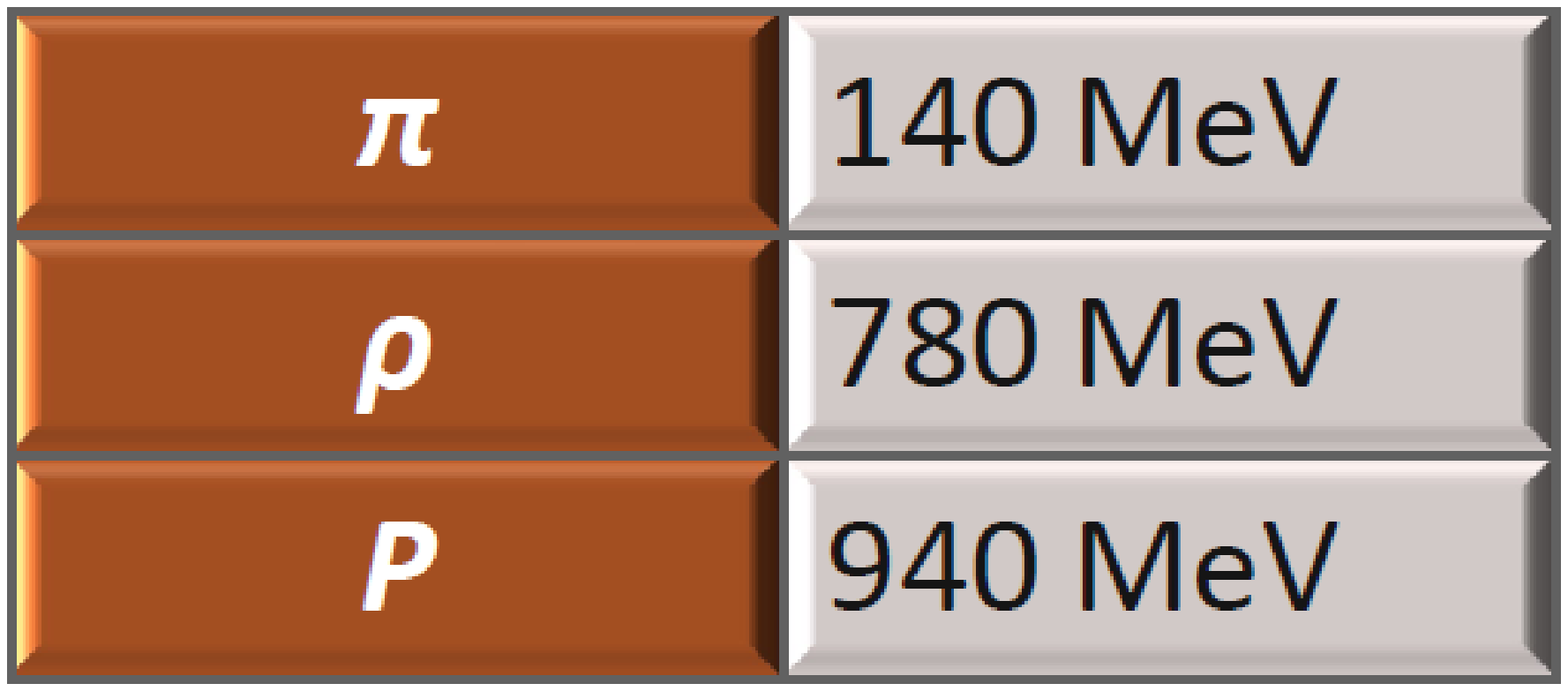}
\hspace*{0.15\textwidth}\includegraphics[clip,width=0.4\textwidth]{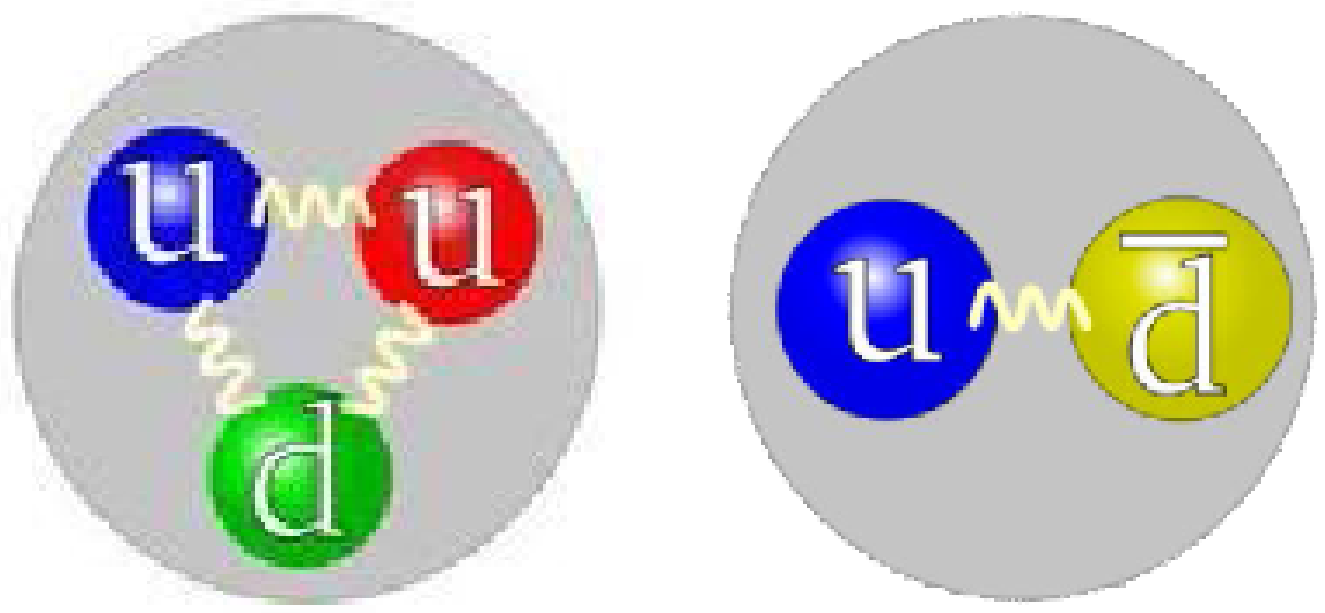}

\caption{\label{F4}
\emph{Left panel} -- the proton ($P$) and $\rho$-meson each have a mass that is typical of hadrons but the pion ($\pi$-meson) is remarkably light, with a mass more typical of leptons.
\emph{Right panel} -- ``eightfold way'' pictures of the proton (left) and pion (right), wherein both are quantum mechanical objects comprised of constituent-quarks bound by a potential.
}
\end{figure}

The discovery of the pion was the beginning of modern particle physics.  Following this came a disentanglement of confusion between the muon and the pion, which have similar masses but vastly different interaction patterns: the muon is a lepton and the pion, a meson.  Then the discovery of \emph{strangeness} \cite{Rochester:1947mi}; and subsequently 28 new hadrons with masses less-than 1\,GeV.  Amongst this collection it was abundantly clear that the pion is too light: it interacts as a hadron but possesses the mass of a lepton, Fig.\,\ref{F4}.

A semblance of order was brought to the hadron spectrum via the ``eightfold way'' \cite{GellMann:1962xb,Ne'eman:1961cd}, a picture based on the Lie group $SU(3)$ wherein the $u$-, $d$- and $s$-quarks are represented by the basis vectors in the fundamental representation.  In such a picture, the pion is a quantum mechanical system comprised from a constituent-quark and constituent-antiquark, interacting via a potential; and the proton is not vastly different: it is three constituent-quarks bound via a symmetry-related potential.  In the usual case this would mean that the mass of the proton is roughly three-times the mass of the constituents; i.e., the constituent-quark mass is $M_U = M_D \approx M_P/3 \simeq 320\,$MeV.  However, the same notion produces $m_\pi \approx (M_U+M_D) = 640\,$MeV, a value that is five-times too large.  The notion cannot quickly be discarded, however, because the $\rho$-meson, which appears to be identical to a pion except that the constituent-quark spins are aligned within the $\rho$, instead of anti-aligned as in the pion, has a mass $770\,{\rm MeV} \approx (M_U+M_D)$.  Furthermore, the properties of the vast majority of ground state hadrons appear to be explained by such simple counting laws.

The correct question therefore appears to be: what is wrong with the pion; or how does one build an almost massless particle from two massive constituent-quarks.  Within quantum mechanics one could naturally tune a potential so that the ground-state is massless but in that case $m_{\rm bound-state}$ rises linearly with the sum of the constituents' masses.  This is not acceptable for the pion because it has long been known that \cite{GellMann:1968rz}
\begin{equation}
\label{GMORprimative}
m_\pi^2 \propto {\cal H}_{\rm mass-term}^1,
\end{equation}
where ${\cal H}_{\rm mass-term}$ is whatever mass-term should be associated with those quarks that actually comprise the pion in the Hamiltonian density which describes them.  It will become clear that in order to resolve the dichotomy of the pion, as a nearly massless bound-state of apparently massive constituents, it is necessary to adopt a theory and an approach in which one can express: a well-defined and valid chiral limit; and an accurate realisation of dynamical chiral symmetry breaking (DCSB).  This is highly nontrivial, impossible in quantum mechanics, and plausible only in asymptotically-free gauge theories.

The flavour index is suppressed in Eq.\,\eqref{LQCD}, but here I note that $M$ represents a current-quark mass matrix:
\begin{equation}
M = \left[
\begin{array}{cccccc}
m_u & 0 & 0 & \ldots & & \\
0 & m_d & 0 & \ldots & & \\
0 & 0 & m_s & \ldots & & \\
\vdots & \vdots & \vdots & m_c & & \\
& & & & m_b & \\
& & & & & m_t
\end{array}\right].
\end{equation}
The current-quark masses are external parameters in QCD.  They are generated by couplings between quark currents and the Higgs boson within the Standard Model.  With six different masses, there are six different couplings.  This mechanism produces the mass values \cite{Nakamura:2010zzi} but doesn't explain them; e.g., no one knows why $m_t \approx 40\,000\,m_u$, although the number of suggested answers is inversely proportional to this level of certainty.

Interacting gauge theories, in which it makes sense to speak of massless fermions, possess a nonperturbative chiral symmetry.  This means that one can set some or all of the current-quark masses to zero, and define and preserve the associated set of Ward-Takahashi identities.  (I will return to this in Sec.\,\ref{sec:inmeson} because the chiral limit can be a complicated concept.)  A related concept is helicity, which is the projection of a particle's spin, $J$, onto its direction of motion:
\begin{equation}
\lambda = J\cdot p\,.
\end{equation}
For a massless fermion, helicity is a Lorentz-invariant spin-observable: $\Lambda = \pm$; i.e., it is parallel or antiparallel to the direction of motion.  This is readily understood: massless particles travel at the speed of light, hence no observer can overtake the particle and thereby view its momentum as having changed sign.

The generator of chiral transformations involves $\gamma_5$ and the simplest chiral rotation is
\begin{equation}
q(x) \to {\rm e}^{i \gamma_5 \theta} \, q(x)\,,\;
\bar q(x) \to \bar q(x)\,{\rm e}^{i \gamma_5 \theta}.
\end{equation}
In connection with a fermion-antifermion bound-state, a rotation through $\theta = \pi/4$ changes a parity-positive state into a parity-negative state: $J^\pm \to J^\mp$; i.e., a particular chiral rotation is equivalent to the operation of parity conjugation  (see Sec.\,3.2 of Ref.\,\cite{Holl:2006ni}).  Therefore, if a theory is chirally symmetric, then its spectrum should be expected to exhibit degenerate parity partners.  (At this point I have overlooked the feature of chiral anomalies in gauge field theories because they have no bearing on this argument.  Section~\ref{flavourless} canvasses aspects of the non-Abelian anomaly.)

Two of the current-quarks in perturbative QCD are light \cite{Leutwyler:2009jg}:
\begin{equation}
\frac{m_u}{m_d} \approx 0.5\,,\; m_d \approx 4\,{\rm MeV}.
\end{equation}
Hence the spectrum of states constituted from $u$- and $d$-quarks should exhibit almost exact parity doubling.  However, the splitting between parity partners is typically greater-than 100-times the current-quark mass-scale; e.g.,
\begin{equation}
\begin{array}{lcc}
{\rm state}\,(J^P)     & {\rm nucleon}\,(\frac{1}{2}^+) & {\rm nucleon}\,(\frac{1}{2}^-) \\
{\rm mass\,(MeV)} & 940 & 1535
\end{array}.
\end{equation}

Plainly, something is happening in QCD: some inherent, dynamical effect is dramatically changing the pattern by which the Lagrangian-density's chiral symmetry is expressed.   The effect is qualitatively different from spontaneous symmetry breaking, $\grave{a}$ \textit{la} the Higgs mechanism.  Nothing is added to QCD.  One has only the original fermion and gauge-boson fields, yet the mass operator produces a spectrum with no sign of the chiral symmetry which characterised the light-quark sector of the Lagrangian density.

The key challenges of real-world QCD are now plain.  Quarks and gluons are apparently confined; viz., no matter how hard one strikes the proton, one cannot liberate an individual quark or gluon.  There is a very unnatural pattern of bound-state masses: whilst the Lagrangian (or perturbative) bare masses are very small, the spectrum does not exhibit a degeneracy between parity partners.  One can express this as a dynamical breakdown of chiral symmetry.  Neither of these phenomena is apparent in QCD's Lagrangian density, yet they are the dominant characteristics of real-world QCD.  The task of hadron physics is to explain the manner through which such complex behaviour emerges from rules so simple that they can be expressed in just a few lines; viz., Eqs.\,\eqref{LQCD2}, (\ref{gluonFST}).

\section{Emergent Phenomena}
\label{sec:Emergent}
Two of the most important hadrons are the neutron and proton, collectively termed the nucleons.  They are characterised by two static properties: electric change and magnetic moment, $\mu_N$.  The charges are obvious: proton, $+1$, and neutron, $0$; however, the magnetic moments are not quantised.  The proton's magnetic moment was discovered by Stern in 1933 \cite{Stern1}, who received the Nobel Prize for Physics in 1943 \cite{Nobel43}: ``for his contribution to the development of the molecular ray method and his discovery of the magnetic moment of the proton.''  The measured magnetic moment of the proton is
\begin{equation}
\mu_p = (1 + 1.79 ) \frac{ e }{2 M},
\end{equation}
where $e$ is the proton's charge and $M$ is its mass.  At the time of discovery, this immediately posed a problem because in Dirac's relativistic quantum mechanics a pointlike fermion with charge $q$ and mass $m$, interacting with an electromagnetic field, has a magnetic moment $\mu= q/[2 m]$ \cite{Dirac:1928hu}.  The actual value is almost three-times Dirac's prediction.

Numerous explanations were proposed but we now know this large discrepancy originates from the finite size of the proton.  This feature of the proton was not uncovered until 1955 \cite{Hofstadter:1955ae}, a discovery that brought Hofstadter a half-share of the 1961 Nobel Prize in physics \cite{Nobel61}: ``for his pioneering studies of electron scattering in atomic nuclei and for his thereby achieved discoveries concerning the structure of the nucleons.''  (The 1961 prize also recognised M\"ossbauer: ``for his researches concerning the resonance absorption of $\gamma$ radiation and his discovery in this connection of the effect which bears his name.'')

The proton is nonpointlike because it is a composite object.  The baryon-number carrying constituents of the proton are quarks, whose existence was confirmed via high-energy electron-proton scattering experiments at the Stanford Linear Accelerator Center (SLAC) in 1968.  This research resulted in Friedman, Kendall and Taylor being awarded the 1990 Nobel Prize in physics \cite{Taylor:1991ew,Kendall:1991np,Friedman:1991nq,Friedman:1991ip}: ``for their pioneering investigations concerning deep inelastic scattering of electrons on protons and bound neutrons, which have been of essential importance for the development of the quark model in particle physics.''

The elastic and inelastic scattering of electrons from an hadronic target are still key tools in hadron physics.  The electron in particular, and leptons in general, are excellent probes because, on the hadron scale at least, they are structureless.  Their interactions are described by a simple current
\begin{equation}
\label{ecurrent}
j^e_\mu(P^\prime,P) = -i e \bar u_e(P^\prime) \Lambda_\mu(P,Q) u_e(P)
\stackrel{\stackrel{\rm Born}{\rm approximation}}{=} -i e \bar u_e(P^\prime)\gamma_\mu u_e(P) \,,
\end{equation}
where the momentum transfer $Q=P^\prime - P$.  Consequently, structures revealed by such experiments can typically be interpreted as properties of the hadron target.

The nucleon's electromagnetic current is
\begin{subequations}
\label{NcurrentA}
\begin{eqnarray}
j^N_\mu(P^\prime,P) &= &i e \, \bar u_N(P^\prime) \Lambda_\mu(P,Q) u_N(P)\\
& = & i e \, \bar u_N(P^\prime)\left[ \gamma_\mu F_{1N}(Q^2)  + \frac{1}{2M}\sigma_{\mu\nu}Q_\nu F_{2N}(Q^2)\right] u_N(P) \,. \label{Ncurrent}
\end{eqnarray}
\end{subequations}
Here, $F_{1N}(Q^2)$ and $F_{2N}(Q^2)$ are, respectively, the nucleons' Dirac and Pauli form factors: $F_1$ is called the Dirac form factor because of its association with the matrix structure of the current for a pointlike fermion; and $F_2$ is the Pauli form factor because it is associated with an additional -- or anomalous Pauli \cite{Pauli:1934xm} -- contribution to the particle's magnetic moment.  The Sachs electric and magnetic form factors are expressed in terms of $F_{1,2}$ as follows:
\begin{subequations}
\label{eq:sachs}
\begin{eqnarray}
G_{EN}(Q^2) &= & F_{1N}(Q^2) - \frac{Q^2}{4 M^2} F_{2N}(Q^2)\,,\\
G_{MN}(Q^2) &= & F_{1N}(Q^2) + F_{2N}(Q^2)\,.
\end{eqnarray}
\end{subequations}
If one is studying a system for which a nonrelativistic limit is a useful or valid approximation, then Fourier transforms of the Sachs form factors produce configuration-space densities that describe the distribution of charge and magnetisation within the composite object.  A comparison between Eqs.\,\eqref{ecurrent} and \eqref{Ncurrent} reveals that for a structureless or simply-structured fermion $F_1(Q^2) \equiv 1$ and $F_2(Q^2) \equiv 0$, and hence $G_E(Q^2) \equiv G_M(Q^2)$.  Thus the distribution of charge and magnetisation is identical within composite but simply-structured fermions.

As observed in Sec.\,\ref{sec:HadronPhysics}, a central goal of nuclear physics is to understand the structure and properties of protons and neutrons, and ultimately atomic nuclei, in terms of the quarks and gluons of QCD.  Confinement and dynamical chiral symmetry breaking, in fact, a fundamental theory of unprecedented complexity means that this aim is far from being achieved.  The struggle with QCD defines the difference between nuclear and particle physics: nuclear physicists try to solve this theory, whereas particle physicists escape to a place where perturbation theory, the last refuge of a scoundrel, is all that is necessary.

Before one can make progress toward this goal, one must first grasp what is meant by ``the quarks and gluons of QCD.''  Is there such a thing as a constituent-gluon or a constituent-quark?  After all, these are the concepts for which Gell-Mann was awarded the Nobel Prize.  Do they or can they be rigorously defined in QCD, perhaps as quasi-particle degrees-of-freedom?  If not, then with what should they be replaced and how should the goal be rigorously interpreted?  Finally, why is it important to achieve this goal?

The last question is readily answered.  Recall the dichotomy of the pion, discussed in connection with Eq.\,\eqref{GMORprimative}: how does one compose an almost massless object from two very massive constituents?  The alternative is to have $m_\pi \approx m_\rho$.  In that case the attractive and repulsive forces in the nucleon-nucleon potential have the same range and there is no intermediate range attraction.  Under these circumstances, it is probable that \cite{Flambaum:2007mj}: the deuteron would not be stable, big-bang nucleosynthesis would not have occurred; $^{12}$C could not have been produced and wouldn't be stable if it were; etc.  These questions matter because their answers explain how we came to exist and ask them.

Neither confinement nor DCSB is apparent in QCD's Lagrangian and yet they play the dominant role in determining the observable characteristics of real-world QCD.  The physics of hadrons is ruled by \emph{emergent phenomena} such as these, which can only be elucidated through the use of nonperturbative methods in quantum field theory.  This is both the greatest novelty and the greatest challenge within the Standard Model.  We must find essentially new ways and means to explain precisely via mathematics the observable content of QCD.

\subsection{Dyson-Schwinger equations}
Beginning with the quark model, many insights into the nature of hadrons have been obtained by using simple models of their structure.  A problem with such studies, however, is that they typically employ degrees-of-freedom, potentials, etc., that have no ascertainable connection with QCD.  At another extreme is the brute-force numerical simulation of lattice-regularised QCD.  This is today a sizeable international endeavour, with the annual five-day lattice conference attracting approximately four-hundred participants.  There are numerous reviews, recent amongst them being Refs.\,\cite{Beane:2008dv,Bazavov:2009bb,Hagler:2009ni,Colangelo:2010et}.

In between, however, there is a compromise.  In 1994 the following observations were made \cite{Roberts:1994dr}: ``As computer technology continues to improve, lattice gauge theory [LGT] will become an increasingly useful means of studying hadronic physics through investigations of discretised quantum chromodynamics [QCD] \ldots\ However, it is equally important to develop other complementary nonperturbative methods based on continuum descriptions.  In particular, with the advent of new accelerators such as CEBAF (VA) and RHIC (NY), there is a need for the development of approximation techniques and models which bridge the gap between short-distance, perturbative QCD and the extensive amount of low- and intermediate-energy phenomenology in a single covariant framework \ldots\ Cross-fertilisation between LGT studies and continuum techniques provides a particularly useful means of developing a detailed understanding of nonperturbative QCD.''  These statements have proved true and a useful compromise is provided by QCD's Dyson-Schwinger equations (DSEs).

\begin{figure}[t]
\vspace*{-1ex}

\includegraphics[clip,width=0.5\textwidth]{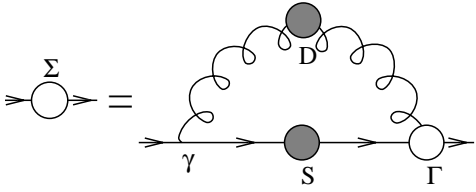}

\caption{\label{fig:gap} In QCD this gap equation describes the manner by which a quark's acquires a momentum-dependent self-energy, $\Sigma(p)$, via interaction with its own gluon field.  This diagram represents a nonlinear integral equation because $S(p)=1/[i\gamma\cdot p + m+ \Sigma(p)]$.  The kernel of that integral equation is composed from the dressed-gluon propagator, $D_{\mu\nu}(k)$, and the dressed-quark-gluon vertex, $\Gamma_\nu(\ell,p)$, both of which satisfy integral equations of their own.  (The quark is represented by the solid line, the gluon by the spring-like line and the quark-gluon vertex by the open circle.)}

\end{figure}

It has long been known \cite{Dyson:1949ha,Schwinger:1951ex,Schwinger:1951hq} that one can derive a system of coupled integral equations relating all the Green functions for a theory, one to another.  The gap equation, illustrated in Fig.\,\ref{fig:gap}, is one such expression.  The DSEs are nonperturbative equivalents in quantum field theory of the Lagrange equations of motion and, amongst their many uses, they are essential in simplifying the general proof of renormalisability of gauge field theories.

The DSEs are a particularly well suited for use in QCD.  At the simplest level they provide a generating tool for perturbation theory.  It therefore follows from asymptotic freedom that in all applications any model-dependence is restricted to the infrared or, equivalently, the long-range domain.  On the other hand, a nonperturbative solution of the DSEs enables the study of: hadrons as composites of dressed-quarks and -gluons; the phenomena of confinement and DCSB; and therefrom an articulation of any connection between them.  The solutions of the DSEs are Schwinger functions and because all cross-sections can be constructed from such $n$-point functions, the DSEs can be used to make predictions for real-world experiments.  One of the merits in this is that assumptions about the infrared form of $n$-point functions can be tested, verified and improved, or rejected in favour of more promising alternatives.  In this mode the DSEs provide a bridge between experiment and theory, and thereby a means by which to use nonperturbative phenomena to chart or at least constrain the infrared behaviour of QCD's $\beta$-function.

Let's return to the \index{Quark propagator} dressed-quark propagator, which is given by the solution of QCD's gap equation, Fig.\,\ref{fig:gap}:
\begin{eqnarray}
\label{Sqdse}
S(p)^{-1} &=& i\gamma\cdot p \,+ m + \Sigma(p)\,,\\
\Sigma(p) &= & \int {d^4\ell\over (2\pi)^4} \, g^2\,D_{\mu\nu}(p-\ell)\, \gamma_\mu\frac{\lambda^a}{2} \frac{1}{i\gamma\cdot \ell A(\ell^2) + B(\ell^2)} \, \Gamma_\nu^a(\ell,p). \label{qdse}
\end{eqnarray}
The most general Poincar\'e covariant solution of this gap equation involves two scalar functions.  There are three common, equivalent expressions:
\begin{equation}
\label{Sgeneral}
S(p) = \frac{1}{i\gamma\cdot p \, A(p^2) + B(p^2)} = \frac{Z(p^2)}{i \gamma\cdot p + M(p^2)} = -i \gamma\cdot p\, \sigma_V(p^2) + \sigma_S(p^2)\,.
\end{equation}
In the second form, $Z(p^2)$ is called the wave-function renormalisation and $M(p^2)$ is the dressed-quark mass function.

Given that a weak coupling expansion of the DSEs produces every diagram in perturbation theory, one can use Eq.\,\eqref{qdse}, and its analogues for the dressed-gluon propagator and dressed-quark gluon vertex, to obtain
\begin{equation}
B_{\rm pert}(p^2) = m \biggL 1 + \frac{\alpha}{\pi} \ln \biggL \frac{m^2}{p^2}\biggR + \ldots \biggR,
\end{equation}
where the ellipsis denotes terms of higher order in $\alpha$ that involve $(\ln [m^2/p^2])^2$ and $(\ln\ln [m^2/p^2])$, etc.  However, at arbitrarily large finite order in perturbation theory it is always true that
\begin{equation}
\label{mto0Bpert}
\lim_{m\to 0} B_{\rm pert}(p^2) \equiv 0.
\end{equation}
(N.B.\ In contrast to QED, a chiral limit is nonperturbatively defined in QCD owing to asymptotic freedom.)  Equation~\eqref{mto0Bpert} means that if one starts with a chirally symmetric theory, then in perturbation theory one also finishes with a chirally symmetric theory: the fermion DSE cannot generate a mass-gap if there is no bare-mass seed in the first place.  Thus \index{Dynamical chiral symmetry breaking (DCSB)} DCSB is \label{impossible} impossible in perturbation theory.  The question is whether this conclusion can ever be avoided; namely, are there circumstances under which it is possible to obtain a nonzero dressed-quark mass function in the chiral limit? \index{Chiral limit}

\subsection{Exploring dynamical mass generation}
\label{sec:njldcsb}
To begin the search for an answer, recall the gap equation, Eq.\,\eqref{qdse}, and consider the expression obtained with the following forms for the dressed-gluon propagator and quark-gluon vertex
\begin{eqnarray}
\label{DnjlGnjl}
g^2 D_{\mu\nu}(p-\ell) =  \delta_{\mu\nu}\, \frac{\alpha_{\rm IR}}{m_G^2}\,,\;
\Gamma_\nu^a(k,p) = \gamma_\nu \frac{\lambda^a}{2}\,,
\end{eqnarray}
wherein $m_G=0.8\,$GeV is a mass-scale, typical of gluons, as we shall see, and $\alpha_{\rm IR}$ is a parameter that specifies the interaction strength.  (One may view the interaction in Eq.\,(\ref{DnjlGnjl}) as being within a class inspired by the Nambu--Jona-Lasinio model \cite{Nambu:1961tp}.  The application to which it is here put most certainly relates to Nambu's half-share in the 2008 Nobel Prize in physics \cite{Nambu:2009zza}: ``for the discovery of the mechanism of spontaneous broken symmetry in subatomic physics.'')  Using Eqs.\,\eqref{DnjlGnjl}, the gap equation becomes
\begin{equation}
 S^{-1}(p) =  i \gamma \cdot p + m +  \frac{16\pi}{3}\frac{\alpha_{\rm IR}}{m_G^2} \int\!\frac{d^4 q}{(2\pi)^4} \,
\gamma_{\mu} \, S(q) \, \gamma_{\mu}\,,   \label{gap-1}
\end{equation}
an expression in which the integral possesses a quadratic divergence, even in the chiral limit.  When the divergence is regularised in a Poincar\'e covariant manner, the solution is \cite{GutierrezGuerrero:2010md}
\begin{equation}
\label{genS}
S(p) = \frac{1}{i \gamma\cdot p + M}\,;
\end{equation}
i.e., $A(p^2)\equiv 1$ and $M(p^2)\equiv M$, a momentum-independent quantity determined by the following one-dimensional nonlinear integral equation:
\begin{equation}
M = m + M\frac{4\alpha_{\rm IR}}{3\pi m_G^2} \int \!ds \, s\, \frac{1}{s+M^2}\,.
\end{equation}
(One arrives at these results via a little matrix algebra involving the Dirac matrices and capitalising on $O(4)$ invariance.)

A heat-kernel-like regularisation procedure is useful in hadron physics applications; viz., one writes \cite{Ebert:1996vx}:
\begin{equation}
\frac{1}{s+M^2} = \int_0^\infty d\tau\,{\rm e}^{-\tau (s+M^2)}
\rightarrow \int_{\tau_{\rm uv}^2}^{\tau_{\rm ir}^2} d\tau\,{\rm e}^{-\tau (s+M^2)}
%
=\frac{{\rm e}^{- (s+M^2)\tau_{\rm uv}^2}-e^{-(s+M^2) \tau_{\rm ir}^2}}{s+M^2} \,, \label{ExplicitRS}
\end{equation}
where $\tau_{\rm ir,uv}$ are, respectively, infrared and ultraviolet regulators.  Since Eq.\,(\ref{DnjlGnjl}) does not define a renormalisable theory, then $\Lambda_{\rm uv}:=1/\tau_{\rm uv}$ cannot be removed but instead plays a dynamical role, setting the scale of all dimensioned quantities.  Using Eq.\,\eqref{ExplicitRS}, the gap equation becomes
\begin{equation}
M = m + M\frac{4\alpha_{\rm IR}}{3\pi m_G^2}\,\,{\cal C}^{\rm iu}(M^2)\,,
\label{gapactual}
\end{equation}
where ${\cal C}^{\rm iu}(M^2)/M^2 = \Gamma(-1,M^2 \tau_{\rm uv}^2) - \Gamma(-1,M^2 \tau_{\rm ir}^2)$, with $\Gamma(\alpha,y)$ being the incomplete gamma-function.

In the current application, one may set $\Lambda_{\rm ir}=0$.  However, when employed in hadron physics phenomenology, a finite value of $\tau_{\rm ir}=:1/\Lambda_{\rm ir}$ implements confinement by ensuring the absence of quark production thresholds in $S$-matrix amplitudes \cite{Krein:1990sf}.  This is apparent from Eq.\,(\ref{ExplicitRS}).  Subsequently, $\Lambda_{\rm uv}$ defines the only mass-scale in a nonrenormalisable model.  Hence one can set $\Lambda_{\rm uv} \equiv 1$ and hereafter merely interpret all other mass-scales as being expressed in units of $\Lambda_{\rm uv}$, in which case ${\cal C}^{\rm 01}(M^2)= M^2 \Gamma(-1,M^2)$.

Now consider Eq.\,(\ref{gapactual}) in the chiral limit; i.e., $m_0=0$:
\begin{equation}
\label{pgapnjl0}
M = M\frac{4\alpha_{\rm IR}}{3\pi \hat m_G^2}\,\,{\cal C}^{\rm 01}(M^2)\,,
\end{equation}
where $\hat m_G = m_G/\Lambda_{\rm uv}$.  One solution of Eq.\,\eqref{pgapnjl0} is obviously $M\equiv 0$.  This is the result that connects smoothly with perturbation theory: one begins with no mass, and no mass is generated.  In this instance the theory is said to realise chiral symmetry in the \index{Chiral symmetry: realisation, Wigner-Weyl mode} Wigner-Weyl mode.

Suppose, on the other hand, that $M\neq 0$ in Eq.\,(\ref{pgapnjl0}).  The expression then makes sense if, and only if, the following equation has a solution:
\begin{equation}
1= \frac{4\alpha_{\rm IR}}{3\pi \hat m_G^2}\,\,{\cal C}^{\rm 01}(M^2)\,.
\end{equation}
${\cal C}^{\rm 01}(M^2)$ is a monotonically decreasing function with maximum value one at $M=0$.  Consequently, $\exists M \neq 0$ solution if, and only if,
\begin{equation}
\label{critmG1}
\frac{\alpha_{\rm IR}}{\pi} > \frac{3}{4} \, \hat m_G^2 \,.
\end{equation}
It is thus apparent that for any gluon mass-scale, $m_G$, there is always a domain of values for $\alpha_{\rm IR}$ upon which a nontrivial solution of the gap equation can be found.  If one supposes $\Lambda_{\rm uv} \sim m_G\,$, then $\exists M \neq 0$ solution for
\begin{equation}
\label{critmG2}
\frac{\alpha_{\rm IR}}{\pi} > \frac{3}{4}.
\end{equation}

This result, derived in a straightforward manner, is remarkable enough to have delivered a Nobel Prize to Nambu.  It reveals the power of a nonperturbative solution to nonlinear equations.  Although one began with a model of massless fermions, the interaction alone has provided those fermions with mass.  This is the phenomenon of dynamical chiral symmetry breaking; namely, the generation of mass \textit{from nothing}; and when it happens, chiral symmetry is said to be realised in the \index{Chiral symmetry: realisation, Nambu-Goldstone mode} Nambu-Goldstone mode.  It is clear from Eqs.\,(\ref{critmG1}), (\ref{critmG2}) that \index{Dynamical chiral symmetry breaking (DCSB)} DCSB is guaranteed to be possible so long as the interaction exceeds a particular minimal strength.  This behaviour is typical of gap equations.

\begin{figure}[t]
\vspace*{-1ex}

\includegraphics[clip,width=0.6\textwidth]{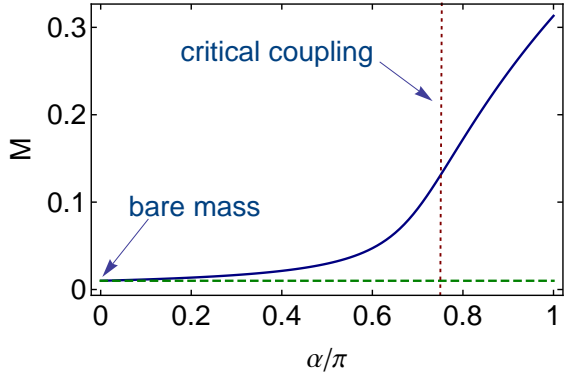}

\caption{\label{fig:NJLgap} Dynamical mass obtained from Eq.\,\protect\eqref{gapactual} as a function of the coupling, $\alpha/\pi$, with bare mass $0.01$, marked by the horizontal dashed line, $\Lambda_{\rm ir}=0$ and $\Lambda_{\rm uv}=m_G$. The vertical dotted line marks the critical coupling in Eq.\,\protect\eqref{critmG2}.}
\end{figure}

Figure~\ref{fig:NJLgap} depicts the $\alpha_{\rm IR}$-dependence of the self-consistent solution to Eq.\,(\ref{gapactual}) obtained with a current-quark mass $m = 0.01$ (measured in units of $\Lambda_{\rm uv}$).  The vertical line marks the critical coupling in Eq.\,\protect\eqref{critmG2}; namely, the value $\alpha_{\rm IR}^{\rm cr}=3/4$ below which a DCSB solution of Eq.\,(\ref{pgapnjl0}) is impossible.  Evidently, for $\alpha_{\rm IR} < \alpha_{\rm IR}^{\rm cr}$, the dressed mass $M \approx m_0$; i.e., the self-consistent solution is well approximated by the perturbative result.  However, a transition takes place for $\alpha \simeq \alpha^{\rm cr}$, and for $\alpha_{\rm IR} > \alpha_{\rm IR}^{\rm cr}$ the dynamical mass is much greater than the bare mass, with $M$ increasing rapidly as the coupling $\alpha_{\rm IR}$ is increased.

\subsection{Dynamical mass and confinement}
\index{Dynamical mass and confinement}
\label{confinement}
One aspect of quark confinement is the absence from the strong interaction spectrum of free-particle-like quarks.  What does the model of Sec.\,\ref{sec:njldcsb} have to contribute in this connection?  Well, whether one works within the domain of the model on which DCSB takes place, or not, the \index{Quark propagator} quark propagator always has the form, Eq.\,\eqref{genS},
\begin{equation}
S^{\rm NJL}(p)=\frac{1}{i\gamma\cdot p \, [A(p^2)=1] + [B(p^2)=M]} =  \frac{- i\gamma\cdot p  + M}{p^2 + M^2}\,,
\end{equation}
where $M$ is constant.  This expression has a pole at $p^2 + M^2 = 0$ and thus is always effectively the propagator for a noninteracting quasiparticle fermion with mass $M$: the fermion propagates as a plane wave.  Hence, while it is true that models of the Nambu--Jona-Lasinio type support DCSB, they do not generally exhibit confinement.  (This is true so long as $\Lambda_{\rm ir}=0$.)

Consider an alternative theory associated with a gap equation defined via the Munczek-Nemirovsky interaction \cite{Munczek:1983dx}
\begin{equation}
\label{mnprop} g^2 D_{\mu\nu}(k) = (2\pi)^4\, \check{G}\,\delta^4(k) \biggL\delta_{\mu\nu} - \frac{k_\mu k_\nu}{k^2}\biggR\,,\; \Gamma_\nu(\ell,p)=\gamma_\nu\,.
\end{equation}
Here $\check{G}$ defines the model's mass-scale and the interaction is a $\delta$-function in momentum space, which is the antithesis to models of the Nambu--Jona-Lasinio type wherein the interaction is instead a $\delta$-function in configuration space.  Using Eq.\,\eqref{mnprop}, the gap equation is
\begin{equation}
i\gamma\cdot p\, A(p^2) + B(p^2)  =  i \gamma\cdot p + m_0 + \check{G}\,\gamma_\mu\,
\frac{-i\gamma\cdot p\, A(p^2) + B(p^2)}{p^2 A^2(p^2) + B^2(p^2)}\,
\gamma_\mu\,,\label{mngap1}
\end{equation}
which yields the following coupled nonlinear algebraic equations:
\begin{eqnarray}
A(p^2) & = & 1 + 2 \, \frac{A(p^2)}{p^2 A^2(p^2) + B^2(p^2)} \, ,\label{AMN}\\
B(p^2) & = & m_0 + 4\, \frac{B(p^2)}{p^2 A^2(p^2) + B^2(p^2)} \,. \label{BMN}
\end{eqnarray}
Equation (\ref{mngap1}) yields an ultraviolet finite model and hence there is no regularisation mass-scale.  In this instance one can therefore refer all dimensioned quantities to the model's mass-scale and set $\check{G}=1$.

Consider the chiral limit of Eq.\,(\ref{BMN}):
\begin{equation}
B(p^2)  =   4\, \frac{B(p^2)}{p^2 A^2(p^2) + B^2(p^2)}\,. \label{BMN0}
\end{equation}
Obviously, like Eq.\,(\ref{pgapnjl0}), this equation admits a trivial solution $B(p^2) \equiv 0$ that is smoothly connected to the perturbative result, but is there another?
The existence of a $B\not\equiv 0$ solution; i.e., a solution that dynamically
breaks chiral symmetry, requires (in units of $\check{G}$)
\begin{equation}
p^2 A^2(p^2) + B^2(p^2) = 4\,.
\end{equation}
Suppose this identity to be satisfied, then its substitution into Eq.~(\ref{AMN}) gives \begin{equation}
A(p^2) - 1 = \frac{1}{2}\,A(p^2) \; \Rightarrow \; A(p^2) \equiv 2\,,
\end{equation}
which in turn entails
\begin{equation}
B(p^2) = 2\,\sqrt{1 - p^2}\,.
\end{equation}

A complete chiral-limit solution is composed subject to the physical requirement that the quark self-energy is real at spacelike momenta, and hence
\begin{eqnarray}
\label{AMNres}
A(p^2) &= & \left\{ \begin{array}{ll}
2\,;\; & p^2\leq 1\\
\frac{1}{2}\left( 1 + \sqrt{1+8/p^2} \right)\,;\; & p^2>1 \end{array} \right.\\
\label{BMNres}
B(p^2) &= & \left\{ \begin{array}{ll}
2 \sqrt{1-p^2}\,;\; & p^2\leq 1\\
0\,; & p^2>1 \,.\end{array} \right.
\end{eqnarray}
In this case both scalar functions characterising the \index{Quark propagator} dressed-quark propagator differ significantly from their free-particle forms and are momentum dependent.  (We will see that this is also true in QCD.)  It is noteworthy that the magnitude of the model's mass-scale plays no role in the appearance of this \index{Dynamical chiral symmetry breaking (DCSB)} DCSB solution of the gap equation; viz., reinstating $\check G$, in the chiral limit one has, $\forall \check G \neq 0$, $B(0) = 2 \check G \neq 0$.  Thus, in models of the Munczek-Nemirovsky type, the interaction is always strong enough to support the generation of mass from nothing.

The DCSB solution of Eqs.\,(\ref{AMNres}), (\ref{BMNres}) is defined and continuous for all values of $p^2$, including timelike momenta, $p^2<0$.  It gives a dressed-quark propagator whose denominator
\begin{equation}
p^2\,A^2(p^2) + B^2(p^2) > 0\,, \; \forall \, p^2\,.
\end{equation}
This is an extraordinary result.  It means that the propagator does not exhibit any free-particle-like poles.  It has long been argued that this feature can be interpreted as a realisation of quark confinement; e.g., \index{Confinement} Refs.\,\cite{Krein:1990sf,Munczek:1983dx,Gribov:1999ui,Stingl:1983pt,Cahill:1988zi}.

\begin{figure}[t]
\includegraphics[clip,width=0.60\textwidth]{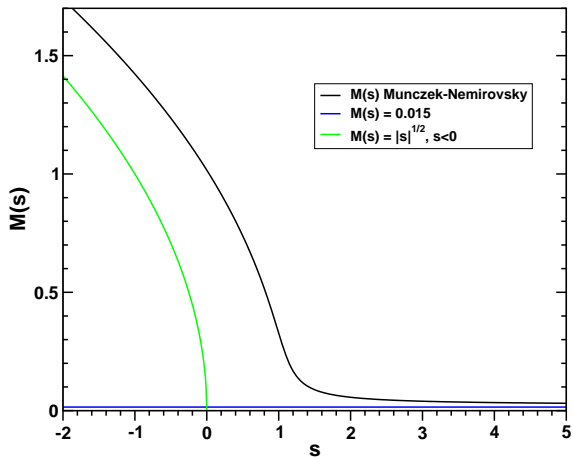}

\caption{\label{MNmassfn} With a bare mass $m=0.015$, the dressed-quark mass function \mbox{$M(s=p^2)$} obtained using the interaction specified by Eq.\,(\ref{mnprop}) -- solid curve.  A free-particle-like mass-pole would occur at that value of $s$ for which the solid and green curves intersect.  As the figure suggests, this never happens in models of the Munczek-Nemirovsky type.  (All dimensioned quantities measured in units of $\check{G}$, the model's mass-scale.)}
\end{figure}

The interaction in Eq.\,\eqref{mnprop} has taken a massless quark and turned it into something which at timelike momenta bears little resemblance to the perturbative quark.  It does that for all nonzero values of the model's mass-scale.  This model exemplifies an intriguing possibility; namely, that all models with quark confinement necessarily exhibit DCSB.  It is obvious from Sec.\,\ref{sec:njldcsb} that the converse is certainly not true.

In the chirally asymmetric case the gap equation yields
\begin{eqnarray}
A(p^2) & = & \frac{2 \,B(p^2)}{ m+ B(p^2) }\,,\\
B(p^2) & = & m + \frac{4\, [m + B(p^2)]^2}{B(p^2) ([m+B(p^2)]^2 + 4 p^2)}\,.
\end{eqnarray}
The second of these is a quartic equation for $B(p^2)$.  It can be solved algebraically.  There are four solutions, obtainable in closed form, only one of which possesses the physically sensible ultraviolet spacelike behaviour: $B(p^2) \to m$ as \mbox{$p^2\to \infty$}.  The physical solution is depicted in Fig.\,\ref{MNmassfn}.  At large spacelike momenta, $M(s=p^2)\to m^+$ and a perturbative analysis is reliable.  That is never the case for $s\lesssim 1$ (in units of $\check{G}$), on which domain $M(s)\gg m$, a defining characteristic of DCSB; and the difference $(M(s) - |s|)$ is always nonzero, a feature that is consistent with confinement.  

\begin{figure}[t]
\includegraphics[clip,width=0.7\textwidth]{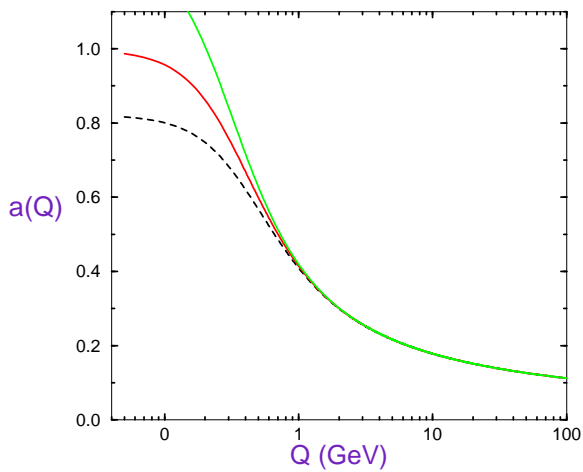}
\vspace*{-2ex}

\caption{\label{generalinteraction} Two classes of effective interaction in the gap equation.  A theory will exhibit dynamical chiral symmetry breaking if, and only if, $a(Q=0) \gtrsim 1$.  Here the ultraviolet behaviour of the interactions is constrained to be that of QCD.}
\end{figure}

I have illustrated two variations on the theme of dynamical mass generation via the gap equation.  A general class of models for asymptotically free theories may be discussed in terms of an effective interaction
\begin{equation}
g^2 D(Q^2) = 4\pi \frac{a(Q^2)}{Q^2}\,,
\end{equation}
where $a(Q^2)$ is such that $g^2 D(Q^2)$ evolves according to QCD's renormalisation group in the ultraviolet.  This situation is depicted in \protect\index{Dynamical chiral symmetry breaking} Fig.\,\ref{generalinteraction}.  The types of effective interaction fall into two classes.  In those for which $a(0) \lesssim 1$ [recall Eq.\,\eqref{critmG2}], the only solution of the gap equation in the chiral limit is \mbox{$M(s) \equiv 0$}.  Whereas, when $a(0)\gtrsim 1$, $M(s)\neq 0$ is possible in the chiral limit and, indeed, corresponds to the energetically favoured ground state \cite{Roberts:1994dr,Cahill:1985mh,Roberts:1985ju,Qin:2010nq}.  In this case $M(0)\neq 0$ is a new, dynamically generated mass-scale.  If large enough, it can explain how a theory that is perturbatively massless nevertheless possesses the spectrum of a massive theory.

It should now be plain that confinement and DCSB are key emergent phenomena in QCD, and that their understanding requires the nonperturbative solution of a fully-fledged relativistic quantum field theory.  Both mathematics and physics are still far from being able to accomplish that.  On the other hand, I have illustrated that confinement and DCSB are expressed in QCD's propagators.  These phenomena are also evident in dressed vertices.  Such nonperturbative modifications should have observable consequences.  The DSEs are a useful analytical and numerical tool for the nonperturbative study of relativistic quantum field theory.  They can be used to show that simple models exhibit DCSB but that DCSB $\not\Rightarrow$ confinement; and conversely, that simple models can exhibit confinement and that confinement $\Rightarrow$ DCSB.  However, the question of what can be established for QCD remains.
	
\subsection{Confinement}
\label{confinementII}
Gluon and quark confinement is an empirical fact: no matter how hard one has thus far struck the proton, or any other hadron, no individual gluon or quark has yet been liberated.  However,  there is no agreed, theoretical definition of light-quark confinement.  One thing is nevertheless certain; namely, true confinement entails quark-hadron duality, which is the statement that all observable consequences of a confining QCD could, in principle, be computed using an hadronic basis.

Confinement is often discussed in terms of Wilson's area law, exposed in an influential article entitled ``Confinement of quarks,'' which introduced the concept of lattice gauge theory \cite{Wilson:1974sk}.  The area law identifies confinement with a linearly rising potential between coloured charges.  Phenomenologists with an attachment to quantum mechanics and string theorists, too, have long been nourished by this result.  Unfortunately, the only quarks in the article are static; i.e., infinitely heavy, so the concept doesn't bear on real-world QCD.  It is therefore worth stating plainly that the potential between infinitely-heavy quarks measured in numerical simulations of quenched lattice-regularised QCD -- the so-called static potential -- is simply \emph{irrelevant} to the question of confinement in the real world, in which light quarks are ubiquitous.  In fact, it is a basic feature of QCD that light-particle creation and annihilation effects are essentially nonperturbative and therefore it is impossible in principle to compute a potential between two light quarks \cite{Bali:2005fu}.

Some consider that a belief in linear potentials and their relevance is justified by the appearance of Regge trajectories in the hadron spectrum; i.e., that associated with a given ground state composite, there are higher mass composite states with larger total angular momentum (spin $J$), all of which lie on a straight line in the plane $(J,M^2)$.  The appearance does not necessarily match reality, however, as observed in Ref.\,\cite{Veltmann:2003}: ``In time the Regge trajectories thus became the cradle of string theory.  Nowadays the Regge trajectories have largely disappeared, not in the least because these higher spin bound states are hard to find experimentally.  At the peak of the Regge fashion (around 1970) theoretical physics produced many papers containing families of Regge trajectories, with the various (hypothetically straight) lines based on one or two points only!''  A detailed analysis supports this view; viz., Ref.\,\cite{Tang:2000tb}: ``Early Chew-Frautschi plots show that meson and baryon Regge trajectories are approximately linear and non-intersecting.  In this paper, we reconstruct all Regge trajectories from the most recent data. Our plots show that meson trajectories are non-linear and intersecting.  We also show that all current meson Regge trajectories models are ruled out by data.''

\begin{figure}[t]
\includegraphics[clip,width=0.70\textwidth]{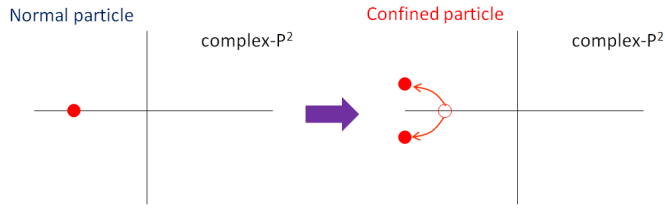}
\caption{\label{fig:confined} \emph{Left panel} -- An observable particle is associated with a pole at timelike-$P^2$.  This becomes a branch point if, e.g., the particle is dressed by photons.  \emph{Right panel} -- When the dressing interaction is confining, the real-axis mass-pole splits, moving into pairs of complex conjugate poles or branch points.  No mass-shell can be associated with a particle whose propagator exhibits such singularity structure.}
\end{figure}

Drawing on a long list of sources; e.g., Refs.\,\cite{Munczek:1983dx,Gribov:1999ui,Stingl:1983pt,Cahill:1988zi}, a perspective on confinement was laid out in Ref.\,\cite{Krein:1990sf}.  As illustrated above and in Fig.\,\ref{fig:confined}, confinement can be related to the analytic properties of QCD's Schwinger functions.  For example, it can be read from the reconstruction theorem \cite{SW80,GJ81} that the only Schwinger functions which can be associated with expectation values in the Hilbert space of observables -- namely, the set of measurable expectation values -- are those that satisfy the axiom of reflection positivity.  This is an extremely tight constraint.  It can be shown to require as a necessary condition that the Fourier transform of the momentum-space Schwinger function is a positive-definite function of its arguments.  This condition suggests a practical confinement test, which can be used with numerical solutions of the DSEs (see, e.g., Sec.\,III.C of Ref.\,\cite{Hawes:1993ef} and Sec.\,IV of Ref.\,\cite{Maris:1995ns}).  The implications and use of reflection positivity are discussed and illustrated in Sec.~2 of Ref.\,\cite{Roberts:2007ji}.

It is notable that any 2-point Schwinger function with an inflexion point at $p^2 > 0$ must breach the axiom of reflection positivity, so that a violation of positivity can be determined by inspection of the pointwise behaviour of the Schwinger function in momentum space (Sec.\,IV.B of Ref.\,\cite{Bashir:2008fk}).


\begin{figure}[t]
\includegraphics[clip,width=0.7\textwidth]{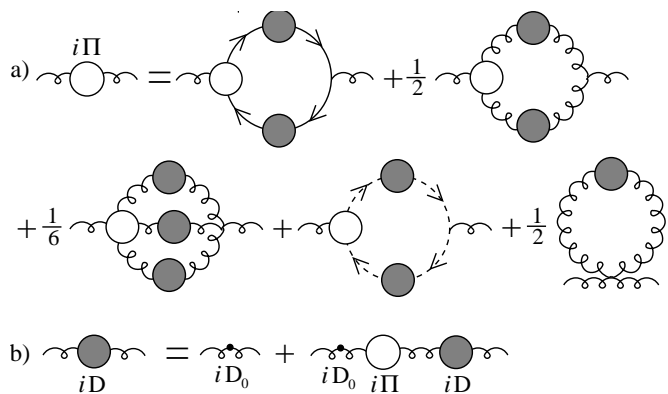}
\vspace*{-2ex}

\caption{\label{fig:gluondse} Dyson-Schwinger equation for the gluon propagator.  The gluon vacuum polarisation tensor $\Pi_{\mu\nu}=(\delta_{\mu\nu} - k_\mu k_\nu/k^2) \Pi(k^2)$ is the gluon's self energy, from which the propagator is obtained: in Landau gauge, $D_{\mu\nu}(k)= (\delta_{\mu\nu} - k_\mu k_\nu/k^2)[1/(k^2[1+\Pi(k^2)])]$ is the propagator obtained (The broken line represents the propagator for the ghost field.)}
\end{figure}

\begin{figure}[t]

\includegraphics[clip,width=0.60\textwidth]{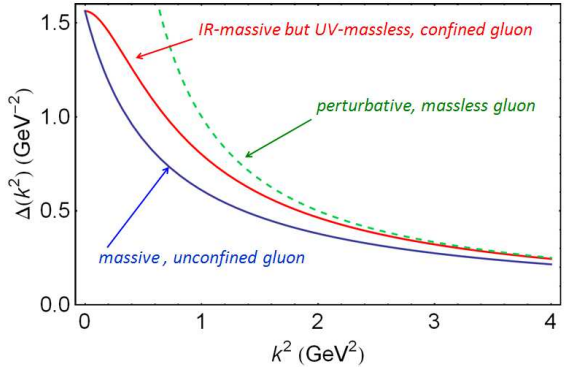}

\caption{\label{fig:gluonrp}
$\Delta(k^2)$, the function that describes dressing of a Landau-gauge gluon propagator, plotted for three distinct cases.
A bare gluon is described by $\Delta(k^2) = 1/k^2$ (the dashed line), which is plainly convex on $k^2\in (0,\infty)$.  Such a propagator has a representation in terms of a non-negative spectral density.
In some theories, interactions generate a mass in the transverse part of the gauge-boson propagator, so that $\Delta(k^2) = 1/(k^2+m_g^2)$, which can also be represented in terms of a non-negative spectral density.
In QCD, however, self-interactions generate a momentum-dependent mass for the gluon, which is large at infrared momenta but vanishes in the ultraviolet \protect\cite{Boucaud:2011ug}.  This is illustrated by the curve labelled ``IR-massive but UV-massless.''  With the generation of a mass-\emph{function}, $\Delta(k^2)$ exhibits an inflexion point and hence cannot be expressed in terms of a non-negative spectral density.
}
\end{figure}

The gluon two-point function satisfies the gap equation depicted in Fig.\,\ref{fig:gluondse}.  Consider then $\Delta(k^2)=1/(k^2[1+\Pi(k^2)])$, which is the single scalar function that describes the dressing of a Landau-gauge gluon propagator.  A large body of work has focused on exposing the behaviour of $\Delta(k^2)$ in the pure Yang-Mills sector of QCD.  These studies are reviewed in Ref.\,\cite{Boucaud:2011ug}.
A connection with the expression and nature of confinement in the Yang-Mills sector is indicated in Fig.\,\ref{fig:gluonrp}.  The appearance of an inflexion point in the two-point function generated by the gluon's momentum-dependent mass-function is impossible to overlook.  Hence this gluon cannot appear in the Hilbert space of observable states.  
At present, DSE and lattice-QCD agree that the gluon is truly described by a momentum-dependent mass function but the infrared scale is only known to within a factor of two; i.e.,
\begin{equation}
m_G^2 := \lim_{k^2\to 0}k^2 \Pi(k^2) \approx 2 - 4\, \Lambda_{\rm QCD}.
\end{equation}

{F}rom the perspective that confinement can be related to the analytic properties of QCD's Schwinger functions, the question of light-quark confinement can be translated into the challenge of charting the infrared behavior of QCD's \emph{universal} $\beta$-function.  (Although this function may depend on the scheme chosen to renormalise the theory, it is unique within a given scheme \protect\cite{Celmaster:1979km}.  Of course, the behaviour of the $\beta$-function on the perturbative domain is well known.)  This is a well-posed problem whose solution is an elemental goal of modern hadron physics and which can be addressed in any framework enabling the nonperturbative evaluation of renormalisation constants.  It is the $\beta$-function that is responsible for the behaviour evident in Fig.\,\ref{fig:gluonrp}, and Fig.\,\ref{gluoncloud} below; and one of the more interesting of contemporary questions is whether it is possible to reconstruct the $\beta$-function, or at least constrain it tightly, given empirical information on the gluon and quark mass functions.

\begin{figure}[t]
\includegraphics[clip,width=0.47\textwidth]{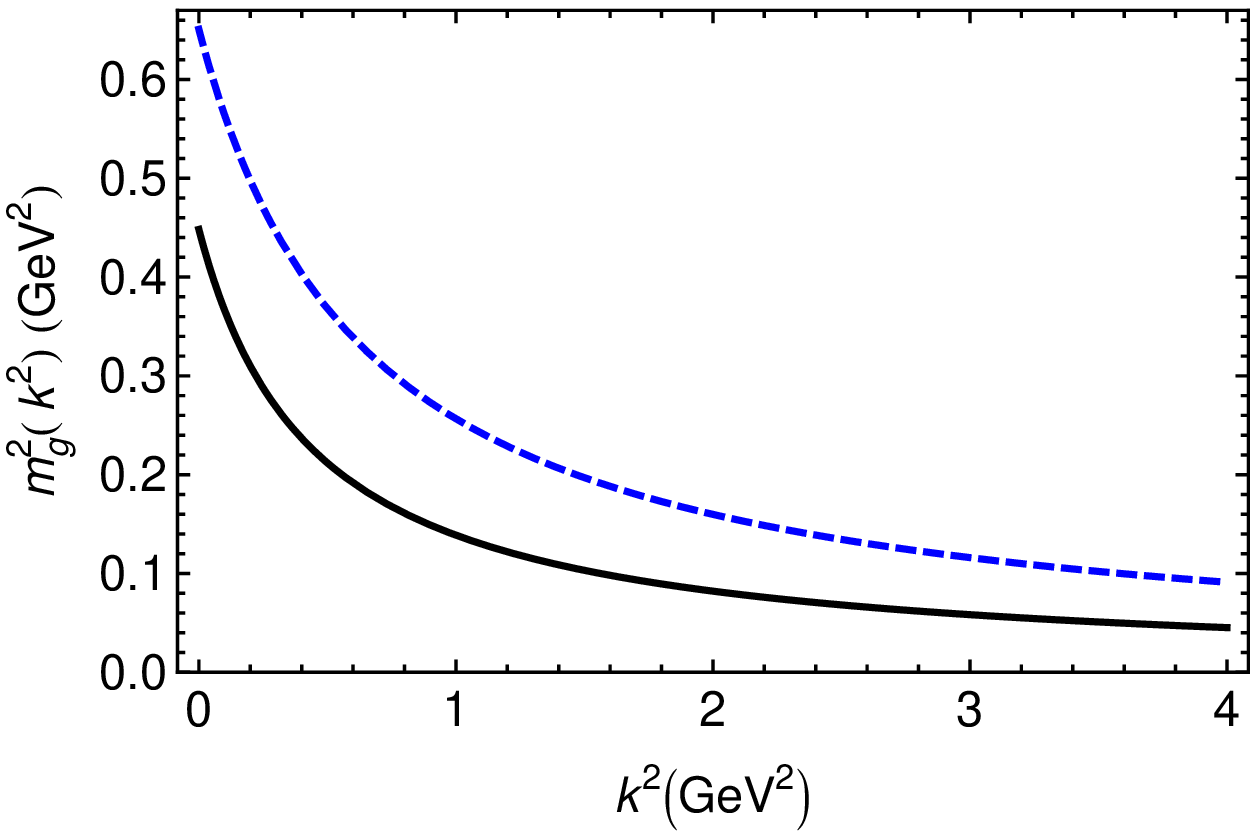}
\includegraphics[clip,width=0.49\textwidth]{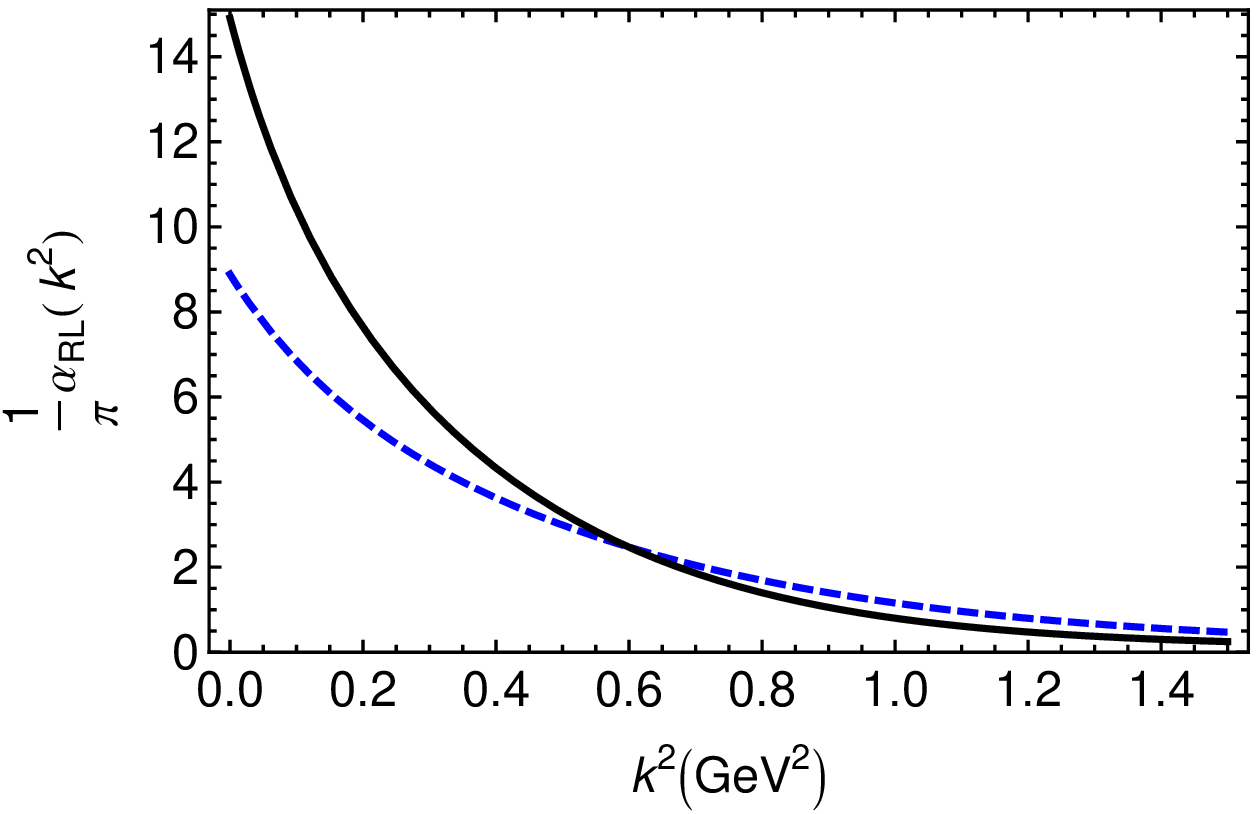}

\caption{\label{fig:gluonrunning}
\emph{Left panel} -- Rainbow-ladder gluon running-mass; and \emph{right panel} -- rainbow-ladder effective running-coupling.  Matching related curves in each panel (solid to solid or dashed to dashed) gives an equivalent description of observables within the rainbow-ladder truncation.
}
\end{figure}

In this connection I note that the DSEs connect the $\beta$-function to experimental observables.  Hence, comparison between DSE computations and observations of the hadron mass spectrum, and elastic and transition form factors, can be used to chart $\beta$-function's long-range behaviour.  Extant studies show that the properties of hadron excited states are a great deal more sensitive to the long-range behaviour of the $\beta$-function than those of ground states.  This is illustrated in Refs.\,\cite{Qin:2011dd,Qin:2011xq}, which through a study of ground-state, radially-excited and exotic scalar-, vector- and flavoured-pseudoscalar-mesons in rainbow-ladder truncation, which is leading order in the most widely used nonperturbative scheme \cite{Munczek:1994zz,Bender:1996bb}, produced the effective coupling and running gluon mass depicted in Fig.\,\ref{fig:gluonrunning}.

\begin{figure}[t]

\includegraphics[clip,width=0.60\textwidth]{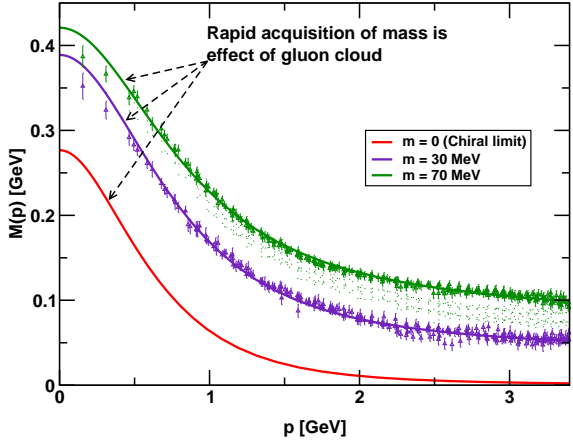}
\caption{\label{gluoncloud} Dressed-quark mass function, $M(p)$: \emph{solid curves} -- DSE results, \protect\cite{Bhagwat:2003vw,Bhagwat:2006tu,Bhagwat:2007vx}, ``data'' -- lattice-QCD simulations \protect\cite{Bowman:2005vx}.  (N.B.\ $m=70\,$MeV is the uppermost curve.  Current-quark mass decreases from top to bottom.)  
The constituent mass arises from a cloud of low-momentum gluons attaching themselves to the current-quark: DCSB is a truly nonperturbative effect that generates a quark mass \emph{from nothing}; namely, it occurs even in the chiral limit, as evidenced by the $m=0$ curve.}
\end{figure}

\subsection{Aper{\c{c}}u}
\label{sec:Apercu}
Whilst confinement is empirically true, Fig.\,\ref{gluoncloud} shows that dynamical chiral symmetry breaking is a fact in QCD.  Contemporary DSE- and lattice-QCD studies agree on the basic features of the mass function for a dressed light-quark.  The mass function is momentum dependent.  For light quarks with nonzero Lagrangian mass it exhibits logarithmic growth as one proceeds toward the infrared from large momenta: this is the perturbative domain upon which asymptotic freedom is evident.  At a momentum scale of approximately $5\Lambda_{\rm QCD}$, however, the rate of increase accelerates dramatically as power-law growth sets in: this is the beginning of the nonperturbative domain, whereupon dynamical chiral symmetry breaking is key to the evolution of a dressed-quark's mass.  As the $m=0$ curve in Fig.\,\ref{gluoncloud} shows, even a quark that is perturbatively massless would acquire a mass in QCD.  This is kindred to the behaviour of the running gluon mass in Fig.\,\ref{fig:gluonrunning}.  The inflexion point at $p \sim \Lambda_{\rm QCD}$, which is predicted by DSE studies and appears to be supported by the lattice results, indicates that beyond this point one has very probably entered the confinement domain.  The inflexion point is connected with the behaviour illustrated in the right panel of Fig.\,\ref{fig:confined}.

The reality of DCSB means the Higgs mechanism is largely irrelevant to the bulk of normal matter in the Universe.  Instead the most important mass generating mechanism for light-quark hadrons is the strong interaction effect of DCSB; e.g., one may identify it as being responsible for 98\% of a proton's mass \cite{Flambaum:2005kc,Holl:2005st}.  There is a caveat; namely, as so often, the pion is exceptional.  Its mass is given by the simple product of two terms, one of which is the ratio of two order parameters for DCSB whilst the other is determined by the current-quark mass (Sec.\,\ref{psmassformula}).  Hence the pion would be massless in the absence of a mechanism that can generate a current-mass for at least one light-quark.  The impact of a massless, strongly-interacting particle on the physics of the Universe would be dramatic.

It is QCD's $\beta$-function that is responsible for the behavior evident in Figs.\,\ref{fig:gluonrp} and \ref{gluoncloud}, and thereby the scale-dependence of the structure and interactions of dressed-gluons and -quarks.  Elucidation of an unambiguous connection between observables and the behaviour evident in Figs.\,\ref{fig:gluonrp} and \ref{gluoncloud} is therefore a promising method for empirically charting the long range behaviour of the strong interaction, something which will not be achieved by theory alone in the foreseeable future.  As described and illustrated in Secs.\,\ref{sec:Charting}, \ref{sec:Baryons} below, this can potentially lead to an explanation and understanding of the origin of the interaction strength at infrared momenta that guarantees DCSB through the gap equation and ties confinement to DCSB.

\section{Gap and Bethe-Salpeter Equations}
\label{gapbse}
In order to proceed it is useful to recapitulate on the best known and simplest DSE.  The Dyson or gap equation describes how quark propagation is influenced by interactions; viz., for a quark of flavour $f$,
\begin{equation}
 S_f(p)^{-1} = Z_2 \,(i\gamma\cdot p + m_f^{\rm bm}) + Z_1 \int^\Lambda_q\!\! g^2 D_{\mu\nu}(p-q)\frac{\lambda^a}{2}\gamma_\mu S_f(q) \frac{\lambda^a}{2}\Gamma^f_\nu(q,p) ,
\label{gendseN}
\end{equation}
where: $D_{\mu\nu}$ is the gluon propagator; $\Gamma^f_\nu$, the quark-gluon vertex; $\int^\Lambda_q$, a symbol that represents a Poincar\'e invariant regularization of the four-dimensional Euclidean integral, with $\Lambda$ the regularization mass-scale; $m_f^{\rm bm}(\Lambda)$, the current-quark bare mass; and $Z_{1,2}(\zeta^2,\Lambda^2)$, respectively, the vertex and quark wave-function renormalisation constants, with $\zeta$ the renormalisation point -- dependence upon which is not usually made explicit.  The gap equation will be familiar from Eq.\,(\ref{Sqdse}) but here I've been more careful with the definitions.

The gap equation's solution is the dressed-quark propagator,
\begin{equation}
 S(p) =
%
\frac{1}{i \gamma\cdot p \, A(p^2,\zeta^2) + B(p^2,\zeta^2)}
= \frac{Z(p^2,\zeta^2)}{i\gamma\cdot p + M(p^2)}\,,
%
\label{SgeneralN}
\end{equation}
which is obtained from Eq.\,(\ref{gendseN}) augmented by a renormalisation condition.  A mass-independent scheme is a useful choice and can be implemented by fixing all renormalisation constants in the chiral limit.  (See, e.g., Ref.\,\cite{Chang:2008ec} and references therein; or Ref.\,\protect\cite{tarrach} for a detailed discussion of renormalisation.)

The mass function, $M(p^2)=B(p^2,\zeta^2)/A(p^2,\zeta^2)$, is independent of the renormalisation point, $\zeta$; and the renormalised current-quark mass,
\begin{equation}
\label{mzeta}
m_f^\zeta = Z_m(\zeta,\Lambda) \, m^{\rm bm}(\Lambda) = Z_4^{-1} Z_2\, m_f^{\rm bm},
\end{equation}
wherein $Z_4$ is the renormalisation constant associated with the Lagrangian's mass-term. Like the running coupling constant, this ``running mass'' is familiar from textbooks.  However, it is not commonly appreciated that $m^\zeta$ is simply the dressed-quark mass function evaluated at one particular deep spacelike point; viz,
\begin{equation}
m_f^\zeta = M_f(\zeta^2)\,.
\end{equation}
The renormalisation-group invariant current-quark mass may be inferred via
\begin{equation}
\hat m_f = \lim_{p^2\to\infty} \left[\frac{1}{2}\ln \frac{p^2}{\Lambda^2_{\rm QCD}}\right]^{\gamma_m} M_f(p^2)\,,
\label{RGIcqmass}
\end{equation}
where $\gamma_m = 12/(33-2 N_f)$ (see Fig.\,\ref{F3}).  The chiral limit is expressed by
\begin{equation}
\hat m_f = 0\,.
\end{equation}
Moreover,
\begin{equation}
\forall \zeta \gg \Lambda_{\rm QCD}, \;
\frac{m_{f_1}^\zeta}{m^\zeta_{f_2}}=\frac{\hat m_{f_1}}{\hat m_{f_2}}\,.
\end{equation}
However, I would like to emphasise that in the presence of DCSB the ratio
\begin{equation}
\frac{m_{f_1}^{\zeta=p^2}}{m^{\zeta=p^2}_{f_2}}=\frac{M_{f_1}(p^2)}{M_{f_2}(p^2)}
\end{equation}
is not independent of $p^2$: in the infrared; i.e., $\forall p^2 \lesssim \Lambda_{\rm QCD}^2$, it then expresses a ratio of constituent-like quark masses, which, for light quarks, are two orders-of-magnitude larger than their current-masses and nonlinearly related to them \cite{Flambaum:2005kc,Holl:2005st}.

The gap equation illustrates the features and flaws of each DSE.  It is a nonlinear integral equation for the dressed-quark propagator and hence can yield much-needed nonperturbative information.  However, the kernel involves the two-point function $D_{\mu\nu}$ and the three-point function $\Gamma^f_\nu$.  The gap equation is therefore coupled to the DSEs satisfied by these functions, which in turn involve higher $n$-point functions.  Hence the DSEs are a tower of coupled integral equations, with a tractable problem obtained only once a truncation scheme is specified.  It is unsurprising that the best known truncation scheme is the weak coupling expansion, which reproduces every diagram in perturbation theory.  This scheme is systematic and valuable in the analysis of large momentum transfer phenomena because QCD is asymptotically free but it precludes any possibility of obtaining nonperturbative information.

Given the importance of DCSB in QCD, it is significant that the dressed-quark propagator features in the axial-vector Ward-Takahashi identity, which expresses chiral symmetry and its breaking pattern:
\begin{equation}
P_\mu \Gamma_{5\mu}^{fg}(k;P) + \, i\,[m_f(\zeta)+m_g(\zeta)] \,\Gamma_5^{fg}(k;P)
= S_f^{-1}(k_+) i \gamma_5 +  i \gamma_5 S_g^{-1}(k_-) \,,
\label{avwtimN}
\end{equation}
where $P=p_1+p_2$ is the total-momentum entering the vertex and $k$ is the relative-momentum between the amputated quark legs.  To be explicit, $k=(1-\eta) p_1 + \eta p_2$, with $\eta \in [0,1]$, and hence $k_+ = p_1 = k + \eta P$, $k_- = p_2 = k - (1-\eta) P$.  In a Poincar\'e covariant approach, such as presented by a proper use of DSEs, no observable can depend on $\eta$; i.e., the definition of the relative momentum.  N.B.\ Sec.\,\ref{flavourless} discusses the important differences encountered in treating flavourless pseudoscalar mesons.

In Eq.\,(\ref{avwtimN}), $\Gamma_{5\mu}^{fg}$ and $\Gamma_5^{fg}$ are, respectively, the amputated axial-vector and pseudoscalar vertices.  They are both obtained from an inhomogeneous Bethe-Salpeter equation (BSE), which is exemplified here using a textbook expression \cite{Salpeter:1951sz}:
\begin{equation}
[\Gamma_{5\mu}(k;P)]_{tu} = Z_2 [\gamma_5 \gamma_\mu]_{tu}+ \int_q^\Lambda [ S(q_+) \Gamma_{5\mu}(q;P) S(q_-) ]_{sr} K_{tu}^{rs}(q,k;P),
\label{bsetextbook}
\end{equation}
in which $K$ is the fully-amputated quark-antiquark scattering kernel, and
the colour-, Dirac- and flavour-matrix structure of the elements in the equation is  denoted by the indices $r,s,t,u$.  N.B.\ By definition, $K$ does not contain quark-antiquark to single gauge-boson annihilation diagrams, nor diagrams that become disconnected by cutting one quark and one antiquark line.

The Ward-Takahashi identity, Eq.\,(\ref{avwtimN}), entails that an intimate relation exists between the kernel in the gap equation and that in the BSE.  (This is another example of the coupling between DSEs.)  Therefore an understanding of chiral symmetry and its dynamical breaking can only be obtained with a truncation scheme that preserves this relation, and hence guarantees Eq.\,(\ref{avwtimN}) without a fine-tuning of model-dependent parameters.  Until 1995--1996 no one had a good idea about how to do this.  Equations were truncated, sometimes with good phenomenological results and sometimes with poor results.  Neither the successes nor the failures could be explained.

\begin{figure}[t]
\vspace*{-1ex}

\includegraphics[clip,width=0.66\textwidth]{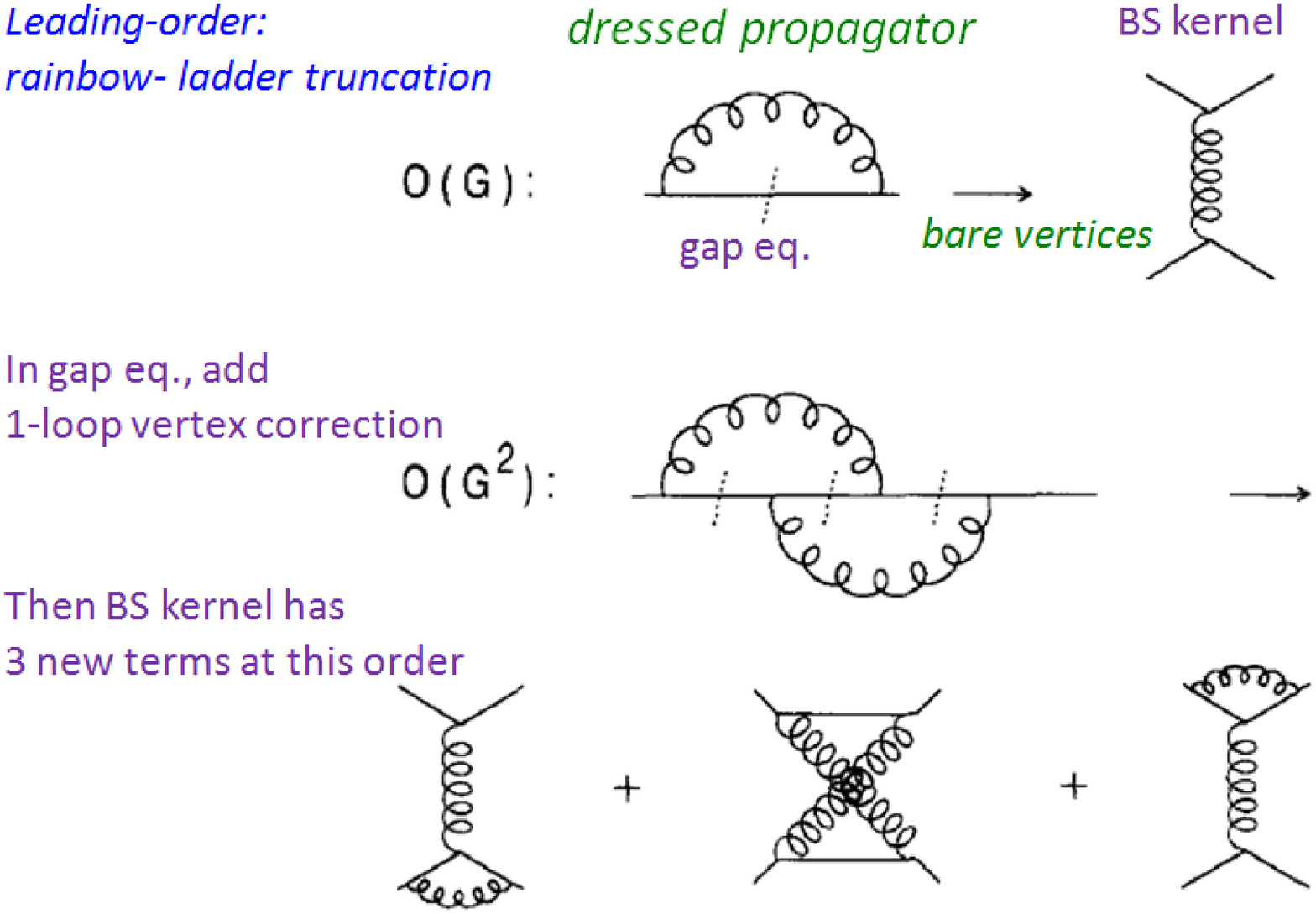}

\caption{\label{FSysTruncation}
Illustration of the procedure used to produce a nonperturbative systematic symmetry-preserving truncation.  It produces a modified skeleton expansion, in which the propagators are fully-dressed but the vertices are constructed term-by-term.}

\end{figure}

That changed with Refs.\,\cite{Munczek:1994zz,Bender:1996bb}, which described a procedure, illustrated in Fig.\,\ref{FSysTruncation}, that generates a Bethe-Salpeter equation from the kernel of any gap equation whose diagrammatic content is known.  That this is possible and achievable systematically is necessary and sufficient to prove numerous exact results in QCD.  The procedure also enables the formulation of practical phenomenological models that can be used to illustrate the exact results and provide predictions for experiment with readily quantifiable errors.

\section{Exact Results in Hadron Physics}
\label{sec:Exact}
Before detailing some of the exact results, it is worth explicating the range of applications; e.g.,
veracious statements about the pion $\sigma$-term \cite{Flambaum:2005kc};
radially-excited and hybrid pseudoscalar mesons \cite{Holl:2004fr,Holl:2005vu};
heavy-light \cite{Ivanov:1998ms} and heavy-heavy mesons \cite{Bhagwat:2006xi};
novel results for the pion susceptibility obtained via analysis of the isovector-pseudoscalar vacuum polarisation \cite{Chang:2009at}, which bear upon the essential content of the so-called ``Mexican hat'' potential that is used in building models for QCD;
and a derivation \cite{Chang:2008sp} of the Weinberg sum rule \cite{Weinberg:1967kj}.

\subsection{Pseudoscalar meson mass formula}
\label{psmassformula}
The first of the results was introduced in Ref.\,\cite{Maris:1997hd}; namely, a mass formula that is exact for flavour non-diagonal pseudoscalar mesons:
\begin{equation}
\label{mrtrelation}
f_{H_{0^-}} m_{H_{0^-}}^2 = (m_{f_1}^\zeta + m_{f_2}^\zeta) \rho_{H_{0^-}}^\zeta,
\end{equation}
where: $m_{f_i}^\zeta$ are the current-masses of the quarks constituting the meson; and
\begin{eqnarray}
f_{H_{0^-}} P_\mu = \langle 0 | \bar q_{f_2} \gamma_5 \gamma_\mu q_{f_1} |H_{0^-}\rangle
& = & Z_2\; {\rm tr}_{\rm CD}
\int_q^\Lambda i\gamma_5\gamma_\mu S_{f_1}(q_+) \Gamma_{H_{0^-}}(q;P) S_{f_2}(q_-)\,, \rule{1.5em}{0ex} \label{fpigen}
\\
i\rho_{H_{0^-}} = -\langle 0 | \bar q_{f_2} i\gamma_5 q_{f_1} |H_{0^-} \rangle & = & Z_4\; {\rm tr}_{\rm CD}
\int_q^\Lambda \gamma_5 S_{f_1}(q_+) \Gamma_{H_{0^-}}(q;P) S_{f_2}(q_-) \,, \rule{1.5em}{0ex}
\label{rhogen}
\end{eqnarray}
where $\Gamma_{H_{0^-}}$ is the pseudoscalar meson's bound-state Bethe-Salpeter amplitude:
\begin{eqnarray}
\nonumber \Gamma_{H_{0^-}}(k;P) &=& \gamma_5 \biggL i E_{H_{0^-}}(k;P) + \gamma\cdot P F_{H_{0^-}}(k;P)  \\
&&  + \gamma\cdot k \, G_{H_{0^-}}(k;P) - \sigma_{\mu\nu} k_\mu P_\nu H_{H_{0^-}}(k;P) \biggR ,
\label{genGpi}
\end{eqnarray}
which is determined from the associated homogeneous BSE; viz., Eq.\,\eqref{bsetextbook} without the inhomogeneity.

Equation~\eqref{fpigen} is the pseudovector projection of the pseudoscalar meson's Bethe-Salpeter wave-function onto the origin in configuration space; namely, the meson's leptonic decay constant, which is the strong interaction contribution to the strength of the meson's weak interaction.  Similarly, Eq.\,\eqref{rhogen} is the pseudoscalar projection of the wave-function onto the origin in configuration space; viz., a pseudoscalar analogue of the meson's leptonic decay constant.  These observations will subsequently become quite important herein.

It is worth emphasising that the quark wave-function and Lagrangian mass renormalisation constants, $Z_{2,4}(\zeta,\Lambda)$, respectively, depend on the gauge parameter in precisely the manner needed to ensure that the right-hand sides of Eqs.\,(\ref{fpigen}), (\ref{rhogen}) are gauge-invariant.  Moreover, $Z_2(\zeta,\Lambda)$ ensures that the right-hand side of Eq.\,(\ref{fpigen}) is independent of both $\zeta$ and $\Lambda$, so that $f_{H_{0^-}}$ is truly an observable; and $Z_4(\zeta,\Lambda)$ ensures that $\rho_{H_{0^-}}^\zeta$ is independent of $\Lambda$ and evolves with $\zeta$ in just the way necessary to guarantee that the product $m^\zeta \rho_{H_{0^-}}^\zeta$ is
renormalisation-point-independent.

In addition, it should be noted that Eq.\,(\ref{mrtrelation}) is valid for every pseudoscalar meson and for any value of the current-quark masses; viz., $\hat m_{f_i} \in [ 0,\infty)$, $i=1,2$.  This includes the chiral limit, in whose neighbourhood Eq.\,(\ref{mrtrelation}) can be shown \cite{Maris:1997hd} to reproduce the familiar Gell-Mann--Oakes--Renner relation \cite{GellMann:1968rz}, Eq.\,\eqref{GMORprimative}.  It also includes arbitrarily large values of one or both current-quark masses.  In the former case; viz., one quark mass fixed and the other becoming very large, so that $m_q/m_Q \ll 1$, then $f_{H_{0^-}} \propto 1/\surd m_{H_{0^-}}$, $\rho_{H_{0^-}} \propto \surd m_{H_{0^-}}$, and one arrives at $m_{H_{0^-}} = m_Q$ \cite{Ivanov:1998ms}.  The mass formula thereby provides the path to the QCD proof of a potential model result.

The axial-vector Ward-Takahashi identity, Eq.\,(\ref{avwtimN}), is a crucial bridge to Eqs.\,(\ref{mrtrelation}) -- (\ref{rhogen}); and on the way one can also prove the following Goldberger-Treiman-like relations \cite{Maris:1997hd}:
\begin{eqnarray}
\label{gtlrelE}
f_{H_{0^-}}^0 E_{H_{0^-}}(k;0) &=& B^0(k^2)\,,\\
\label{gtlrelF}
F_R(k;0) + 2 f_{H_{0^-}}^0 F_{H_{0^-}}(k;0) &=& A^0(k^2)\,,\\
\label{gtlrelG}
G_R(k;0) + 2 f_{H_{0^-}}^0 G_{H_{0^-}}(k;0) &=& \frac{d}{dk^2}A^0(k^2)\,,\\
\label{gtlrelH}
H_R(k;0) + 2 f_{H_{0^-}}^0 H_{H_{0^-}}(k;0) &=& 0\,,
\end{eqnarray}
wherein the superscript indicates that the associated quantity is evaluated in the chiral limit, and $F_R$, $G_R$, $H_R$ are analogues in the inhomogeneous axial-vector vertex of the scalar functions in the $H_{0^-}$-meson's Bethe-Salpeter amplitude.

These identities are of critical importance in QCD.
The first, Eq.\,(\ref{gtlrelE}), can be used to prove that a massless pseudoscalar meson appears in the chiral-limit spectrum if, and only if, chiral symmetry is dynamically broken.  Moreover, it exposes the fascinating consequence that the solution of the two-body pseudoscalar bound-state problem is almost completely known once the one-body problem is solved for the dressed-quark propagator, with the relative momentum within the bound-state identified unambiguously with the momentum of the dressed-quark.  This latter emphasises that Goldstone's theorem has a pointwise expression in QCD.
The remaining three identities are also important because they show that a pseudoscalar meson \emph{must} contain components of pseudovector origin.  This result overturned a misapprehension of twenty-years standing; namely, that only $E_{H_{0^-}}(k;0)$ is nonzero \cite{Delbourgo:1979me}.  These pseudovector components materially influence the observable properties of pseudoscalar mesons \cite{Nguyen:2011jy,Roberts:2010rn,GutierrezGuerrero:2010md,Maris:1998hc,Roberts:2011wy}, something I will illustrate in Sec.\,\ref{sec:LookingDeeper}, as do their analogues in other mesons \cite{Maris:1999nt,Maris:1999bh,Maris:2000sk}.

I have explained that pseudoscalar mesons hold a special place in QCD, Sec.\,\ref{sec:QCNP}.  There are three such states, composed of $u$,$d$-quarks, in the hadron spectrum with masses below $2\,$GeV \cite{Nakamura:2010zzi}: $\pi(140)$; $\pi(1300)$; and $\pi(1800)$.  Of these, the pion [$\pi(140)$] is naturally well known and much studied.  The other two are observed, e.g., as resonances in the coherent production of three-pion final states via pion-nucleus collisions \cite{Bellini:1982ec}.  In the context of a quantum mechanical constituent-quark-model Hamiltonian, these mesons are often viewed as the first three members of a $Q\bar Q$ $n\, ^1\!S_0$ trajectory, where $n$ is the principal quantum number; i.e., the $\pi(140)$ is viewed as the $S$-wave ground state and the others are its first two radial excitations.  By this reasoning the properties of the $\pi(1300)$ and $\pi(1800)$ are likely to be sensitive to details of the long-range part of the quark-quark interaction because the constituent-quark wave-functions will possess material support at large interquark separation.  There is a caveat here; namely, in comparison with $\pi(1300)$, the $\pi(1800)$ is narrow, with a width of $210$MeV \protect\cite{Nakamura:2010zzi}, and has a decay pattern that may be consistent with its interpretation as a \emph{hybrid} meson in constituent-quark models \cite{Barnes:1996ff}.\footnote{It is worth providing some definitions.  \emph{Exotic mesons} -- states whose quantum numbers cannot be supported by quantum mechanical quark-antiquark systems; and \emph{hybrid mesons} -- states with quark-model quantum numbers but a non-quark-model decay pattern.  Both systems are suspected to possess ``constituent gluon'' content, which translates into a statement that they are expected to have a large overlap with interpolating fields that explicitly contain gluon fields.}
This picture has the constituent-quarks' spins aligned to produce $S_{Q\bar Q} = 1$ with $J=0$ obtained by coupling $S_{Q\bar Q}$ to a spin-$1$ excitation of the confinement potential.  Nevertheless, from this perspective, too, the properties of $\pi(1800)$ are sensitive to the interaction responsible for light-quark confinement.  Hence the development of an understanding of the properties of these three pseudoscalars might provide information about \index{Confinement} light-quark confinement; and, as has been emphasised, developing an understanding of confinement is one of the most important goals in mathematics and physics.

\begin{figure}[t]

\includegraphics[clip,width=0.66\textwidth]{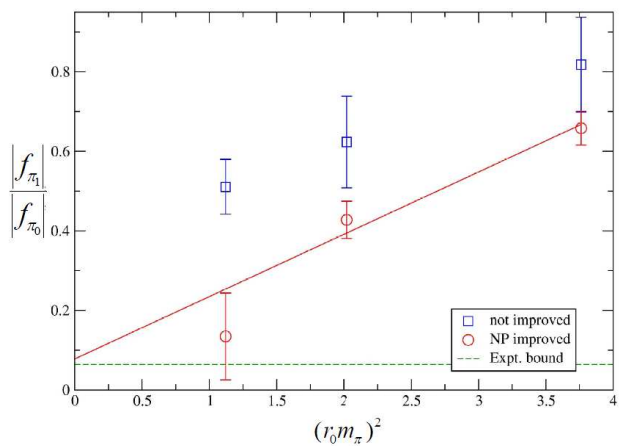}

\caption{\label{fig:fRadial}
Lattice-QCD results for the ratio of the decay constants for the first-excited- and ground-state pseudoscalar mesons as a function of the pion mass squared.  (Lattice parameters: volume\,$=16^3\times 32$; $\beta = 5.2$, spacing $a\simeq 0.1\,$fm, two flavours of degenerate sea quarks; Wilson gauge action and clover fermions.)
The ``not improved'' results were obtained from a fermion action with poor chiral symmetry properties.  In this case $|f_{\pi_1}/f_{\pi_0}|\approx 0.4$, consistent with expectations based on quantum mechanics.
The ``improved'' results were obtained through implementation of the full ALPHA method for the nonperturbative improvement of the fermion action, which greatly improves the simulation's chiral symmetry properties.  In this case, $|f_{\pi_1}/f_{\pi_0}|\approx 0.01$.  (N.B. The sign of the ratio was not determined in the simulation but is discussed in  Ref.\,\protect\cite{Qin:2011xq}.)}
\end{figure}

The puissance of Eq.\,(\ref{mrtrelation}) is further emphasised by the fact that it is also applicable here \cite{Holl:2004fr,Holl:2005vu}.  The formula holds at each pole common to the pseudoscalar and axial-vector vertices and therefore it also impacts upon the properties of non ground state pseudoscalar mesons.  I will work with a label $n\geq 0$ for the pseudoscalar mesons: $\pi_n$, with $n=0$ denoting the ground state, $n=1$ the state with the next lowest mass, and so on.  Plainly, by assumption, $m_{\pi_{n\neq 0}}>m_{\pi_0}$ , and hence $m_{\pi_{n\neq 0}} > 0$ in the chiral limit.  Moreover, the ultraviolet behaviour of the quark-antiquark scattering kernel in QCD guarantees that Eq.\,(\ref{rhogen}) is cutoff independent.  Thus
\begin{equation}
0 < \rho_{\pi_{n}}^0(\zeta):= \lim_{\hat m\to 0} \rho_{\pi_{n}}(\zeta) < \infty \,, \; \forall \, n\,,
\end{equation}
where $\hat m$ is defined in Eq.\,\eqref{RGIcqmass}.  Hence, it is a necessary consequence of chiral symmetry and its dynamical breaking in QCD; viz., Eq.\,(\ref{mrtrelation}), that \index{Leptonic decay constant: pseudoscalar excited states}
\begin{equation}
\label{fpinzero}
f_{\pi_n}^0 \equiv 0\,, \forall \, n\geq 1\,.
\end{equation}
This is the statement that Goldstone modes are the only pseudoscalar mesons to possess a nonzero leptonic decay constant in the chiral limit when chiral symmetry is dynamically broken.  The decay constants of all other pseudoscalar mesons on this trajectory, e.g., radial excitations, vanish: DCSB impacts upon every pseudoscalar meson.  On the other hand, in the absence of DCSB the leptonic decay constant of each such pseudoscalar meson vanishes in the chiral limit; i.e, Eq.\,(\ref{fpinzero}) is true $\forall n \geq 0$: absent DCSB, then all pseudoscalar mesons decouple from the weak interaction.

{F}rom the perspective of quantum mechanics, Eq.\,(\ref{fpinzero}) is a surprising fact.  The leptonic decay constant for $S$-wave states is typically proportional to the wave-function at the origin.  Compared with the ground state, this is smaller for
an excited state because the wave-function is broader in configuration space and wave-functions are normalised.  However, it is a modest effect.  For example, consider the $e^+e^-$ decay of vector mesons.  A calculation in relativistic quantum mechanics based on light-front dynamics \cite{deMelo:2005cy} yields $|f_{\rho_1}/f_{\rho_0}| = 0.5$, consistent with the value inferred from experiment and DSEs in rainbow-ladder truncation \cite{Qin:2011dd}: $|f_{\rho_1}/f_{\rho_0}| = 0.45$.  Thus, it is not uncommon for Eq.\,(\ref{fpinzero}) to be perceived as ``remarkable'' or ``unbelievable.''  Notwithstanding this, in connection with the pion's first radial excitation, the value of $f_{\pi_1}= -2\,$MeV predicted in Ref.\,\cite{Holl:2004fr} is consistent with experiment \cite{Diehl:2001xe} and simulations of lattice-QCD \cite{McNeile:2006qy}, as illustrated in Fig.\,\ref{fig:fRadial}.  It is now recognised that the suppression of $f_{\pi_1}$ is a useful benchmark, which can be used to tune and validate lattice QCD techniques that try to determine the properties of excited states mesons.

\subsection{Flavourless pseudoscalar mesons}
\label{flavourless}
\begin{figure}[t]

\includegraphics[clip,width=0.5\textwidth]{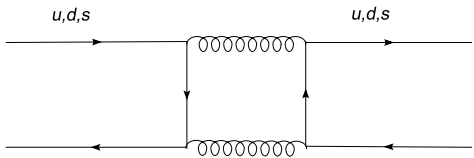}

\caption{\label{fig:nonanomaly}
This simple flavour-mixing diagram is immaterial to the resolution of the $\eta$-$\eta^\prime$ conundrum, as is any collection of processes for which the figure may serve as a skeleton diagram.  (Straight lines denote quarks and springs denote gluons.)
}
\end{figure}

In connection with electric-charge-neutral pseudoscalar mesons, Eq.\,(\ref{mrtrelation}) is strongly modified owing to the non-Abelian anomaly.  This entails that whilst the classical action associated with QCD is invariant under $U_A(N_f)$ (non-Abelian axial transformations generated by  $\lambda_0 \gamma_5$, where $\lambda_0 \propto{\rm diag}[1,\ldots ,1_{N_f}]$), the quantum field theory is not.  The modification is particularly important to properties of $\eta$ and $\eta^\prime$ mesons.  The latter is obviously a peculiar pseudoscalar meson because its mass is far greater than that of any other light-quark pseudoscalar meson; e.g., $m_{\eta^\prime} = 1.75\, m_{\eta}$.  N.B.\ The diagram depicted in Fig.\,\ref{fig:nonanomaly} is often cited as central to a solution of the $\eta$-$\eta^\prime$ puzzle.  However, as will become clear below, whilst it does contribute to flavour-mixing, the process is immaterial in resolving the $\eta$-$\eta^\prime$ conundrum.

The correct mass formula for flavourless pseudoscalars follows from consideration of the complete $U_A(N_f)$ Ward-Takahashi identity:
\begin{equation}
%
%
%
P_\mu \Gamma_{5\mu}^a(k;P) = {\cal S}^{-1}(k_+) i \gamma_5 {\cal F}^a
+ i \gamma_5 {\cal F}^a {\cal S}^{-1}(k_-)
- 2 i {\cal M}^{ab}\Gamma_5^b(k;P)  - {\cal A}^a(k;P)\,,
\label{avwtiG}
\end{equation}
which generalises Eq.\,(\ref{avwtimN}).  In Eq.\,(\ref{avwtiG}),
$\{{\cal F}^a | \, a=0,\ldots,N_f^2-1\}$ are the generators of $U(N_f)$ in the fundamental representation, orthonormalised according to tr${\cal F}^a {\cal F}^b= \frac{1}{2}\delta^{ab}$;
the dressed-quark propagator ${\cal S}=\,$diag$[S_u,S_d,S_s,S_c,S_b,\ldots]$ is matrix-valued;
and
\begin{equation}
{\cal M}^{ab} = {\rm tr}_F \left[ \{ {\cal F}^a , {\cal M}^\zeta \} {\cal F}^b \right],
\end{equation}
where ${\cal M}^\zeta$ is a matrix of renormalised current-quark masses and the trace is over flavour indices.

The final term in the last line of Eq.\,(\ref{avwtiG}) expresses the non-Abelian axial anomaly.  It can be written
\begin{eqnarray}
\label{amputate}
{\cal A}^a(k;P) &=&  {\cal S}^{-1}(k_+) \,\delta^{a0}\, {\cal A}_U(k;P) {\cal S}^{-1}(k_-)\,,\\
{\cal A}_U(k;P) &=& \!\!  \int\!\! d^4xd^4y\, e^{i(k_+\cdot x - k_- \cdot y)} N_f \left\langle  {\cal F}^0\,q(x)  \, {\cal Q}(0) \,   \bar q(y)
\right\rangle, \label{AU}
\end{eqnarray}
and since ${\cal A}^{a=0}(k;P)$ is a pseudoscalar, it has the general form
\begin{equation}
{\cal A}^0(k;P) = {\cal F}^0\gamma_5 \left[ i {\cal E}_{\cal A}(k;P) + \gamma\cdot P {\cal F}_{\cal A}(k;P)  +\, \gamma\cdot k \, k\cdot P {\cal G}_{\cal A}(k;P) + \sigma_{\mu\nu} k_\mu P_\nu {\cal H}_{\cal A}(k;P)\right].
\end{equation}
The matrix element in Eq.\,(\ref{AU}) represents an operator expectation value in full QCD; the operation in Eq.\,(\ref{amputate}) amputates the external quark lines; and
\begin{equation}
{\cal Q}(x) = i \frac{\alpha_s }{4 \pi} {\rm tr}_{C}\left[ \epsilon_{\mu\nu\rho\sigma} F_{\mu\nu} F_{\rho\sigma}(x)\right]  \label{topQ} = \partial_\mu K_\mu(x)
\end{equation}
is the topological charge density operator, where the trace is over colour indices and $F_{\mu\nu}=\frac{1}{2}\lambda^a F_{\mu\nu}^a$ is the matrix-valued gluon field strength tensor.  It is plain and important that only ${\cal A}^{a=0}$ is nonzero.  N.B.\ While ${\cal Q}(x)$ is gauge invariant, the associated Chern-Simons current, $K_\mu$, is not.  Thus in QCD no physical state can couple to $K_\mu$ and hence no state which appears in the observable spectrum can contribute to a resolution of the so-called $U_A(1)$-problem; i.e., physical states cannot play any role in ensuring that the $\eta^\prime$ is not a Goldstone mode.

As described in Sec.\,\ref{psmassformula}, if one imagines there are $N_f$ massless quarks, then DCSB is a necessary and sufficient condition for the $a\neq 0$ components of Eq.\,(\ref{avwtiG}) to guarantee the existence of $N_f^2-1$ massless bound-states of a dressed-quark and -antiquark.  However, owing to Eq.\,(\ref{amputate}), $a=0$ in Eq.\,(\ref{avwtiG}) requires special consideration.  One case is easily covered; viz., it is clear that if ${\cal A}^{0} \equiv 0$, then the $a=0$ component of Eq.\,(\ref{avwtiG}) is no different to the others and there is an additional massless bound-state in the chiral limit.

On the other hand, the large disparity between the mass of the $\eta^\prime$-meson and the octet pseudoscalars suggests that ${\cal A}^{0} \neq 0$ in real-world QCD.  If one carefully considers that possibility, then the Goldberger-Treiman relations in Eqs.\,(\ref{gtlrelE}) -- (\ref{gtlrelH}) become \cite{Bhagwat:2007ha}
\begin{eqnarray}
\label{ewti}
2 f_{H_{0^-}}^0 E_{BS}(k;0) &= & 2 B^{0}(k^2) - {\cal E}_{\cal A}(k;0),\\
\label{fwti}
F_R^0(k;0) + 2 f_{H_{0^-}}^0 F_{BS}(k;0) & = & A^{0}(k^2) - {\cal F}_{\cal A}(k;0),\\
G_R^0(k;0) + 2 f_{H_{0^-}}^0 G_{BS}(k;0) & = & 2 \frac{d}{dk^2}A^{0}(k^2) - {\cal G}_{\cal A}(k;0),\\
\label{hwti}
H_R^0(k;0) + 2 f_{H_{0^-}}^0 H_{BS}(k;0) & = & - {\cal H}_{\cal A}(k;0),
\end{eqnarray}
It follows that the relationship
\begin{equation}
\label{calEB}
{\cal E}_{\cal A}(k;0) = 2 B^{0}(k^2) \,,
\end{equation}
is necessary and sufficient to guarantee that $\Gamma_{5\mu}^0(k;P)$, the flavourless pseudoscalar vertex, does not possess a massless pole in the chiral limit; i.e., that there are only $N_f^2-1$ massless Goldstone bosons.  Now, in the chiral limit, $B^{0}(k^2) \neq 0 $ if, and only if, chiral symmetry is dynamically broken.   Hence, the absence of an additional massless bound-state is only assured through the existence of an intimate connection between DCSB and an expectation value involving the topological charge density.

This critical connection is further highlighted by the following result, obtained through a few straightforward manipulations of Eqs.\,(\ref{avwtiG}), (\ref{amputate}) and (\ref{AU}):
\begin{eqnarray}
\langle \bar q q \rangle_\zeta^0 = - \lim_{\hat m \to 0} \kappa_{H_{0^-}}^\zeta(\hat m)
& = & -\lim_{\Lambda\to \infty}Z_4(\zeta^2,\Lambda^2)\, {\rm tr}_{\rm CD}\int^\Lambda_q\!
S^{0}(q,\zeta)  \\
& = &
\mbox{\footnotesize $\displaystyle \frac{N_f}{2}$} \int d^4 x\, \langle \bar q(x) i\gamma_5  q(x) {\cal Q}(0)\rangle^0.
\end{eqnarray}
The absence of a Goldstone boson in the $a=0$ channel is only guaranteed if this explicit identity between the chiral-limit in-meson condensate and a mixed vacuum polarisation involving the topological charge density is satisfied.  (I will elucidate the concept of in-meson condensates \cite{Maris:1997tm,Brodsky:2009zd,Brodsky:2010xf,Chang:2011mu} in Sec.\,\ref{sec:inmeson}.)

Mass formulae valid for all pseudoscalar mesons have also been obtained \cite{Bhagwat:2007ha}
\begin{equation}
\label{newmass}
%
f_{H_{0^-}}^a m_{H_{0^-}}^2 = 2\,{\cal M}^{ab} \rho_{H_{0^-}}^b + \delta^{a0} \, n_{H_{0^-}}\,,
\end{equation}
where
\begin{eqnarray}
\label{fpia} f_{H_{0^-}}^a \,  P_\mu &=& Z_2\,{\rm tr} \int^\Lambda_q
{\cal F}^a \gamma_5\gamma_\mu\, \chi_{H_{0^-}}(q;P) \,, \\
\label{cpres} i  \rho_{H_{0^-}}^a\!(\zeta)  &=& Z_4\,{\rm tr}
\int^\Lambda_q {\cal F}^a \gamma_5 \, \chi_{H_{0^-}}(q;P)\,,\\
n_{H_{0^-}} &=& \mbox{\footnotesize $\displaystyle \sqrt{\frac{N_f}{2}}$} \, \nu_{H_{0^-}} \,, \; \nu_{H_{0^-}}= \langle 0 | {\cal Q} | H_{0^-}\rangle \,.
\end{eqnarray}
For charged pseudoscalar mesons, Eq.\,(\ref{newmass}) is equivalent to Eq.\,(\ref{mrtrelation}), but the novelty of Eq.\,(\ref{newmass}) is what it expresses for flavourless pseudoscalars.  To illustrate, consider the case of a $U(N_f=3)$-symmetric mass matrix, in which all $N_f=3$ current-quark masses assume the single value $m^\zeta$, then this formula yields:
\begin{equation}
\label{etapchiral}
m_{\eta^\prime}^2 f_{\eta^\prime}^0 = n_{\eta^\prime} + 2 m^\zeta\rho_{\eta^\prime}^{0 \zeta} \,.
\end{equation}
Plainly, the $\eta^\prime$ is split from the Goldstone modes so long as $n_{\eta^\prime} \neq 0$.  Numerical simulations of lattice-QCD have confirmed this identity \protect\cite{Bardeen:2000cz,Ahmad:2005dr}.

It is important to elucidate the physical content of $n_{\eta^\prime}$.  Returning to the definition:
\begin{equation}
\nu_{\eta^\prime}= \mbox{\footnotesize $\displaystyle \sqrt{\frac{3}{2}}$} \, \langle 0 | {\cal Q} | \eta^\prime \rangle \,,
\end{equation}
it is readily seen to be another type of in-meson condensate.  It is analogous to those that will be discussed in Sec.\,\ref{sec:inmeson} but in this case the hadron-to-vacuum transition amplitude measures the topological content of the $\eta^\prime$.  One may therefore state that the $\eta^\prime$ is split from the Goldstone modes so long as its wave-function possesses nonzero topological content.  This is plainly very different to requiring that the QCD vacuum is topologically nontrivial.

\begin{figure}[t]
\vspace*{-25ex}

\includegraphics[clip,width=0.7\textwidth]{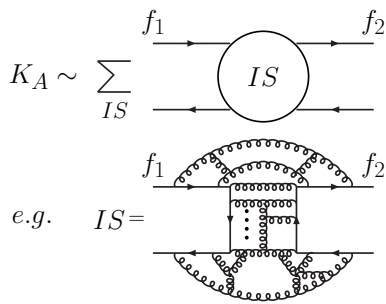}
\vspace*{-25ex}

\caption{\label{etaglue} An illustration of the nature of the contribution to the Bethe-Salpeter kernel associated with the non-Abelian anomaly.  All terms have the ``hairpin'' structure illustrated in the lower panel.  No finite sum of such intermediate states is sufficient.  (Straight lines denote quarks, with $f_1$ and $f_2$ independent, and springs denote gluons.)}
\end{figure}

Within QCD the properties of the $\eta^\prime$ can be computed via the BSE, just like other mesons.  A nonzero value of $n_{\eta^\prime}$ can be achieved with a Bethe-Salpeter kernel that contains the contribution depicted in Fig.\,\ref{etaglue} because one may argue from Eqs.\,(\ref{AU}) and (\ref{topQ}) that an anomaly-related contribution to a meson's Bethe-Salpeter kernel cannot contain external quark or antiquark lines that are connected to the incoming lines: purely gluonic configurations must mediate, as illustrated in Fig.\,\ref{etaglue}.  Furthermore, it is straightforward to see that no finite sum of gluon exchanges can serve this purpose.  Indeed, consider any one such single contribution in the chiral limit.  It will be proportional to the total momentum and hence vanish for $P=0$, in conflict with Eq.\,(\ref{etapchiral}).  This lies behind the need for something like the Kogut-Susskind \emph{ghost}; i.e., the coupling of a massless axial-vector gauge-like field to the Chern-Simons current, which does not appear in the particle spectrum of QCD because the current is not gauge invariant.  (See Ref.\,\protect\cite{Kogut:1974kt} and Sec.\,5.1 of Ref.\,\cite{Christos:1984tu}.)

It is argued \cite{Witten:1979vv,Veneziano:1979ec} that in QCD with $N_c$ colours,
\begin{equation}
n_{\eta^\prime} \sim \frac{1}{\sqrt{N_c}}\,,
\end{equation}
and it can be seen to follow from the gap equation, the homogeneous BSE and Eqs.\,(\ref{fpia}), (\ref{cpres}) that
\begin{equation}
f_{\eta^\prime}^0 \sim \sqrt{N_c} \sim \rho_{\eta^\prime}^0(\zeta)\,.
\end{equation}
One thus obtains
\begin{equation}
m_{\eta^\prime}^2 =  \frac{n_{\eta^\prime}}{f_{\eta^\prime}^0} + 2 m(\zeta) \frac{\rho_{\eta^\prime}^0(\zeta)}{f_{\eta^\prime}^0} \,.
\end{equation}
The first term vanishes in the limit $N_c\to \infty$ while the second remains finite.  Subsequently taking the chiral limit, the $\eta^\prime$ mass approaches zero in the manner characteristic of all Goldstone modes.  (N.B.\ One must take the limit $N_c\to \infty$ before the chiral limit because the procedures do not commute \cite{Narayanan:2004cp}.)  These results are realised in the effective Lagrangian of Ref.\,\cite{DiVecchia:1979bf} in a fashion that is consistent with all the constraints of the anomalous Ward identity.  It is noteworthy that this is not true of the so-called 't\,Hooft determinant \protect\cite{Christos:1984tu,Crewther:1977ce,Crewther:1978zz}.

Implications of the mass formula in Eq.\,(\ref{newmass}) were exemplified in Ref.\,\cite{Bhagwat:2007ha} using an elementary dynamical model that includes a one-parameter \emph{Ansatz} for that part of the Bethe-Salpeter kernel related to the non-Abelian anomaly, an illustration of which is provided in Fig.\,\ref{etaglue}.  The study compares ground-state pseudoscalar- and vector-mesons constituted from all known quarks, excluding the $t$-quark.  Amongst the notable results is a prediction for the mixing angles between neutral mesons; e.g.,
\begin{equation}
\label{valmixing}
\theta_\eta = -15.4^\circ\,,\;
\theta_{\eta^\prime} = -15.7^\circ\,.
\end{equation}
N.B.\ There are necessarily two mixing angles, with each determined at the appropriate pole position in the inhomogeneous vertex.  It is interesting that the angles are approximately equal and compare well with the value inferred from a single mixing angle analysis \cite{Bini:2007zza} $\theta = -13.3^\circ \pm 1.0^\circ$.

It is worth explicating the nature of the flavour-induced difference between the $\pi^0$ and $\pi^\pm$ masses.  If one ignores mixing with mesons containing other than $u,d$-quarks; viz., works solely within $SU(N_f=2)$, then $m_{\pi^0}-m_{\pi^+}=-0.04\,$MeV.  On the other hand, the full calculation yields $m_{\pi^0}-m_{\pi^+}=-0.4\,$MeV, a factor of ten greater, and one obtains a $\pi^0$-$\eta$ mixing angle, whose value at the neutral pion mass shell is
\begin{equation}
\theta_{\pi \eta}(m_{\pi^0}^2)=1.2^\circ.
\end{equation}
For comparison, Ref.\,\cite{Green:2003qw} infers a mixing angle of $0.6^\circ \pm 0.3^\circ$ from a $K$-matrix analysis of the process $p\, d \rightarrow\, ^3$He$\,\pi^0$.  Plainly, mixing with the $\eta$-meson is the dominant non-electromagnetic contribution to the $\pi^\pm$-$\pi^0$ mass splitting.  The analogous angle at the $\eta$ mass-shell is
\begin{equation}
\theta_{\pi \eta}(m_{\eta}^2)=1.3^\circ.
\end{equation}

The angles in Eq.\,(\ref{valmixing}) correspond to
\begin{eqnarray}
\label{pi0f}
|\pi^0\rangle & \sim & 0.72 \, \bar u u - 0.69 \, \bar d d - 0.013 \, \bar s s\,, \\
\label{pi8f}
|\eta\rangle & \sim & 0.53\, \bar u u + 0.57 \, \bar d d - 0.63 \, \bar s s\,, \\
\label{pi9f}
|\eta^\prime\rangle & \sim & 0.44\, \bar u u + 0.45 \, \bar d d + 0.78 \, \bar s s \,.
\end{eqnarray}
Evidently, in the presence of a sensible amount of isospin breaking, the $\pi^0$ is still predominantly characterised by ${\cal F}^3$ but there is a small admixture of $\bar ss$.  It is found in Ref.\,\cite{Bhagwat:2007ha} that mixing with the $\pi^0$ has a similarly modest impact on the flavour content of the $\eta$ and $\eta^\prime$.  It's effect on their masses is far less.

\section{Condensates are Confined within Hadrons}
\label{sec:inmeson}
Dynamical chiral symmetry breaking and its connection with the generation of hadron masses was first considered in Ref.\,\cite{Nambu:1961tp}.  The effect was represented as a vacuum phenomenon.  Two essentially inequivalent classes of ground-state were identified in the mean-field treatment of a meson-nucleon field theory: symmetry preserving (Wigner phase); and symmetry breaking (Nambu phase).  Notably, within the symmetry breaking class, each of an uncountable infinity of distinct configurations is related to every other by a chiral rotation.  This is arguably the origin of the concept that strongly-interacting quantum field theories possess a nontrivial vacuum.

\begin{figure}[t]
\vspace*{-1ex}

\includegraphics[clip,width=0.66\textwidth]{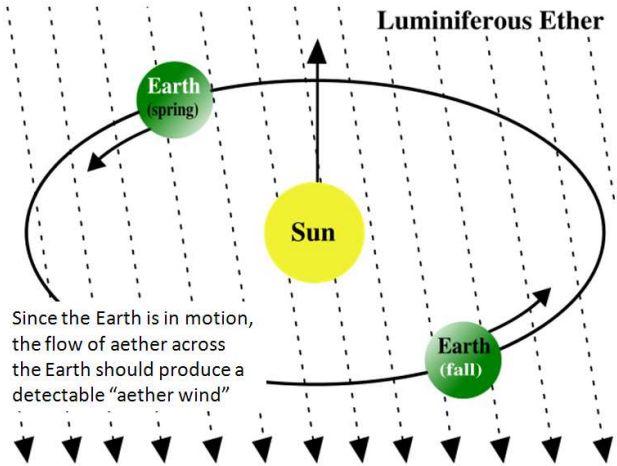}

\caption{\label{fig:aether} Physics theories of the late 19th century postulated that, just as water waves must have a medium to move across (water), and audible sound waves require a medium to move through (such as air or water), so also light waves require a medium, the ``luminiferous aether.''  This was apparently unassailable logic until, of course, one of the most famous failed experiments in the history of science to date \protect\cite{MichelsonMorley}.
}

\end{figure}

With the introduction of the parton model for the description of deep inelastic scattering (DIS), this notion was challenged via an argument \cite{Casher:1974xd} that DCSB can be realised as an intrinsic property of hadrons, instead of through a nontrivial vacuum exterior to the observable degrees of freedom.  Such a perspective is tenable because the essential ingredient required for dynamical symmetry breaking in a composite system is the existence of a divergent number of constituents and DIS provided evidence for the existence within every hadron of a sea of low-momentum partons.  This view has, however, received scant attention.  On the contrary, the introduction of QCD sum rules as a theoretical artifice to estimate nonperturbative strong-interaction matrix elements entrenched the belief that the QCD vacuum is characterised by numerous distinct, spacetime-independent condensates.  Faith in empirical vacuum condensates might be compared with an earlier misguided conviction that the universe was filled with a luminiferous aether, Fig.\,\ref{fig:aether}.

Notwithstanding the prevalence of the belief in empirical vacuum condensates, it does lead to problems; e.g., entailing, as explained below, a cosmological constant that is $10^{46}$-times greater than that which is observed \cite{Brodsky:2009zd,Turner:2001yu}.  This unwelcome consequence is partly responsible for reconsideration of the possibility that the so-called vacuum condensates are in fact an intrinsic property of hadrons.  Namely, in a confining theory -- and confinement is essential to this view -- condensates are not constant, physical mass-scales that fill all spacetime; instead, they are merely mass-dimensioned parameters that serve a practical purpose in some theoretical truncation schemes but otherwise do not have an existence independent of hadrons \cite{Brodsky:2009zd,Brodsky:2010xf,Chang:2011mu,%
Burkardt:1998dd,Brodsky:2008be,Glazek:2011vg}.

To account for the gross overestimate, recall that in Sec.\,\ref{sec:introduction} I remarked that the expansion of the universe is accelerating \cite{Turner:2001yu}.  One piece of evidence is provided by the observations of type Ia supernovae reported in Refs.\,\cite{1538-3881-116-3-1009,0004-637X-517-2-565}; and another by measurements of the composition of the universe, which point to a missing energy component with negative pressure.  To explain the latter, the curvature of the universe may be characterised by a mass-energy density, $\rho_U$.  There is a critical value of this density for which the universe is flat: $\rho_F$.  Observations of the cosmic microwave background's power spectrum, e.g., Ref.\,\cite{Netterfield:2001yq}, indicate that $\Omega_0=\rho_U/\rho_F=1 \pm 0.04$.  In a flat universe, the matter density, $\Omega_M$, and the energy density must sum to the critical density.  However, matter only contributes about one-third of the critical density, $\Omega_M=0.33\pm0.04$.  Thus two-thirds of the critical density is missing.

In order to have escaped detection, the missing energy must be smoothly distributed.  In order not to interfere with the formation of structure (by inhibiting the growth of density perturbations) the energy density in this component must change more slowly than matter (so that it was subdominant in the past).  The universe's accelerated expansion can be accommodated in general relativity through the cosmological constant, $\Lambda$, and observations determine an associated density, $\rho_{\Lambda}^{\rm obs}$.  (Recall that Einstein introduced the repulsive effect of the cosmological constant in order to balance the attractive gravity of matter so that a static universe was possible.  However, he promptly discarded it after the discovery of the expansion of the Universe.)

It has been suggested that the advent of quantum field theory makes consideration of the cosmological constant obligatory not optional \cite{Turner:2001yu}.  Indeed, the only possible covariant form for the energy of the (quantum) vacuum; viz.,
\begin{equation}
T_{\mu\nu}^{\rm VAC} = \rho_{\rm VAC}\, \delta_{\mu\nu}
\end{equation}
is mathematically equivalent to the cosmological constant.  The vacuum is \cite{Turner:2001yu} ``\ldots a perfect fluid and precisely spatially uniform \ldots'' so that ``Vacuum energy is almost the perfect candidate for dark energy.''  Now, if the ground state of QCD is truly expressed in a nonzero spacetime-independent expectation value $\langle\bar q q\rangle$, then the energy difference between the symmetric and broken phases is of order $M_{\rm QCD} \sim 0.3\,$GeV, as indicated by Fig.\,\ref{gluoncloud}.  One obtains therefrom:
\begin{equation}
\rho_\Lambda^{\rm QCD} = 10^{46} \rho_\Lambda^{\rm obs}.
\end{equation}
In fact, the discrepancy is far greater if the Higgs vacuum expectation value is treated in a similar manner.

This mismatch has been called ``the greatest embarrassment in theoretical physics.''  However, it vanishes if one discards the notion that condensates have a physical existence, which is independent of the hadrons that express QCD's asymptotically realisable degrees of freedom \cite{Brodsky:2009zd}; namely, if one accepts that such condensates are merely mass-dimensioned parameters in one or another theoretical truncation scheme.  This appears mandatory in a confining theory \cite{Brodsky:2010xf,Chang:2011mu}, a perspective one may embed in a broader context by considering just what is observable in quantum field theory \cite{Weinberg:1978kz}: ``\ldots although individual quantum field theories have of course a good deal of content, quantum field theory itself has no content beyond analyticity, unitarity, cluster decomposition and symmetry.''  If QCD is a confining theory, then the principle of cluster decomposition is only realised for colour singlet states \cite{Krein:1990sf} and all observable consequences of the theory, including its ground state, can be expressed via an hadronic basis.  This is quark-hadron duality.

It is worthwhile to recapitulate upon the arguments in Refs.\,\cite{Brodsky:2010xf,Chang:2011mu}.  To begin, note that Eq.\,(\ref{fpigen}) is the exact expression in QCD for the leptonic decay constant of a pseudoscalar meson.  It is a property of the pion and, as consideration of the integral expression reveals, it can be described as the pseudovector projection of the pion's Bethe-Salpeter wavefunction onto the origin in configuration space.  Note that the product $\psi = S \Gamma S$ is called the Bethe-Salpeter wavefunction because, when a nonrelativistic limit can validly be performed, the quantity $\psi$ at fixed time becomes the quantum mechanical wavefunction for the system under consideration.  (N.B.\ In the neighborhood of the chiral limit, a value for $f_{H_{0^-}}$ can be estimated via either of two approximation formulae \protect\cite{Cahill:1985mh,Pagels:1979hd,Chang:2009zb}.  These formulae both illustrate and emphasize the role of $f_{H_{0^-}}$ as an order parameter for DCSB.)

If chiral symmetry were not dynamically broken, then in the neighborhood of the chiral limit $f_{H_{0^-}} \propto \hat m$ \cite{Holl:2004fr}.  Of course, chiral symmetry is dynamically broken in QCD \cite{Bhagwat:2003vw,Bhagwat:2006tu,Bhagwat:2007vx,Bowman:2005vx} and so for the ground-state pseudoscalar
\begin{equation}
\lim_{\hat m\to 0} f_{H_{0^-}}(\hat m) = f^0_{H_{0^-}} \neq 0\,.
\end{equation}
Taken together, these last two observations express the fact that $f_{H_{0^-}}$, which is an intrinsic property of the pseudoscalar meson, is a \emph{bona fide} order parameter for DCSB.  An analysis within chiral perturbation theory \cite{Bijnens:2006zp} suggests that the chiral limit value, $f^0_{H_{0^-}}$, is $\sim 5$\% below the measured value of 92.4\,MeV; and efficacious DSE studies give a 3\% chiral-limit reduction~\cite{Maris:1997tm}.

Now, Eq.\,(\ref{rhogen}) is kindred to Eq.\,(\ref{fpigen}); it is the expression in quantum field theory which describes the \emph{pseudoscalar} projection of the pseudoscalar meson's Bethe-Salpeter wavefunction onto the origin in configuration space.  It is thus truly just another type of pseudoscalar meson decay constant.  

In this connection it is therefore notable that one may rigorously define an ``in-meson'' condensate; viz.\,\cite{Maris:1997hd,Maris:1997tm}:
\begin{equation}
\label{inpiqbq}
-\langle \bar q_{f_2} q_{f_1} \rangle^\zeta_{H_{0^-}} \equiv -
f_{H_{0^-}} \langle 0 | q_{f_2} \gamma_5 q_{f_1} |H_{0^-} \rangle
= f_{H_{0^-}} \rho_{H_{0^-}}^\zeta =: \kappa_{H_{0^-}}^\zeta(\hat m)\,.
\end{equation}
Now, using Eq.\,(\ref{gtlrelE}), one finds \cite{Maris:1997hd}
\begin{equation}
\lim_{\hat m\to 0} \kappa^\zeta_{H_{0^-}}(\hat m)
=
Z_4 \, {\rm tr}_{\rm CD}\int^\Lambda \!\!\!\! \mbox{\footnotesize $\displaystyle\frac{d^4 q}{(2\pi)^4}$} S^0(q;\zeta) =  -\langle \bar q q \rangle_\zeta^0\,.
\label{qbqpiqbq0}
\end{equation}
Hence the so-called vacuum quark condensate is, in fact, the chiral-limit value of the in-meson condensate; i.e., it describes a property of the chiral-limit pseudoscalar meson.  One can therefore argue that this condensate is no more a property of the ``vacuum'' than the pseudoscalar meson's chiral-limit leptonic decay constant.  Moreover, Ref.\,\cite{Langfeld:2003ye} establishes the equivalence of all three definitions of the so-called vacuum quark condensate: a constant in the operator product expansion \cite{Lane:1974he,Politzer:1976tv}; via the Banks-Casher formula \cite{Banks:1979yr}; and the trace of the chiral-limit dressed-quark propagator.  Hence, they are all related to the in-meson condensate via Eq.\,\eqref{qbqpiqbq0} and none is defined essentially in connection with the vacuum.


It is worth remarking that in the presence of confinement it is impossible to write a valid nonperturbative definition of a single quark or gluon annihilation operator; and therefore impossible to rigorously define a second quantised vacuum (ground state) for QCD upon a foundation of gluon and quark (quasiparticle) operators.   To do so would be to answer the question: What is the state that is annihilated by an operator which is - as appears at present - unknowable?  However, with the assumptions that confinement is absolute and that it entails quark-hadron duality, the question changes completely.  In this case, the nonperturbative Hamiltonian of observable phenomena in QCD is diagonalised by colour-singlet states alone.  The ground state of this nonperturbative strong-interaction Hamiltonian is the state with zero hadrons.  One may picture the creation and annihilation operators for such states as rigorously defined via smeared sources on a spacetime lattice.  The ground-state is defined with reference to such operators, employing, e.g., the Gell-Mann - Low theorem \cite{GellMann:1951rw}, which is applicable in this case because there are well-defined asymptotic states and associated annihilation and creation operators.

In learning that the so-called vacuum quark condensate is actually the chiral-limit value of an in-pion property, some respond as follows.
The electromagnetic radius of any hadron which couples to pseudoscalar mesons must diverge in the chiral limit.  This long-known effect arises because the propagation of \emph{massless} on-shell colour-singlet pseudoscalar mesons is undamped \cite{Beg:1973sc,Pervushin:1974nm,Gasser:1983yg,Alkofer:1993gu}.
Therefore, does not each pion grow to fill the universe; so that, in this limit, the in-pion condensate reproduces the conventional paradigm?

Confinement, again, vitiates this objection.  Both DSE- and lattice-QCD studies indicate that confinement entails dynamical mass generation for both gluons and quarks, see Secs.\,\ref{confinement}, \ref{confinementII}.  The dynamical gluon and quark masses remain large in the limit of vanishing current-quark mass.  In fact, the dynamical masses are almost independent of the current-quark mass in the neighbourhood of the chiral limit.  It follows that for any hadron the quark-gluon containment-radius does not diverge in the chiral limit.  Instead, it is almost insensitive to the magnitude of the current-quark mass because the dynamical masses of the hadron's constituents are frozen at large values; viz., $2 - 3\,\Lambda_{\rm QCD}$.  These considerations show that the divergence of the electromagnetic radius does not correspond to expansion of a condensate from within the pion but rather to the copious production and subsequent propagation of composite pions, each of which contains a condensate whose value is essentially unchanged from its nonzero current-quark mass value within a containment-domain whose size is similarly unaffected.

There is more to be said in connection with the definition and consequences of a chiral limit.  Plainly, the existence of strongly-interacting massless composites would have an enormous impact on the evolution of the universe; and it is naive to imagine that one can simply set $\hat m_{u,d}=0$ and consider a circumscribed range of manageable consequences whilst ignoring the wider implications for hadrons, the Standard Model and beyond.  For example, with all else held constant, Big Bang Nucleosynthesis is very sensitive to the value of the pion-mass \cite{Flambaum:2007mj}.  We are fortunate that the absence of quarks with zero current-quark mass has produced a universe in which we exist so that we may carefully ponder the alternative.

The discussion of Ref.\,\cite{Brodsky:2010xf} was restricted to pseudoscalar mesons.  It is expanded in Ref.\,\cite{Chang:2011mu} via a demonstration that the in-pseudoscalar-meson condensate can be represented through the pseudoscalar-meson's scalar form factor at zero momentum transfer.  With the aid of a mass formula for scalar mesons, revealed therein, the analogue was shown to be true for in-scalar-meson condensates.  As argued, the concept is readily extended to all hadrons so that, via the zero momentum transfer value of any hadron's scalar form factor, one can readily extract the value for a quark condensate in that hadron which is a measure of dynamical chiral symmetry breaking.

Given that quark condensates are an intrinsic property of hadrons, one arrives at a new paradigm, as observed in the popular science press \cite{Courtland:2010zz}: ``EMPTY space may really be empty.  Though quantum theory suggests that a vacuum should be fizzing with particle activity, it turns out that this paradoxical picture of nothingness may not be needed.  A calmer view of the vacuum would also help resolve a nagging inconsistency with dark energy, the elusive force thought to be speeding up the expansion of the universe.''  In connection with the cosmological constant, putting QCD condensates back into hadrons reduces the mismatch between experiment and theory by a factor of $10^{46}$.  If technicolour-like theories are the correct scheme for extending the Standard Model \cite{Andersen:2011yj}, then the impact of the notion of in-hadron condensates is far greater still. 

\section{Many Facets of DCSB}
\label{sec:Facets}
%
The importance and interconnection of confinement and DCSB are summarised in Sec.\,\ref{sec:Apercu}; and some of the profound implications of DCSB for pseudoscalar mesons are detailed in Sec.\,\ref{sec:Exact}.  The latter could be proved owing to the existence of at least one systematic nonperturbative symmetry-preserving DSE truncation scheme (Fig.\,\ref{FSysTruncation} and the associated discussion).  On the other hand, the practical application of this particular scheme has numerous shortcomings.
For example, at leading-order (rainbow-ladder) the truncation is accurate for ground-state vector- and electrically-charged pseudoscalar-mesons because corrections in these channels largely cancel, owing to parameter-free preservation of the Ward-Takahashi identities.  However, they do not cancel in other channels \cite{Roberts:1996jx,Roberts:1997vs,Bender:2002as,Bhagwat:2004hn}.  Hence studies based on the rainbow-ladder truncation, or low-order improvements thereof, have usually provided poor results for scalar- and axial-vector-mesons \cite{Burden:1996nh,Watson:2004kd,Maris:2006ea,Cloet:2007pi,Fischer:2009jm,%
Krassnigg:2009zh}, produced masses for exotic states that are too low in comparison with other estimates \cite{Qin:2011dd,Qin:2011xq,Burden:1996nh,Cloet:2007pi,Krassnigg:2009zh}, and exhibit gross sensitivity to model parameters for tensor-mesons \cite{Krassnigg:2010mh} and excited states \cite{Qin:2011dd,Qin:2011xq,Holl:2004fr,Holl:2004un}.  In these circumstances one must conclude that physics important to these states is omitted.
One anticipates therefore that significant qualitative advances in understanding the essence of QCD could be made with symmetry-preserving kernel \emph{Ans\"atze} that express important additional nonperturbative effects, which are impossible to capture in any finite sum of contributions.  Such an approach has recently become available \cite{Chang:2009zb} and will be summarised below.  It is hoped that the mathematics of Fa\`{a} di Bruno Hopf algebras will assist hadron physics practitioners to improve further upon this.

\subsection{DCSB in the Bethe-Salpeter kernel}
\label{sec:building}
\begin{figure}[t]
\includegraphics[width=0.55\textwidth]{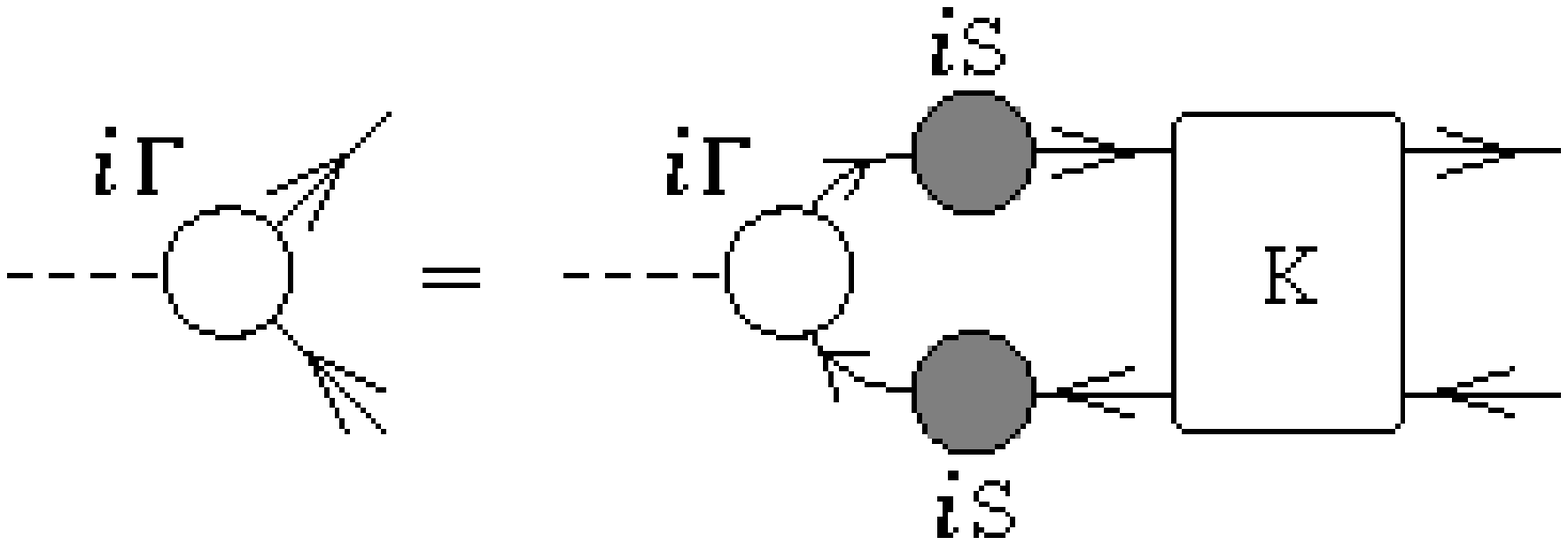}

\includegraphics[width=0.75\textwidth]{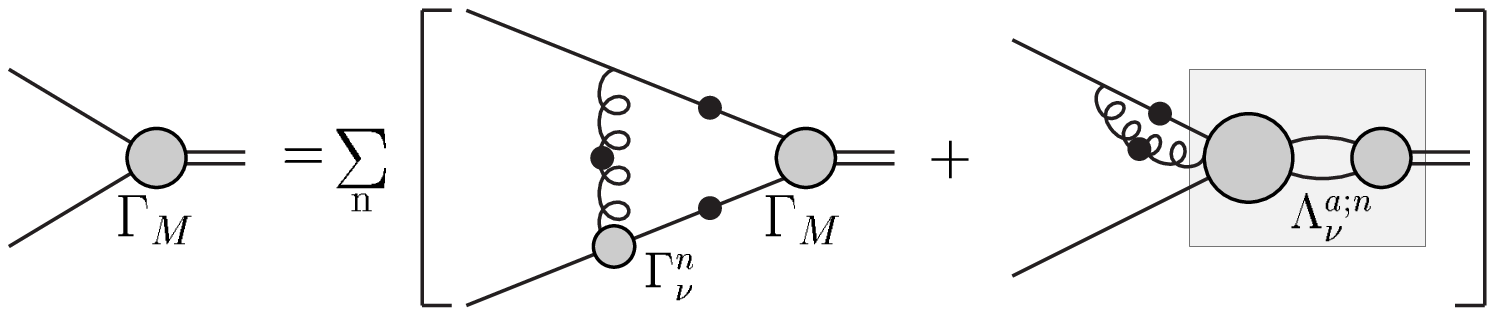}
\caption{\label{detmoldkernel}
Omitting the inhomogeneity, the \emph{upper panel} illustrates the textbook form of the Bethe-Salpeter equation, Eq.\,(\protect\ref{bsetextbook}), whereas the \emph{lower panel} depicts the form expressed in Eq.\,(\protect\ref{genbse}).  The reversal of the total-momentum's flow is immaterial here.
N.B.\ In any symmetry-preserving truncation, beyond the leading-order identified in Ref.\,\protect\cite{Bender:1996bb}; i.e., rainbow-ladder, the Bethe-Salpeter kernel is nonplanar even if the vertex in the gap equation is planar \protect\cite{Bender:2002as}.  This is illustrated in Fig.\,\protect\ref{FSysTruncation}.}
\end{figure}

In order to illustrate the decisive importance of DCSB in the Bethe-Salpeter kernel, consider, e.g., flavoured pseudoscalar and axial-vector mesons, which appear as poles in the inhomogeneous BSE for the axial-vector vertex, $\Gamma_{5\mu}^{fg}$, where $f,g$ are flavour labels.  An exact form of that equation is ($k$, $q$ are relative momenta, $P$ is the total momentum flowing into the vertex, and $q_\pm = q\pm P/2$, etc.)
\begin{eqnarray}
\nonumber
\Gamma_{5\mu}^{fg}(k;P) &=& Z_2 \gamma_5\gamma_\mu - \int_q^\Lambda g^2 D_{\alpha\beta}(k-q) \frac{\lambda^a}{2}\,\gamma_\alpha S_f(q_+) \Gamma_{5\mu}^{fg}(q;P) S_g(q_-) \frac{\lambda^a}{2}\,\Gamma_\beta^g(q_-,k_-) \\
&  + &\int^\Lambda_q g^2D_{\alpha\beta}(k-q)\, \frac{\lambda^a}{2}\,\gamma_\alpha S_f(q_+) \frac{\lambda^a}{2} \Lambda_{5\mu\beta}^{fg}(k,q;P), \rule{1em}{0ex} \label{genbse}
\end{eqnarray}
where $\Lambda_{5\mu\beta}^{fg}$ is a 4-point Schwinger function.  [The pseudoscalar vertex satisfies an analogue of Eq.\,(\ref{genbse}).]  This form of the BSE was first written in Ref.\,\cite{Bender:2002as} and is illustrated in the lower-panel of Fig.\,\ref{detmoldkernel}.  The diagrammatic content of the right-hand-side is completely equivalent to that of Eq.\,(\ref{bsetextbook}), which is depicted in the upper-panel of the figure.  However, in striking qualitative opposition to that textbook equation, Eq.\,(\ref{genbse}) partly embeds the solution vertex in the four-point function, $\Lambda$, whilst simultaneously explicating a part of the effect of the dressed-quark-gluon vertex.  This has the invaluable consequence of enabling the derivation of both an integral equation for the new Bethe-Salpeter kernel, $\Lambda$, in which the driving term is the dressed-quark-gluon vertex \cite{Bender:2002as}, and a Ward-Takahashi identity relating $\Lambda$ to that vertex \cite{Chang:2009zb}.  No similar equations have yet been found for $K$ and hence the textbook form of the BSE, whilst tidy, is very limited in its capacity to expose the effects of DCSB in bound-state physics.

As emphasised above, no study of light-quark hadrons is dependable if it fails to comply with the axial-vector Ward-Takahashi identity, Eq.\,(\ref{avwtimN}).  The condition
\begin{equation}
P_\mu \Lambda_{5\mu\beta}^{fg}(k,q;P) + i [m_f(\zeta)+m_g(\zeta)] \Lambda_{5\beta}^{fg}(k,q;P)= \Gamma_\beta^f(q_+,k_+) \, i\gamma_5+ i\gamma_5 \, \Gamma_\beta^g(q_-,k_-) \,, \label{LavwtiGamma}
\end{equation}
where $\Lambda_{5\beta}^{fg}$ is the analogue of $\Lambda_{5\mu\beta}^{fg}$ in the pseudoscalar equation, is necessary and sufficient to ensure the Ward-Takahashi identity is satisfied by the solution of Eqs.\,(\ref{gendseN}) and (\ref{genbse}) \cite{Chang:2009zb}.

Consider Eq.\,(\ref{LavwtiGamma}).  Rainbow-ladder is the lead\-ing-or\-der term in the systematic DSE truncation scheme of Refs.\,\cite{Munczek:1994zz,Bender:1996bb}.  It corresponds to $\Gamma_\nu^f=\gamma_\nu$, in which case Eq.\,(\ref{LavwtiGamma}) is solved by $\Lambda_{5\mu\beta}^{fg}\equiv 0 \equiv \Lambda_{5\beta}^{fg}$.  This is the solution that indeed provides the rainbow-ladder forms of Eq.\,(\ref{genbse}).  Such consistency will be apparent in any valid systematic term-by-term improvement of the rainbow-ladder truncation.

However, since the two-point functions of elementary excitations are strongly modified in the infrared, one must accept that the same is generally true for three-point functions; i.e., the vertices.  Hence the bare vertex will be a poor approximation to the complete result unless there are extenuating circumstances.  This is readily made apparent, for with a dressed-quark propagator of the form in Eq.\,\eqref{Sgeneral}, one finds immediately that the Ward-Takahashi identity is breached; viz.,
\begin{equation}
P_\mu i \gamma_\mu \neq S^{-1}(k+P/2) - S^{-1}(k-P/2)\,,
\end{equation}
and the violation is significant whenever and wherever the mass function in Fig.\,\ref{gluoncloud} is large.  This was actually realised early on, with studies of the fermion--gauge-boson vertex in Abelian gauge theories \cite{Ball:1980ay} that have inspired numerous ensuing analyses.  The importance of this dressing to the reliable computation of hadron physics observables was exposed in Refs.\,\cite{Frank:1994mf,Roberts:1994hh}, insights from which have subsequently been exploited effectively; e.g., Refs.\,\cite{Chang:2011vu,Maris:1997hd,Maris:2000sk,Eichmann:2008ef,Cloet:2008re,Chang:2010hb,%
Eichmann:2011vu,Chang:2011ei,Wilson:2011rj}.

The most important feature of the perturbative or bare vertex is that it cannot cause spin-flip transitions; namely, it is an helicity conserving interaction.  However, one must expect that DCSB introduces nonperturbatively generated structures that very strongly break helicity conservation.  These contributions will be large when the dressed-quark mass-function is large.  Conversely, they will vanish in the ultraviolet; i.e., on the perturbative domain.  The exact form of the vertex contributions is still the subject of study but their existence is model-independent.

Critical now is a realisation that Eq.\,(\ref{LavwtiGamma}) is far more than just a device for checking a truncation's consistency.  For, just as the vector Ward-Takahashi identity has long been used to build \emph{Ans\"atze} for the dressed-quark-photon vertex \cite{Roberts:1994dr,Ball:1980ay,Kizilersu:2009kg,Bashir:2011dp}, Eq.\,(\ref{LavwtiGamma}) provides a tool for constructing a symmetry-preserving kernel of the BSE that is matched to any reasonable form for the dressed-quark-gluon vertex which appears in the gap equation.  With this powerful capacity, Eq.\,(\ref{LavwtiGamma}) achieves a goal that has been sought ever since the Bethe-Salpeter equation was introduced \cite{Salpeter:1951sz}.  As will become apparent, it produces a symmetry-preserving kernel that promises to enable the first reliable Poincar\'e invariant calculation of the spectrum of mesons with masses larger than 1\,GeV.

The utility of Eq.\,(\ref{LavwtiGamma}) was illustrated in Ref.\,\cite{Chang:2009zb} through an application to ground state pseudoscalar and scalar mesons composed of equal-mass $u$- and $d$-quarks.  To this end, it was supposed that in Eq.\,(\ref{gendseN}) one employs an \emph{Ansatz} for the quark-gluon vertex which satisfies
\begin{equation}
P_\mu i \Gamma_\mu^f(k_+,k_-) = {\cal B}(P^2)\left[ S_f^{-1}(k_+) - S_f^{-1}(k_-)\right] , \label{wtiAnsatz}
\end{equation}
with ${\cal B}$ flavour-independent.  (N.B.\ While the true quark-gluon vertex does not satisfy this identity, owing to the form of the Slavnov-Taylor identity which it does satisfy, it is plausible that a solution of Eq.\,(\protect\ref{wtiAnsatz}) can provide a reasonable pointwise approximation to the true vertex \cite{Bhagwat:2004kj}.)  Given Eq.\,(\ref{wtiAnsatz}), then Eq.\,(\ref{LavwtiGamma}) entails ($l=q-k$)
\begin{equation}
i l_\beta \Lambda_{5\beta}^{fg}(k,q;P) =
{\cal B}(l^2)\left[ \Gamma_{5}^{fg}(q;P) - \Gamma_{5}^{fg}(k;P)\right], \label{L5beta}
\end{equation}
with an analogous equation for $P_\mu l_\beta i\Lambda_{5\mu\beta}^{fg}(k,q;P)$.  This identity can be solved to obtain
\begin{equation}
\Lambda_{5\beta}^{fg}(k,q;P)  :=  {\cal B}((k-q)^2)\, \gamma_5\,\overline{ \Lambda}_{\beta}^{fg}(k,q;P) \,, \label{AnsatzL5a}
\end{equation}
with, using an obvious analogue of Eq.\,(\ref{genGpi}),
\begin{eqnarray}
\nonumber
\lefteqn{\overline{ \Lambda}_{\beta}^{fg}(k,q;P) =
2 \ell_\beta \, [ i \Delta_{E_5}(q,k;P)+ \gamma\cdot P \Delta_{F_5}(q,k;P)] + \gamma_\beta \, \Sigma_{G_5}(q,k;P) }\\
&&   + \, 2 \ell_\beta \,  \gamma\cdot\ell\, \Delta_{G_5}(q,k;P)+[ \gamma_\beta,\gamma\cdot P]
\Sigma_{H_5}(q,k;P) + 2 \ell_\beta  [ \gamma\cdot\ell ,\gamma\cdot P]  \Delta_{H_5}(q,k;P) \,,\rule{2em}{0ex} \label{AnsatzL5b}
\end{eqnarray}
where $\ell=(q+k)/2$, $\Sigma_{\Phi}(q,k;P) = [\Phi(q;P)+\Phi(k;P)]/2$ and $\Delta_{\Phi}(q,k;P) = [\Phi(q;P)-\Phi(k;P)]/[q^2-k^2]$.

Now, given any \emph{Ansatz} for the quark-gluon vertex that satisfies Eq.\,(\ref{wtiAnsatz}), then the pseudoscalar analogue of Eq.\,(\ref{genbse}), and Eqs.\,(\ref{gendseN}), (\ref{AnsatzL5a}), (\ref{AnsatzL5b}) provide a symmetry-preserving closed system whose solution predicts the properties of pseudoscalar mesons.
The relevant scalar meson equations are readily derived.  With these systems one can anticipate, elucidate and understand the influence on hadron properties of the rich nonperturbative structure expected of the fully-dressed quark-gluon vertex in QCD: in particular, that of the dynamically generated dressed-quark mass function, whose impact is quashed at any finite order in the truncation scheme of Ref.\,\cite{Bender:1996bb}, or any kindred scheme.

To proceed one need only specify the gap equation's kernel because, as noted above, the BSEs are completely defined therefrom.  To complete the illustration \cite{Chang:2009zb} a simplified form of the effective interaction in Ref.\,\cite{Maris:1997tm} was employed and two vertex \emph{Ans\"atze} were compared; viz., the bare vertex $\Gamma_\mu^g = \gamma_\mu$, which defines the rainbow-ladder truncation of the DSEs and omits vertex dressing; and the Ball-Chiu (BC) vertex \cite{Ball:1980ay}, which nonperturbatively incorporates some of the vertex dressing associated with DCSB:
\begin{equation}
i\Gamma^g_\mu(q,k)  =
i\Sigma_{A^g}(q^2,k^2)\,\gamma_\mu+
2 \ell_\mu \biggL i\gamma\cdot \ell\,
\Delta_{A^g}(q^2,k^2) + \Delta_{B^g}(q^2,k^2)\biggR \!.
\label{bcvtx}
\end{equation}

\begin{figure}[t]
\vspace*{-3ex}

\begin{minipage}[t]{\textwidth}
\begin{minipage}[t]{0.49\textwidth}
\leftline{\includegraphics[clip,width=1.0\textwidth]{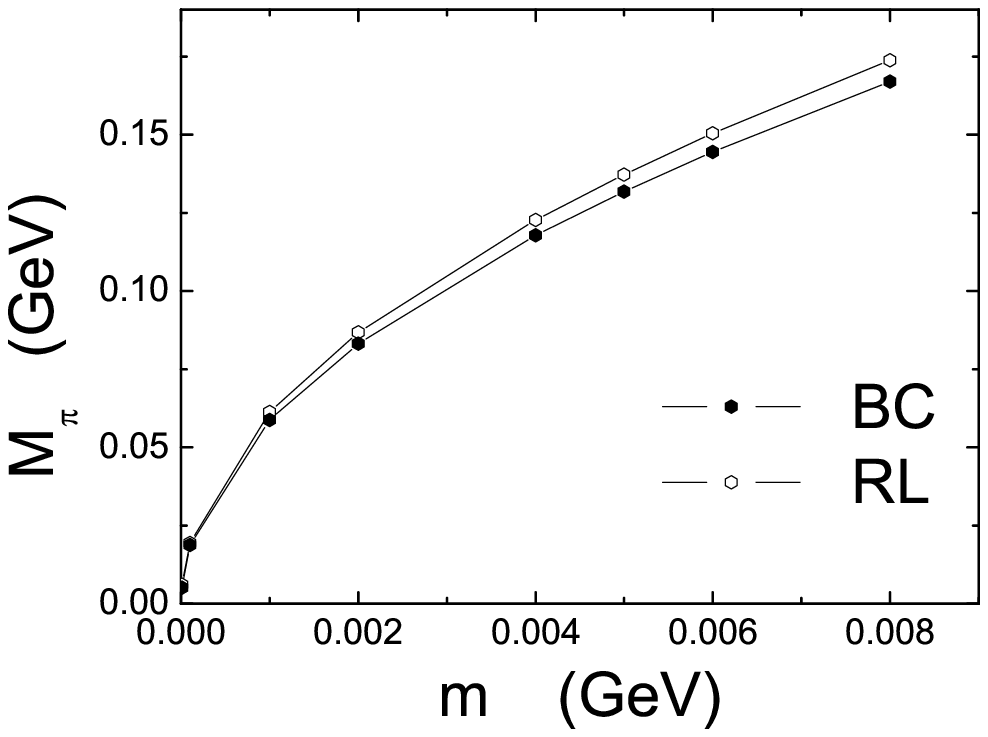}}
\end{minipage}
\begin{minipage}[t]{0.49\textwidth}
\leftline{\includegraphics[clip,width=1.0\textwidth]{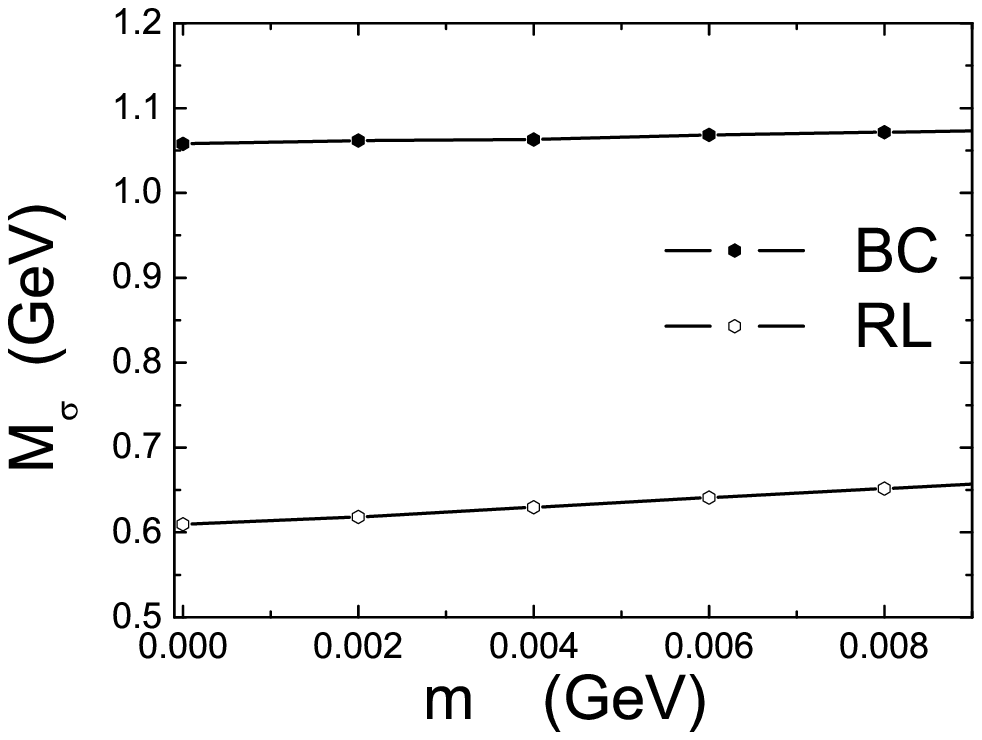}}
\end{minipage}\vspace*{3ex}
\end{minipage}
\vspace*{-4ex}



\caption{\label{massDlarge} Dependence of pseudoscalar (left panel) and scalar (right) meson masses on the current-quark mass, $m$.  The Ball-Chiu vertex (BC) result is compared with the rainbow-ladder (RL) result.  (Figure adapted from Ref.\,\protect\cite{Chang:2009zb}.)}
\end{figure}

A particular novelty of the study is that one can calculate the current-quark-mass-dependence of meson masses using a symmetry-preserving DSE truncation whose diagrammatic content is unknown. That dependence is depicted in Fig.\,\ref{massDlarge} and compared with the rainbow-ladder result.  The $m$-dependence of the pseudoscalar meson's mass provides numerical confirmation of the algebraic fact that the axial-vector Ward-Takahashi identity is preserved by both the rainbow-ladder truncation and the BC-consistent \emph{Ansatz} for the Bethe-Salpeter kernel.  The figure also shows that the axial-vector Ward-Takahashi identity and DCSB conspire to shield the pion's mass from material variation in response to dressing the quark-gluon vertex \cite{Roberts:2007jh,Bender:2002as,Bhagwat:2004hn}.

As noted in Ref.\,\cite{Chang:2009zb}, a rainbow-ladder kernel with realistic interaction strength yields
\begin{equation}
\label{epsilonRL}
\varepsilon_\sigma^{\rm RL} := \frac{2 M(0) - m_\sigma }{2 M(0)}
\rule[-2.5ex]{0.1ex}{6ex}_{\rm RL}
= (0.3 \pm 0.1)\,,
\end{equation}
which can be contrasted with the value obtained using the BC-consistent Bethe-Salpeter kernel; viz.,
\begin{equation}
\label{epsilonBC}
\varepsilon_\sigma^{\rm BC} \lesssim 0.1\,.
\end{equation}
Plainly, significant additional repulsion is present in the BC-consistent truncation of the scalar BSE.

\begin{figure}[t]

\includegraphics[clip,width=0.66\textwidth]{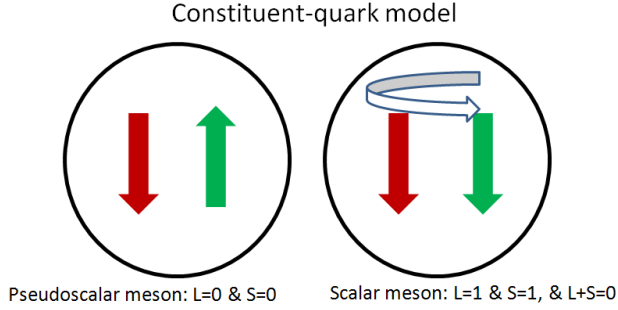}

\caption{\label{fig:ScalarCQM}
In constituent-quark-like models, pseudoscalar mesons are $^1\!S_0$ states -- constituent spins antiparallel and zero orbital angular momentum; and scalar mesons are $^3\!P_0$ states -- constituent spins parallel and one unit of orbital angular momentum.  Hence a scalar is a spin and orbital excitation of a pseudoscalar meson.}

\end{figure}

Scalar mesons are commonly identified as $^3\!P_0$ states, see Fig.\,\ref{fig:ScalarCQM}.  This assignment expresses a constituent-quark-model perspective, from which a $J^{PC}=0^{++}$ fermion-antifermion bound-state must have the constituents' spins aligned and one unit of constituent orbital angular momentum.  Hence a scalar is a spin and orbital excitation of a pseudoscalar meson.  Of course, no constituent-quark-model can be connected systematically with QCD.  Nevertheless, the presence of orbital angular momentum in a hadron's rest frame is a necessary consequence of Poincar\'e covariance and the momentum-dependent vector-boson-exchange character of QCD \cite{Roberts:2007ji,Bhagwat:2006xi,Bhagwat:2006pu}, so there is a realisation in QCD of the quark-model anticipation.

Extant studies of realistic corrections to the rainbow-ladder truncation show that they reduce hyperfine splitting \cite{Bhagwat:2004hn}.  Hence, with the comparison between Eqs.\,(\ref{epsilonRL}) and (\ref{epsilonBC}) one has a clear indication that in a Poincar\'e covariant treatment the BC-consistent truncation magnifies spin-orbit splitting, an effect which can be attributed to the influence of the quark's dynamically-enhanced scalar self-energy \cite{Roberts:2007ji} in the Bethe-Salpeter kernel.

\subsection{Quark anomalous magnetic moments}
\label{spectrum2}
It was conjectured in Ref.\,\cite{Chang:2009zb} that a full realisation of DCSB in the Bethe-Salpeter kernel will have a big impact on mesons with mass greater than 1\,GeV.  Moreover, that it can overcome a longstanding failure of theoretical hadron physics.  Namely, no extant continuum hadron spectrum calculation is believable because all symmetry preserving studies produce a splitting between vector and axial-vector mesons that is far too small: just one-quarter of the experimental value (see, e.g., Refs.\,\cite{Watson:2004kd,Maris:2006ea,Cloet:2007pi,Fischer:2009jm}).  Significant developments have followed that conjecture \cite{Chang:2010hb,Chang:2011ei} and will now be related.

As described in Sec.\,\ref{sec:Emergent}, when using Dirac's relativistic quantum mechanics, a fermion with charge $q$ and mass $m$, interacting with an electromagnetic field, has a magnetic moment $\mu= q/[2 m]$ \cite{Dirac:1928hu}.  For the electron, this prediction held true for twenty years, until improvements in experimental techniques enabled the discovery of a small deviation \cite{Foley:1948zz}, with the moment increased by a multiplicative factor: $1.00119\pm 0.00005$.  This correction was explained by the first systematic computation using renormalised quantum electrodynamics (QED) \cite{Schwinger:1948iu}:
\begin{equation}
\label{anommme}
\frac{q}{2m} \to \left(1 + \frac{\alpha}{2\pi}\right) \frac{q}{2m}\,,
\end{equation}
where $\alpha$ is QED's fine structure constant.  The agreement with experiment established quantum electrodynamics as a valid tool.  The correction defines the electron's \emph{anomalous magnetic moment}, which is now known with extraordinary precision and agrees with theory at O$(\alpha^5)$ \cite{Mohr:2008fa}.

The fermion-photon coupling in QED is described by:
\begin{equation}
\label{QEDinteraction}
\int d^4\! x\, i q \,\bar\psi(x) \gamma_\mu \psi(x)\,A_\mu(x)\,,
\end{equation}
where $\psi(x)$, $\bar\psi(x)$ describe the fermion field and $A_\mu(x)$ describes the photon.  As I explained in Sec.\,\ref{sec:Emergent}, this interaction generates the following electromagnetic current for an on-shell Dirac fermion ($k=p_f -p_i$),
\begin{equation}
\label{ecurrent2}
i q \, \bar u(p_f) \left[ \gamma_\mu F_1(k^2)+ \frac{1}{2 m} \sigma_{\mu\nu} k_\nu F_2(k^2)\right] u(p_i)\,,
\end{equation}
where: $F_1(k^2)$, $F_2(k^2)$ are form factors; and $u(p)$, $\bar u(p)$ are electron spinors.  Using their Euclidean space definition, one can derive a Gordon-identity; viz., with $2 \ell=p_f + p_i$,
\begin{equation}
\label{Gordon}
2 m \, \bar u(p_f) i \gamma_\mu u(p_i) = \bar u(p_f)\left[ 2 \ell_\mu + i \sigma_{\mu\nu} k_\nu \right]u(p_i)\,.
\end{equation}
With this rearrangement one sees that for massive fermions the interaction can be decomposed into two terms: the first describes the spin-independent part of the fermion-photon interaction, and is common to spin-zero and spin-half particles, whilst the second expresses the spin-dependent, helicity flipping part.
Moreover, one reads from Eqs.\,(\ref{ecurrent2}) and (\ref{Gordon}) that a point-particle in the absence of radiative corrections has $F_1 \equiv 1$ and $F_2 \equiv 0$, and hence Dirac's value for the magnetic moment.  The anomalous magnetic moment in Eq.\,(\ref{anommme}) corresponds to $F_2(0) = \alpha/2\pi$.

One infers from Eq.\,(\ref{Gordon}) that an anomalous contribution to the magnetic moment can be associated with an additional interaction term:
\begin{equation}
\label{anominteraction}
\int d^4\! x\, \mbox{\small $\frac{1}{2}$} q \, \bar \psi(x) \sigma_{\mu\nu} \psi(x) F_{\mu\nu}(x)\,,
\end{equation}
where $F_{\mu\nu}(x)$ is the gauge-boson field strength tensor.  This term is invariant under local $U(1)$ gauge transformations but is not generated by minimal substitution in the action for a free Dirac field.

Consider the effect of the global chiral transformation $\psi(x) \to \exp(i \theta\gamma_5) \psi(x)$.  The term in Eq.\,(\ref{QEDinteraction}) is invariant.  However, the interaction of Eq.\,(\ref{anominteraction}) is not.  These observations facilitate the understanding of a general result: $F_2\equiv 0$ for a massless fermion in a quantum field theory with chiral symmetry realized in the Wigner mode; i.e., when the symmetry is not dynamically broken.  A firmer conclusion can be drawn.  For $m=0$ it follows from Eq.\,(\ref{Gordon}) that Eq.\,(\ref{QEDinteraction}) does not mix with the helicity-flipping interaction of Eq.\,(\ref{anominteraction}) and hence a massless fermion does not possess a measurable magnetic moment.

A reconsideration of Ref.\,\cite{Schwinger:1948iu} reveals no manifest conflict with these facts.  The perturbative expression for $F_2(0)$ contains a multiplicative numerator factor of $m$ and the usual analysis of the denominator involves steps that are only valid for $m\neq 0$.  Fundamentally, there is no conundrum because QED is not an asymptotically free theory and hence, as mentioned in Sec.\,\ref{sec:QCD}, does not have a well-defined nonperturbative chiral limit.

On the other hand, in QCD the chiral limit is rigorously defined nonperturbatively \cite{Maris:1997tm}.  (It remains to be seen whether the theory thus obtained is meaningful, as indicated in the antepenultimate paragraph of Sec.\,\ref{sec:inmeson}.)  The analogue of Schwinger's one-loop calculation can then be carried out to find an anomalous \emph{chromo}-magnetic moment for the quark.  There are two diagrams in this case: one similar in form to that in QED; and another owing to the gluon self-interaction.  One reads from Ref.\,\cite{Davydychev:2000rt} that the perturbative result vanishes in the chiral limit.  However, Fig.\,\ref{gluoncloud} demonstrates that chiral symmetry is dynamically broken in QCD and one must therefore ask whether this affects the chromomagnetic moment.

Of course, it does; and it is now known that this is signalled by the appearance of $\Delta_{B^g}$ in Eq.\,(\ref{bcvtx}).  If one writes the quark-gluon vertex as
\begin{equation}
i \Gamma_\mu(p_f,p_i;k) = \lambda_1(p_f,p_i;k)\, i \gamma_\mu
+ 2 \ell_\mu \biggL i\gamma\cdot \ell\,
\lambda_2(p_f,p_i;k) + \lambda_3(p_f,p_i;k) \biggR  + \ldots \,,
\end{equation}
then contemporary simulations of lattice-regularised QCD \cite{Skullerud:2003qu} and DSE studies \cite{Bhagwat:2004kj} agree that
\begin{equation}
\label{alpha3B}
\lambda_3(p,p;0) \approx \frac{d}{dp^2} B(p^2,\zeta)
\end{equation}
and also on the form of $\lambda_1(p,p;0)$, which is functionally similar to $A(p^2,\zeta)$.  However, owing to non-orthogonality of the tensors accompanying $\lambda_1$ and $\lambda_2$, it is difficult to obtain a lattice signal for $\lambda_2$.  One must therefore consider the DSE prediction for $\lambda_2$ in Ref.\,\cite{Bhagwat:2004kj} more reliable.

As pointed out above, perturbative massless-QCD conserves helicity so the quark-gluon vertex cannot perturbatively have a term with the helicity-flipping characteristics of $\lambda_3$.  Equation~(\ref{alpha3B}) is thus remarkable, showing that the dressed-quark-gluon vertex contains at least one chirally-asymmetric component whose origin and size owe solely to DCSB; and Sec.\,\ref{sec:building} illustrates that $\lambda_3$ has a material impact on the hadron spectrum.

This reasoning is extended in Ref.\,\cite{Chang:2010hb}: massless fermions in gauge field theories cannot possess an anomalous chromo/electro-magnetic moment because the term that describes it couples left- and right-handed fermions; however, if chiral symmetry is strongly broken dynamically, then the fermions should also posses large anomalous magnetic moments.  Such an effect is expressed in the dressed-quark-gluon vertex via a term
\begin{equation}
\label{qcdanom1}
\Gamma_\mu^{\rm acm_5} (p_f,p_i;k) = \sigma_{\mu\nu} k_\nu \, \tau_5(p_f,p_i,k)\,.
\end{equation}

\begin{figure}[t]
\includegraphics[width=0.66\textwidth]{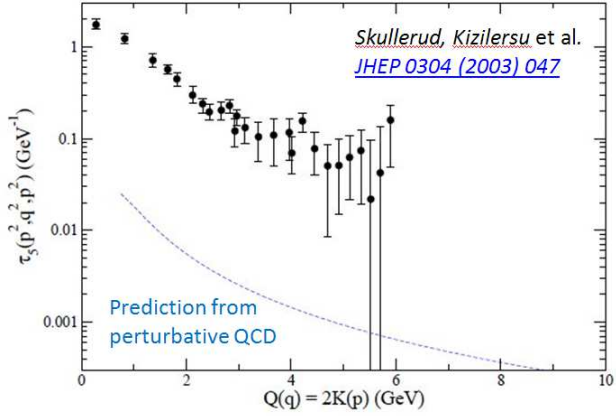}
\caption{\label{latticetau5}
Direct anomalous chromomagnetic moment contribution to the dressed-quark-gluon vertex computed in quenched-QCD with current-quark mass $m\sim 100\,$MeV \protect\cite{Skullerud:2003qu}.  The one-loop perturbative result is shown for comparison.
Plainly, the nonperturbative result is two orders-of-magnitude larger than the perturbative computation.  This level of magnification is typical of DCSB; e.g., with a current-quark mass of 4\,MeV, one obtains $M(p^2=0)\sim 400\,$MeV.
(Figure adapted from Ref.\,\protect\cite{Skullerud:2003qu}.)}
\end{figure}

That QCD generates a strongly momentum-dependent chromomagnetic form factor in the quark-gluon vertex, $\tau_5$, with a large DCSB-component, is confirmed in Ref.\,\cite{Skullerud:2003qu}.  Only a particular kinematic arrangement was readily accessible in that lattice simulation but this is enough to learn that, at the current-quark mass considered: $\tau_5$ is roughly two orders-of-magnitude larger than the perturbative form (see Fig.\,\ref{latticetau5}); and
\begin{equation}
\label{boundtau5}
\forall p^2>0: \; |\tau_5(p,-p;2 p)| \gtrsim |\lambda_3(p,p;0)|\,.
\end{equation}
The magnitude of the lattice result is consistent with instanton-liquid model estimates \cite{Kochelev:1996pv,Diakonov:2002fq}.

This large chromomagnetic moment is likely to have a broad impact on the properties of light-quark systems \cite{Diakonov:2002fq,Ebert:2005es}.  In particular, as will be illustrated in Sec.\,\ref{sec:a1rho}, it can explain the longstanding puzzle of the mass splitting between the $a_1$- and $\rho$-mesons in the hadron spectrum \cite{Chang:2011ei}.  Here a different novelty will be elucidated; viz., the manner in which the quark's chromomagnetic moment generates a quark anomalous \emph{electro}magnetic moment.  This demonstration is only possible now that the method of Ref.\,\cite{Chang:2009zb} is available.  It was accomplished \cite{Chang:2010hb} using the same simplification of the effective interaction in Ref.\,\cite{Maris:1997tm} that produced Figs.\,\ref{massDlarge}.

In order to understand the vertex \emph{Ansatz} used in Ref.\,\cite{Chang:2010hb}, it is necessary to return to perturbation theory.  As mentioned above Eq.\,\eqref{alpha3B}, one can determine from Ref.\,\cite{Davydychev:2000rt} that at leading-order in the coupling, $\alpha_s$, the three-gluon vertex does not contribute to the QCD analogue of Eq.\,(\ref{anommme}) and the Abelian-like diagram produces the finite and negative correction $(-\alpha_s/[12 \pi])$.
The complete cancellation of ultraviolet divergences occurs only because of the dynamical generation of another term in the transverse part of the quark-gluon vertex; namely,
\begin{equation}
\Gamma_\mu^{\rm acm_4}(p_f,p_i) = [ \ell_\mu^{\rm T} \gamma\cdot  k + i \gamma_\mu^{\rm T} \sigma_{\nu\rho}\ell_\nu k_\rho] \tau_4(p_f,p_i)\,,
\end{equation}
with $T_{\mu\nu} = \delta_{\mu\nu} - k_\mu k_\nu/k^2$, $a_\mu^{\rm T} := T_{\mu\nu}a_\nu$. (N.B.\ The tensor denominated $\Gamma_\mu^4$ here is labelled $T^8$ in Refs.\,\cite{Kizilersu:2009kg,Bashir:2011dp}.)

Cognisant of this, one may use a simple \emph{Ansatz} to express the dynamical generation of an anomalous chromomagnetic moment via the dressed-quark gluon vertex; viz.,
\begin{eqnarray}
\label{ourvtx}
\tilde\Gamma_\mu(p_f,p_i)  & = & \Gamma_\mu^{\rm BC}(p_f,p_i) +
\Gamma_\mu^{\rm acm}(p_f,p_i)\,,\\
\Gamma_\mu^{\rm acm}(p_f,p_i) &=& \Gamma_\mu^{\rm acm_4}(p_f,p_i) + \Gamma_\mu^{\rm acm_5}(p_f,p_i)\,,
\label{ourvtxacm}
\end{eqnarray}
with $\tau_5(p_f,p_i) =  (-7/4)\Delta_B(p_f^2,p_i^2)$ and
\begin{equation}
\tau_4(p_f,p_i) = {\cal F}(z) \bigg[  \frac{1-2\eta}{M_E}\Delta_B(p_f^2,p_i^2) - \Delta_A(p_f^2,p_i^2) \bigg]. \label{tau4}
\end{equation}
The damping factor ${\cal F}(z)=(1- \exp(-z))/z$, $z=(p_i^2 + p_f^2- 2 M_E^2)/\Lambda_{\cal F}^2$, $\Lambda_{\cal F}=1\,$GeV, simplifies numerical analysis but is otherwise irrelevant; and $M_E=\{ p| p>0, p^2 = M^2(p^2)\}$ is the Euclidean constituent-quark mass.
A confined quark does not possess a mass-shell (Sec.\,\ref{sec:Emergent}).  Hence, one cannot unambiguously assign a single value to its anomalous magnetic moment.  One can nonetheless compute a magnetic moment distribution.  At each value of $p^2$, spinors can be defined to satisfy the free-particle Euclidean Dirac equation with mass $m\to M(p^2)=:\varsigma$, so that
\begin{equation}
\bar u(p_f;\varsigma) \, \Gamma_\mu( p_f,p_i;k)\,  u(p_i;\varsigma)
= \bar u(p_f) [ F_1(k^2) \gamma_\mu + \frac{1}{2 \varsigma} \,\sigma_{\mu \nu} k_\nu F_2(k^2)] u(p_i)
\label{GenSpinors}
\end{equation}
and then, from Eqs.\,(\ref{ourvtx}) -- (\ref{tau4}),
\begin{equation}
\label{kappaacm}
\kappa^{\rm acm}(\varsigma) = \frac{ - 2 \varsigma \, \eta \delta_B^{\varsigma}}
    {\sigma_A^{\varsigma} - 2 \varsigma^2 \delta_A^{\varsigma}+ 2 \varsigma \delta_B^{\varsigma} }\,,
\end{equation}
where $\sigma_A^{\varsigma} = \Sigma_A(\varsigma,\varsigma)$, $\delta_A^{\varsigma} = \Delta_A(\varsigma,\varsigma)$, etc.  The numerator's simplicity owes to a premeditated cancellation between $\tau_4$ and $\tau_5$, which replicates the one at leading-order in perturbation theory.
Where a comparison of terms is possible, this vertex \emph{Ansatz} is semi-quantitatively in agreement with Refs.\,\cite{Skullerud:2003qu,Bhagwat:2004kj}.  However, the presence and understanding of the role of $\Gamma_\mu^{\rm acm_4}$ is a novel contribution by Ref.\,\cite{Chang:2010hb}.  N.B.\ It is apparent from Eq.\,(\ref{kappaacm}) that $\kappa^{\rm acm} \propto m^2$ in the absence of DCSB, so that $\kappa^{\rm acm}/[2m]\to 0$ in the chiral limit.

The BSE for the quark-photon vertex can be written following the method of Ref.\,\cite{Chang:2009zb}.  Since the method guarantees preservation of the Ward-Takahashi identities, the general form of the solution is
\begin{eqnarray}
\Gamma_\mu^\gamma(p_f,p_i) & = & \Gamma_\mu^{\rm BC}(p_f,p_i) + \Gamma_\mu^{\rm T}(p_f,p_i)\,,\\
\nonumber
\Gamma_\mu^{\rm T}(p_f,p_i) & = &
\gamma_\mu^{\rm T} \hat F_1
+ \sigma_{\mu\nu} k_\nu \hat F_2
+ T_{\mu\rho} \sigma_{\rho\nu} \ell_\nu \,\ell\cdot k\, \hat F_3
+ [ \ell_\mu^{\rm T} \gamma\cdot  k + i \gamma_\mu^{\rm T} \sigma_{\nu\rho}\ell_\nu k_\rho] \hat F_4
\\
 & & \,- i \ell_\mu^{\rm T} \hat F_5+ \ell_\mu^{\rm T} \gamma\cdot k \, \ell \cdot  k\, \hat F_6 - \ell_\mu^{\rm T} \gamma\cdot \ell \, \hat F_7
%
+ \ell_\mu^T \sigma_{\nu\rho} \ell_\nu k_\rho \hat F_8\,,
\end{eqnarray}
where $\{\hat F_i|i=1,\ldots,8\}$ are scalar functions of Lorentz-invariants constructed from $p_f$, $p_i$, $k$.  The Ward-Takahashi identity is plainly satisfied; viz., 
\begin{equation}
k_\mu i \Gamma_\mu(p_f,p_i) = k_\mu i \Gamma_\mu^{\rm BC} (p_f,p_i)
= S^{-1}(p_f) - S^{-1}(p_i)\,.
\end{equation}

\begin{figure}[t]
\vspace*{-3ex}

\begin{minipage}[t]{\textwidth}
\begin{minipage}[t]{0.5\textwidth}
\leftline{\includegraphics[clip,width=1.0\textwidth]{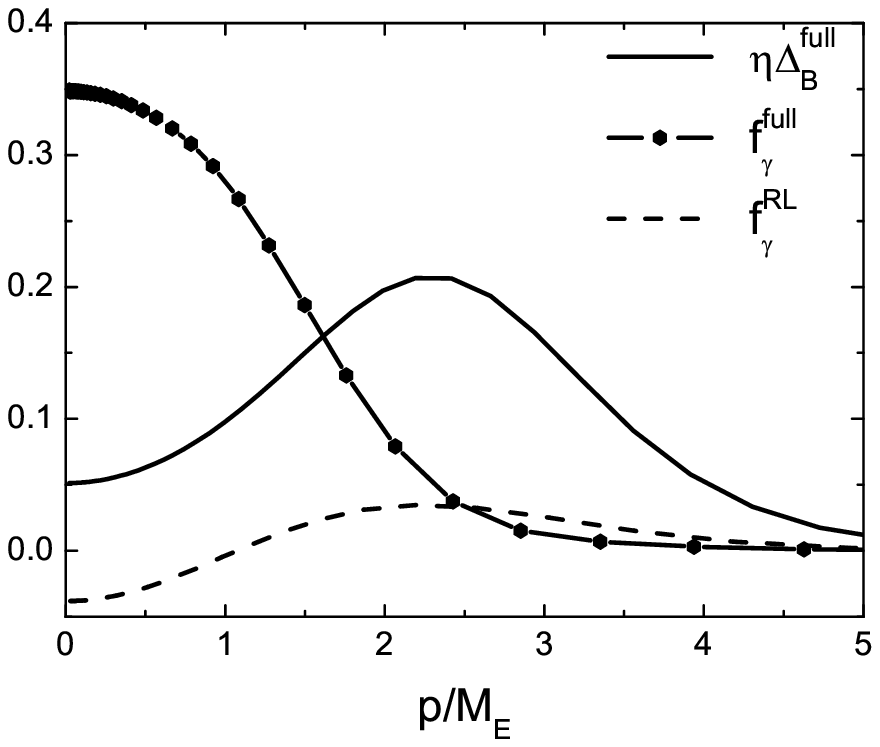}}
\end{minipage}
\begin{minipage}[t]{0.5\textwidth}
\leftline{\includegraphics[clip,width=1.05\textwidth]{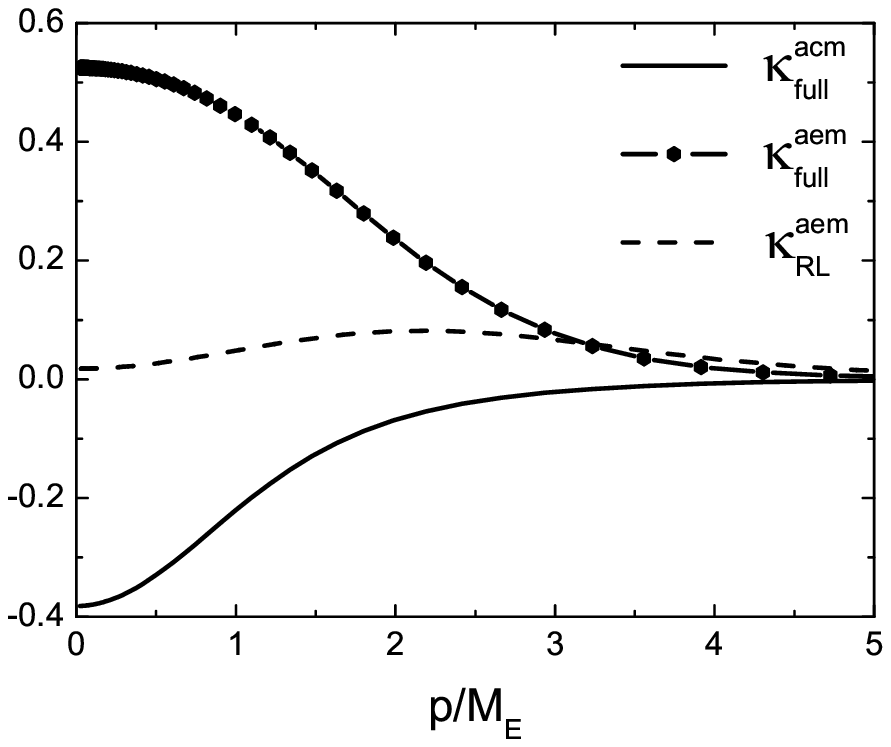}}
\end{minipage}
\end{minipage}



\caption{\label{figACM}
\emph{Left panel} -- $f_\gamma$ (GeV$^{-1}$) in Eq.\,(\protect\ref{fgamma})
cf.\ $(-7/4)\Delta_B(p^2,p^2)$, both computed using Eq.\,(\protect\ref{ourvtx}) and the same simplification of the interaction in Ref.\,\protect\cite{Maris:1997tm}.
\emph{Right panel} -- Anomalous chromo- and electro-magnetic moment distributions for a dressed-quark, computed using Eq.\,(\protect\ref{kappavalue}).
The dashed-curve in both panels is the rainbow-ladder (RL) truncation result.}
\end{figure}

Figure~\ref{figACM} depicts the results obtained for the quark's anomalous electromagnetic moment form factor
\begin{equation}
f_\gamma(p) :=  \lim_{p_f\to p}\frac{-1}{12\,k^2}{\rm tr}\, \sigma_{\mu\nu} k_\mu \Gamma_\nu^\gamma(p_f,p)
=  \hat F_2+ \frac{1}{3}p^2 \hat F_8\,. \label{fgamma}
\end{equation}
The result is evidently sizable.
It is worth reiterating that $f_\gamma$ is completely nonperturbative: in the chiral limit, at any finite order in perturbation theory, $f_\gamma\equiv 0$.  For contrast the figure also displays the result computed in the rainbow-ladder truncation of QCD's DSEs. As the leading-order in a systematic but stepwise symmetry-preserving scheme \cite{Bender:1996bb}, this truncation only partially expresses DCSB: it is exhibited by the dressed-quark propagator but not present in the quark-gluon vertex.  In this case $f_\gamma$ is nonzero but small.  These are artefacts of the truncation that cannot be remedied at any finite order of the procedure in Ref.\,\cite{Bender:1996bb} or a kindred scheme.

Employing Eq.\,(\ref{GenSpinors}), in connection with the dressed-quark-photon vertex,
one can write an expression for the quark's anomalous electromagnetic moment distribution
\begin{equation}
\label{kappavalue}
\kappa(\varsigma)=\frac{2 \varsigma \hat F_{2} + 2 \varsigma^2 \hat F_4  +\Lambda_{\kappa}(\varsigma)}
{\sigma_{A}^{\varsigma} + \hat F_{1}-\Lambda_{\kappa}(\varsigma)}\,,
\end{equation}
where: $\Lambda_{\kappa}(\varsigma)= 2\varsigma^{2}\delta_{A}^\varsigma-2 \varsigma \delta_{B}^\varsigma -\varsigma \hat F_5 - \varsigma^2 \hat F_7$; and the $\hat F_i$ are evaluated at $p_f^2=p_i^2=M(p_f^2)^2=:\varsigma^2$, $k^2=0$.  Plainly, $\kappa(\varsigma)\equiv 0$ in the chiral limit when chiral symmetry is not dynamically broken.  Moreover, as a consequence of asymptotic freedom, $\kappa(\varsigma) \to 0$ rapidly with increasing momentum.
The computed distribution is depicted in Fig.\,\ref{figACM}.  It yields Euclidean mass-shell values:
\begin{equation}
\begin{array}{llll}
& M_{\rm full}^E = 0.44\,{\rm GeV},& \kappa_{\rm full}^{\rm acm}= -0.22\,, &
\kappa_{\rm full}^{\rm aem}= 0.45\,\\[1ex]
{\rm cf}. & M_{\rm RL}^E = 0.35\,{\rm GeV}, & \kappa_{\rm RL}^{\rm acm}= 0\,, & \kappa_{\rm RL}^{\rm aem}= 0.048 .
\end{array}
\end{equation}

It is thus apparent that DCSB produces a dressed light-quark with a momentum-dependent anomalous chromomagnetic moment, which is large at infrared momenta.  Significant amongst the consequences is the generation of an anomalous electromagnetic moment for the dressed light-quark with commensurate size but opposite sign.  (N.B.\ This result was anticipated in Ref.\,\protect\cite{Bicudo:1998qb}, which argued that DCSB usually triggers the generation of a measurable anomalous magnetic moment for light-quarks.)
The infrared dimension of both moments is determined by the Euclidean constituent-quark mass.  This is two orders-of-magnitude greater than the physical light-quark current-mass, which sets the scale of the perturbative result for both these quantities.

There are two more notable features; namely, the rainbow-ladder truncation, and low-order stepwise improvements thereof, underestimate these effects by an order of magnitude; and both the $\tau_4$ and $\tau_5$ terms in the dressed-quark-gluon vertex are indispensable for a realistic description of hadron phenomena.  Whilst a simple interaction was used to illustrate these outcomes, they are robust.

These results are stimulating a reanalysis of hadron elastic and transition electromagnetic form factors \cite{Wilson:2011rj,Chang:2011tx}, and the hadron spectrum, results of which will be described below.
Furthermore, given the magnitude of the muon ``$g_\mu-2$ anomaly'' and its assumed importance as an harbinger of physics beyond the Standard Model \cite{Jegerlehner:2009ry}, it might also be worthwhile to make a quantitative estimate of the contribution to $g_\mu-2$ from the quark's DCSB-induced anomalous moments following, e.g., the computational pattern for the hadronic light-by-light scattering component of the photon polarization tensor indicated in Ref.\,\cite{Goecke:2011pe}.

\subsection{\mbox{\boldmath $a_1$}-\mbox{\boldmath $\rho$} mass splitting}
\label{sec:a1rho}
The analysis in Ref.\,\cite{Chang:2009zb} enables one to construct a symmetry-preserving kernel for the BSE given any form for $\Gamma_\mu$.  Owing to the importance of symmetries in forming the spectrum of a quantum field theory, this is a pivotal advance.  One may now use all information available, from any reliable source, to construct the best possible vertex \emph{Ansatz}.  The last section illustrated that this enables one to incorporate crucial nonperturbative effects, which any finite sum of contributions is incapable of capturing, and thereby prove that DCSB generates material, momentum-dependent anomalous chromo- and electro-magnetic moments for dressed light-quarks.

The vertex described in Sec.\,\ref{spectrum2} contains a great deal of information about DCSB.  It is the best motivated \emph{Ansatz} to date, has stimulated a detailed reanalysis of the quark-photon coupling \cite{Bashir:2011dp}, and may be used in the calculation of the masses of ground-state spin-zero and -one light-quark mesons in order to illuminate the impact of DCSB on the hadron spectrum.  This analysis expands significantly on the discussion of scalar and pseudoscalar mesons in Sec.\,\ref{sec:building}.

A prediction for the spectrum therefore follows once the gap equation's kernel is specified and the Ward-Identity solved for $\Lambda_{5\mu\beta}^{fg}$.  In the pseudoscalar and axial-vector channels the Ward-Takahashi identity for the Bethe-Salpeter kernel is solved by
\begin{eqnarray}
\nonumber
2 \Lambda_{5\beta(\mu)} &= & [\tilde \Gamma_{\beta}(q_{+},k_{+})+\gamma_{5}\tilde \Gamma_{\beta}(q_{-},k_{-})
\gamma_{5}] \frac{1}{S^{-1}(k_{+})+S^{-1}(-k_{-})}\Gamma_{5(\mu)}(k;P)\nonumber\\
&+&\Gamma_{5(\mu)}(q;P)\frac{1}{S^{-1}(-q_{+})+S^{-1}(q_{-})}
[\gamma_{5}\tilde\Gamma_{\beta}(q_{+},k_{+})\gamma_{5}
+\tilde\Gamma_{\beta}(q_{-},k_{-})], \rule{2em}{0ex}
\end{eqnarray}
where $\tilde \Gamma$ is the chosen \emph{Ansatz} for the quark-gluon vertex.  Kernels for other channels are readily constructed.

\begin{figure}[t]

\includegraphics[clip,width=0.67\textwidth]{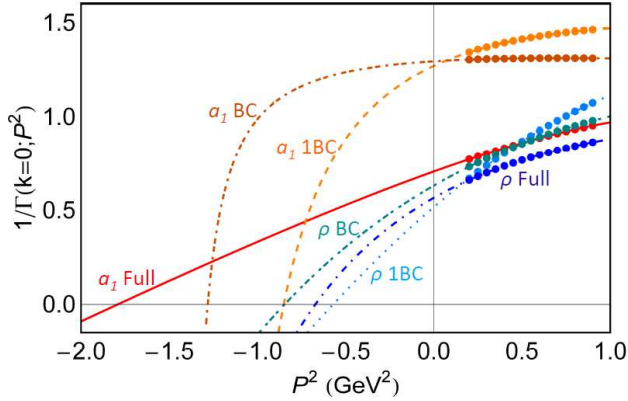}

\caption{\label{F1a1}
Illustration of the procedure used to determine meson masses,
which is fully described in Ref.\,\cite{Bhagwat:2007rj} and analogous to the method used in lattice-QCD.
\emph{Solid curve} -- $a_1$-meson, nonperturbative kernel; \emph{dot-dash-dash} -- $a_1$, kernel derived from Eq.\,(\protect\ref{bcvtx}) only (Ball-Chiu, BC); and \emph{dash} -- $a_1$, kernel derived from just the first term in Eq.\,(\ref{bcvtx}) (1BC, a minimal renormalization improvement \protect\cite{Bloch:2002eq} of the leading-order -- RL, rainbow-ladder -- kernel \protect\cite{Bender:1996bb}).
\emph{Dot-dash curve} -- $\rho$-meson, nonperturbative kernel; \emph{Dot-dash-dot} -- $\rho$, BC-kernel; and \emph{dotted} -- $\rho$, 1BC-kernel.
\emph{Points} -- values of $1/\Gamma(k=0;P^2)$ in the given channel computed with the kernel described.  Pad\'e approximants are constructed in each case; and the location of the zero is identified with $(-m_{\rm meson}^2)$.}
\end{figure}

Reference~\protect\cite{Chang:2011ei} computes ground-state masses using the interaction described in Ref.\,\cite{Qin:2011dd}, which produced Fig.\,\ref{fig:gluonrunning}; the vertex model explicated in Sec.\,\ref{spectrum2}; and the method for solving the inhomogeneous Bethe-Salpeter equation that is detailed in Ref.\,\cite{Bhagwat:2007rj}, which ensures one need only solve the gap and Bethe-Salpeter equations at spacelike momenta.  This simplifies the numerical problem.  To explain, the inhomogeneous BSE is solved for the complete Bethe-Salpeter amplitude in a particular channel on a domain of spacelike total-momenta, $P^2>0$.  Any bound-state in that channel appears as a pole in the solution at $P^2=-m_{\rm meson}^2$.  Denoting the leading Chebyshev moment of the amplitude's dominant Dirac structure by $\Gamma(k;P)$, where $k$ is the relative momentum, then $1/\Gamma(k=0;P^2)$ exhibits a zero at $(-m_{\rm meson}^2)$.  The location of that zero is determined via extrapolation of a Pad\'e approximant to the spacelike-behavior of $1/\Gamma(k=0;P^2)$.  This is illustrated for the $\rho$- and $a_1$-channels in Fig.\,\ref{F1a1}.

\begin{table}[t]
\begin{center}
\caption{\label{tablemasses}
%
Col.~1: Spectrum obtained with the full nonperturbative Bethe-Salpeter kernels described herein, which express effects of DCSB.
The method of Ref.\,\protect\cite{Bhagwat:2007rj} was used:  the error reveals the sensitivity to varying the order of Pad\'e approximant.
Col.~2 -- 
Experimental values; computed, except $m_\sigma$, from isospin mass-squared averages \protect\cite{Nakamura:2010zzi}.
%
Col.~3 -- Masses determined from the inhomogeneous BSE at leading-order in the DSE truncation scheme of Ref.\,\protect\cite{Bender:1996bb} using the interaction in Ref.\,\cite{Chang:2009zb} (with this simple kernel, the Pad\'e error is negligible);
and Col.~4 -- results in Ref.\,\protect\cite{Alkofer:2002bp}, obtained directly from the homogeneous BSE at the same order of truncation.
%
}
\begin{tabular*}
{\hsize}
{|l@{\extracolsep{0ptplus1fil}}
|l@{\extracolsep{0ptplus1fil}}
|l@{\extracolsep{0ptplus1fil}}
|l@{\extracolsep{0ptplus1fil}}
|l@{\extracolsep{0ptplus1fil}}|}\hline
\rule{0em}{3ex}
    & Ref.\,\protect\cite{Chang:2011ei} & Expt.~ & \emph{RL-Pad\'e}~ & \emph{RL-direct}~ \\\hline
$m_\pi$   & $0.138 $~ & 0.138 & 0.138~ & 0.137~ \\
$m_\rho$  & $0.817 \pm 0.016$~ & 0.777 & 0.754~ & 0.758~ \\
$m_\sigma$& $0.90 \pm 0.05$  & $0.4$ -- $1.2$~ & 0.645~ & 0.645~ \\
$m_{a_1}$ & $1.30 \pm 0.11$~ & $1.24 \pm 0.04$~ & 0.938~ & 0.927~  \\
$m_{b_1}$ & $1.15 \pm 0.07$~ & $1.21 \pm 0.02$~ & $0.904 $~ & 0.912~ \\
$m_{a_1}-m_\rho$~ & $0.48 \pm 0.12$ & $0.46 \pm 0.04$ & 0.18 & 0.17 \\
$m_{b_1}-m_\rho$~ & $0.33 \pm 0.09$ & $0.43 \pm 0.02$ & 0.15 & 0.15 \\\hline
\end{tabular*}
\end{center}
\end{table}

A full set of results is listed in Table~\ref{tablemasses}, wherein
the level of agreement between Cols.~3 and 4 illustrates the efficacy of the method used to compute masses: no difference is greater than 1\%.
Next consider $m_\sigma$ and compare Cols.~1--3.  It is an algebraic result that in the RL-truncation of QCD's DSEs, $m_\sigma \approx 2 M$, where $M$ is a constituent-like quark mass \cite{Roberts:2011cf}.  On the other hand, incorporating the quark mass function into the Bethe-Salpeter kernel via $\Gamma_\mu^{\rm BC}$ generates a strong spin-orbit interaction, which significantly boosts $m_\sigma$ \cite{Chang:2009zb}.  This feature is evidently unaffected by the inclusion of $\Gamma_\mu^{\rm acm}$; i.e., those terms associated with a dressed-quark anomalous chromomagnetic moment.
Since terms associated with pion final-state interactions were deliberately omitted from the nonperturbative kernel derived in Ref.\,\cite{Chang:2011ei}, it is noteworthy that $m_\sigma$ in Col.~1 matches estimates for the mass of the dressed-quark-core component of the $\sigma$-meson obtained using unitarised chiral perturbation theory \cite{Pelaez:2006nj,RuizdeElvira:2010cs}.  N.B.\ In addition to providing a width, such final-state interactions necessarily reduce the real part of the mass \cite{Holl:2005st,Cloet:2008fw,Chang:2009ae}.

Now compare the entries in Rows~2, 4--6.  The $\rho$- and $a_1$-mesons have been known for more than thirty years and are typically judged to be parity-partners; i.e., they would be degenerate if chiral symmetry were manifest in QCD.  Plainly, they are not, being split by roughly $450\,$MeV (i.e., $> m_\rho/2$).  It is suspected that this large splitting owes to DCSB.  Hitherto, however, no symmetry-preserving bound-state treatment could explain it.  This is illustrated by Cols.~3, 4, which show that whilst a good estimate of $m_\rho$ is readily obtained at leading-order in the systematic DSE truncation scheme of Ref.\,\cite{Bender:1996bb}, the axial-vector masses are much underestimated.  The flaw persists at next-to-leading-order \cite{Watson:2004kd,Fischer:2009jm}.

The analysis in Ref.\,\cite{Chang:2011ei} points to a remedy for this longstanding failure.  Using the Poincar\'e-covariant, symmetry preserving formulation of the meson bound-state problem enabled by Ref.\,\cite{Chang:2009zb}, with nonperturbative kernels for the gap and Bethe-Salpeter equations, which incorporate effects of DCSB that are impossible to capture in any step-by-step procedure for improving upon the rainbow-ladder truncation, it provides realistic estimates of axial-vector meson masses.
In obtaining these results, Ref.\,\cite{Chang:2011ei} showed that the vertex \emph{Ansatz} used most widely in studies of DCSB, $\Gamma_\mu^{BC}$, is inadequate as a tool in hadron physics.  Used alone, it increases both $m_\rho$ and $m_{a_1}$ but yields $m_{a_1}-m_\rho=0.21\,$GeV, qualitatively unchanged from the rainbow-ladder-like result (see Fig.\,\ref{F1a1}).
A good description of axial-vector mesons is only achieved by including interactions derived from $\Gamma_\mu^{\rm acm}$; i.e., connected with the dressed-quark anomalous chromomagnetic moment \cite{Chang:2010hb}.  Moreover, used alone, neither term in $\Gamma_\mu^{\rm acm}$, Eq.\,\eqref{ourvtxacm}, can produce a satisfactory result.  The full vertex \emph{Ansatz} and the associated gap and Bethe-Salpeter kernels described in Sec.\,\ref{spectrum2} are the minimum required.

Row~5 contains additional information.  The leading-covariant in the $b_1$-meson channel is $\gamma_5 k_\mu$.  The appearance of $k_\mu$ suggests that dressed-quark orbital angular momentum will play a significant role in this meson's structure, even more so than in the $a_1$-channel for which the dominant covariant is $\gamma_5\gamma_\mu$.
(N.B.\ In a simple quark-model, constituent spins are parallel within the $a_1$ but antiparallel within the $b_1$.  Constituents of the $b_1$ may therefore become closer, so that spin-orbit repulsion can exert a greater influence.)
This expectation is borne out by the following: with the full kernel, $m_{b_1}$ is far more sensitive to the interaction's momentum-space range parameter than any other state, decreasing rapidly as the interaction's spatial-variation is increasingly suppressed.
%

The results reviewed in this subsection rest on an \emph{Ansatz} for the quark-gluon vertex and whilst the best available information was used in its construction, improvement is nonetheless possible.  That will involve elucidating the role of Dirac covariants in the quark-gluon vertex which have not yet been considered, as in Ref.\,\cite{Bashir:2011dp}, and of resonant contributions; viz., meson loop effects that give widths to some of the states considered.  In cases for which empirical width-to-mass ratios are $\lesssim 25$\%, one might judge that such contributions can reliably be obtained via bound-state perturbation theory \cite{Pichowsky:1999mu}.  Contemporary studies indicate that these effects reduce bound-state masses but the reduction can uniformly be compensated by a modest inflation of the interaction's mass-scale \cite{Roberts:2011cf,Eichmann:2008ae}, so that the masses in Table~\ref{tablemasses} are semiquantitatively unchanged.  The case of the $\sigma$-meson is more complicated.  However, the prediction of a large mass for this meson's dressed-quark core leaves sufficient room for a strong reduction by resonant contributions \cite{Pelaez:2006nj,RuizdeElvira:2010cs}.

This section reviewed a continuum framework for computing and explaining the meson spectrum, which combines a veracious description of pion properties with estimates for masses of light-quark mesons heavier than $m_\rho$.
(A contemporary lattice-QCD perspective on this problem may be drawn from Refs.\,\protect\cite{Dudek:2011bn,Engel:2011aa}.)
The method therefore offers the promise of a first reliable Poincar\'e-invariant, symmetry-preserving computation of the spectrum of light-quark hybrids and exotics; i.e., those putative states which are impossible to construct in a quantum mechanics based upon constituent-quark degrees-of-freedom.  So long as the promise is promptly fulfilled, the approach will provide predictions to guide the forthcoming generation of facilities.

\section{Probing the Hadronic Interior}
\label{sec:LookingDeeper}
\subsection{Elastic form factors}
Form factors have long been recognised as a basic tool for elucidating bound-state properties and are of particular value in hadron physics because they provide information on hadron structure as a function of $Q^2$, the squared momentum-transfer:
small-$Q^2$ is the nonperturbative domain; and large-$Q^2$ is the perturbative domain. Nonperturbative methods in hadron physics must explain their behaviour from $Q^2=0$ into and through the transition domain, whereupon the behaviour is currently being measured.  Experimental and theoretical studies of hadron electromagnetic form factors have made rapid and significant progress during the last several years, including new data in the timelike region, and material gains have been made in studying the pion form factor.

\begin{figure}[t]
\includegraphics[width=0.70\textwidth]{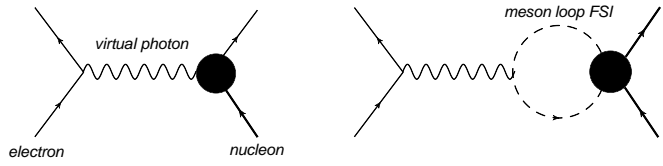}
\caption{\label{fig:PionFSI}
Apart from the direct contribution (left image), electron scattering from a nucleon may also proceed via meson-nucleon final state interactions (right image), which have a measurable effect.
}
\end{figure}

Despite this, many urgent questions remain unanswered.  For example:
how can one use experiment to chart the long-range behaviour of the $\beta$-function in QCD;
given the low mass of the pion and its strong coupling to protons and neutrons, how can one discern features produced by final-state interactions and disentangle them from the intrinsic properties of the target, Fig.\,\ref{fig:PionFSI};
and at which momentum-transfer does the transition from nonperturbative-QCD to perturbative-QCD take place?
Overviews of the recent status of experiment and theory are provided in Refs.\,\cite{Arrington:2006zm,Perdrisat:2006hj}.  However, with the intense interest and investment in this field, the status changes rapidly.

\subsubsection{Pion Form Factor}
\begin{figure}[t]
\includegraphics[width=0.80\textwidth]{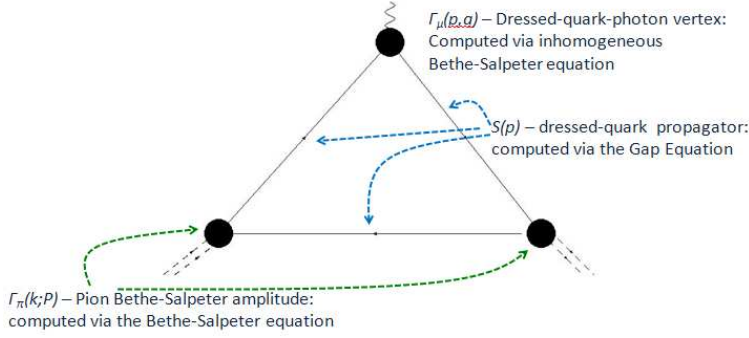}
\caption{\label{pionFFGIA}
Generalised impulse approximation to the electromagnetic charged-pion form factor \protect\cite{Roberts:1994hh}.
As depicted, the relevant Ward-Takahashi identities are satisfied so long as each element in the computation is obtained from the appropriate DSE at leading-order in the truncation scheme of Ref.\,\protect\cite{Bender:1996bb}.
}
\end{figure}

As an illustration, I will consider the electromagnetic charged-pion form factor. Measurement of this form factor is not simple because the pion is an unstable meson with a short lifetime -- roughly 30ns -- and hence one cannot construct a stable pion target.  On the other hand, since the pion is often described as a two body system, many theorists have pretended that a reliable computation of the form factor is easy.  However, owing to the intimate connection between the pion and DCSB, that is not the case and instead one must employ a tool that is capable of veraciously expressing that relationship.  The DSEs fulfill that requirement and a quantitative prediction was obtained \cite{Maris:2000sk} by combining the: dressed-rainbow gap equation, Eq.\,\eqref{gendseN} with $\Gamma_\nu^f\to \gamma_\nu$; dressed-ladder Bethe-Salpeter equation, Eq.\,\eqref{bsetextbook} with the kernel in the top row of Fig.\,\ref{FSysTruncation}; and dressed-impulse approximation to the form factor.  For an incoming pion with momentum $p_1=K-Q/2$ and an outgoing pion with momentum $p_2=K+Q/2$, the latter is provided by
\begin{eqnarray}
\nonumber
iK_\mu F_\pi^{\rm em}(Q^2) & = & N_c \int\frac{d^4t}{(2\pi)^4}{\rm tr_D} \Bigg[ \Gamma_{\pi}(t;-p_2)\\
&& \times S(t+p_2) \Gamma_{\mu}(t+p_2,t+p_1)  S(t+p_1) \; \Gamma_{\pi}(t;p_1) \; S(t) \Bigg],
\label{KF}
\end{eqnarray}
where the trace is over spinor indices alone.  This expression is represented by the diagram in Fig.\,\ref{pionFFGIA}.

\begin{figure}[t] 
\includegraphics[clip,height=0.45\textheight,angle=-90]{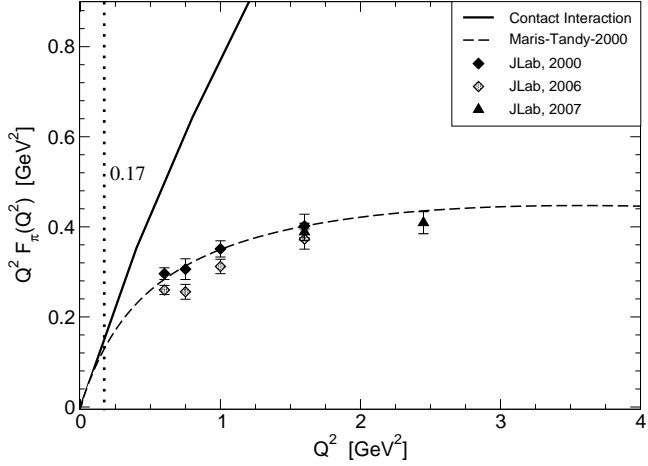}
\caption{\label{FpiUV} $Q^2 F^{\rm em}_{\pi}(Q^2)$ --
\underline{Dashed curve}: DSE prediction \protect\cite{Maris:2000sk}, which employed a momentum-dependent renormalisation-group-improved gluon exchange interaction and produces $r_\pi=0.67\,$fm cf.\ experiment \cite{Nakamura:2010zzi} $0.672 \pm 0.008\,$fm.  Owing to the interaction's momentum-dependence, this computation involves a dressed-quark mass-function which evolves in the chiral limit as $\sim 1/p^2$ for $p^2 \gg \Lambda_{\rm QCD}^2$.
\underline{Solid curve}: $Q^2 F^{\rm em}_{\pi}(Q^2)$ obtained with a symmetry-preserving DSE treatment of a momentum-independent one-gluon exchange interaction, Sec.\,\protect\ref{sec:njldcsb} and Refs.\,\protect\cite{GutierrezGuerrero:2010md,Roberts:2011wy}.  In this computation, the computed dressed-quark mass is momentum independent; viz., $M(p^2) \equiv M$.
For $Q^2>0.17\,$GeV$^2\approx M^2$, marked by the vertical \emph{dotted line}, the contact interaction result for $F^{\rm em}_{\pi}(Q^2)$ differs from that in  Ref.\,\protect\cite{Maris:2000sk} by more than 20\%.
The data are from Refs.\,\protect\cite{Volmer:2000ek,Horn:2006tm,Tadevosyan:2007yd}.
}
\end{figure}

The impulse approximation to the form factor is completely determined once the interaction in the gap equation is specified; viz., a choice made for $Z_1 g^2 D_{\mu\nu}(k) \Gamma_\nu(q,p)$ in Eq.\,\eqref{gendseN}.  That employed in Ref.\,\cite{Maris:2000sk} grew from the ground laid by a series of studies \cite{Munczek:1983dx,Maris:1999nt,Maris:1997tm,Burden:1991gd,Jain:1993qh,Frank:1995uk,Burden:1996vt}.  It combines a one-parameter model for the infrared behaviour of the interaction with renormalisation-group-improved one-gluon exchange at ultraviolet momenta so that the results of perturbative QCD are recovered.  In connection with ground-state vector and flavoured pseudoscalar mesons, this interaction correlates more than forty observables with a root-mean-square relative error of 15\% \cite{Maris:2003vk} and has hitherto provided the most efficacious tool for JLab physics.  It is corrected and updated in Ref.\,\cite{Qin:2011dd}. The pion for factor obtained as a parameter-free prediction in Ref.\,\cite{Maris:2000sk}  was confirmed by subsequent experiments at JLab, as illustrated by the dashed curve in Fig.\,\ref{FpiUV}.

Another of the reasons I focus on $F^{\rm em}_{\pi}(Q^2)$ is the existence of a prediction that $Q^2 F_{\pi}(Q^2)\approx\,$constant for $Q^2\gg m_\pi^2$ in a theory whose interaction is mediated by massless vector-bosons.  To be explicit \cite{Farrar:1979aw,Efremov:1979qk,Lepage:1980fj}:
\begin{equation}
\label{FpiUVpQCD}
Q^2 F_{\pi}(Q^2) \stackrel{Q^2\gg m_\pi^2}{\simeq} 16 \pi f_\pi^2 \alpha(Q^2),
\end{equation}
which takes the value $0.13\,$GeV$^2$ at $Q^2=10\,$GeV$^2$ if one uses the one-loop result for the strong running-coupling $\alpha(Q^2=10\,{\rm GeV}^2)\approx 0.3$.  The verification of this prediction is a strong motivation for modern experiment \cite{Volmer:2000ek,Horn:2006tm,Tadevosyan:2007yd}, which can also be viewed as an attempt to constrain and map experimentally the pointwise behaviour of the exchange interaction that binds the pion.

Section~\ref{psmassformula} details some extraordinary consequences of DCSB, amongst them the Goldberger-Treiman relations of Eqs.\,(\ref{gtlrelE}) -- (\ref{gtlrelH}).  Of these, Eqs.\,(\ref{gtlrelF}) and (\ref{gtlrelG}) entail that the pion possesses components of pseudovector origin which alter the asymptotic form of $F_{\pi}^{\rm em}(Q^2)$ by a multiplicative factor of $Q^2$ cf.\ the result obtained in their absence \cite{Maris:1998hc}.  Here, in connection with Eq.\,\eqref{FpiUVpQCD}, Fig.\,\ref{gluoncloud} and Eqs.\,(\ref{gtlrelE}) -- (\ref{gtlrelH}) are critical.

\begin{figure}[t] 
\includegraphics[clip,height=0.30\textheight]{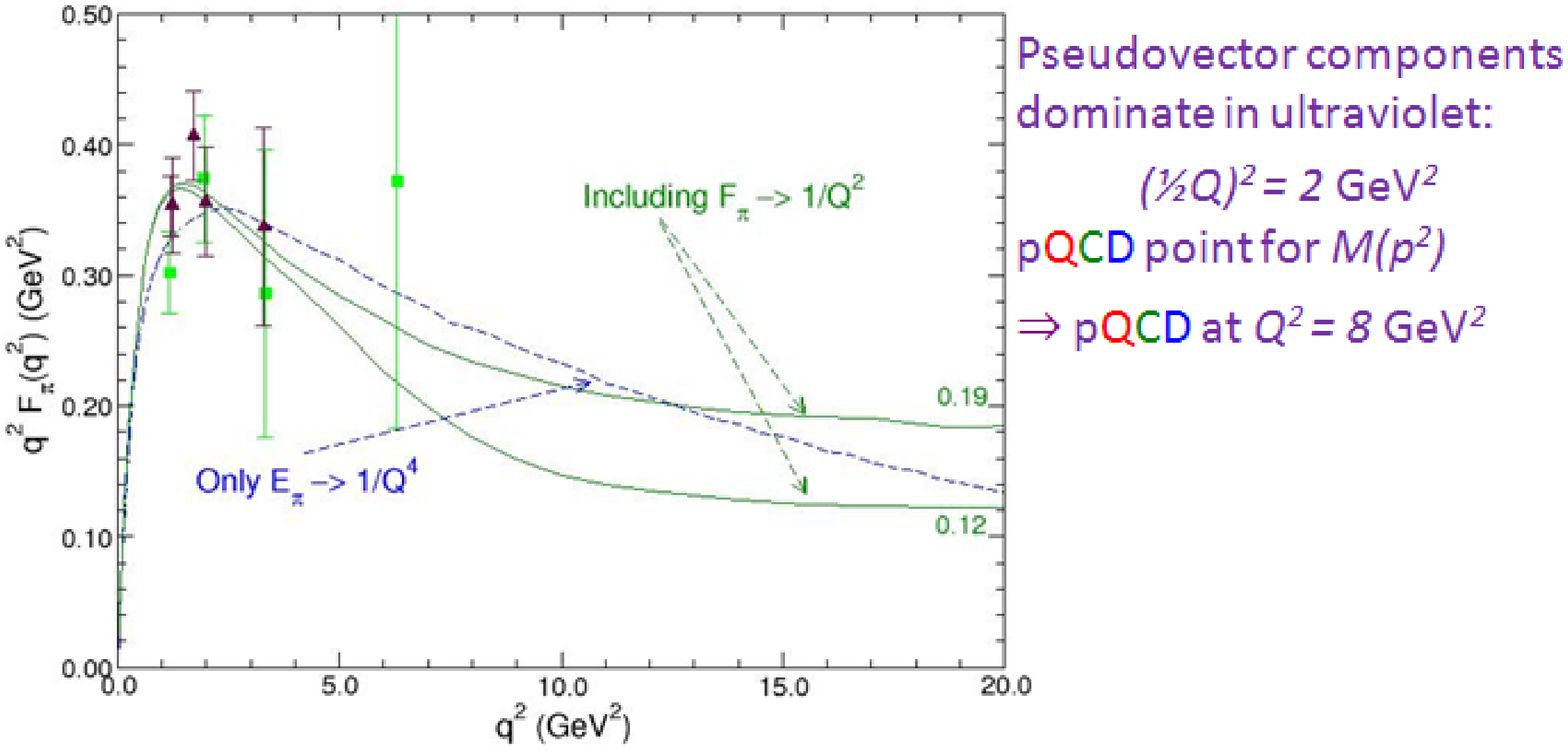}
\caption{\label{FpiQ2MR} Influence of the pion's pseudovector components on the asymptotic behaviour of $Q^2 F^{\rm em}_{\pi}(Q^2)$.  The existence of such components is a necessary consequence of DCSB \cite{Maris:1997hd,Maris:1998hc}.
\underline{Dashed curve} -- Result obtained if one ignores the pseudovector components, $F_\pi(Q^2) \sim 1/Q^4$; and \underline{solid curves} -- result produced via inclusion of the pseudovector parts of the pion's Bethe-Salpeter amplitude, $Q^2 F_\pi(Q^2) \approx\,$constant.
In the latter case the two curves result from slightly different assumptions about the evolution of the strong running-coupling from the infrared to the ultraviolet, something which again exposes the sensitivity of this observable to the momentum-dependence of the strong-interaction's $\beta$-function.
}
\end{figure}

In the electromagnetic elastic scattering process, the momentum transfer, $Q$, is primarily shared equally between the pion's constituents because the bound-state Bethe-Salpeter amplitude is peaked at zero relative momentum.  Thus, one can consider $k\sim Q/2$.
The Goldberger-Trieman-like relations express a mapping between the relative momentum of the pion's constituents and the one-body momentum of dressed-quark; and the momentum dependence of the dressed-quark mass function is well-described by perturbation theory when $k^2>2\,$GeV$^2$.
Hence, one should expect a perturbative-QCD analysis of the pion form factor to be valid for $k^2=Q^2/4 \gtrsim 2\,$GeV$^2$; i.e.,
\begin{equation}
F_\pi^{\rm em}(Q^2) \approx F_\pi^{\rm em\, pQCD}(Q^2) \; \mbox{for} \; Q^2\gtrsim 8\,{\rm GeV}^2.
\end{equation}
This explains the result in Ref.\,\cite{Maris:1998hc}, which is reproduced in Fig.\,\ref{FpiQ2MR}.
A similar argument for baryons suggests that the nucleon form factors should be perturbative for $Q^2\gtrsim 18\,$GeV$^2$.

\subsection{Deep inelastic scattering}
\label{FF2}
As mentioned in Sec.\,\ref{sec:Emergent}, quarks were discovered in deep inelastic scattering (DIS) experiments at SLAC, performed during the period 1966-1978.  DIS is completely different to elastic scattering.  In this process, one disintegrates the target instead of keeping only those events in which it remains intact.  On a well-defined kinematic domain; namely, the Bjorken limit:
\begin{equation}
q^2 \to \infty\,,\; P\cdot q \to -\infty\,,\; x_{\rm Bjorken}:=-\frac{q^2}{2 P\cdot q} = {\rm fixed},
\end{equation}
where $P$ is the target's four-momentum and $q$ is the momentum transfer, the cross-section can rigorously be interpreted as a measurement of the momentum-fraction probability distribution for quarks and gluons within the target hadron: $q(x)$, $g(x)$.  These quantities describe the probability that a quark/gluon within the target will carry a fraction $x$ of the bound-state's momentum, as defined in the infinite-momentum or light-front frame.  (The light-front formulation of quantum field theory is built upon Dirac's front form of relativistic quantum dynamics \cite{Dirac:1949cp}.)

The past forty years have seen a tremendous effort to deduce the parton distribution
functions (PDFs) of the most accessible hadrons -- the proton, neutron and pion.  There are many reasons for this long sustained and thriving interest \cite{Holt:2010vj} but in large part it is motivated by the suspected process-independence of the usual parton distribution functions and hence an ability to unify many hadronic processes through their computation.  In connection with uncovering the essence of the strong interaction, the behaviour of the valence-quark distribution functions at large Bjorken-$x$ is most relevant.
Furthermore, an accurate determination of the behavior of distribution functions in the valence region is also important to high-energy physics.  Particle discovery experiments and Standard Model tests with colliders are only possible if the QCD background is completely understood.  QCD evolution, apparent in the so-called scaling violations by parton distribution functions,\footnote{DGLAP evolution is described in Sec.IID of Ref.\,\cite{Holt:2010vj}.  The evolution equations are derived in perturbative QCD and determine the rate of change of parton densities when the energy-scale chosen for their definition is varied.}
entails that with increasing center-of-mass energy, the support at large-$x$ in the distributions evolves to small-$x$ and thereby contributes materially to the collider background.
N.B.\ The nucleon PDFs are now fairly well determined for $x\lesssim 0.8$ but the pion and kaon PDFs remain poorly known on the entire domain of $x$.

\subsubsection{Pion and kaon valence-quark distributions}
\label{sec:uvx}
Owing to the dichotomous nature of Goldstone bosons, understanding the valence-quark distribution functions in the pion and kaon is of great importance.  Moreover, given the large value of the ratio of $s$-to-$u$ current-quark masses, a comparison between the pion and kaon structure functions offers the chance to chart effects of explicit chiral symmetry breaking on the structure of would-be Goldstone modes.  There is also the prediction \cite{Ezawa:1974wm,Farrar:1975yb} that a theory in which the quarks interact via $1/k^2$-vector-boson exchange will produce valence-quark distribution functions for which
\begin{equation}
\label{pQCDuvx}
q_{\rm v}(x) \propto (1-x)^{2+\gamma} \,,\; x\gtrsim 0.85\,,
\end{equation}
where $\gamma\gtrsim 0$ is an anomalous dimension that grows with increasing momentum transfer.  (See Sec.VI.B.3 of Ref.\,\cite{Holt:2010vj} for a detailed discussion.)

\begin{figure}[t] 
\includegraphics[clip,height=0.15\textheight]{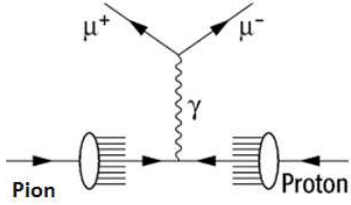}
\caption{\label{piNDY} $\pi N$ Drell-Yan process, in which, e.g., a valence antiquark from the pion annihilates with a valence quark in the nucleon to produce a $\mu^+ \mu^-$ pair.
}
\end{figure}

Owing to the absence of pseudoscalar meson targets, experimental knowledge of the parton structure of the pion and kaon arises primarily from pionic or kaonic Drell-Yan processes, illustrated in Fig.\,\ref{piNDY}, involving nucleons in heavy nuclei \cite{Badier:1980jq,Badier:1983mj,Betev:1985pg,Conway:1989fs,Wijesooriya:2005ir}.  Theoretically, given that DCSB plays a crucial role in connection with pseudoscalar mesons, one must employ an approach that realistically expresses this phenomenon.  The DSEs therefore provide a natural framework: studies of the pion and kaon exist and will be reviewed here.  The first \cite{Hecht:2000xa} computed pion PDFs, using efficacious parametrisations of both the Bethe-Salpeter amplitude and dressed-quark propagators \cite{Roberts:1994hh,Burden:1995ve,ElBennich:2011py}.  The second \cite{Nguyen:2011jy} employed direct, numerical DSE solutions in the computation of the pion and kaon PDFs, adapting the approach employed in successful predictions of electromagnetic form factors \cite{Holl:2005vu,Maris:1999bh,Maris:2000sk,Maris:2002mz}; and also studied the ratio $u_K(x)/u_\pi(x)$ in order to elucidate aspects of the influence of an hadronic environment.

In rainbow-ladder truncation, one obtains the pion's valence-quark distribution from
\begin{eqnarray}
u_\pi(x) = -\frac{1}{2} \int \frac{d^4 \ell}{(2\pi)^4}  {\rm tr}_{\rm cd}\, \left[ \Gamma_\pi(\ell,-P) \,S_u(\ell)\, \Gamma^n(\ell;x) \, S_u(\ell)\, \Gamma_\pi(\ell,P)\, S_d(\ell-P) \right] ,
\label{Eucl_pdf_LR_Ward}
\end{eqnarray}
wherein the Bethe-Salpeter amplitudes and dressed-quark propagators are discussed above and $\Gamma^n(\ell;x)$ is a generalization of the dressed-quark-photon vertex, describing a dressed-quark scattering from a zero momentum photon.  It satisfies a BSE (here with a rainbow-ladder kernel) with the inhomogeneous term  $i\gamma\cdot n \, \delta(\ell \cdot n - x P\cdot n)$.  In Eq.\,\eqref{Eucl_pdf_LR_Ward}, $n_\mu$ is a light-like vector satisfying \mbox{$n^2 = 0$}.
In choosing rainbow-ladder truncation one implements a precise parallel to the symmetry-preserving treatment of the pion charge form factor at \mbox{$Q^2 = 0 $}, wherein the vector current is conserved by use of ladder dynamics at all three vertices and rainbow dynamics for all three quark propagators \cite{Maris:1998hc,Maris:1999bh,Maris:2000sk,Roberts:1994hh}.   Equation~(\ref{Eucl_pdf_LR_Ward}) ensures automatically that
\begin{equation}
 \langle x_f^0 \rangle := \int_0^1 dx \,  q^v_f(x) = 1\; \mbox{for}\; f = u, \bar d\,,
\end{equation}
since $ \int dx \, \Gamma^n(\ell;x)$ gives the Ward-identity vertex and the Bethe-Salpeter amplitudes are canonically normalised.

\begin{figure}[t]
\includegraphics[clip,width=0.6\textwidth]{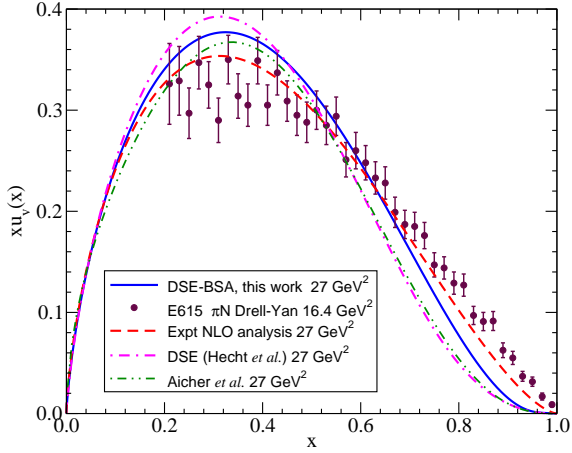}

\caption{ Pion valence quark distribution function evolved to (5.2~GeV)$^2$.  \emph{Solid curve} -- full DSE calculation \protect\cite{Nguyen:2011jy}; \emph{dot-dashed curve} -- semi-phenomenological DSE-based calculation in Ref.\,\protect\cite{Hecht:2000xa}; \emph{filled circles} -- experimental data from Ref.\,\protect\cite{Conway:1989fs}, at scale (4.05\,{\rm GeV})$^2$;
\emph{dashed curve} -- NLO reanalysis of the experimental data \protect\cite{Wijesooriya:2005ir};
and \emph{dot-dot-dashed curve} -- NLO reanalysis of experimental data with inclusion of soft-gluon resummation \protect\cite{Aicher:2010cb}.
\label{fig:pi_DSE}}
\end{figure}

Figure~\ref{fig:pi_DSE} displays the DSE results for the pion's valence $u$-quark distribution, evolved from a resolving scale $Q_0^2=(0.6\,$GeV)$^2$ to $Q^2 = (5.2~{\rm GeV})^2$ using leading-order DGLAP evolution (Sec.IID of Ref.\,\cite{Holt:2010vj}), and a comparison with $\pi N$ Drell-Yan data \cite{Conway:1989fs} at a scale  $Q^2 \sim (4.05~{\rm GeV})^2$, inferred via a leading-order analysis.  The computation's resolving scale, $Q_0$, was fixed by matching the $\langle x^n\rangle^\pi$ moments for $n=1,2,3$ to an experimental analysis at (2\,{\rm GeV})$^2$ \cite{Sutton:1991ay}.

It is notable that at $Q_0$ the DSE results yield
\begin{equation}
\label{momcons}
2 \, \langle x \rangle^\pi_{Q_0} = 0.7\,,\;
2 \, \langle x \rangle^K_{Q_0} = 0.8\,.
\end{equation}
(For comparison, the parametrised valence-like pion parton distributions of Ref.\,\protect\cite{Gluck:1998xa} yield a gluon momentum fraction of $\langle x_g\rangle^\pi_{Q_0=0.51} = 0.3$.)
In each case the remainder of the hadron's momentum is carried by gluons, which effect binding of the meson bound state and are invisible to the electromagnetic probe.  Some fraction of the hadron's momentum is carried by gluons at all resolving scales unless the hadron is a point particle \cite{Holt:2010vj}.  Indeed, it is a simple algebraic exercise to demonstrate that the only non-increasing, convex function which can produce $\langle x^0\rangle =1$ and $\langle x \rangle = \frac{1}{2}$, is the distribution $u(x)=1$, which is uniquely connected with a pointlike meson; viz., a meson whose Bethe-Salpeter amplitude is momentum-independent.  Thus Eqs.\,(\ref{momcons}) are an essential consequence of momentum conservation.

Whilst the DSE results in Fig.\,\ref{fig:pi_DSE} are both consistent with Eq.\,(\ref{pQCDuvx}); i.e., they produce algebraically the precise behaviour predicted by perturbative QCD, on the valence-quark domain -- which is uniquely sensitive to the behaviour of the dressed-quark mass-function, $M(p^2)$ -- it is evident that they disagree markedly with the Drell-Yan data reported in Ref.\,\cite{Conway:1989fs}.  This tension was long seen as a crucial mystery for a QCD description of the lightest and subtlest hadron \cite{Holt:2010vj}.  Its re-emergence with Ref.\,\cite{Hecht:2000xa} motivated a NLO reanalysis of the Drell-Yan data \cite{Wijesooriya:2005ir}, the result of which is also displayed in Fig.\,\ref{fig:pi_DSE}.  At NLO the extracted PDF is softer at high-$x$ but the discrepancy nevertheless remains.
To be precise, Ref.\,\cite{Wijesooriya:2005ir} determined a high-$x$ exponent of $\beta \simeq 1.5$ whereas the exponents produced by the DSE studies \cite{Hecht:2000xa,Nguyen:2011jy} are, respectively, $2.1$ and $2.4$ at the common model scale.  They do not allow much room for a harder PDF at high-$x$.\footnote{Here, ``hard'' means pointlike but can also identify a process with large momentum transfer.  On the other hand, ''soft'' is associated with the structure of diffuse composite objects or processes with small momentum transfer.}

Following the highlighting of this discrepancy in Ref.\,\cite{Holt:2010vj}, a resolution of the conflict between data and well-constrained theory was proposed.  In Ref.\,\cite{Aicher:2010cb} a long-overlooked effect was incorporated; namely, ``soft-gluon resummation.'' With the inclusion of this next-to-leading-logarithmic threshold resummation effect in the calculation of the Drell-Yan cross section, a considerably softer valence-quark distribution was obtained at high-$x$.
This is readily understood.  The Drell-Yan cross-section factorises into two pieces: one hard and the other soft.  The soft piece involves the PDF and the hard piece is calculable in perturbation theory.  Adding additional interactions to the latter, which are important at large-$x$; viz., soft gluons, provides greater strength in the hard piece on the valence-quark domain.  Hence a description of the data is obtained with a softer PDF.
Indeed, the distribution obtained thereby matches precisely the expectations based on perturbative-QCD and obtained using DSEs.  This is evident in a comparison between the \emph{dash-dot} and \emph{dash-dot-dot} curves in Fig.\,\ref{fig:pi_DSE}.
This outcome again emphasises the predictive power and strength of using a single internally-consistent, well-constrained framework to correlate and unify the description of hadron observables.

\begin{figure}[t]
\includegraphics[clip,width=0.66\textwidth]{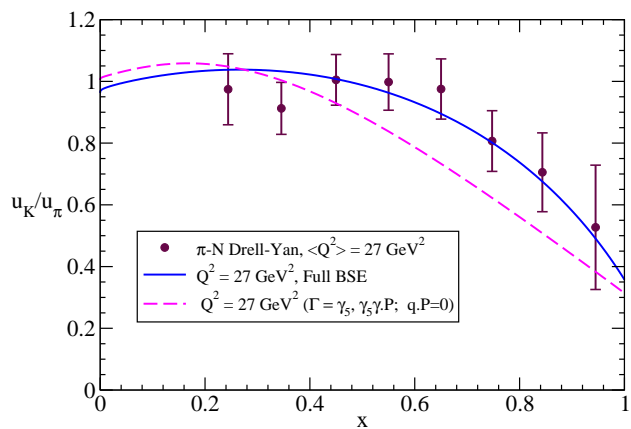}

\caption{\label{fig:pi_DSE_ratio}
DSE prediction for the ratio of $u$-quark distributions in the kaon and pion \protect\cite{Holt:2010vj,Nguyen:2011jy}.  The full Bethe-Salpeter amplitude produces the \emph{solid} curve; a reduced BSE vertex produces the \emph{dashed} curve.  The reduced amplitude retains only the invariants and amplitudes  involving pseudoscalar and axial vector Dirac matrices, and ignores dependence on the variable $q\cdot P$.  These are part of the reductions that occur in a pointlike treatment of pseudoscalar mesons.  The experimental data is from \protect\cite{Badier:1980jq,Badier:1983mj}.}
\end{figure}

The ratio $u_K/u_\pi$ measures the effect of the local hadronic environment.  In the kaon, the $u$-quark is bound with a heavier partner than in the pion ($m_s \approx 25 m_d$) and this should cause $u_K(x)$ to peak at lower-$x$ than $u_\pi(x)$.
In fact, one finds that the $s$-quark distribution peaks at a value of $x$ which is just 15\% larger than that of the $u$-quark.  Hence, even though DIS is a high-$Q^2$ process, constituent-quark-like mass-scales explain this shift: $M_s/M_u \approx 1.25$.
The DSE prediction of $u_K/u_\pi$ \cite{Holt:2010vj,Nguyen:2011jy} is shown in Fig.\,\ref{fig:pi_DSE_ratio} along with available Drell-Yan data \cite{Badier:1980jq,Badier:1983mj}.  The parameter-free DSE result agrees well with the data.
I note that
\begin{equation}
\frac{u_K(0)}{u_\pi(0)} \stackrel{\mbox{\footnotesize\rm DGLAP:}Q^2\to \infty}{\to} 1\,;
\end{equation}
viz, the ratio approaches one under evolution to larger resolving scales owing to the increasingly large population of sea-quarks produced thereby \cite{Chang:2010xs}.  On the other hand, the value at $x=1$ is a fixed-point under evolution:
\begin{equation}
\forall Q_1^2>Q_0^2,\;
\frac{u_K(1)}{u_\pi(1)}\rule[-2.5ex]{0.1ex}{6ex}_{Q_1^2}
\stackrel{\mbox{\footnotesize\rm DGLAP:}Q_0^2\to  Q_1^2}{=}
\frac{u_K(1)}{u_\pi(1)} \rule[-2.5ex]{0.1ex}{6ex}_{Q_0^2}
= \frac{u_K(1)}{u_\pi(1)} \rule[-2.5ex]{0.1ex}{6ex}_{Q_0^2}
\end{equation}
i.e., it is the same at every value of the resolving scale $Q^2$, and is therefore a persistent probe of nonperturbative dynamics \cite{Holt:2010vj}.

With Ref.\,\cite{Nguyen:2011jy} a significant milestone was achieved; viz., unification of the computation of distribution functions that arise in analyses of deep inelastic scattering with that of numerous other properties of pseudoscalar mesons, including meson-meson scattering \cite{Bicudo:2001aw,Bicudo:2001jq} and the successful prediction of electromagnetic elastic and transition form factors.  The results confirm the large-$x$ behavior of distribution functions predicted by the QCD parton model; provide a good account of the $\pi$-$N$ Drell-Yan data for $u_\pi(x)$; and a parameter-free prediction for the ratio $u_K(x)/u_\pi(x)$ that agrees with extant data, showing a strong environment-dependence of the $u$-quark distribution.  The new Drell-Yan experiment running at FNAL is capable of validating this comparison, as is the COMPASS~II experiment at CERN.  Such an experiment should be done so that complete understanding of QCD's Goldstone modes can be claimed. 

\section{Charting the Interaction between Light-Quarks}
\label{sec:Charting}
Let us now return to some remarks from Sec.\,\ref{confinementII}.  Namely, confinement can be related to the analytic properties of QCD's Schwinger functions so that the question of light-quark confinement may be translated into the challenge of charting the infrared behavior of QCD's universal $\beta$-function.  This is a well-posed problem whose solution is an elemental goal of modern hadron physics \cite{Qin:2011dd,Brodsky:2010ur,Aguilar:2010gm}.  The answer provides QCD's running coupling. Naturally, the behaviour of the $\beta$-function on the perturbative domain is well known; indeed, to fourth-loop order \cite{Nakamura:2010zzi}.

The DSEs provide a mapping between experimental observables and the pointwise behaviour of QCD's $\beta$-function, including the infrared domain.  Hence, comparison between computations and observations -- of, e.g., the hadron mass spectrum, and elastic and transition form factors -- can be used to chart the $\beta$-function's long-range behaviour, wherefrom arises the pattern of chiral symmetry breaking.  Whilst extant studies show the static properties of hadron excited states to be more sensitive to the long-range behaviour of the $\beta$-function than those of ground states \cite{Qin:2011dd,Qin:2011xq}, dynamical properties of ground states possess quite some discriminating power.

I will now illustrate these remarks by explicating the impact of differing assumptions about the behaviour of the Bethe-Salpeter kernel on a few selected hadron properties.  In this way we will see that it is possible to build a stock of material which can be used to identify unambiguous signals in experiment for the pointwise behaviour of: the interaction between light-quarks; the light-quarks' mass-function; and other similar quantities.  Whilst these are particular qualities, taken together they will enable a characterisation of the nonperturbative behaviour of the theory underlying strong interaction phenomena.

\begin{table}[t]
\caption{\emph{Row 1} -- Contact-interaction meson-related results obtained with \protect\cite{Roberts:2011wy} $\alpha_{\rm IR}/\pi=0.93$ in Eq.\,\eqref{DnjlGnjl}, and (in GeV): $m=0.007$ in Eq.\,\eqref{gap-1}, and $\Lambda_{\rm ir} = 0.24\,$, $\Lambda_{\rm uv}=0.905$ in Eq.\,\eqref{ExplicitRS}.  $\kappa_\pi$ is the in-pion condensate \protect\cite{Brodsky:2010xf,Chang:2011mu}; and $f_{\pi,\rho}$ are the mesons' leptonic decay constants.
\emph{Row 2} -- Same quantities computed using an interaction based on renormalisation-group-improved one-gluon exchange \protect\cite{Maris:1999nt}.
Empirical values are $\kappa_\pi \approx (0.22\,{\rm GeV})^3$ and \protect\cite{Nakamura:2010zzi} $f_\pi=0.092\,$GeV, $f_\rho=0.153\,$GeV.
\label{Table:cf}
}
\begin{center}
\begin{tabular*}
{\hsize}
{
l@{\extracolsep{0ptplus1fil}}
|c@{\extracolsep{0ptplus1fil}}
c@{\extracolsep{0ptplus1fil}}
c@{\extracolsep{0ptplus1fil}}
c@{\extracolsep{0ptplus1fil}}
c@{\extracolsep{0ptplus1fil}}
c@{\extracolsep{0ptplus1fil}}}\hline
interaction~ & $M$ & $\kappa_\pi^{1/3}$ & $m_\pi$ & $m_\rho$ & $f_\pi$ & $f_\rho$ \\\hline
%
%
contact  & 0.37 & 0.24 & 0.14 & 0.93 & 0.101 & 0.13 \\
QCD-like  & 0.34 & 0.24 & 0.14 & 0.74 & 0.093 & 0.15 \\\hline
\end{tabular*}
\end{center}
\end{table}

\begin{figure}[t]
\includegraphics[width=0.70\textwidth]{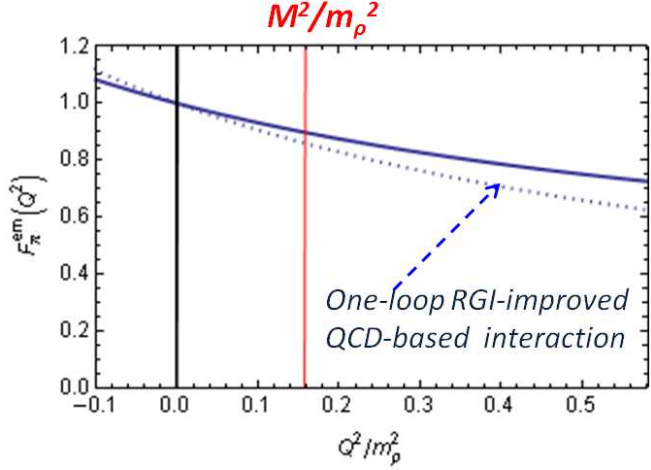}
\caption{\label{fig:FpiemCvQCD}
\emph{Solid curve} -- Contact interaction result for the charged pion's electromagnetic form factor \protect\cite{Roberts:2011wy}; and dotted curve, result obtained with renormalisation-group-improved one-gluon exchange \protect\cite{Maris:2000sk}.
There is no meaningful difference for $Q^2< M^2$.
}
\end{figure}

Consider therefore the interaction kernel specified by Eq.\,\eqref{DnjlGnjl}, and the collection of DSEs derived therefrom via the confining, symmetry-preserving regularisation scheme indicated in Sec.\,\ref{sec:njldcsb}.  A raft of studies -- static properties of $\pi$- and $\rho$-meson \cite{GutierrezGuerrero:2010md,Roberts:2011wy,Roberts:2011cf}, and those of the neutron and proton \cite{Wilson:2011rj,Roberts:2011cf} -- have shown that contact-interaction results are not realistically distinguishable from those obtained with renormalisation-group-improved one-gluon exchange for processes involving momentum transfers $Q^2<M^2$, where $M$ is the dressed-quark mass in Eq.\,\eqref{genS}.  This is exemplified in Table~\ref{Table:cf} and Fig.\,\ref{fig:FpiemCvQCD}

The picture changes dramatically, however, for processes involving larger momentum transfers.  This may again be illustrated via the charged pion electromagnetic form factor.  With a symmetry preserving regularisation of the interaction in Eq.\,\eqref{DnjlGnjl}, no meson's Bethe-Salpeter amplitude can depend on relative momentum.  Hence Eq.\,(\ref{genGpi}) becomes
\begin{equation}
\Gamma_\pi(P) = \gamma_5 \biggL i E_\pi(P) + \frac{1}{M} \gamma\cdot P F_\pi(P) \biggR ,
\end{equation}
the dressed-quark propagator is given in Eq.\,\eqref{genS}, and the explicit form of the model's ladder BSE is
\begin{equation}
\label{bsefinal0}
\biggerL
\begin{array}{c}
E_\pi(P)\\
F_\pi(P)
\end{array}
\biggerR
= \frac{4 \alpha_{\rm IR}}{3\pi m_G^2}
\biggerL
\begin{array}{cc}
{\cal K}_{EE} & {\cal K}_{EF} \\
{\cal K}_{FE} & {\cal K}_{FF}
\end{array}\biggerR
\biggerL
\begin{array}{c}
E_\pi(P)\\
F_\pi(P)
\end{array}
\biggerR,
\end{equation}
where, with $m=0=P^2$, anticipating the Goldstone character of the pion, anticipating the Goldstone character of the pion,
\begin{equation}
\label{pionKernel}
{\cal K}_{EE}  =  {\cal C}(M^2;\tau_{\rm ir}^2,\tau_{\rm uv}^2)\,, \;
{\cal K}_{EF}  =  0\,, \;
2 {\cal K}_{FE} = {\cal C}_1(M^2;\tau_{\rm ir}^2,\tau_{\rm uv}^2) \,, \;
{\cal K}_{FF} = - 2 {\cal K}_{FE}\,,
\end{equation}
and ${\cal C}_1(z) = - z {\cal C}^\prime(z)$, where I have suppressed the dependence on $\tau_{\rm ir,uv}$.  The solution of Eq.\,(\ref{bsefinal0}) gives the pion's chiral-limit Bethe-Salpeter amplitude, which, for the computation of observables, should be normalised canonically; viz.,
\begin{equation}
%
 P_\mu = N_c\, {\rm tr} \int\! \frac{d^4q}{(2\pi)^4}\Gamma_\pi(-P)
 \frac{\partial}{\partial P_\mu} S(q+P) \, \Gamma_\pi(P)\, S(q)\,. \label{Ndef}
\end{equation}
Hence, in the chiral limit,
\begin{equation}
1 = \frac{N_c}{4\pi^2} \frac{1}{M^2} \, {\cal C}_1(M^2;\tau_{\rm ir}^2,\tau_{\rm uv}^2)
E_\pi [ E_\pi - 2 F_\pi],
\label{Norm0}
\end{equation}
and the pion's leptonic decay constant is
\begin{equation}
\label{fpi0}
f_\pi^0 = \frac{N_c}{4\pi^2} \frac{1}{M} {\cal C}_1(M^2;\tau_{\rm ir}^2,\tau_{\rm uv}^2)  [ E_\pi - 2 F_\pi ]\,.
\end{equation}

If one has preserved Eq.\,(\ref{avwtimN}), then, for $m=0$ in the neighbourhood of $P^2=0$, the solution of the axial-vector BSE has the form:
\begin{equation}
\Gamma_{5\mu}(k_+,k) = \frac{P_\mu}{P^2} \, 2 f_\pi^0 \, \Gamma_\pi(P) + \gamma_5\gamma_\mu F_R(P)
\end{equation}
and the following subset of Eqs.\,(\ref{gtlrelE}) -- (\ref{gtlrelH}) will hold:
\begin{equation}
\label{GTI}
f_\pi^0 E_\pi = M \,,\; 2\frac{F_\pi}{E_\pi}+ F_R = 1\,.
\end{equation}
Hence $F_\pi(P)$ is necessarily nonzero in a vector exchange theory, irrespective of the pointwise behaviour of the interaction.  In fact, for $m=0$, $E_\pi(P) = 3.568$, $F_\pi(P)=0.459$, so that $F_\pi(P)$ has a measurable impact on the value of $f_\pi$. 

More importantly, it has a big impact on the form factor, which may be computed using Eq.\,\eqref{KF}.  This yields the solid curve in Fig.\,\ref{FpiUV}, which exposes that a contact interaction, Eq.\,(\ref{DnjlGnjl}), generates
\begin{equation}
\label{Fpiconstant}
F_\pi^{\rm em}(Q^2 \to\infty) =\,{\rm constant.}
\end{equation}
Equation~(\ref{Fpiconstant}) should not come as a surprise: with a symmetry-preserving regularisation of the interaction in Eq.\,(\ref{DnjlGnjl}), the pion's Bethe-Salpeter amplitude cannot depend on the constituent's relative momentum.  This is characteristic of a pointlike particle, which must have a hard form factor.  Notwithstanding this, for more than twenty-years numerous practitioners held to the notion that the pion form factor could not distinguish between a contact interaction and QCD.  Indeed, this misapprehension still prevails in other contexts, as explained in Ref.\,\cite{Roberts:2010rn}.

This illustration emphasises that when a momentum-independent vector-exchange interaction is regularised in a symmetry-preserving manner, the results are directly comparable with experiment, computations based on well-defined and systematically-improvable truncations of QCD's DSEs \cite{Maris:2000sk}, and perturbative QCD.  In this context it will be apparent that a contact interaction, whilst capable of describing pion static properties well, Table\,\ref{Table:cf}, generates a momentum-independent dressed-quark mass, which entails a form factor whose evolution with $Q^2$ deviates markedly from experiment for $Q^2>0.17\,$GeV$^2\approx M^2$ and produces asymptotic power-law behaviour, Eq.\,(\ref{Fpiconstant}), in serious conflict with QCD \cite{Farrar:1979aw,Efremov:1979qk,Lepage:1980fj}.
Plainly, experiment is very sensitive to the momentum-dependence of the interaction between light-quarks, and the raft of computations reported in Refs.\,\cite{Roberts:2010rn,GutierrezGuerrero:2010md,Roberts:2011wy,Wilson:2011rj} therefore provide important benchmarks for use in experimentally determining QCD's dressed-quark mass function and therefrom constraining the $\beta$-function's infrared behaviour.

It is important now to strive toward the reliable computation of properties of meson excited states, and the spectrum of hybrid and exotic mesons, states which are hypothetically possible in QCD but have not yet been seen experimentally.  As mentioned above, hybrids are defined to be states that possess normal constituent-quark model meson quantum numbers but non-quark-model decay patterns whilst exotics are described by quantum numbers not possible for quantum mechanical quark-antiquark systems.  If such states exist, then it is suspected that they will possess ``constituent gluon'' content; i.e., they are expected to have a large overlap with interpolating fields that explicitly contain gluon fields.  This being the case, their existence would completely eliminate the distinction between matter- and force-fields that has held since Maxwell's time.  Given such a composition, however, it is possible that a Bethe-Salpeter equation treatment is inadequate to understand their properties.  One might instead have to work with a three-body Faddeev equation that describes the interactions between two dressed-quarks and a dressed-gluon.\footnote{Glueballs, states with no valence quark content at all, are found in numerical simulations of lattice-regularised QCD \protect\cite{Richards:2010ck}.  However, the light-quark current-mass is four-times larger than the physical value.  Such predictions will not be firm until they are produced with more realistic masses.}   This is a daunting prospect.  If one is going to work with the Faddeev equation, then it is certainly best to first study systems about which a good deal is known experimentally; viz, baryons in QCD.

\section{Describing Baryons and Mesons Simultaneously}
\label{sec:Baryons}
While a symmetry-preserving description of mesons is essential, it is only part of the physics that nonperturbative QCD must describe because Nature also presents us with baryons: light-quarks in three-particle composites.  An explanation of the spectrum of baryons and the nature of interactions between them is basic to understanding the Standard Model.  The present and planned experimental programmes at JLab, and other facilities worldwide, are critical elements in this effort.

No approach to QCD is comprehensive if it cannot provide a unified explanation of both mesons and baryons.  We have explained that DCSB is a keystone of the Standard Model, which is evident in the momentum-dependence of the dressed-quark mass function -- Fig.\,\ref{gluoncloud}: it is just as important to baryons as it is to mesons.  Since constituent-quark-like models cannot incorporate the momentum-dependent dressed-quark mass-function, they are not a viable tool for use in this programme.  The DSEs furnish the only extant continuum framework that can simultaneously connect both meson and baryon observables with this basic feature of QCD, having provided, e.g., a direct correlation of meson and baryon properties via a single interaction kernel, which preserves QCD's one-loop renormalisation group behaviour and can systematically be improved.  This is evident in the preceding sections and their combination with Refs.\,\cite{Eichmann:2008ef,Eichmann:2011vu,Eichmann:2008ae,Eichmann:2011ej}.

In order to illustrate this programme, I will focus on nucleon electromagnetic form factors, which were introduced and described in connection with Eqs.\,\eqref{NcurrentA} -- \eqref{eq:sachs}.  As noted thereabouts, for a structureless or simply-structured fermion $F_1(Q^2) \equiv 1$ and $F_2(Q^2) \equiv 0$, so that $G_E(Q^2) \equiv G_M(Q^2)$ and the distribution of charge and magnetisation is identical.  This was believed to be the case for the proton until 1999.  In that year, enabled by the high luminosity of the accelerator at JLab, a new method was employed to measure the ratio $G_E^p(Q^2)/G_M^p(Q^2)$ \cite{Jones:1999rz}.  The result astonished the community: whilst $G_E^p(Q^2)/G_M^p(Q^2)\approx 1$ for $Q^2< 1\,$GeV$^2$, $G_E^p(Q^2)/G_M^p(Q^2)$ is a rapidly decreasing function for $Q^2>1\,$GeV$^2$ (see Fig.\,\ref{fig:GEGMp}).  How is this to be understood?

\subsection{Faddeev equation}
\label{sec:FE}
In quantum field theory a baryon appears as a pole in a six-point quark Green function.  The residue is proportional to the baryon's Faddeev amplitude, which is obtained from a Poincar\'e covariant Faddeev equation that sums all possible exchanges and interactions that can take place between three dressed-quarks.  A tractable Faddeev equation for baryons \cite{Cahill:1988dx} is founded on the observation that an interaction which describes colour-singlet mesons also generates nonpointlike quark-quark (diquark) correlations in the colour-$\bar 3$ (antitriplet) channel \cite{Cahill:1987qr}.  The dominant correlations for ground state octet and decuplet baryons are scalar ($0^+$) and axial-vector ($1^+$) diquarks because, for example, the associated mass-scales are smaller than the baryons' masses \cite{Burden:1996nh,Maris:2002yu} and their parity matches that of these baryons.  It follows that only they need be retained in approximating the quark-quark scattering matrix which appears as part of the Faddeev equation \cite{Eichmann:2008ef,Cloet:2008re,Roberts:2011cf}.  On the other hand, pseudoscalar ($0^-$) and vector ($1^-$) diquarks dominate in the parity-partners of ground state octet and decuplet baryons \cite{Roberts:2011cf}.  The DSE approach treats mesons and baryons on the same footing and, in particular, enables the impact of DCSB to be expressed in the prediction of baryon properties.

It is important to appreciate that diquarks do not appear in the strong interaction spectrum \cite{Bender:1996bb,Bender:2002as,Bhagwat:2004hn}.  However, the attraction between quarks in this channel justifies a picture of baryons in which two quarks within a baryon are always correlated as a colour-$\bar 3$ diquark pseudoparticle, and binding is effected by the iterated exchange of roles between the bystander and diquark-participant quarks.   Here it is important to emphasise strongly that QCD supports \emph{nonpointlike} diquark correlations \cite{Roberts:2011wy,Maris:2004bp}.  Hence models that employ pointlike diquark degrees of freedom have little connection with QCD.

\begin{figure}[t]
\includegraphics[clip,width=0.65\textwidth]{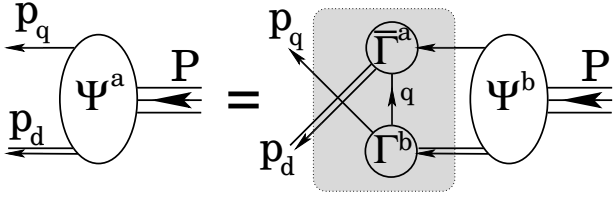}
\caption{\label{fig:Faddeev} Diagrammatic representation of a Poincar\'e covariant Faddeev equation for a baryon.  $\Psi$ is the Faddeev amplitude for a baryon of total momentum $P= p_q + p_d$.  It expresses the relative momentum correlation between the dressed-quark and -diquarks within the baryon.
The shaded region demarcates the kernel of the Faddeev equation, in which: the \emph{single line} denotes the dressed-quark propagator, $\Gamma$ is the diquark Bethe-Salpeter amplitude; and the \emph{double line} is the diquark propagator.}
\end{figure}

The Faddeev equation, illustrated in Fig.\,\ref{fig:Faddeev}, is a linear homogeneous matrix equation, in many respects similar to a Bethe-Salpeter equation.  Its solution is the nucleon's Poincar\'e-covariant Faddeev amplitude, which describes quark-diquark relative motion within the nucleon.
The composite nature of the diquark correlations, and their dynamical breakup and reformation through dressed-quark exchange, pictured in Fig.\,\ref{fig:Faddeev}, is crucial to maintaining fermion statistics for the nucleon bound-state.
Furthermore, owing to the critical importance of both scalar- and axial-vector-diquark correlations, the nucleon's rest-frame wave-function possesses $S$-, $P$- and  $D$-wave correlations; i.e., a nucleon should \emph{a priori} be expected to contain significant dressed-quark orbital angular momentum.
This is verified in Ref.\,\cite{Cloet:2007pi}, which shows that in the nucleon's
rest frame just 37\% of the total spin of the nucleon is contained within components of
the Faddeev amplitude which possess zero quark orbital angular momentum.

It is worthwhile expressing one Faddeev equation concretely and I choose that for the $\Delta$-resonance, a $J=3/2$ bound-state of three valence light-quarks.  It is possible to obtain a simple yet extremely informative equation if one employs the interaction presented in Eq.\,\eqref{DnjlGnjl} and an additional simplification; i.e., representing the quark exchanged between the diquarks as
\begin{equation}
S^{\rm T}(k) \to \frac{g_\Delta^2}{M}\,,
\label{staticexchange}
\end{equation}
where $g_\Delta=1.56$ \cite{Roberts:2011cf}.  This is a variant of the so-called ``static approximation,'' which itself was introduced in Ref.\,\cite{Buck:1992wz} and has subsequently been used in studying a range of nucleon properties \cite{Bentz:2007zs}.  In combination with diquark correlations generated by Eq.\,(\ref{DnjlGnjl}), whose Bethe-Salpeter amplitudes are momentum-independent, Eq.\,(\ref{staticexchange}) generates Faddeev equation kernels which themselves are independent of the external relative momentum variable.  The dramatic simplifications which this produces are the merit of Eq.\,(\ref{staticexchange}).

Following this through, one derives \cite{Roberts:2011cf}
\begin{eqnarray}
1 &=& 8 \frac{g_\Delta^2}{M } \frac{E_{qq_{1^+}}^2}{m_{qq_{1^+}}^2}
\int\frac{d^4\ell^\prime}{(2\pi)^4} \int_0^1 d\alpha\,
\frac{(m_{qq_{1^+}}^2 + (1-\alpha)^2 m_\Delta^2)(\alpha m_\Delta + M)}
{[\ell^{^\prime 2} + \sigma_\Delta(\alpha,M,m_{qq_{1^+}},m_\Delta)]^2}\\
\nonumber
&=& \frac{g_\Delta^2}{M}\frac{E_{qq_{1^+}}^2}{m_{qq_{1^+}}^2}\frac{1}{2\pi^2}\\
&& \times
\int_0^1 d\alpha\, (m_{qq_{1^+}}^2 + (1-\alpha)^2 m_\Delta^2)(\alpha m_\Delta + M)\overline{\cal C}^{\rm iu}_1(\sigma_\Delta(\alpha,M,m_{qq_{1^+}},m_\Delta)),
\label{DeltaFaddeev}
\end{eqnarray}
where
$E_{qq_{1^+}}$ is the canonically-normalised axial-vector diquark Bethe-Salpeter amplitude,
$m_{qq_{1^+}}$ is the computed mass for this correlation,
\begin{equation}
\sigma_\Delta(\alpha,M,m_{qq_{1^+}},m_\Delta)=(1-\alpha)\, M^2 + \alpha \, m_{qq_{1^+}} - \alpha (1-\alpha)\, m_\Delta^2,
\end{equation}
and  $\overline{\cal C}^{\rm iu}_1(z) = {\cal C}^{\rm iu}_1(z)/z$ [see text around Eqs.\,\eqref{gapactual}, \eqref{pionKernel}].
Equation~\eqref{DeltaFaddeev} is an eigenvalue problem whose solution yields the mass for the dressed-quark-core of the $\Delta$-resonance, $m_\Delta$.  It is one dimensional because only the axial-vector diquark correlation contributes to the structure of the $\Delta$.  (The nucleon, on the other hand, is constituted from scalar- and axial-vector-diquarks and presents a five-dimensional eigenvalue problem.)

\begin{figure}[t]
\centerline{\includegraphics[clip,width=0.95\textwidth]{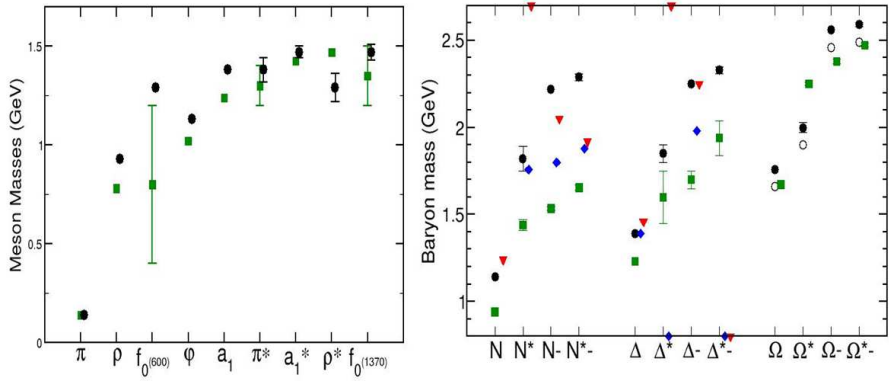}}
\caption{\label{fig:Fig2}
Comparison between DSE-computed hadron masses (\emph{filled circles}) and: bare baryon masses from Ref.\,\protect\cite{Suzuki:2009nj} (\emph{filled diamonds}) and Ref.\,\protect\cite{Gasparyan:2003fp} (\emph{filled triangles}); and experiment \protect\cite{Nakamura:2010zzi}, \emph{filled-squares}.
For the coupled-channels models a symbol at the lower extremity indicates that no associated state is found in the analysis, whilst a symbol at the upper extremity indicates that the analysis reports a dynamically-generated resonance with no corresponding bare-baryon state.
In connection with $\Omega$-baryons the \emph{open-circles} represent a shift downward in the computed results by $100\,$MeV.  This is an estimate of the effect produced by pseudoscalar-meson loop corrections in $\Delta$-like systems at a $s$-quark current-mass.
}
\end{figure}

With experience, one can look at Eq.\,\eqref{DeltaFaddeev} and see that increasing the current-quark mass will boost the mass of the bound-state.  This is just one of the Faddeev equations in Ref.\,\cite{Roberts:2011cf}.  In fact, building on lessons from meson studies \cite{Chang:2011vu}, a unified spectrum of $u,d$-quark hadrons was obtained therein using the symmetry-preserving regularization of a vector$\,\times\,$vector contact interaction that I have briefly described herein.  Reference~\cite{Roberts:2011cf} reports a study that simultaneously correlates the masses of meson and baryon ground- and excited-states within a single framework.  In comparison with relevant quantities, the computation produces $\overline{\mbox{rms}}$=13\%, where $\overline{\mbox{rms}}$ is the root-mean-square-relative-error$/$degree-of freedom.  As evident in Fig.\,\ref{fig:Fig2}, the prediction uniformly overestimates the PDG values of meson and baryon masses \cite{Nakamura:2010zzi}.  Given that the employed truncation deliberately omitted meson-cloud effects in the Faddeev kernel -- analogues of those in the right image of Fig.\,\ref{fig:PionFSI} -- this is a good outcome, since inclusion of such contributions acts to reduce the computed masses.

\begin{figure}[t]
\includegraphics[clip,width=0.75\textwidth]{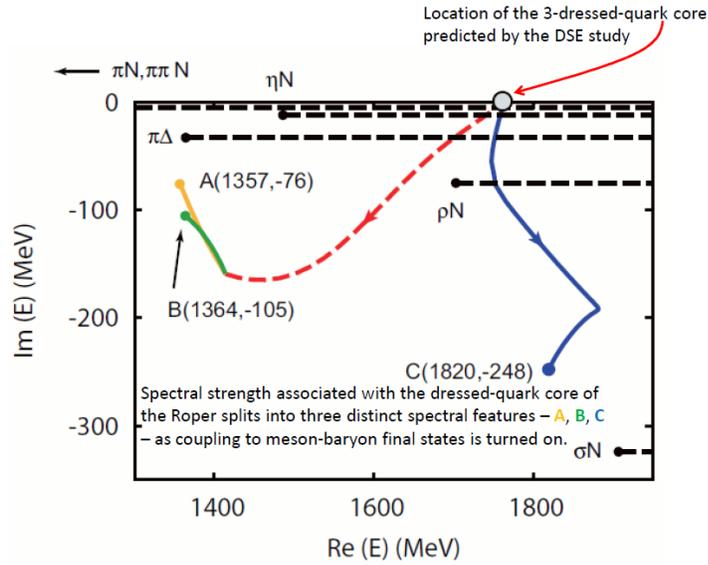}

\caption{\label{ebac}
The Excited Baryon Analysis Center (EBAC) examined the $P_{11}$-channel and found that the two poles associated with the Roper resonance and the next higher resonance were all associated with the same seed dressed-quark state.  Coupling to the continuum of meson-baryon final states induces multiple observed resonances from the same bare state.  In EBAC's analysis, all PDG-identified resonances were found to consist of a core state plus meson-baryon components.  (Adapted from Ref.\,\protect\cite{Suzuki:2009nj}.)}

\end{figure}

Following this line of reasoning, a striking result is agreement between the DSE-computed baryon masses \cite{Roberts:2011cf} and the bare masses employed in modern coupled-channels models of pion-nucleon reactions \cite{Suzuki:2009nj,Gasparyan:2003fp}, see Fig.\,\ref{fig:Fig2} and also Ref.\,\cite{Wilson:2011rj}.  The Roper resonance is very interesting.  The DSE study \cite{Roberts:2011cf} produces an excitation of the nucleon at $1.82\pm0.07\,$GeV.  This state is predominantly a radial excitation of the quark-diquark system, with both the scalar- and axial-vector diquark correlations in their ground state.  Its predicted mass lies precisely at the value determined in the analysis of Ref.\,\cite{Suzuki:2009nj}.  This is significant because for almost 50 years the ``Roper resonance'' has defied understanding.  Discovered in 1963, it appears to be an exact copy of the proton except that its mass is 50\% greater.  The mass was the problem: hitherto it could not be explained by any symmetry-preserving QCD-based tool.  That has now changed.  Combined, see Fig.\,\ref{ebac}, Refs.\,\cite{Roberts:2011cf,Suzuki:2009nj} demonstrate that the Roper resonance is indeed the proton's first radial excitation, and that its mass is far lighter than normal for such an excitation because the Roper obscures its dressed-quark-core with a dense cloud of pions and other mesons.  Such feedback between QCD-based theory and reaction models is critical now and for the foreseeable future, especially since analyses of experimental data on nucleon-resonance electrocouplings suggest strongly that this structure is typical; i.e., most low-lying $N^\ast$-states can best be understood as an internal quark-core dressed additionally by a meson cloud \cite{Aznauryan:2011td,Gothe:2011up}.

Additional analysis \cite{Wilson:2011rj} suggests a fascinating new feature of the Roper.  To elucidate, I focus first on the nucleon, whose Faddeev amplitude describes a ground-state that is dominated by its scalar diquark component (78\%).  The axial-vector component is significantly smaller but nevertheless important.  This heavy weighting of the scalar diquark component persists in solutions obtained with more sophisticated Faddeev equation kernels (see, e.g., Table~2 in Ref.\,\cite{Cloet:2008re}).  From a perspective provided by the nucleon's parity partner and the radial excitation of that state, in which the scalar and axial-vector diquark probabilities are \cite{Roberts:2011ym} 51\%-49\% and 43\%-57\%, respectively, the scalar diquark component of the ground-state nucleon actually appears to be unnaturally large.

One can nevertheless understand the structure of the nucleon.  As with so much else, the composition of the nucleon is intimately connected with dynamical chiral symmetry breaking.  In a two-color version of QCD, the scalar diquark is a Goldstone mode, just like the pion \cite{Roberts:1996jx}.  (This is a long-known result of Pauli-G\"ursey symmetry \cite{Pauli:1957,Gursey:1958}.)  A memory of this persists in the three-color theory and is evident in many ways.  Amongst them, through a large value of the canonically normalized Bethe-Salpeter amplitude and hence a strong quark$+$quark$-$diquark coupling within the nucleon.  (A qualitatively identical effect explains the large value of the $\pi N$ coupling constant.) There is no such enhancement mechanism associated with the axial-vector diquark.  Therefore the scalar diquark dominates the nucleon.

With the Faddeev equation treatment described herein, the effect on the Roper is dramatic: orthogonality of the ground- and excited-states forces the Roper to be constituted almost entirely (81\%) from the axial-vector diquark correlation.  It is important to check whether this outcome survives with a Faddeev equation kernel built from a momentum-dependent interaction.

\subsection{Nucleon form factors}
With masses and Faddeev amplitudes in hand, it is possible to compute baryon electromagnetic form factors.  For the nucleon, studies of the Faddeev equation exist that are based on the one-loop renormalisation-group-improved interaction that was used efficaciously in the study of mesons \cite{Eichmann:2008ef,Cloet:2008re}.  These studies retain the scalar and axial-vector diquark correlations, for the reasons explained in Sec.\,\ref{sec:FE}.

\begin{figure}[t]
\begin{minipage}[t]{\textwidth}
\begin{minipage}[t]{0.4\textwidth}
\leftline{\includegraphics[width=0.90\textwidth]{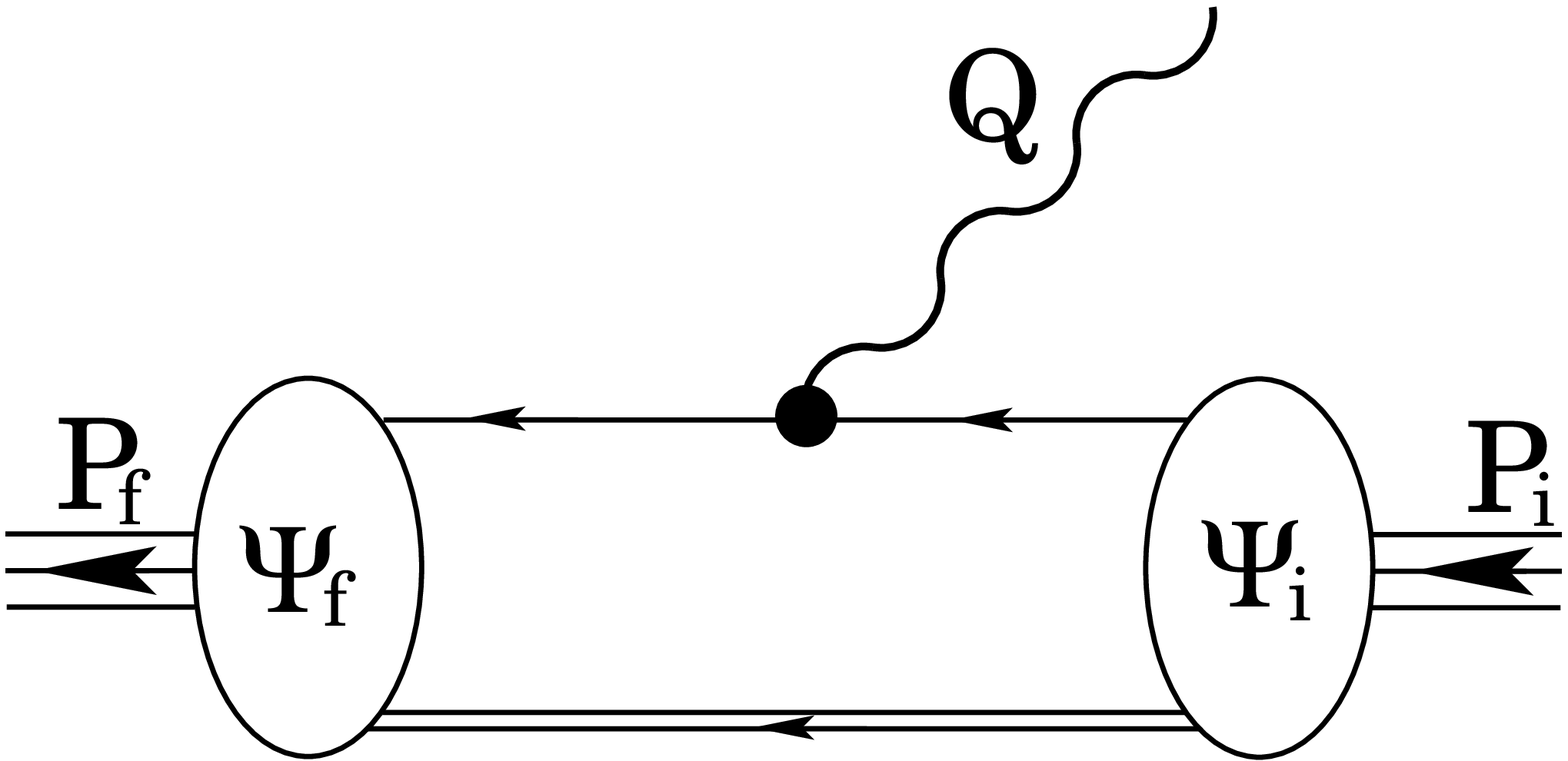}}
\end{minipage}
\begin{minipage}[t]{0.4\textwidth}
\rightline{\includegraphics[width=0.90\textwidth]{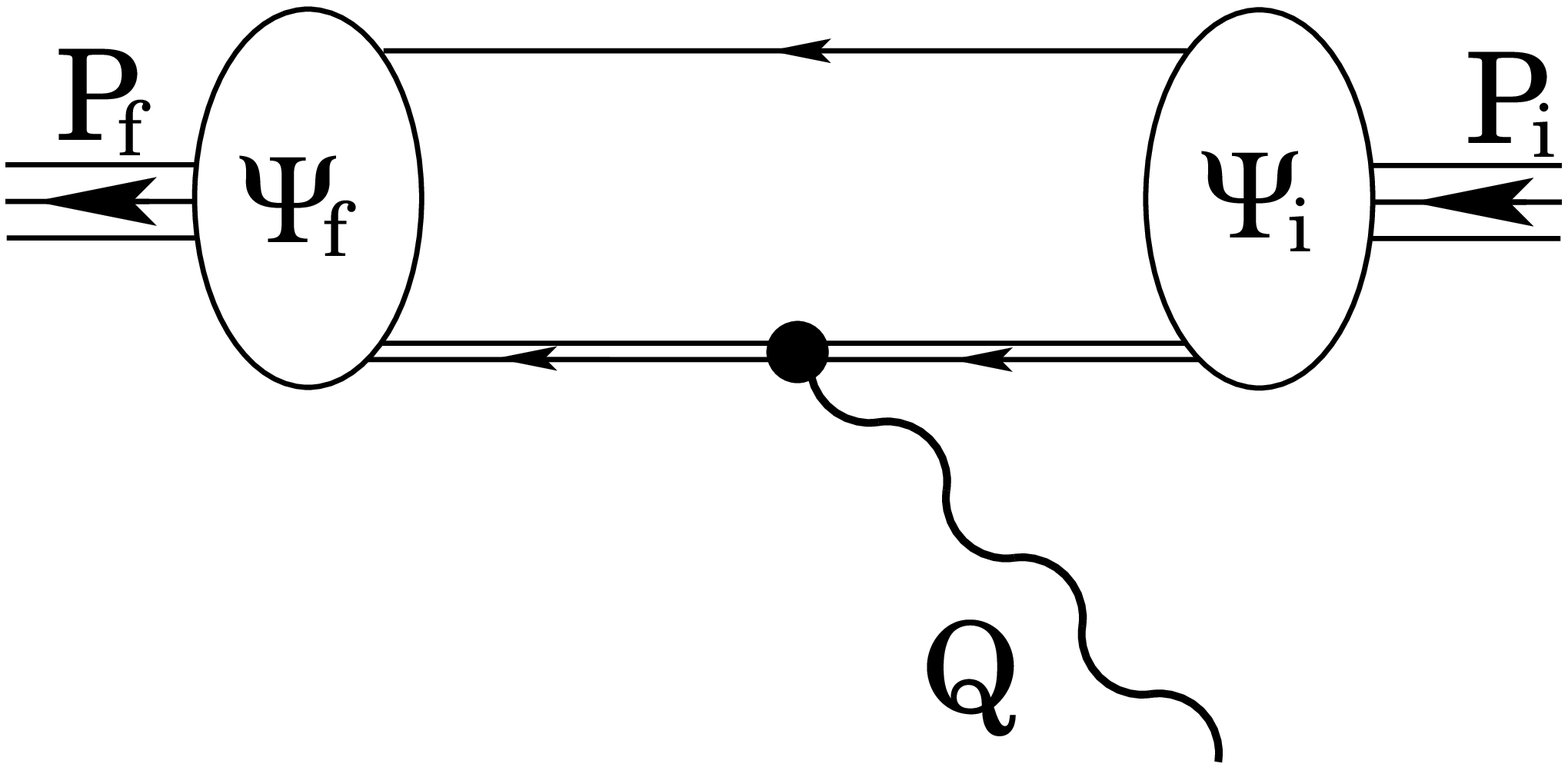}}
\end{minipage}\vspace*{3ex}

\begin{minipage}[t]{0.4\textwidth}
\leftline{\includegraphics[width=0.90\textwidth]{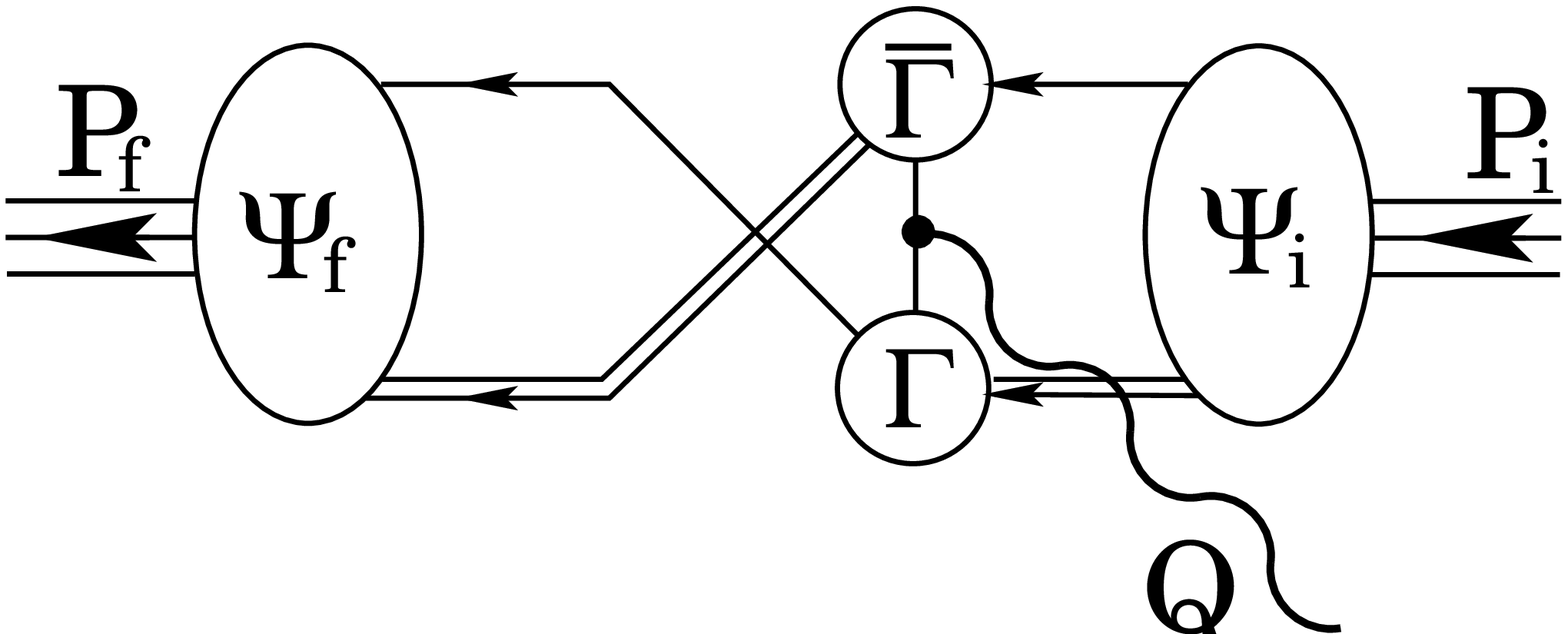}}
\end{minipage}
\begin{minipage}[t]{0.4\textwidth}
\rightline{\includegraphics[width=0.90\textwidth]{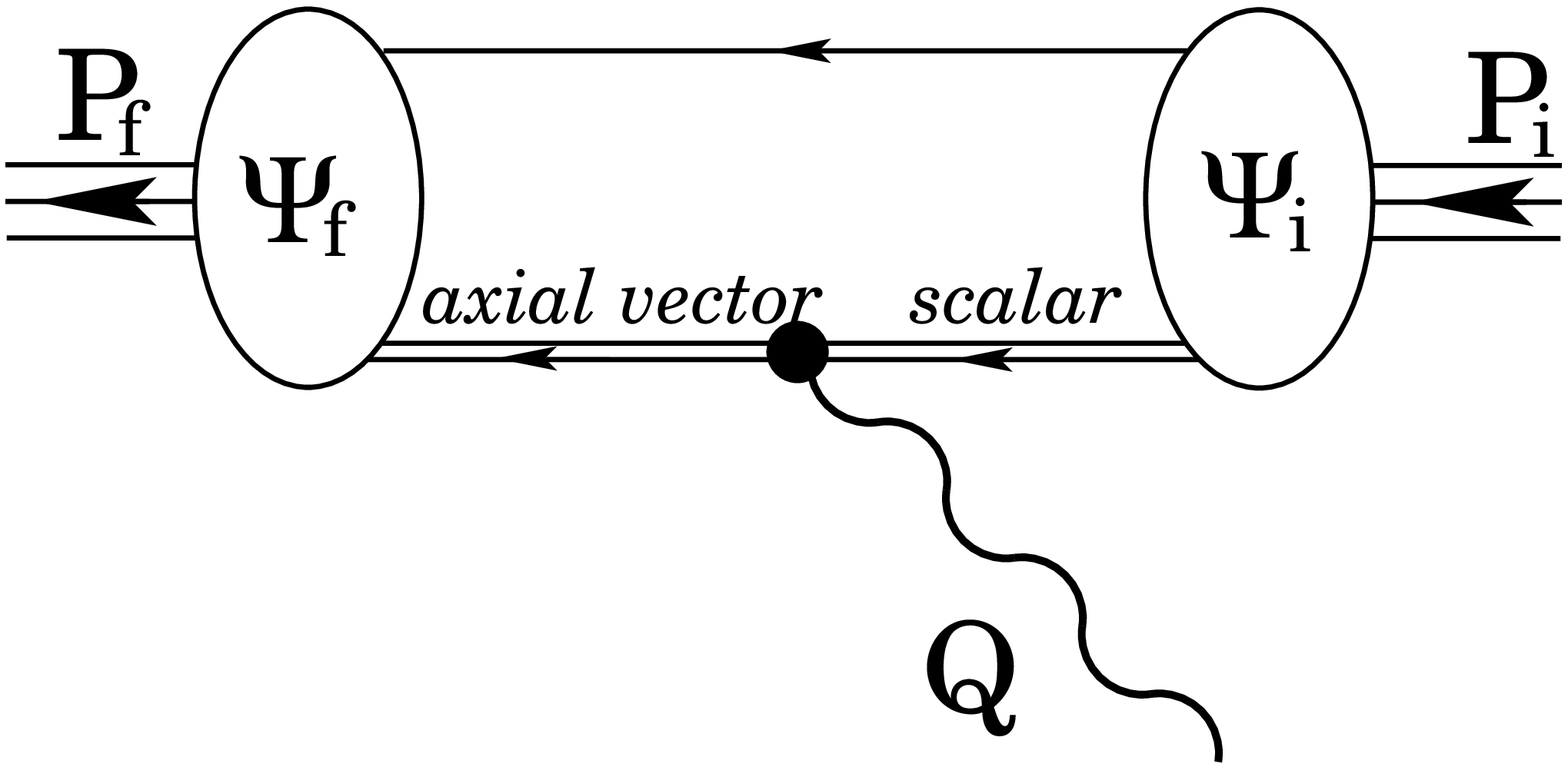}}
\end{minipage}\vspace*{3ex}

\begin{minipage}[t]{0.4\textwidth}
\leftline{\includegraphics[width=0.90\textwidth]{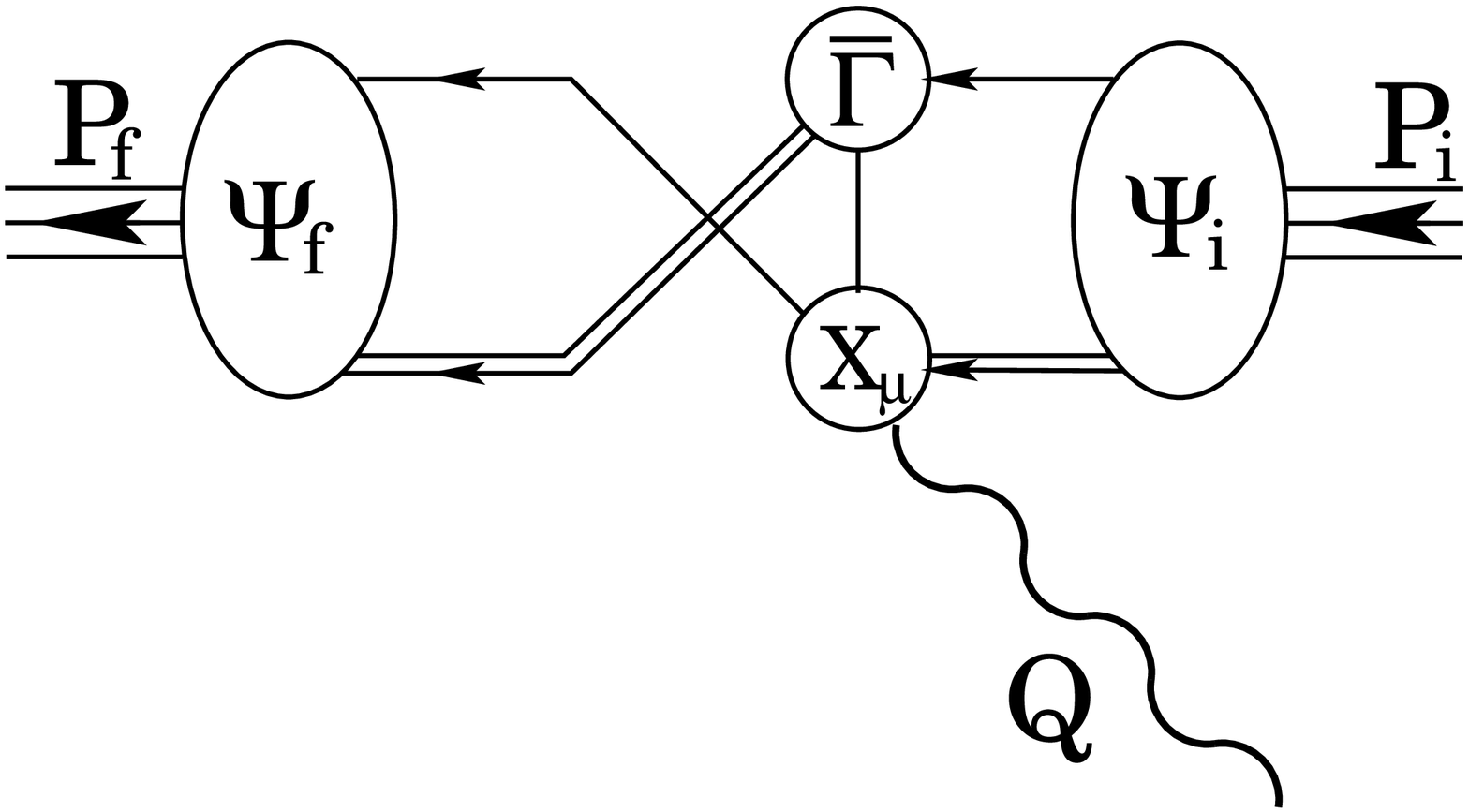}}
\end{minipage}
\begin{minipage}[t]{0.4\textwidth}
\rightline{\includegraphics[width=0.90\textwidth]{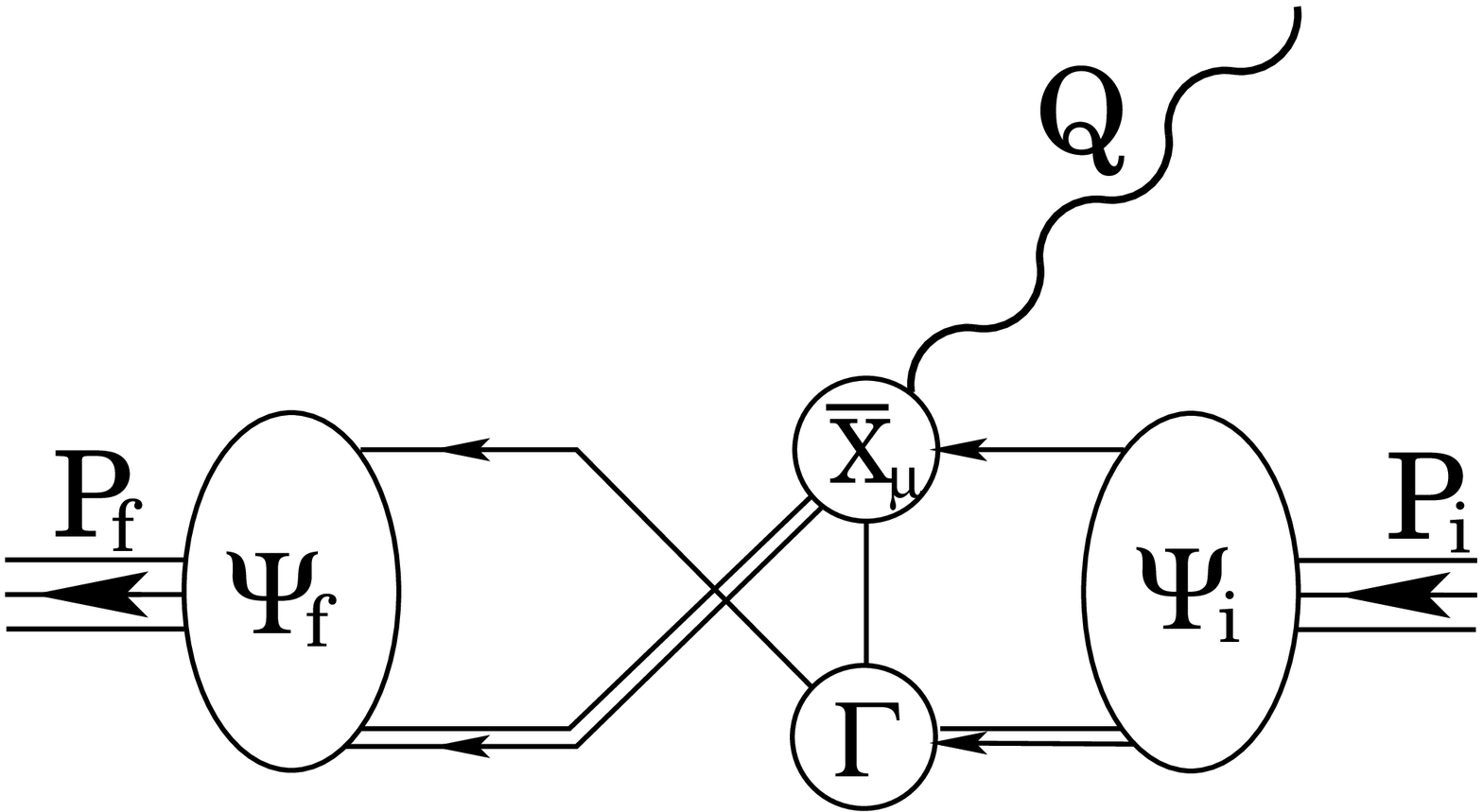}}
\end{minipage}
\end{minipage}
\caption{\label{fig:vertex} Vertex which ensures a conserved current for on-shell nucleons described by the Faddeev amplitudes, $\Psi_{i,f}$, obtained from the equation depicted in Fig.\,\protect\ref{fig:Faddeev}.  The photon probing the nucleon introduces momentum $Q$ and is represented by the wiggly line; the single line represents $S(p)$, the dressed-quark propagator, and the double line, the diquark propagator; and $\Gamma$ is the diquark Bethe-Salpeter amplitude.  The vertex in Diagram~4 (second row, right column) represents an electromagnetically induced transition between the scalar and axial-vector diquarks; and the vertices in row~3 are ``seagull terms,'' which appear as partners to Diagram~3 and arise because binding in the nucleons' Faddeev equations is, in general, effected by the exchange of a momentum-carrying dressed-quark between \textit{nonpointlike} diquark correlations \cite{Oettel:1999gc}.}
\end{figure}

A nonpointlike composite nucleon must interact with the photon via a sophisticated current, whose form is constrained by vector Ward-Takahashi identities.  This continues a thread that pervades these notes.  For the bound-state described by the Faddeev equation in Fig.\,\ref{fig:Faddeev} that current is described in Ref.\,\cite{Oettel:1999gc} and depicted in Fig.\,\ref{fig:vertex}: Diagrams~4--6 represent eight-dimensional integrals, which can be evaluated using Monte-Carlo techniques.

The pattern of the computation should now be clear.  In principle, one specifies an interaction kernel for the gap equation and solves for the dressed-quark propagator.  This completes the specification of the Bethe-Salpeter kernel, so one may compute the masses and amplitudes for the diquark correlations.  With all these quantities determined, the Faddeev equation is defined and can be solved for a baryon's mass and amplitude.  In combination with the current, Fig.\,\ref{fig:vertex}, it is a straightforward numerical task to compute the form factors.

\begin{figure}[t]
\begin{minipage}[t]{\textwidth}
\begin{minipage}[t]{0.5\textwidth}
\leftline{\includegraphics[width=0.95\textwidth]{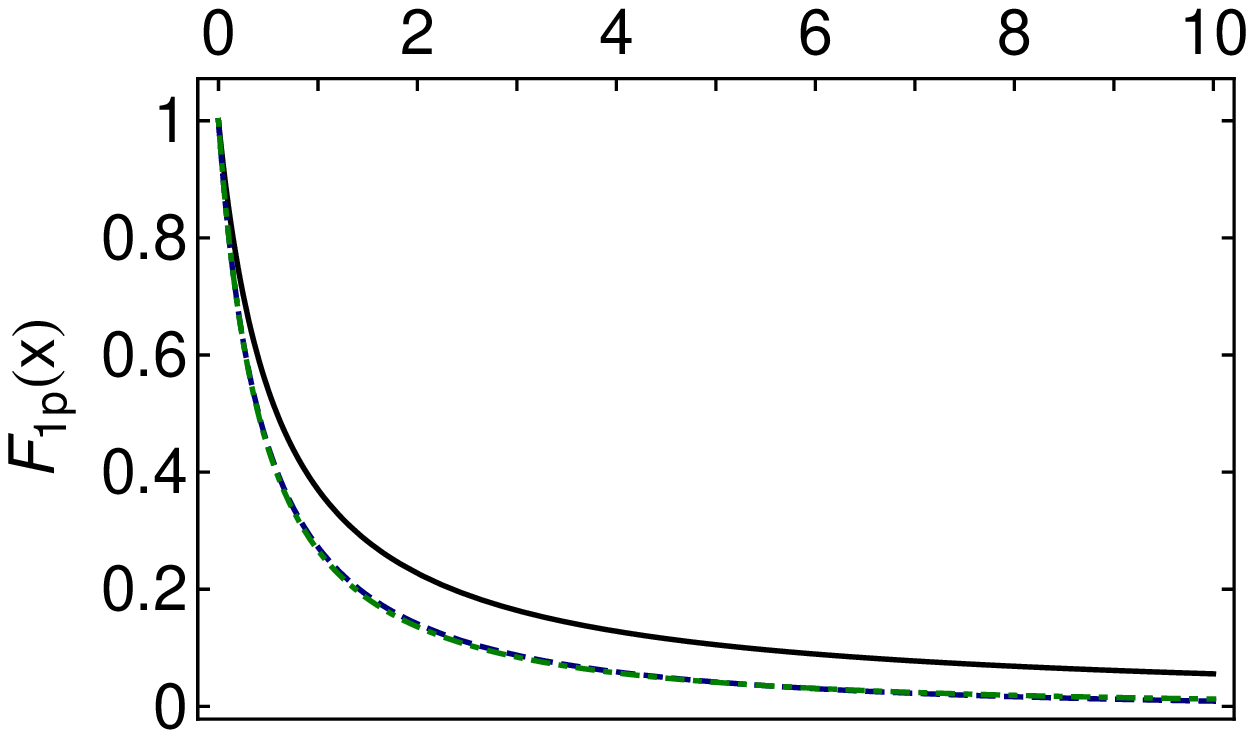}}
\end{minipage}
\begin{minipage}[t]{0.5\textwidth}
\rightline{\includegraphics[width=0.97\textwidth]{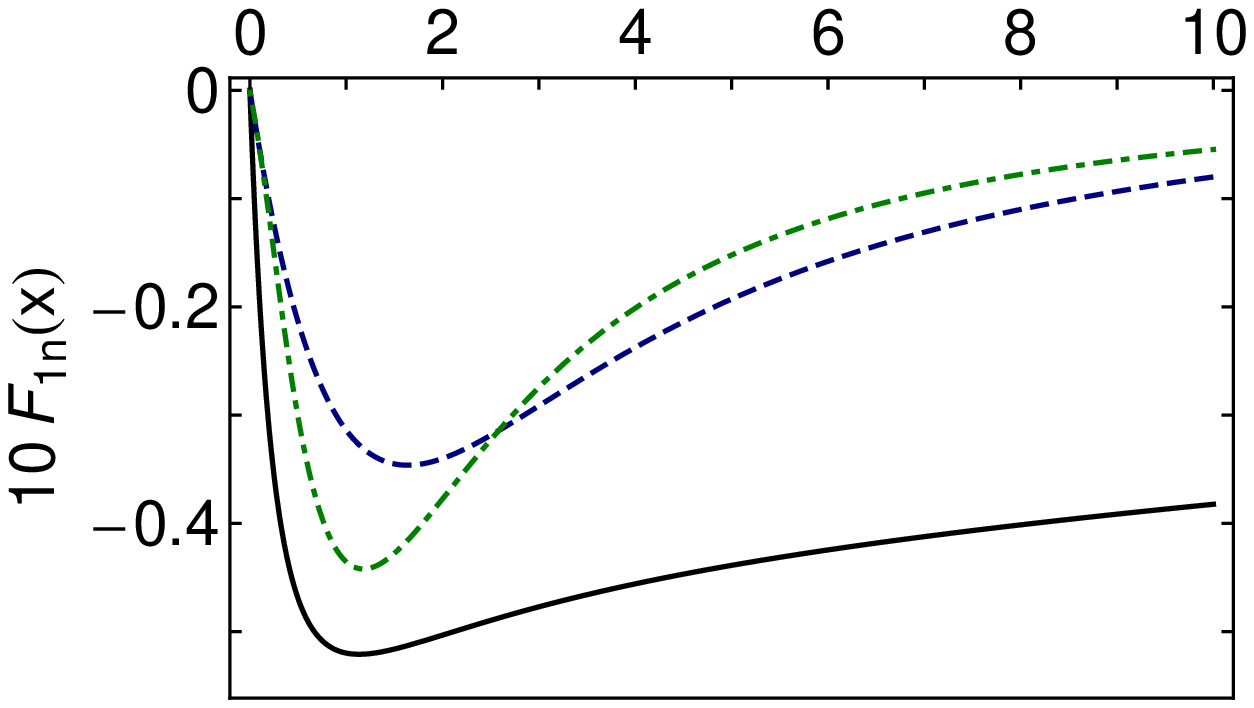}}
\end{minipage}\vspace*{-3ex}

\begin{minipage}[t]{0.5\textwidth}
\leftline{\includegraphics[width=0.95\textwidth]{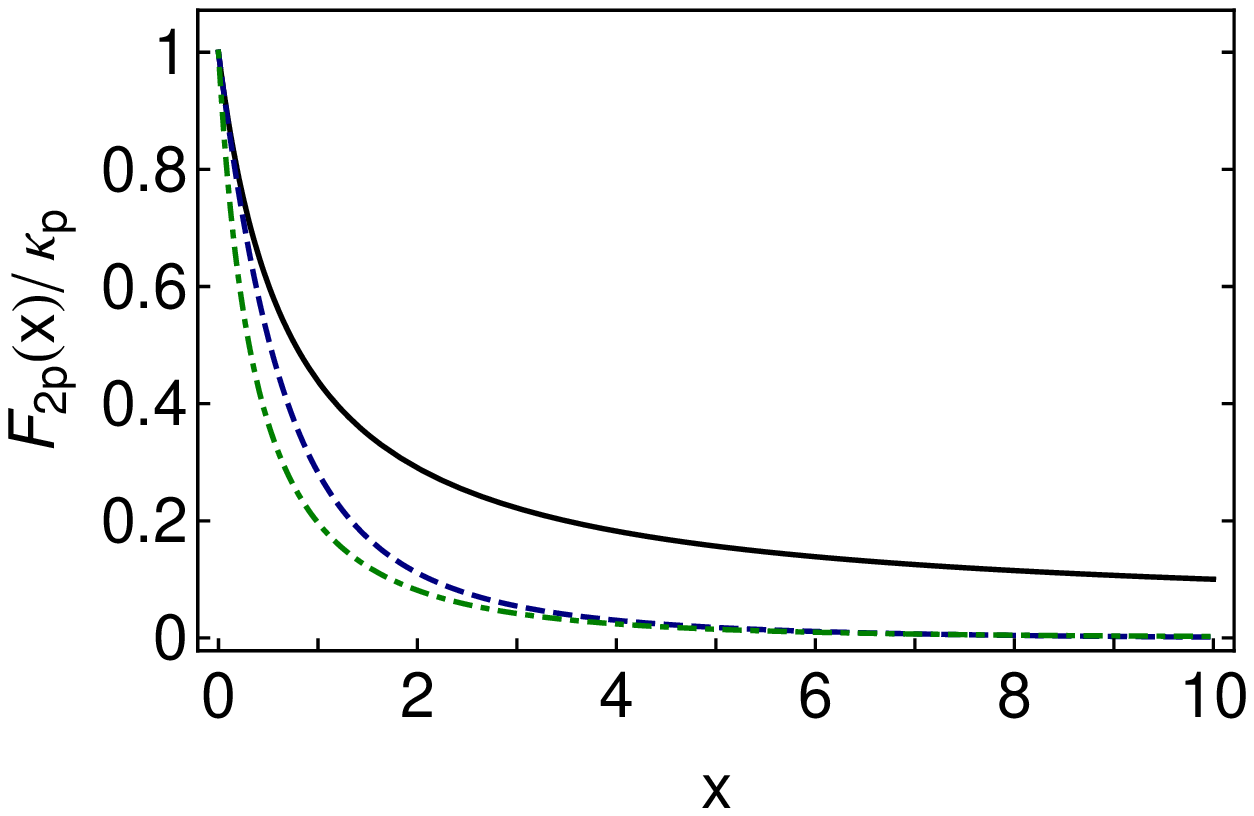}}
\end{minipage}
\begin{minipage}[t]{0.5\textwidth}
\rightline{\includegraphics[width=0.97\textwidth]{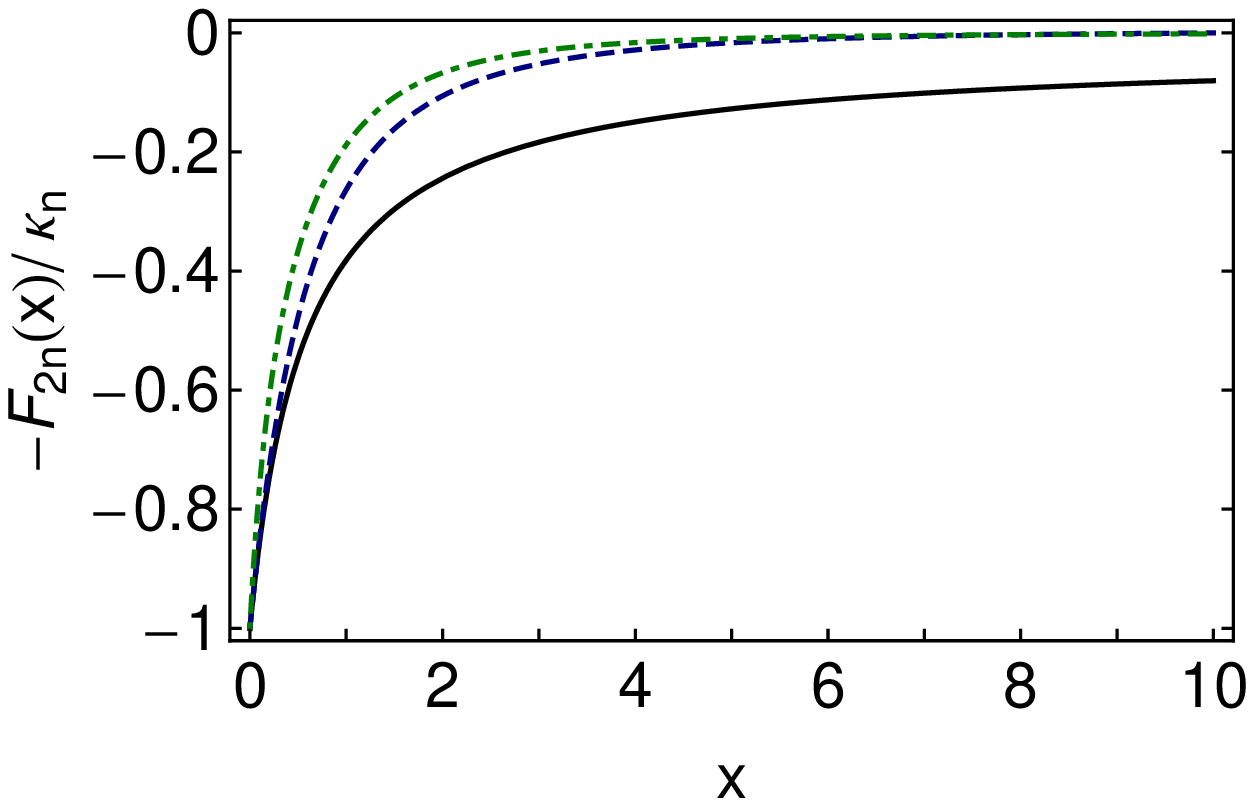}}
\end{minipage}
\end{minipage}
\caption{\label{fig:ProtonF1F2} Nucleon Dirac (upper panel) and Pauli (lower panel) form factors, as a function of $x=Q^2/m_N^2$ -- proton on left and neutron on right.
\emph{dashed curve} -- result obtained in Ref.\,\protect\cite{Cloet:2008re}, which employed QCD-like momentum-dependence for the dressed-quark propagators and diquark Bethe-Salpeter amplitudes in solving the Faddeev equation;
\emph{Solid curve} -- result obtained using a symmetry-preserving regularisation of a contact-interaction \protect\cite{Wilson:2011rj} and hence a dressed-quark mass-function and diquark Bethe-Salpeter amplitudes that are momentum-independent;
and \emph{dot-dashed curve} -- a parametrisation of experimental data \protect\cite{Kelly:2004hm}.}
\end{figure}

Some of the steps indicated here can be simplified by employing algebraic parametrisations of one or more elements (propagators and/or amplitudes).  Such parametrisations, based upon numerical solutions of the gap- and Bethe-Salpeter equations, have long been employed efficaciously \cite{Roberts:1994hh,Burden:1995ve,Roberts:1993ks}.  Following that path, Ref.\,\cite{Cloet:2008re} produced a comprehensive survey of nucleon electromagnetic form factors, the quality of which is characterised by the results depicted in Fig.\,\ref{fig:ProtonF1F2}.  With these results, Ref.\,\cite{Cloet:2008re} unified the computation of meson and nucleon form factors, and also their valence-quark distribution functions (see Sec.\,\ref{sec:VPDF}).

Figure~\ref{fig:ProtonF1F2} also depicts nucleon form factors obtained \cite{Wilson:2011rj} with a symmetry-preserving DSE-treatment of the contact interaction in Eq.\,\eqref{DnjlGnjl}.  These form factors characterise a nucleon that is constructed from diquarks whose Bethe-Salpeter amplitudes are momentum-independent and dressed-quarks with a momentum-independent mass-function, which inputs to the Faddeev equation yield a bound-state described by a momentum-independent Faddeev amplitude.  This last is the hallmark of a pointlike composite particle and explains the hardness of the computed form factors, which is evident in Figs.\,\ref{fig:ProtonF1F2}.  The hardness contrasts starkly with results obtained from a momentum-dependent Faddeev amplitude produced by dressed-quark propagators and diquark Bethe-Salpeter amplitudes with QCD-like momentum-dependence; and with experiment.  As I have remarked above, evidence for a connection between the momentum-dependence of each of these elements and the behaviour of QCD's $\beta$-function is accumulating; e.g., Refs.\,\cite{Roberts:2010rn,GutierrezGuerrero:2010md,Roberts:2011wy,Maris:2000sk,%
Eichmann:2008ef,Eichmann:2011vu,Bhagwat:2006pu}.
The comparisons in Figs.\,\ref{fig:ProtonF1F2} add to this evidence, in connection here with readily accessible observables, and support the view, presented herein, that experiment is a sensitive probe of the running of the $\beta$-function to infrared momenta.

\begin{figure}[t]
\includegraphics[clip,width=0.55\textwidth]{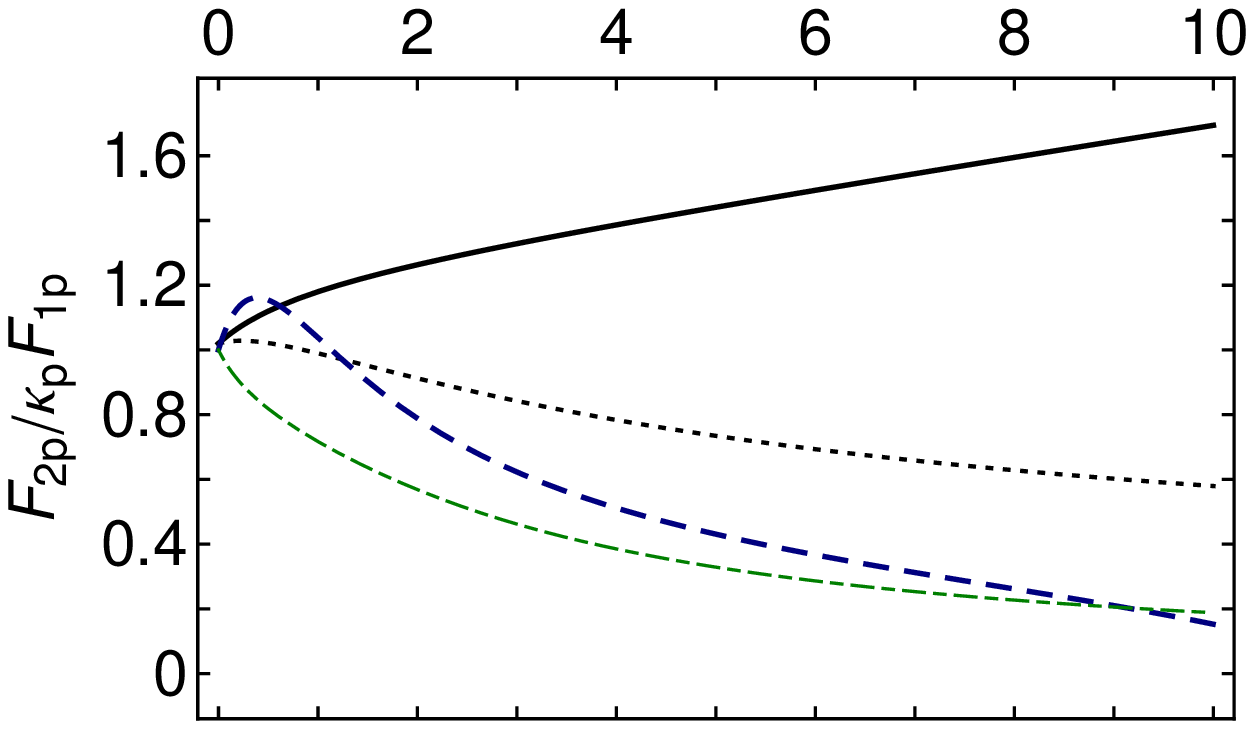}
\\[-12.4ex]

\includegraphics[clip,width=0.55\textwidth]{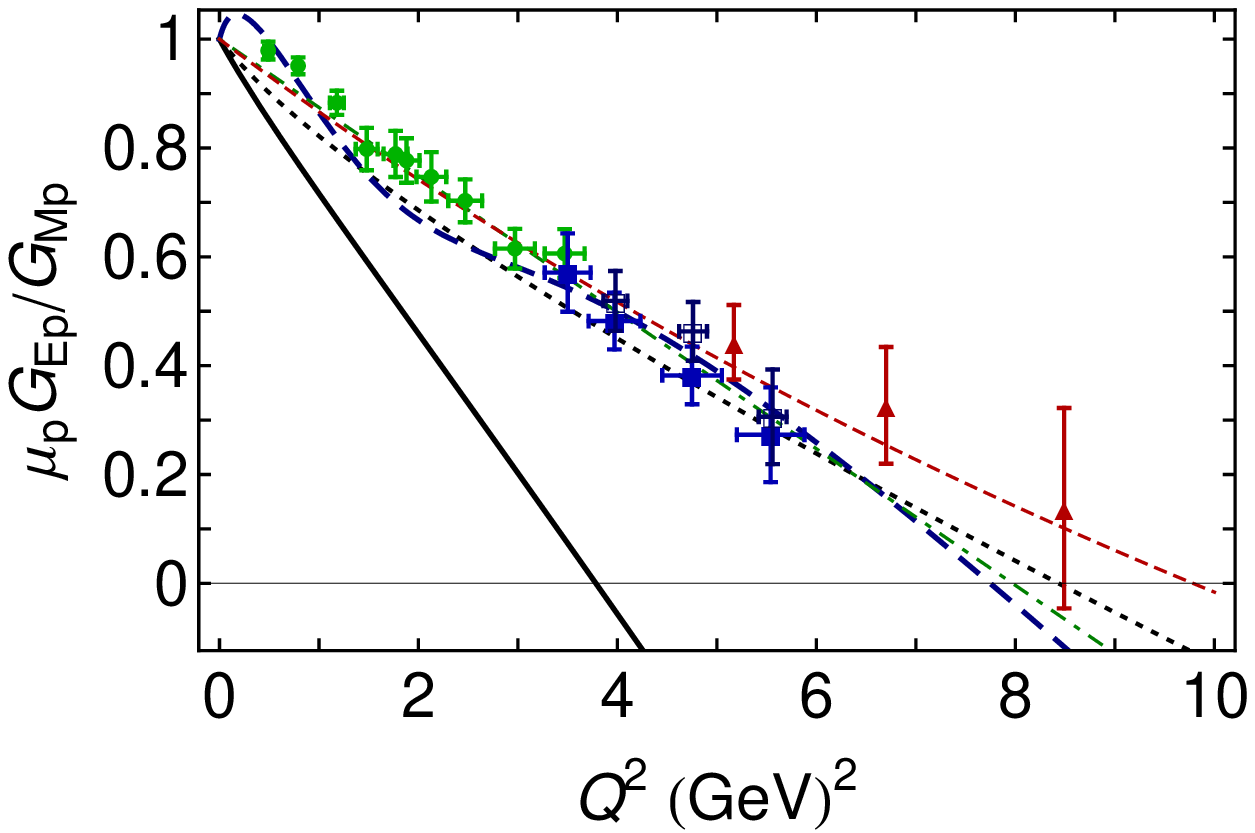}
\caption{\label{fig:GEGMp}
\emph{Upper panel}: Normalised ratio of proton Pauli and Dirac form factors.  \emph{Solid curve} -- contact interaction \protect\cite{Wilson:2011rj}; \emph{long-dashed curve} -- result from Ref.\,\protect\cite{Chang:2011tx}, which employed QCD-like momentum-dependence for the dressed-quark propagators and diquark Bethe-Salpeter amplitudes; \emph{long-dash-dotted curve} -- drawn from parametrisation of experimental data in Ref.\,\protect\cite{Kelly:2004hm}; and \emph{dotted curve} -- softened contact-interaction result, described in connection with Eq.\,\eqref{SoftF2p}.
\emph{Lower panel}: Normalised ratio of proton Sachs electric and magnetic form factors.  \emph{Solid curve} and \emph{long-dashed curve}, as above; \emph{dot-dashed curve} -- linear fit to data in Refs.\,\protect\cite{Jones:1999rz,Gayou:2001qd,Qattan:2004ht,Punjabi:2005wq,Puckett:2010ac}, constrained to one at $Q^2=0$; \emph{short-dashed curve} -- $[1,1]$-Pad\'e fit to that data; and \emph{dotted curve} -- softened contact-interaction result, described in connection with Eq.\,\eqref{SoftF2p}.
In addition, we have represented a selection of data explicitly: filled-squares \protect\cite{Gayou:2001qd}; circles \protect\cite{Punjabi:2005wq}; up-triangles \protect\cite{Puckett:2010ac}; and open-squares \protect\cite{Puckett:2011xg}.
}
\end{figure}

This returns us to the ratio $G_E^p(Q^2)/G_M^p(Q^2)$ described in the introduction to this section and depicted in Fig.\,\ref{fig:GEGMp}.  One observes that, independent of the nature of the interaction, computations of this ratio exhibit a zero.  Its location, however, is very sensitive to the interaction, although not to the electromagnetic size of the diquark correlations \cite{Cloet:2008re}.  (I note that the DSE prediction of the same ratio for the neutron has been confirmed by recent experiments \cite{Riordan:2010id}, as also is the trend of the nucleon's flavour separated Dirac and Pauli form factors \cite{Wilson:2011rj,Riordan:2010id,Cates:2011pz}.)

In order to assist in explaining the origin and location of a zero in the Sachs form factor ratio, in the top panel of Fig.\,\ref{fig:GEGMp} I depict the ratio of Pauli and Dirac form factors: both the actual contact-interaction result and that obtained when the Pauli form factor is artificially ``softened;'' viz.,
\begin{equation}
\label{SoftF2p}
F_{2p}(Q^2) \to \frac{F_{2p}(Q^2)}{1+Q^2/(4 m_N^2)}\,.
\end{equation}
As observed in Ref.\,\cite{Bloch:2003vn}, a softening of the proton's Pauli form factor has the effect of shifting the zero to larger values of $Q^2$.  In fact, if $F_{2p}$ becomes soft quickly enough, then the zero disappears completely.

The Pauli form factor is a gauge of the distribution of magnetisation within the proton. Ultimately, this magnetisation is carried by the dressed-quarks and influenced by correlations amongst them, which are expressed in the Faddeev wave-function.
If the dressed-quarks are described by a momentum-independent mass-function, then they behave as Dirac particles with constant Dirac values for their magnetic moments and produce a hard Pauli form factor.
%
Alternatively, suppose that the dressed-quarks possess a momentum-dependent mass-function, which is large at infrared momenta but vanishes as their momentum increases.  At small momenta they will then behave as constituent-like particles with a large magnetic moment, but their mass and magnetic moment will drop toward zero as the probe momentum grows.  (N.B.\ As described in Sec.\,\ref{spectrum2}, massless fermions do not possess a measurable magnetic moment.)  Such dressed-quarks will produce a proton Pauli form factor that is large for $Q^2 \sim 0$ but drops rapidly on the domain of transition between nonperturbative and perturbative QCD, to give a very small result at large-$Q^2$.  The precise form of the $Q^2$-dependence will depend on the evolving nature of the angular momentum correlations between the dressed-quarks.
From this perspective, existence, and location if so, of the zero in $\mu_p G_{Ep}(Q^2)/G_{Mp}(Q^2)$ are a fairly direct measure of the location and width of the transition region between the nonperturbative and perturbative domains of QCD as expressed in the momentum-dependence of the dressed-quark mass-function.

One may expect that a mass-function which rapidly becomes partonic -- namely, is very soft -- will not produce a zero; has seen that a constant mass-function produces a zero at a small value of $Q^2$, and knows that a mass-function which resembles that obtained in the best available DSE studies \cite{Qin:2011dd,Bhagwat:2006tu} and via lattice-QCD simulations \cite{Bowman:2005vx}, produces a zero at a location that is consistent with extant data.  There is plainly an opportunity here for very constructive feedback between future experiments and theory.  It is anticipated that experiments at JLab in the 12\,GeV era will establish conclusively whether or not this ratio possesses a zero.  The result will assist greatly in refining understanding of the dressed-quark mass function and therefrom QCD's $\beta$-function.

\subsection{Valence-quark distributions at $x=1$}
\label{sec:VPDF}
Before closing this section I would like to exploit a connection between the $Q^2=0$ values of elastic form factors and the Bjorken-$x=1$ values of the dimensionless structure functions of deep inelastic scattering, $F_2^{n,p}(x)$.  First recall that the $x=1$ value of a structure function is invariant under the evolution equations, Sec.\,\ref{sec:uvx}.  Hence the value of
\begin{equation}
\label{dvuv1}
\left. \frac{d_v(x)}{u_v(x)}\right|_{x\to 1}\rule{-0.5em}{0ex}, \;\mbox{where} \rule{1em}{0ex}
\frac{d_v(x)}{u_v(x)} =
\frac{4 \frac{F_2^n(x)}{F_2^p(x)} - 1}{4- \frac{F_2^n(x)}{F_2^p(x)}},
\end{equation}
is a scale-invariant feature of QCD and a discriminator between models.  Next, when Bjorken-$x$ is unity, then $Q^2+2P\cdot Q=0$; i.e., one is dealing with elastic scattering.  Therefore, in the neighbourhood of $x=1$ the structure functions are determined by the target's elastic form factors.
The ratio in Eq.\,\eqref{dvuv1} expresses the relative probability of finding a $d$-quark carrying all the proton's light-front momentum compared with that of a $u$-quark doing the same or, equally, owing to invariance under evolution, the relative probability that a $Q^2=0$ probe either scatters from a $d$-quark or a $u$-quark; viz.,
\begin{equation}
\label{dvuvF1}
\left. \frac{d_v(x)}{u_v(x)}\right|_{x\to 1} = \frac{P_{1}^{p,d}}{P_{1}^{p,u}}.
\end{equation}
%

In constituent-quark models with $SU(6)$-symmetric spin-flavour wave-functions the right-hand-side of Eq.\,\eqref{dvuvF1} is $1/2$ because there is nothing to distinguish between the wave-functions of $u$- and $d$-quarks, and the proton is constituted from $u$-quarks and one $d$-quark.  On the other hand, when a Poincar\'e-covariant Faddeev equation is employed to describe the nucleon,
\begin{equation}
\label{dvuvF1result}
\frac{P_{1}^{p,d}}{P_{1}^{p,u}} =
\frac{\frac{2}{3} P_1^{p,a} + \frac{1}{3} P_1^{p,m}}
{P_1^{p,s}+\frac{1}{3} P_1^{p,a} + \frac{2}{3} P_1^{p,m}},
\end{equation}
where I have used the notation of Ref.\,\cite{Cloet:2008re}.  Namely,
$P_1^{p,s}=F_{1p}^s(Q^2=0)$ is the contribution to the proton's charge arising from diagrams with a scalar diquark component in both the initial and final state: $u[ud]\otimes \gamma \otimes u[ud]$.  The diquark-photon interaction is far softer than the quark-photon interaction and hence this diagram contributes solely to $u_v$ at $x=1$.
$P_1^{p,a}=F_{1p}^a(Q^2=0)$, is the kindred axial-vector diquark contribution; viz., $2 d\{uu\}\otimes \gamma\otimes d\{uu\}+u\{ud\} \otimes\gamma \otimes u\{ud\}$.  At $x=1$ this contributes twice as much to $d_v$ as it does to $u_v$.
$P_1^{p,m}=F_{1p}^m(Q^2=0)$, is the contribution to the proton's charge arising from diagrams with a different diquark component in the initial and final state.  The existence of this contribution relies on the exchange of a quark between the diquark correlations and hence it contributes twice as much to $u_v$ as it does to $d_v$.  If one uses the ``static approximation'' to the nucleon form factor, Eq.\,\eqref{staticexchange}, as with the treatment of the contact-interaction in Ref.\,\cite{Wilson:2011rj}, then $P_1^{p,m}\equiv 0$.  It is plain from Eq.\,\eqref{dvuvF1result} that $d_v/u_v=0$ in the absence of axial-vector diquark correlations; i.e., in scalar-diquark-only models of the nucleon, which were once common and, despite their weaknesses, still too often employed.

Using the probabilities presented in Refs.\,\cite{Cloet:2008re,Wilson:2011rj}, respectively, one obtains:
\begin{equation}
\label{compdvonuv}
\begin{array}{l|ccccc}
 & P_1^{p,s} & P_1^{p,a} & P_1^{p,m} & \frac{d_v}{u_v} & \frac{F_2^n}{F_2^p} \\\hline
M(p^2) & 0.60 & 0.25 & 0.15 & 0.28 & 0.49\\
%
%
\mbox{M=constant} &  0.78 & 0.22 & 0\rule{1.2em}{0ex} & 0.18 & 0.41\\
\end{array}\;,
\end{equation}
Both rows in Eq.\,\eqref{compdvonuv} are consistent with $d_v/u_v= 0.23\pm 0.09$ (90\% confidence level, $F_2^n/F_2^p = 0.45 \pm 0.08$) inferred recently via consideration of electron-nucleus scattering at $x>1$ \cite{Hen:2011rt}.  On the other hand, this is also true of the result obtained through a naive consideration of the isospin and helicity structure of a proton's light-front quark wave function at $x\sim 1$, which leads one to expect that $d$-quarks are five-times less likely than $u$-quarks to possess the same helicity as the proton they comprise; viz., $d_v/u_v=0.2$ \cite{Farrar:1975yb}.  Plainly, contemporary experiment-based analyses do not provide a particularly discriminating constraint.  Future experiments with a tritium target should help \cite{Holt:2010zz}, emphasising again the critical interplay between experiment and theory in elucidating the nature of the strong interaction.

\section{Epilogue}
QCD is the most interesting part of the Standard Model and Nature's only example of an essentially nonperturbative fundamental theory.  Whilst confinement remains a puzzle, it is recognised that dynamical chiral symmetry breaking (DCSB) is a fact in QCD.  It is manifest in dressed-propagators and vertices, and, amongst other things, it is responsible for:
the transformation of the light current-quarks in QCD's Lagrangian into heavy constituent-like quarks, in terms of which order was first brought to the hadron spectrum;
the unnaturally small values of the masses of light-quark pseudoscalar mesons and the $\eta$-$\eta^\prime$ splitting;
the unnaturally strong coupling of pseudoscalar mesons to light-quarks -- $g_{\pi \bar q q} \approx 4.3$;
and the unnaturally strong coupling of pseudoscalar mesons to the lightest baryons -- $g_{\pi \bar N N} \approx 12.8 \approx 3 g_{\pi \bar q q}$.

Herein I have illustrated the dramatic impact that DCSB has upon observables in hadron physics.  A ``smoking gun'' for DCSB is the behaviour of the dressed-quark mass function.  The momentum dependence manifest in Fig.\,\ref{gluoncloud} is an essentially quantum field theoretical effect.  Exposing and elucidating its consequences therefore requires a nonperturbative and symmetry-preserving approach, where the latter means preserving Poincar\'e covariance, chiral and electromagnetic current-conservation, etc.  The Dyson-Schwinger equations (DSEs) provide such a framework.  I have explained the nature of some of the experimental and theoretical studies that are underway which can potentially identify observable signals of $M(p^2)$ and thereby confirm and explain the mechanism responsible for the vast bulk of visible mass in the Universe.

Along the way I have described a number of exact results proved in QCD using the DSEs, amongst them:
\begin{itemize}
\item Light-quark confinement is a dynamical phenomenon, which cannot in principle be expressed via a potential;
\item Goldstone's theorem is fundamentally an expression of equivalence between the one-body problem and the two-body problem in the pseudoscalar channel;
\item quarks are not Dirac particles -- they possess anomalous chromo- and electro-magnetic moments which are large at infrared momenta;
\item and gluons are nonperturbatively massive, being described by a mass-function which is large in the infrared but diminishes with power-law behaviour in the ultraviolet.
\end{itemize}
Numerous items could be added to this list, some of which are described above.

There are many reasons why this is an exciting time.  As per my brief, I have focused on one.  Namely, through the DSEs, one is unifying phenomena as apparently diverse as: the hadron spectrum; hadron elastic and transition form factors, from small- to large-$Q^2$; and parton distribution functions.  The key is an understanding of both the fundamental origin of visible mass and the far-reaching consequences of the mechanism responsible; namely, DCSB.  Through continuing feedback between experiment and theory, these studies should lead us to an explanation of confinement, the phenomenon that makes nonperturbative QCD the most interesting piece of the Standard Model.  They might also provide an understanding of nonperturbative physics that enables the formulation of a realistic extension of that model.

\subsection*{Acknowledgments}
I am grateful to the organisers for the opportunity to be involved in this conference on
\emph{Dyson-Schwinger Equations and Fa\`{a} di Bruno Hopf Algebras in Physics and Combinatorics (DSFdB2011)}, for the financial support that enabled my participation and, above all, for their kindness and hospitality.
The original material described in this contribution was drawn from collaborations and discussions with A.~Bashir, S.\,J.~Brodsky, L.~Chang, C.~Chen, H.~Chen, I.\,C.~Clo\"et, B.~El-Bennich, X.~Guti\'{e}rrez-Guerrero, R.\,J.~Holt, M.\,A.~Ivanov, Y.-x.~Liu, V.~Mokeev, T.~Nguyen, S.-x.~Qin, H.\,L.\,L.~Roberts, R.~Shrock, P.\,C.~Tandy and D.\,J.~Wilson;
and the work was also supported by
U.\,S.\ Department of Energy, Office of Nuclear Physics, contract no.~DE-AC02-06CH11357.
%


\bibliographystyle{ieeetr}
\bibliography{CDRobertsIRMA}

\begin{thebibliography}{100}

\bibitem{Connes:1999yr}
A.~Connes and D.~Kreimer, ``{Renormalization in quantum field theory and the
  Riemann-Hilbert problem. 1. The Hopf algebra structure of graphs and the main
  theorem},'' {\em Commun. Math. Phys.}, vol.~210, pp.~249--273, 2000.

\bibitem{vanBaalen:2009hu}
G.~van Baalen, D.~Kreimer, D.~Uminsky, and K.~Yeats, ``{The QCD beta-function
  from global solutions to Dyson-Schwinger equations},'' {\em Annals Phys.},
  vol.~325, pp.~300--324, 2010.

\bibitem{Roberts:1994dr}
C.~D. Roberts and A.~G. Williams, ``{Dyson-Schwinger equations and their
  application to hadronic physics},'' {\em Prog. Part. Nucl. Phys.}, vol.~33,
  pp.~477--575, 1994.

\bibitem{Roberts:2000aa}
C.~D. Roberts and S.~M. Schmidt, ``{Dyson-Schwinger equations: Density,
  temperature and continuum strong QCD},'' {\em Prog. Part. Nucl. Phys.},
  vol.~45, pp.~S1--S103, 2000.

\bibitem{Maris:2003vk}
P.~Maris and C.~D. Roberts, ``{Dyson-Schwinger equations: A tool for hadron
  physics},'' {\em Int. J. Mod. Phys.}, vol.~E12, pp.~297--365, 2003.

\bibitem{Pennington:2005be}
M.~R. Pennington, ``{Swimming with quarks},'' {\em J. Phys. Conf. Ser.},
  vol.~18, pp.~1--73, 2005.

\bibitem{Holl:2006ni}
A.~H{\"o}ll, C.~D. Roberts, and S.~V. Wright, ``{Hadron physics and
  Dyson-Schwinger equations},'' nucl-th/0601071.

\bibitem{Fischer:2006ub}
C.~S. Fischer, ``{Infrared properties of QCD from Dyson-Schwinger equations},''
  {\em J. Phys.}, vol.~G32, pp.~R253--R291, 2006.

\bibitem{Roberts:2007jh}
C.~D. Roberts, M.~S. Bhagwat, A.~H{\"o}ll, and S.~V. Wright, ``{Aspects of
  hadron physics},'' {\em Eur. Phys. J. ST}, vol.~140, pp.~53--116, 2007.

\bibitem{Roberts:2007ji}
C.~D. Roberts, ``{Hadron Properties and Dyson-Schwinger Equations},'' {\em
  Prog. Part. Nucl. Phys.}, vol.~61, pp.~50--65, 2008.

\bibitem{Holt:2010vj}
R.~J. Holt and C.~D. Roberts, ``{Distribution Functions of the Nucleon and Pion
  in the Valence Region},'' {\em Rev. Mod. Phys.}, vol.~82, pp.~2991--3044,
  2010.

\bibitem{Chang:2010jq}
L.~Chang and C.~D. Roberts, ``{Hadron Physics: The Essence of Matter},'' {\em
  AIP Conf. Proc.}, vol.~1361, pp.~91--114, 2011.

\bibitem{Swanson:2010pw}
E.~S. Swanson, ``{A Primer on Functional Methods and the Schwinger-Dyson
  Equations},'' {\em AIP Conf. Proc.}, vol.~1296, pp.~75--121, 2010.

\bibitem{Chang:2011vu}
L.~Chang, C.~D. Roberts, and P.~C. Tandy, ``{Selected highlights from the study
  of mesons},'' {\em Chin. J. Phys.}, vol.~49, pp.~955--1004, 2011.

\bibitem{Boucaud:2011ug}
P.~Boucaud {\em et~al.}, ``{The Infrared Behaviour of the Pure Yang-Mills Green
  Functions},'' arXiv:1109.1936 [hep-ph].

\bibitem{GellMann:1964nj}
M.~Gell-Mann, ``{A Schematic Model of Baryons and Mesons},'' {\em Phys. Lett.},
  vol.~8, pp.~214--215, 1964.

\bibitem{Zweig:2010jf}
G.~Zweig, ``{Memories of Murray and the Quark Model},'' {\em Int.J.Mod.Phys.},
  vol.~A25, pp.~3863--3877, 2010.

\bibitem{Nobel65}
S.~Lundqvist~(Editor), {\em Nobel Lectures in Physics (1963-1970)},
  pp.~121--180.
\newblock World Scientific, Singapore, 1998.

\bibitem{Nobel79}
S.~Lundqvist~(Editor), {\em Nobel Lectures in Physics (1971-1980)},
  pp.~485--560.
\newblock World Scientific, Singapore, 1994.

\bibitem{Politzer:1973fx}
H.~Politzer, ``{Reliable Perturbative Results for Strong Interactions?},'' {\em
  Phys. Rev. Lett.}, vol.~30, pp.~1346--1349, 1973.

\bibitem{Politzer:1974fr}
H.~Politzer, ``{Asymptotic Freedom: An Approach to Strong Interactions},'' {\em
  Phys. Rept.}, vol.~14, pp.~129--180, 1974.

\bibitem{Gross:1973id}
D.~Gross and F.~Wilczek, ``{Ultraviolet Behavior of Nonabelian Gauge
  Theories},'' {\em Phys. Rev. Lett.}, vol.~30, pp.~1343--1346, 1973.

\bibitem{Davis:1968cp}
J.~Davis, Raymond, D.~S. Harmer, and K.~C. Hoffman, ``{Search for neutrinos
  from the sun},'' {\em Phys.Rev.Lett.}, vol.~20, pp.~1205--1209, 1968.

\bibitem{Cleveland:1998nv}
B.~Cleveland, T.~Daily, J.~Davis, Raymond, J.~R. Distel, K.~Lande, {\em
  et~al.}, ``{Measurement of the solar electron neutrino flux with the
  Homestake chlorine detector},'' {\em Astrophys.J.}, vol.~496, pp.~505--526,
  1998.

\bibitem{Ahmad:2001an}
Q.~Ahmad {\em et~al.}, ``{Measurement of the rate of $\nu_e + d \to p + p +
  e^-$ interactions produced by $^8B$ solar neutrinos at the Sudbury Neutrino
  Observatory},'' {\em Phys.Rev.Lett.}, vol.~87, p.~071301, 2001.

\bibitem{Pontecorvo:1957cp}
B.~Pontecorvo, ``{Mesonium and anti-mesonium},'' {\em Sov. Phys. JETP}, vol.~6,
  p.~429, 1957.

\bibitem{Maki:1962mu}
Z.~Maki, M.~Nakagawa, and S.~Sakata, ``{Remarks on the unified model of
  elementary particles},'' {\em Prog. Theor. Phys.}, vol.~28, pp.~870--880,
  1962.

\bibitem{Pontecorvo:1967fh}
B.~Pontecorvo, ``{Neutrino Experiments and the Problem of Conservation of
  Leptonic Charge},'' {\em Sov. Phys. JETP}, vol.~26, pp.~984--988, 1968.

\bibitem{Sakharov:1967dj}
A.~D. Sakharov, ``{Violation of CP Invariance, c Asymmetry, and Baryon
  Asymmetry of the Universe},'' {\em Pisma Zh. Eksp. Teor. Fiz.}, vol.~5,
  pp.~32--35, 1967.

\bibitem{Murayama:2007ek}
H.~Murayama, ``{Physics Beyond the Standard Model and Dark Matter},'' {\em
  arXiv:0704.2276 [hep-ph]}, pp.~1--61, 2007.

\bibitem{Hilbert:1900}
D.~Hilbert, ``{Mathematical Problems},'' {\em Bull.\ Amer.\ Math.\ Soc.},
  vol.~8, pp.~437--479.

\bibitem{1538-3881-116-3-1009}
A.~G. Riess {\em et~al.}, ``Observational evidence from supernovae for an
  accelerating universe and a cosmological constant,'' {\em The Astronomical
  Journal}, vol.~116, no.~3, p.~1009, 1998.

\bibitem{0004-637X-517-2-565}
S.~Perlmutter {\em et~al.}, ``Measurements of \mbox{$\Omega$} and
  \mbox{$\Lambda$} from 42 high-redshift supernovae,'' {\em The Astrophysical
  Journal}, vol.~517, no.~2, p.~565, 1999.

\bibitem{Rakow:1990jv}
P.~E.~L. Rakow, ``{Renormalization group flow in QED: An investigation of the
  SchwingeR-Dyson equations},'' {\em Nucl. Phys.}, vol.~B356, pp.~27--45, 1991.

\bibitem{Reenders:1999bg}
M.~Reenders, ``{On the nontriviality of Abelian gauged Nambu-Jona-Lasinio
  models in four dimensions},'' {\em Phys. Rev.}, vol.~D62, p.~025001, 2000.

\bibitem{Jaffe:Clay}
A.~M. Jaffe, ``{The Millennium Grand Challenge in Mathematics},'' {\em Notices
  of the American Mathematical Society}, vol.~53, no.~6, pp.~652--660.

\bibitem{Wilson:1974sk}
K.~G. Wilson, ``{Confinement of quarks},'' {\em Phys. Rev.}, vol.~D10,
  pp.~2445--2459, 1974.

\bibitem{1962hep..conf..845O}
L.~B. {Okun}, ``{The theory of weak interaction},'' in {\em 1962 International
  Conference on High-Energy Physics at CERN} ({J.~Prentki}, ed.), p.~845, 1962.

\bibitem{Keister:1991sb}
B.~D. Keister and W.~N. Polyzou, ``{Relativistic Hamiltonian dynamics in
  nuclear and particle physics},'' {\em Adv. Nucl. Phys.}, vol.~20,
  pp.~225--479, 1991.

\bibitem{Coester:1992cg}
F.~Coester, ``{Null plane dynamics of particles and fields},'' {\em Prog. Part.
  Nucl. Phys.}, vol.~29, pp.~1--32, 1992.

\bibitem{Hecht:2000xa}
M.~B. Hecht, C.~D. Roberts, and S.~M. Schmidt, ``{Valence-quark distributions
  in the pion},'' {\em Phys. Rev.}, vol.~C63, p.~025213, 2001.

\bibitem{Aicher:2010cb}
M.~Aicher, A.~Schafer, and W.~Vogelsang, ``{Soft-Gluon Resummation and the
  Valence Parton Distribution Function of the Pion},'' {\em Phys.\ Rev.\
  Lett.}, vol.~105, p.~252003, 2010.

\bibitem{Nguyen:2011jy}
T.~Nguyen, A.~Bashir, C.~D. Roberts, and P.~C. Tandy, ``{Pion and kaon
  valence-quark parton distribution functions},'' {\em Phys. Rev.}, vol.~C83,
  p.~062201(R), 2011.

\bibitem{Roberts:2010rn}
H.~L.~L. Roberts, C.~D. Roberts, A.~Bashir, L.~X. Guti{\'e}rrez-Guerrero, and
  P.~C. Tandy, ``{Abelian anomaly and neutral pion production},'' {\em Phys.
  Rev.}, vol.~C82, p.~065202, 2010.

\bibitem{Symanzik69}
K.~Symanzik, ``{Euclidean quantum field theory},'' in {\em Local Quantum
  Theory} ({R. Jost}, ed.), Academic, New York, 1969.
\newblock Proceedings of the International School of Physics \mbox{``Enrico
  Fermi''}, Course XLV.

\bibitem{SW80}
R.~F. Streater and A.~S. Wightman, {\em PCT, Spin and Statistics, and All
  That}.
\newblock Addison-Wesley, Reading, Mass, 1980, 3rd~ed.

\bibitem{GJ81}
J.~Glimm and A.~Jaffee, {\em Quantum Physics. A Functional Point of View}.
\newblock Springer-Verlag, New York, 1981.

\bibitem{SE82}
E.~Seiler, {\em Gauge Theories as a Problem of Constructive Quantum Theory and
  Statistical Mechanics}.
\newblock Springer-Verlag, New York, 1982.

\bibitem{Brambilla:2010cs}
N.~Brambilla {\em et~al.}, ``{Heavy quarkonium: progress, puzzles, and
  opportunities},'' {\em Eur. Phys. J.}, vol.~C71, p.~1534, 2011.

\bibitem{Lattes:1947mw}
C.~Lattes, H.~Muirhead, G.~Occhialini, and C.~Powell, ``{Processes involving
  charged mesons},'' {\em Nature}, vol.~159, pp.~694--697, 1947.

\bibitem{Rochester:1947mi}
G.~D. Rochester and C.~C. Butler, ``{Evidence for the existence of new unstable
  elementary particles},'' {\em Nature}, vol.~160, pp.~855--857, 1947.

\bibitem{GellMann:1962xb}
M.~Gell-Mann, ``{Symmetries of baryons and mesons},'' {\em Phys. Rev.},
  vol.~125, pp.~1067--1084, 1962.
\newblock See also {``The Eightfold Way: A Theory of Strong Interaction
  Symmetry,''} DOE Technical Report TID-12608, 1961.

\bibitem{Ne'eman:1961cd}
Y.~Ne'eman, ``{Derivation of strong interactions from a gauge invariance},''
  {\em Nucl. Phys.}, vol.~26, pp.~222--229, 1961.

\bibitem{GellMann:1968rz}
M.~Gell-Mann, R.~J. Oakes, and B.~Renner, ``{Behavior of current divergences
  under SU(3) x SU(3)},'' {\em Phys. Rev.}, vol.~175, pp.~2195--2199, 1968.

\bibitem{Nakamura:2010zzi}
K.~Nakamura {\em et~al.}, ``{Review of particle physics},'' {\em J. Phys.},
  vol.~G37, p.~075021, 2010.

\bibitem{Leutwyler:2009jg}
H.~Leutwyler, ``{Light quark masses},'' {\em PoS}, vol.~CD09, p.~005, 2009.

\bibitem{Stern1}
R.~Frisch and O.~Stern, ``{{\"U}ber die magnetische Ablenkung von
  Wasserstoffmolekülen und das magnetische Moment des Protons},'' {\em Zeits.
  f. Physik}, vol.~85, pp.~4--16, 1933.

\bibitem{Nobel43}
N.~Foundation, {\em Nobel Lectures in Physics (1942-1962)}, pp.~3--24.
\newblock World Scientific, Singapore, 1998.

\bibitem{Dirac:1928hu}
P.~A.~M. Dirac, ``{The Quantum theory of electron},'' {\em Proc. Roy. Soc.
  Lond.}, vol.~A117, pp.~610--624, 1928.

\bibitem{Hofstadter:1955ae}
R.~Hofstadter and R.~W. McAllister, ``{Electron scattering from the proton},''
  {\em Phys. Rev.}, vol.~98, pp.~217--218, 1955.

\bibitem{Nobel61}
N.~Foundation, {\em Nobel Lectures in Physics (1942-1962)}, pp.~555--604.
\newblock World Scientific, Singapore, 1998.

\bibitem{Taylor:1991ew}
R.~E. Taylor, ``{Deep inelastic scattering: The Early years},'' {\em Rev. Mod.
  Phys.}, vol.~63, pp.~573--595, 1991.

\bibitem{Kendall:1991np}
H.~W. Kendall, ``{Deep inelastic scattering: Experiments on the proton and the
  observation},'' {\em Rev. Mod. Phys.}, vol.~63, pp.~597--614, 1991.

\bibitem{Friedman:1991nq}
J.~I. Friedman, ``{Deep inelastic scattering: Comparisons with the quark
  model},'' {\em Rev. Mod. Phys.}, vol.~63, pp.~615--629, 1991.

\bibitem{Friedman:1991ip}
J.~I. Friedman, H.~W. Kendall, and R.~E. Taylor, ``{Deep inelastic scattering:
  Acknowledgements},'' {\em Rev. Mod. Phys.}, vol.~63, p.~629, 1991.

\bibitem{Pauli:1934xm}
W.~Pauli and V.~F. Weisskopf, ``{On quantization of the scalar relativistic
  wave equation. (In German)},'' {\em Helv. Phys. Acta}, vol.~7, pp.~709--731,
  1934.

\bibitem{Flambaum:2007mj}
V.~V. Flambaum and R.~B. Wiringa, ``{Dependence of nuclear binding on hadronic
  mass variation},'' {\em Phys. Rev.}, vol.~C76, p.~054002, 2007.

\bibitem{Beane:2008dv}
S.~R. Beane, K.~Orginos, and M.~J. Savage, ``{Hadronic Interactions from
  Lattice QCD},'' {\em Int. J. Mod. Phys.}, vol.~E17, pp.~1157--1218, 2008.

\bibitem{Bazavov:2009bb}
A.~Bazavov {\em et~al.}, ``{Full nonperturbative QCD simulations with 2+1
  flavors of improved staggered quarks},'' {\em Rev. Mod. Phys.}, vol.~82,
  pp.~1349--1417, 2010.

\bibitem{Hagler:2009ni}
P.~H{\"a}gler, ``{Hadron structure from lattice quantum chromodynamics},'' {\em
  Phys. Rept.}, vol.~490, pp.~49--175, 2010.

\bibitem{Colangelo:2010et}
G.~Colangelo {\em et~al.}, ``{Review of lattice results concerning low energy
  particle physics},'' {\em Eur. Phys. J.}, vol.~C71, p.~1695, 2011.

\bibitem{Dyson:1949ha}
F.~J. Dyson, ``{The S matrix in quantum electrodynamics},'' {\em Phys. Rev.},
  vol.~75, pp.~1736--1755, 1949.

\bibitem{Schwinger:1951ex}
J.~S. Schwinger, ``{On the Green's functions of quantized fields. 1},'' {\em
  Proc. Nat. Acad. Sci.}, vol.~37, pp.~452--455, 1951.

\bibitem{Schwinger:1951hq}
J.~S. Schwinger, ``{On the Green's functions of quantized fields. 2},'' {\em
  Proc. Nat. Acad. Sci.}, vol.~37, pp.~455--459, 1951.

\bibitem{Nambu:1961tp}
Y.~Nambu and G.~Jona-Lasinio, ``{Dynamical Model of Elementary Particles Based
  on an Analogy with Superconductivity. 1.},'' {\em Phys.Rev.}, vol.~122,
  pp.~345--358, 1961.

\bibitem{Nambu:2009zza}
Y.~Nambu, ``{Nobel Lecture: Spontaneous symmetry breaking in particle physics:
  A case of cross fertilization},'' {\em Int. J. Mod. Phys.}, vol.~A24,
  pp.~2371--2377, 2009.

\bibitem{GutierrezGuerrero:2010md}
L.~X. Guti{\'e}rrez-Guerrero, A.~Bashir, I.~C. Clo{\"e}t, and C.~D. Roberts,
  ``{Pion form factor from a contact interaction},'' {\em Phys. Rev.},
  vol.~C81, p.~065202, 2010.

\bibitem{Ebert:1996vx}
D.~Ebert, T.~Feldmann, and H.~Reinhardt, ``{Extended NJL model for light and
  heavy mesons without q anti-q thresholds},'' {\em Phys. Lett.}, vol.~B388,
  pp.~154--160, 1996.

\bibitem{Krein:1990sf}
C.~D. Roberts, A.~G. Williams, and G.~Krein, ``{On the implications of
  confinement},'' {\em Int. J. Mod. Phys.}, vol.~A7, pp.~5607--5624, 1992.

\bibitem{Munczek:1983dx}
H.~J. Munczek and A.~M. Nemirovsky, ``{The Ground State q anti-q Mass Spectrum
  in QCD},'' {\em Phys. Rev.}, vol.~D28, pp.~181--186, 1983.

\bibitem{Gribov:1999ui}
V.~N. Gribov, ``{The theory of quark confinement},'' {\em Eur. Phys. J.},
  vol.~C10, pp.~91--105, 1999.

\bibitem{Stingl:1983pt}
M.~Stingl, ``{A schematic model of mesons based on analytic propagators},''
  {\em Phys. Rev.}, vol.~D29, p.~2105, 1984.

\bibitem{Cahill:1988zi}
R.~T. Cahill, ``{Hadronization of QCD},'' {\em Austral. J. Phys.}, vol.~42,
  pp.~171--186, 1989.

\bibitem{Cahill:1985mh}
R.~T. Cahill and C.~D. Roberts, ``{Soliton Bag Models of Hadrons from QCD},''
  {\em Phys. Rev.}, vol.~D32, p.~2419, 1985.

\bibitem{Roberts:1985ju}
C.~D. Roberts and R.~T. Cahill, ``{Dynamically selected vacuum field
  configuration in massless QED},'' {\em Phys. Rev.}, vol.~D33, p.~1755, 1986.

\bibitem{Qin:2010nq}
S.-x. Qin, L.~Chang, H.~Chen, Y.-x. Liu, and C.~D. Roberts, ``{Phase diagram
  and critical endpoint for strongly- interacting quarks},'' {\em Phys. Rev.
  Lett.}, vol.~106, p.~172301, 2011.

\bibitem{Bali:2005fu}
G.~S. Bali, H.~Neff, T.~Duessel, T.~Lippert, and K.~Schilling, ``{Observation
  of string breaking in QCD},'' {\em Phys. Rev.}, vol.~D71, p.~114513, 2005.

\bibitem{Veltmann:2003}
M.~J.~G. Veltman, {\em Facts and Mysteries in Elementary Particle Physics}.
\newblock World Scientific, Singapore, 2003.

\bibitem{Tang:2000tb}
A.~Tang and J.~W. Norbury, ``{Properties of Regge trajectories},'' {\em Phys.
  Rev.}, vol.~D62, p.~016006, 2000.

\bibitem{Hawes:1993ef}
F.~T. Hawes, C.~D. Roberts, and A.~G. Williams, ``{Dynamical chiral symmetry
  breaking and confinement with an infrared vanishing gluon propagator},'' {\em
  Phys. Rev.}, vol.~D49, pp.~4683--4693, 1994.

\bibitem{Maris:1995ns}
P.~Maris, ``{Confinement and complex singularities in QED in three-
  dimensions},'' {\em Phys. Rev.}, vol.~D52, pp.~6087--6097, 1995.

\bibitem{Bashir:2008fk}
A.~Bashir, A.~Raya, I.~C. Clo{\"e}t, and C.~D. Roberts, ``{Regarding
  confinement and dynamical chiral symmetry breaking in QED3},'' {\em Phys.
  Rev.}, vol.~C78, p.~055201, 2008.

\bibitem{Celmaster:1979km}
W.~Celmaster and R.~J. Gonsalves, ``{The Renormalization Prescription
  Dependence of the QCD Coupling Constant},'' {\em Phys. Rev.}, vol.~D20,
  p.~1420, 1979.

\bibitem{Qin:2011dd}
S.-x. Qin, L.~Chang, Y.-x. Liu, C.~D. Roberts, and D.~J. Wilson, ``{Interaction
  model for the gap equation},'' {\em Phys. Rev.}, vol.~C84, p.~042202(R),
  2011.

\bibitem{Qin:2011xq}
S.-x. Qin, L.~Chang, Y.-x. Liu, C.~D. Roberts, and D.~J. Wilson, ``{Commentary
  on rainbow-ladder truncation for excited states and exotics},''
  arXiv:1109.3459 [nucl-th].

\bibitem{Munczek:1994zz}
H.~J. Munczek, ``{Dynamical chiral symmetry breaking, Goldstone's theorem and
  the consistency of the Schwinger-Dyson and Bethe- Salpeter Equations},'' {\em
  Phys. Rev.}, vol.~D52, pp.~4736--4740, 1995.

\bibitem{Bender:1996bb}
A.~Bender, C.~D. Roberts, and L.~Von~Smekal, ``{Goldstone Theorem and Diquark
  Confinement Beyond Rainbow- Ladder Approximation},'' {\em Phys. Lett.},
  vol.~B380, pp.~7--12, 1996.

\bibitem{Bhagwat:2003vw}
M.~Bhagwat, M.~Pichowsky, C.~Roberts, and P.~Tandy, ``{Analysis of a quenched
  lattice QCD dressed quark propagator},'' {\em Phys.Rev.}, vol.~C68,
  p.~015203, 2003.

\bibitem{Bhagwat:2006tu}
M.~S. Bhagwat and P.~C. Tandy, ``{Analysis of full-QCD and quenched-QCD lattice
  propagators},'' {\em AIP Conf. Proc.}, vol.~842, pp.~225--227, 2006.

\bibitem{Bhagwat:2007vx}
M.~S. Bhagwat, I.~C. Clo{\"e}t, and C.~D. Roberts, ``{Covariance, Dynamics and
  Symmetries, and Hadron Form Factors},'' pp.~112--120, {in Proceedings of the
  \emph{Workshop on Exclusive Reactions at High Momentum Transfer}, Newport
  News, Virginia, 21-24 May 2007, Eds.\ A.~Radyushkin and P.~Stoler (World
  Scientific, Singapore, 2007)}.

\bibitem{Bowman:2005vx}
P.~O. Bowman {\em et~al.}, ``{Unquenched quark propagator in Landau gauge},''
  {\em Phys. Rev.}, vol.~D71, p.~054507, 2005.

\bibitem{Flambaum:2005kc}
V.~V. Flambaum, A.~Holl, P.~Jaikumar, C.~D. Roberts, and S.~V. Wright, ``{Sigma
  terms of light-quark hadrons},'' {\em Few Body Syst.}, vol.~38, pp.~31--51,
  2006.

\bibitem{Holl:2005st}
A.~H{\"o}ll, P.~Maris, C.~D. Roberts, and S.~V. Wright, ``{Schwinger functions
  and light-quark bound states, and sigma terms},'' {\em Nucl. Phys. Proc.
  Suppl.}, vol.~161, pp.~87--94, 2006.

\bibitem{Chang:2008ec}
L.~Chang {\em et~al.}, ``{Chiral susceptibility and the scalar Ward
  identity},'' {\em Phys. Rev.}, vol.~C79, p.~035209, 2009.

\bibitem{tarrach}
P.~Pascual and R.~Tarrach, {\em QCD: Renormalization for the Practitioner}.
\newblock Springer-Verlag, Berlin, 1984.
\newblock Lecture Notes in Physics \textbf{194}.

\bibitem{Salpeter:1951sz}
E.~E. Salpeter and H.~A. Bethe, ``{A Relativistic equation for bound state
  problems},'' {\em Phys. Rev.}, vol.~84, pp.~1232--1242, 1951.

\bibitem{Holl:2004fr}
A.~H{\"o}ll, A.~Krassnigg, and C.~D. Roberts, ``{Pseudoscalar meson radial
  excitations},'' {\em Phys. Rev.}, vol.~C70, p.~042203, 2004.

\bibitem{Holl:2005vu}
A.~H{\"o}ll, A.~Krassnigg, P.~Maris, C.~D. Roberts, and S.~V. Wright,
  ``{Electromagnetic properties of ground and excited state pseudoscalar
  mesons},'' {\em Phys. Rev.}, vol.~C71, p.~065204, 2005.

\bibitem{Ivanov:1998ms}
M.~A. Ivanov, Y.~L. Kalinovsky, and C.~D. Roberts, ``{Survey of heavy-meson
  observables},'' {\em Phys. Rev.}, vol.~D60, p.~034018, 1999.

\bibitem{Bhagwat:2006xi}
M.~S. Bhagwat, A.~Krassnigg, P.~Maris, and C.~D. Roberts, ``{Mind the gap},''
  {\em Eur. Phys. J.}, vol.~A31, pp.~630--637, 2007.

\bibitem{Chang:2009at}
L.~Chang {\em et~al.}, ``{Vacuum pseudoscalar susceptibility},'' {\em Phys.
  Rev.}, vol.~C81, p.~032201, 2010.

\bibitem{Chang:2008sp}
L.~Chang, Y.-x. Liu, W.-m. Sun, and H.-s. Zong, ``{Revisiting the Vector and
  Axial-vector Vacuum Susceptibilities},'' {\em Phys. Lett.}, vol.~B669,
  pp.~327--330, 2008.

\bibitem{Weinberg:1967kj}
S.~Weinberg, ``{Precise relations between the spectra of vector and axial
  vector mesons},'' {\em Phys. Rev. Lett.}, vol.~18, pp.~507--509, 1967.

\bibitem{Maris:1997hd}
P.~Maris, C.~D. Roberts, and P.~C. Tandy, ``{Pion mass and decay constant},''
  {\em Phys. Lett.}, vol.~B420, pp.~267--273, 1998.

\bibitem{Delbourgo:1979me}
R.~Delbourgo and M.~D. Scadron, ``{Proof of the Nambu-Goldstone realization for
  vector gluon quark theories},'' {\em J. Phys.}, vol.~G5, p.~1621, 1979.

\bibitem{Maris:1998hc}
P.~Maris and C.~D. Roberts, ``{Pseudovector components of the pion, pi0 -->
  gamma gamma, and F(pi)(q**2)},'' {\em Phys. Rev.}, vol.~C58, pp.~3659--3665,
  1998.

\bibitem{Roberts:2011wy}
H.~L.~L. Roberts, A.~Bashir, L.~X. Guti{\'e}rrez-Guerrero, C.~D. Roberts, and
  D.~J. Wilson, ``{pi- and rho-mesons, and their diquark partners, from a
  contact interaction},'' {\em Phys. Rev.}, vol.~C83, p.~065206, 2011.

\bibitem{Maris:1999nt}
P.~Maris and P.~C. Tandy, ``{Bethe-Salpeter study of vector meson masses and
  decay constants},'' {\em Phys. Rev.}, vol.~C60, p.~055214, 1999.

\bibitem{Maris:1999bh}
P.~Maris and P.~C. Tandy, ``{The quark photon vertex and the pion charge
  radius},'' {\em Phys. Rev.}, vol.~C61, p.~045202, 2000.

\bibitem{Maris:2000sk}
P.~Maris and P.~C. Tandy, ``{The pi, K+, and K0 electromagnetic form
  factors},'' {\em Phys. Rev.}, vol.~C62, p.~055204, 2000.

\bibitem{Bellini:1982ec}
G.~Bellini {\em et~al.}, ``{Evidence for new \mbox{$0^-$} S resonances in the
  \mbox{$\pi^+ \pi^- \pi^-$} systems},'' {\em Phys. Rev. Lett.}, vol.~48,
  pp.~1697--1700, 1982.

\bibitem{Barnes:1996ff}
T.~Barnes, F.~E. Close, P.~R. Page, and E.~S. Swanson, ``{Higher Quarkonia},''
  {\em Phys. Rev.}, vol.~D55, pp.~4157--4188, 1997.

\bibitem{deMelo:2005cy}
J.~P. B.~C. de~Melo, T.~Frederico, E.~Pace, and G.~Salme, ``{Space-like and
  time-like pion electromagnetic form factor and Fock state components within
  the light-front dynamics},'' {\em Phys. Rev.}, vol.~D73, p.~074013, 2006.

\bibitem{Diehl:2001xe}
M.~Diehl and G.~Hiller, ``{New ways to explore factorization in b decays},''
  {\em JHEP}, vol.~06, p.~067, 2001.

\bibitem{McNeile:2006qy}
C.~McNeile and C.~Michael, ``{The decay constant of the first excited pion from
  lattice QCD},'' {\em Phys. Lett.}, vol.~B642, pp.~244--247, 2006.

\bibitem{Bhagwat:2007ha}
M.~S. Bhagwat, L.~Chang, Y.-X. Liu, C.~D. Roberts, and P.~C. Tandy, ``{Flavour
  symmetry breaking and meson masses},'' {\em Phys. Rev.}, vol.~C76, p.~045203,
  2007.

\bibitem{Maris:1997tm}
P.~Maris and C.~D. Roberts, ``{{$\pi$} and {$K$} meson Bethe-Salpeter
  amplitudes},'' {\em Phys. Rev.}, vol.~C56, pp.~3369--3383, 1997.

\bibitem{Brodsky:2009zd}
S.~J. Brodsky and R.~Shrock, ``{Condensates in Quantum Chromodynamics and the
  Cosmological Constant},'' {\em Proc. Nat. Acad. Sci.}, vol.~108, pp.~45--50,
  2011.

\bibitem{Brodsky:2010xf}
S.~J. Brodsky, C.~D. Roberts, R.~Shrock, and P.~C. Tandy, ``{Essence of the
  vacuum quark condensate},'' {\em Phys. Rev.}, vol.~C82, p.~022201, 2010.

\bibitem{Chang:2011mu}
L.~Chang, C.~D. Roberts, and P.~C. Tandy, ``{Expanding the concept of in-hadron
  condensates},'' arXiv:1109.2903 [nucl-th].

\bibitem{Bardeen:2000cz}
W.~A. Bardeen, A.~Duncan, E.~Eichten, and H.~Thacker, ``{Anomalous chiral
  behavior in quenched lattice QCD},'' {\em Phys. Rev.}, vol.~D62, p.~114505,
  2000.

\bibitem{Ahmad:2005dr}
S.~Ahmad, J.~T. Lenaghan, and H.~B. Thacker, ``{Coherent topological charge
  structure in CP(N-1) models and QCD},'' {\em Phys. Rev.}, vol.~D72,
  p.~114511, 2005.

\bibitem{Kogut:1974kt}
J.~B. Kogut and L.~Susskind, ``{How to solve the \mbox{$\eta \to 3 \pi$}
  problem by seizing the vacuum},'' {\em Phys. Rev.}, vol.~D11, p.~3594, 1975.

\bibitem{Christos:1984tu}
G.~A. Christos, ``{Chiral symmetry and the U(1) problem},'' {\em Phys. Rept.},
  vol.~116, pp.~251--336, 1984.

\bibitem{Witten:1979vv}
E.~Witten, ``{Current algebra theorems for the U(1) Goldstone boson},'' {\em
  Nucl. Phys.}, vol.~B156, p.~269, 1979.

\bibitem{Veneziano:1979ec}
G.~Veneziano, ``{U(1) without instantons},'' {\em Nucl. Phys.}, vol.~B159,
  pp.~213--224, 1979.

\bibitem{Narayanan:2004cp}
R.~Narayanan and H.~Neuberger, ``{Chiral symmetry breaking at large {$N_c$}},''
  {\em Nucl. Phys.}, vol.~B696, pp.~107--140, 2004.

\bibitem{DiVecchia:1979bf}
P.~Di~Vecchia, ``{An effective Lagrangian with no U(1) problem in
  \mbox{CP$^{n-1}$} models and QCD},'' {\em Phys. Lett.}, vol.~B85, p.~357,
  1979.

\bibitem{Crewther:1977ce}
R.~J. Crewther, ``{Chirality Selection Rules and the U(1) Problem},'' {\em
  Phys. Lett.}, vol.~B70, p.~349, 1977.

\bibitem{Crewther:1978zz}
R.~J. Crewther, ``{Effects of Topological Charge in Gauge Theories},'' {\em
  Acta Phys. Austriaca Suppl.}, vol.~19, pp.~47--153, 1978.

\bibitem{Bini:2007zza}
C.~Bini, ``{Recent KLOE results on hadron physics},'' {\em Eur. Phys. J.},
  vol.~A31, pp.~446--450, 2007.

\bibitem{Green:2003qw}
A.~M. Green and S.~Wycech, ``{On \mbox{$\eta \pi$} mixing close to the
  \mbox{$\eta$}-He threshold},'' {\em Phys. Rev.}, vol.~C68, p.~061601, 2003.

\bibitem{Casher:1974xd}
A.~Casher and L.~Susskind, ``{Chiral magnetism (or magnetohadrochironics)},''
  {\em Phys. Rev.}, vol.~D9, pp.~436--460, 1974.

\bibitem{MichelsonMorley}
A.~A. Michelson and E.~W. Morley, ``{On the Relative Motion of the Earth and
  the Luminiferous Ether},'' {\em American Journal of Science}, vol.~34,
  pp.~333--345, 1887.

\bibitem{Turner:2001yu}
M.~S. Turner, ``{Dark energy and the new cosmology},'' astro-ph/0108103.

\bibitem{Burkardt:1998dd}
M.~Burkardt, ``{Dynamical vertex mass generation and chiral symmetry breaking
  on the light-front},'' {\em Phys. Rev.}, vol.~D58, p.~096015, 1998.

\bibitem{Brodsky:2008be}
S.~J. Brodsky and R.~Shrock, ``{Maximum Wavelength of Confined Quarks and
  Gluons and Properties of Quantum Chromodynamics},'' {\em Phys. Lett.},
  vol.~B666, pp.~95--99, 2008.

\bibitem{Glazek:2011vg}
S.~D. Glazek, ``{Reinterpretation of gluon condensate in dynamics of hadronic
  constituents},'' {\em Acta Phys. Polon.}, vol.~B42, pp.~1933--2010, 2011.

\bibitem{Netterfield:2001yq}
C.~B. Netterfield {\em et~al.}, ``{A measurement by BOOMERANG of multiple peaks
  in the angular power spectrum of the cosmic microwave background},'' {\em
  Astrophys. J.}, vol.~571, pp.~604--614, 2002.

\bibitem{Weinberg:1978kz}
S.~Weinberg, ``{Phenomenological Lagrangians},'' {\em Physica}, vol.~A96,
  p.~327, 1979.

\bibitem{Pagels:1979hd}
H.~Pagels and S.~Stokar, ``{The Pion Decay Constant, Electromagnetic
  Form-Factor and Quark Electromagnetic Selfenergy in QCD},'' {\em Phys. Rev.},
  vol.~D20, p.~2947, 1979.

\bibitem{Chang:2009zb}
L.~Chang and C.~D. Roberts, ``{Sketching the Bethe-Salpeter kernel},'' {\em
  Phys. Rev. Lett.}, vol.~103, p.~081601, 2009.

\bibitem{Bijnens:2006zp}
J.~Bijnens, ``{Chiral Perturbation Theory Beyond One Loop},'' {\em Prog. Part.
  Nucl. Phys.}, vol.~58, pp.~521--586, 2007.

\bibitem{Langfeld:2003ye}
K.~Langfeld, H.~Markum, R.~Pullirsch, C.~D. Roberts, and S.~M. Schmidt,
  ``{Concerning the quark condensate},'' {\em Phys. Rev.}, vol.~C67, p.~065206,
  2003.

\bibitem{Lane:1974he}
K.~D. Lane, ``{Asymptotic Freedom and Goldstone Realization of Chiral
  Symmetry},'' {\em Phys. Rev.}, vol.~D10, p.~2605, 1974.

\bibitem{Politzer:1976tv}
H.~D. Politzer, ``{Effective Quark Masses in the Chiral Limit},'' {\em Nucl.
  Phys.}, vol.~B117, p.~397, 1976.

\bibitem{Banks:1979yr}
T.~Banks and A.~Casher, ``{Chiral Symmetry Breaking in Confining Theories},''
  {\em Nucl. Phys.}, vol.~B169, p.~103, 1980.

\bibitem{GellMann:1951rw}
M.~Gell-Mann and F.~Low, ``{Bound states in quantum field theory},'' {\em Phys.
  Rev.}, vol.~84, pp.~350--354, 1951.

\bibitem{Beg:1973sc}
M.~A.~B. Beg and A.~Zepeda, ``{Pion radius and isovector nucleon radii in the
  limit of small pion mass},'' {\em Phys. Rev.}, vol.~D6, pp.~2912--2918, 1972.

\bibitem{Pervushin:1974nm}
V.~N. Pervushin and M.~K. Volkov, ``{Low-Energy Scattering of Massive Pions},''
  {\em Sov. J. Nucl. Phys.}, vol.~20, pp.~408--413, 1975.
\newblock [Yad.Fiz.20:762-774,1974].

\bibitem{Gasser:1983yg}
J.~Gasser and H.~Leutwyler, ``{Chiral Perturbation Theory to One Loop},'' {\em
  Annals Phys.}, vol.~158, p.~142, 1984.

\bibitem{Alkofer:1993gu}
R.~Alkofer, A.~Bender, and C.~D. Roberts, ``{Pion loop contribution to the
  electromagnetic pion charge radius},'' {\em Int. J. Mod. Phys.}, vol.~A10,
  pp.~3319--3342, 1995.

\bibitem{Courtland:2010zz}
R.~Courtland, ``{Void that is truly empty solves dark energy puzzle},'' {\em
  New Sci.}, vol.~207N2776, p.~10, 2010.

\bibitem{Andersen:2011yj}
J.~R. Andersen {\em et~al.}, ``{Discovering Technicolor},'' {\em Eur. Phys. J.
  Plus}, vol.~126, p.~81, 2011.

\bibitem{Roberts:1996jx}
C.~D. Roberts, {\em \mbox{\rm in} Como 1996, Quark confinement and the hadron
  spectrum II}, ch.~{Confinement, diquarks and Goldstone's theorem},
  pp.~224--230.
\newblock World Scientific, Singapore, 1996.
\newblock Eds.\, N.~Brambilla and G.~M.~Prosperi.

\bibitem{Roberts:1997vs}
C.~D. Roberts, ``{Nonperturbative effects in QCD at finite temperature and
  density},'' {\em Phys. Part. Nucl.}, vol.~30, pp.~223--257, 1999.

\bibitem{Bender:2002as}
A.~Bender, W.~Detmold, C.~D. Roberts, and A.~W. Thomas, ``{Bethe-Salpeter
  equation and a nonperturbative quark gluon vertex},'' {\em Phys. Rev.},
  vol.~C65, p.~065203, 2002.

\bibitem{Bhagwat:2004hn}
M.~S. Bhagwat, A.~H{\"o}ll, A.~Krassnigg, C.~D. Roberts, and P.~C. Tandy,
  ``{Aspects and consequences of a dressed-quark-gluon vertex},'' {\em Phys.
  Rev.}, vol.~C70, p.~035205, 2004.

\bibitem{Burden:1996nh}
C.~J. Burden, L.~Qian, C.~D. Roberts, P.~C. Tandy, and M.~J. Thomson,
  ``{Ground-state spectrum of light-quark mesons},'' {\em Phys. Rev.},
  vol.~C55, pp.~2649--2664, 1997.

\bibitem{Watson:2004kd}
P.~Watson, W.~Cassing, and P.~C. Tandy, ``{Bethe-Salpeter meson masses beyond
  ladder approximation},'' {\em Few Body Syst.}, vol.~35, pp.~129--153, 2004.

\bibitem{Maris:2006ea}
P.~Maris, ``{Hadron Physics and the Dyson--Schwinger Equations of QCD},'' {\em
  AIP Conf. Proc.}, vol.~892, pp.~65--71, 2007.

\bibitem{Cloet:2007pi}
I.~C. Clo{\"e}t, A.~Krassnigg, and C.~D. Roberts, ``{Dynamics, Symmetries and
  Hadron Properties},'' arXiv:0710.5746 [nucl-th].
\newblock In {\it Proceedings} {\it of} \emph{11th} \emph{International}
  \emph{Conference} \emph{on} {Meson-Nucleon Physics and} {\it the} {\it
  Structure} \emph{of} \emph{the} \emph{Nucleon} \emph{(MENU 2007)},
  J{\"u}lich, Germany, 10-14 Sep 2007, eds.\ H.~Machner and S.~Krewald, paper
  125.

\bibitem{Fischer:2009jm}
C.~S. Fischer and R.~Williams, ``{Probing the gluon self-interaction in light
  mesons},'' {\em Phys. Rev. Lett.}, vol.~103, p.~122001, 2009.

\bibitem{Krassnigg:2009zh}
A.~Krassnigg, ``{Survey of J=0,1 mesons in a Bethe-Salpeter approach},'' {\em
  Phys. Rev.}, vol.~D80, p.~114010, 2009.

\bibitem{Krassnigg:2010mh}
A.~Krassnigg and M.~Blank, ``{A covariant study of tensor mesons},'' {\em Phys.
  Rev.}, vol.~D83, p.~096006, 2011.

\bibitem{Holl:2004un}
A.~H{\"o}ll, A.~Krassnigg, C.~D. Roberts, and S.~V. Wright, ``{On the
  complexion of pseudoscalar mesons},'' {\em Int. J. Mod. Phys.}, vol.~A20,
  pp.~1778--1784, 2005.

\bibitem{Ball:1980ay}
J.~S. Ball and T.-W. Chiu, ``{Analytic Properties of the Vertex Function in
  Gauge Theories. 1},'' {\em Phys. Rev.}, vol.~D22, p.~2542, 1980.

\bibitem{Frank:1994mf}
M.~R. Frank, ``{Nonperturbative aspects of the quark - photon vertex},'' {\em
  Phys. Rev.}, vol.~C51, pp.~987--998, 1995.

\bibitem{Roberts:1994hh}
C.~D. Roberts, ``{Electromagnetic pion form-factor and neutral pion decay
  width},'' {\em Nucl. Phys.}, vol.~A605, pp.~475--495, 1996.

\bibitem{Eichmann:2008ef}
G.~Eichmann, I.~C. Clo{\"e}t, R.~Alkofer, A.~Krassnigg, and C.~D. Roberts,
  ``{Toward unifying the description of meson and baryon properties},'' {\em
  Phys. Rev.}, vol.~C79, p.~012202, 2009.

\bibitem{Cloet:2008re}
I.~C. Clo{\"e}t, G.~Eichmann, B.~El-Bennich, T.~Kl{\"a}hn, and C.~D. Roberts,
  ``{Survey of nucleon electromagnetic form factors},'' {\em Few Body Syst.},
  vol.~46, pp.~1--36, 2009.

\bibitem{Chang:2010hb}
L.~Chang, Y.-X. Liu, and C.~D. Roberts, ``{Dressed-quark anomalous magnetic
  moments},'' {\em Phys. Rev. Lett.}, vol.~106, p.~072001, 2011.

\bibitem{Eichmann:2011vu}
G.~Eichmann, ``{Nucleon electromagnetic form factors from the covariant Faddeev
  equation},'' {\em Phys. Rev.}, vol.~D84, p.~014014, 2011.

\bibitem{Chang:2011ei}
L.~Chang and C.~D. Roberts, ``{Tracing masses of ground-state light-quark
  mesons},'' arXiv:1104.4821 [nucl-th].
\newblock {Tracing masses of ground-state light-quark mesons}.

\bibitem{Wilson:2011rj}
D.~J. Wilson, I.~C. Clo{\"e}t, L.~Chang, and C.~D. Roberts, ``{Nucleon and
  Roper electromagnetic elastic and transition form factors},'' arXiv:1112.2212
  [nucl-th].
\newblock {Nucleon and Roper electromagnetic elastic and transition form
  factors}.

\bibitem{Kizilersu:2009kg}
A.~K{\i}z{\i}lers{\"u} and M.~R. Pennington, ``{Building the Full
  Fermion-Photon Vertex of QED by Imposing Multiplicative Renormalizability of
  the Schwinger-Dyson Equations for the Fermion and Photon Propagators},'' {\em
  Phys. Rev.}, vol.~D79, p.~125020, 2009.

\bibitem{Bashir:2011dp}
A.~Bashir, R.~Bermudez, L.~Chang, and C.~D. Roberts, ``{Dynamical chiral
  symmetry breaking and the fermion--gauge- boson vertex},'' arXiv:1112.4847
  [nucl-th].

\bibitem{Bhagwat:2004kj}
M.~S. Bhagwat and P.~C. Tandy, ``{Quark-gluon vertex model and lattice-QCD
  data},'' {\em Phys. Rev.}, vol.~D70, p.~094039, 2004.

\bibitem{Bhagwat:2006pu}
M.~S. Bhagwat and P.~Maris, ``{Vector meson form factors and their quark-mass
  dependence},'' {\em Phys. Rev.}, vol.~C77, p.~025203, 2008.

\bibitem{Foley:1948zz}
H.~M. Foley and P.~Kusch, ``{On the Intrinsic Moment of the Electron},'' {\em
  Phys. Rev.}, vol.~73, pp.~412--412, 1948.

\bibitem{Schwinger:1948iu}
J.~S. Schwinger, ``{On Quantum electrodynamics and the magnetic moment of the
  electron},'' {\em Phys. Rev.}, vol.~73, pp.~416--417, 1948.

\bibitem{Mohr:2008fa}
P.~J. Mohr, B.~N. Taylor, and D.~B. Newell, ``{CODATA Recommended Values of the
  Fundamental Physical Constants: 2006},'' {\em Rev. Mod. Phys.}, vol.~80,
  pp.~633--730, 2008.

\bibitem{Davydychev:2000rt}
A.~I. Davydychev, P.~Osland, and L.~Saks, ``{Quark gluon vertex in arbitrary
  gauge and dimension},'' {\em Phys. Rev.}, vol.~D63, p.~014022, 2001.

\bibitem{Skullerud:2003qu}
J.~I. Skullerud, P.~O. Bowman, A.~K{\i}z{\i}lers{\"u}, D.~B. Leinweber, and
  A.~G. Williams, ``{Nonperturbative structure of the quark gluon vertex},''
  {\em JHEP}, vol.~04, p.~047, 2003.

\bibitem{Kochelev:1996pv}
N.~I. Kochelev, ``{Anomalous quark chromomagnetic moment induced by
  instantons},'' {\em Phys. Lett.}, vol.~B426, pp.~149--153, 1998.

\bibitem{Diakonov:2002fq}
D.~Diakonov, ``{Instantons at work},'' {\em Prog. Part. Nucl. Phys.}, vol.~51,
  pp.~173--222, 2003.

\bibitem{Ebert:2005es}
D.~Ebert, R.~N. Faustov, and V.~O. Galkin, ``{Masses and electroweak properties
  of light mesons in the relativistic quark model},'' {\em Eur. Phys. J.},
  vol.~C47, pp.~745--755, 2006.

\bibitem{Bicudo:1998qb}
P.~J.~A. Bicudo, J.~E. F.~T. Ribeiro, and R.~Fernandes, ``{The anomalous
  magnetic moment of quarks},'' {\em Phys. Rev.}, vol.~C59, pp.~1107--1112,
  1999.

\bibitem{Chang:2011tx}
L.~Chang, I.~C. Clo{\"e}t, C.~D. Roberts, and H.~L.~L. Roberts, ``{T(r)opical
  Dyson-Schwinger Equations},'' {\em AIP Conf. Proc.}, vol.~1354, pp.~110--117,
  2011.

\bibitem{Jegerlehner:2009ry}
F.~Jegerlehner and A.~Nyffeler, ``{The Muon g-2},'' {\em Phys. Rept.},
  vol.~477, pp.~1--110, 2009.

\bibitem{Goecke:2011pe}
T.~Goecke, C.~S. Fischer, and R.~Williams, ``{Leading-order calculation of
  hadronic contributions to the muon $g-2$ using the Dyson-Schwinger
  approach},'' {\em Phys. Lett.}, vol.~B704, pp.~211--217, 2011.

\bibitem{Bhagwat:2007rj}
M.~S. Bhagwat, A.~Hoell, A.~Krassnigg, C.~D. Roberts, and S.~V. Wright,
  ``{Schwinger functions and light-quark bound states},'' {\em Few Body Syst.},
  vol.~40, pp.~209--235, 2007.

\bibitem{Bloch:2002eq}
J.~C.~R. Bloch, ``{Multiplicative renormalizability and quark propagator},''
  {\em Phys. Rev.}, vol.~D66, p.~034032, 2002.

\bibitem{Alkofer:2002bp}
R.~Alkofer, P.~Watson, and H.~Weigel, ``{Mesons in a Poincare covariant
  Bethe-Salpeter approach},'' {\em Phys. Rev.}, vol.~D65, p.~094026, 2002.

\bibitem{Roberts:2011cf}
H.~L.~L. Roberts, L.~Chang, I.~C. Clo{\"e}t, and C.~D. Roberts, ``{Masses of
  ground and excited-state hadrons},'' {\em Few Body Syst.}, vol.~51,
  pp.~1--25, 2011.

\bibitem{Pelaez:2006nj}
J.~R. Pelaez and G.~Rios, ``{Nature of the {$f_0(600)$} from its {$N_c$}
  dependence at two loops in unitarized Chiral Perturbation Theory},'' {\em
  Phys. Rev. Lett.}, vol.~97, p.~242002, 2006.

\bibitem{RuizdeElvira:2010cs}
J.~Ruiz~de Elvira, J.~R. Pelaez, M.~R. Pennington, and D.~J. Wilson, ``{Chiral
  Perturbation Theory, the ${1/N_c}$ expansion and Regge behaviour determine
  the structure of the lightest scalar meson},'' {\em Phys. Rev.}, vol.~D84,
  p.~096006, 2011.

\bibitem{Cloet:2008fw}
I.~C. Clo{\"e}t and C.~D. Roberts, ``{Form Factors and Dyson-Schwinger
  Equations},'' {\em PoS}, vol.~LC2008, p.~047, 2008.

\bibitem{Chang:2009ae}
L.~Chang, I.~C. Clo{\"e}t, B.~El-Bennich, T.~Klahn, and C.~D. Roberts,
  ``{Exploring the light-quark interaction},'' {\em Chin. Phys.}, vol.~C33,
  pp.~1189--1196, 2009.

\bibitem{Pichowsky:1999mu}
M.~A. Pichowsky, S.~Walawalkar, and S.~Capstick, ``{Meson-loop contributions to
  the rho omega mass splitting and rho charge radius},'' {\em Phys. Rev.},
  vol.~D60, p.~054030, 1999.

\bibitem{Eichmann:2008ae}
G.~Eichmann, R.~Alkofer, I.~C. Clo{\"e}t, A.~Krassnigg, and C.~D. Roberts,
  ``{Perspective on rainbow-ladder truncation},'' {\em Phys. Rev.}, vol.~C77,
  p.~042202, 2008.

\bibitem{Dudek:2011bn}
J.~J. Dudek, ``{The lightest hybrid meson supermultiplet in QCD},'' {\em Phys.
  Rev.}, vol.~D84, p.~074023, 2011.

\bibitem{Engel:2011aa}
G.~P. Engel, C.~B. Lang, M.~Limmer, D.~Mohler, and A.~Schaefer, ``{QCD with two
  light dynamical chirally improved quarks: Mesons},'' arXiv:1112.1601
  [hep-lat].

\bibitem{Arrington:2006zm}
J.~Arrington, C.~D. Roberts, and J.~M. Zanotti, ``{Nucleon electromagnetic form
  factors},'' {\em J. Phys.}, vol.~G34, pp.~S23--S52, 2007.

\bibitem{Perdrisat:2006hj}
C.~F. Perdrisat, V.~Punjabi, and M.~Vanderhaeghen, ``{Nucleon electromagnetic
  form factors},'' {\em Prog. Part. Nucl. Phys.}, vol.~59, pp.~694--764, 2007.

\bibitem{Burden:1991gd}
C.~J. Burden, C.~D. Roberts, and A.~G. Williams, ``{Singularity structure of a
  model quark propagator},'' {\em Phys. Lett.}, vol.~B285, pp.~347--353, 1992.

\bibitem{Jain:1993qh}
P.~Jain and H.~J. Munczek, ``{q anti-q bound states in the Bethe-Salpeter
  formalism},'' {\em Phys. Rev.}, vol.~D48, pp.~5403--5411, 1993.

\bibitem{Frank:1995uk}
M.~R. Frank and C.~D. Roberts, ``{Model gluon propagator and pion and rho meson
  observables},'' {\em Phys. Rev.}, vol.~C53, pp.~390--398, 1996.

\bibitem{Burden:1996vt}
C.~J. Burden and D.~S. Liu, ``{A non-perturbative treatment of heavy quarks and
  mesons},'' {\em Phys. Rev.}, vol.~D55, pp.~367--376, 1997.

\bibitem{Volmer:2000ek}
J.~Volmer {\em et~al.}, ``{New results for the charged pion electromagnetic
  form- factor},'' {\em Phys. Rev. Lett.}, vol.~86, pp.~1713--1716, 2001.

\bibitem{Horn:2006tm}
T.~Horn {\em et~al.}, ``{Determination of the Charged Pion Form Factor at
  {$Q^2=1.60$} and {$2.45 \,({\rm GeV/c})^2$}},'' {\em Phys. Rev. Lett.},
  vol.~97, p.~192001, 2006.

\bibitem{Tadevosyan:2007yd}
V.~Tadevosyan {\em et~al.}, ``{Determination of the pion charge form factor for
  {$Q^2=0.60- 1.60\,{\rm GeV}^2$}},'' {\em Phys. Rev.}, vol.~C75, p.~055205,
  2007.

\bibitem{Farrar:1979aw}
G.~R. Farrar and D.~R. Jackson, ``{The Pion Form-Factor},'' {\em Phys. Rev.
  Lett.}, vol.~43, p.~246, 1979.

\bibitem{Efremov:1979qk}
A.~V. Efremov and A.~V. Radyushkin, ``{Factorization and Asymptotical Behavior
  of Pion Form- Factor in QCD},'' {\em Phys. Lett.}, vol.~B94, pp.~245--250,
  1980.

\bibitem{Lepage:1980fj}
G.~P. Lepage and S.~J. Brodsky, ``{Exclusive Processes in Perturbative Quantum
  Chromodynamics},'' {\em Phys. Rev.}, vol.~D22, p.~2157, 1980.

\bibitem{Dirac:1949cp}
P.~A.~M. Dirac, ``{Forms of Relativistic Dynamics},'' {\em Rev. Mod. Phys.},
  vol.~21, pp.~392--399, 1949.

\bibitem{Ezawa:1974wm}
Z.~F. Ezawa, ``{Wide-Angle Scattering in Softened Field Theory},'' {\em Nuovo
  Cim.}, vol.~A23, pp.~271--290, 1974.

\bibitem{Farrar:1975yb}
G.~R. Farrar and D.~R. Jackson, ``{Pion and Nucleon Structure Functions Near
  x=1},'' {\em Phys. Rev. Lett.}, vol.~35, p.~1416, 1975.

\bibitem{Badier:1980jq}
J.~Badier {\em et~al.}, ``{Measurement of the {$K^- / \pi^-$} structure
  function ratio using the Drell-Yan process},'' {\em Phys. Lett.}, vol.~B93,
  p.~354, 1980.

\bibitem{Badier:1983mj}
J.~Badier {\em et~al.}, ``{Experimental determination of the {$\pi$}-meson
  structure functions by the Drell-Yan mechanism},'' {\em Z. Phys.}, vol.~C18,
  p.~281, 1983.

\bibitem{Betev:1985pg}
B.~Betev {\em et~al.}, ``{Observation of anomalous scaling violation in muon
  pair production by 194-GeV/c {$\pi$}-tungsten interactions},'' {\em Z.
  Phys.}, vol.~C28, p.~15, 1985.

\bibitem{Conway:1989fs}
J.~S. Conway {\em et~al.}, ``{Experimental study of muon pairs produced by
  252-GeV pions on tungsten},'' {\em Phys. Rev.}, vol.~D39, pp.~92--122, 1989.

\bibitem{Wijesooriya:2005ir}
K.~Wijesooriya, P.~E. Reimer, and R.~J. Holt, ``{The pion parton distribution
  function in the valence region},'' {\em Phys. Rev.}, vol.~C72, p.~065203,
  2005.

\bibitem{Burden:1995ve}
C.~J. Burden, C.~D. Roberts, and M.~J. Thomson, ``{Electromagnetic Form Factors
  of Charged and Neutral Kaons},'' {\em Phys. Lett.}, vol.~B371, pp.~163--168,
  1996.

\bibitem{ElBennich:2011py}
B.~El-Bennich, G.~Krein, L.~Chang, C.~D. Roberts, and D.~J. Wilson, ``{Flavor
  SU(4) breaking between effective couplings},'' {\em Phys. Rev.}, vol.~D,
  2011.
\newblock \emph{Rapid Comm., in press}, arXiv:1111.364 [nucl-th].

\bibitem{Maris:2002mz}
P.~Maris and P.~C. Tandy, ``{Electromagnetic transition form factors of light
  mesons},'' {\em Phys. Rev.}, vol.~C65, p.~045211, 2002.

\bibitem{Sutton:1991ay}
P.~J. Sutton, A.~D. Martin, R.~G. Roberts, and W.~J. Stirling, ``{Parton
  distributions for the pion extracted from Drell-Yan and prompt photon
  experiments},'' {\em Phys. Rev.}, vol.~D45, pp.~2349--2359, 1992.

\bibitem{Gluck:1998xa}
M.~Gluck, E.~Reya, and A.~Vogt, ``{Dynamical parton distributions revisited},''
  {\em Eur. Phys. J.}, vol.~C5, pp.~461--470, 1998.

\bibitem{Chang:2010xs}
L.~Chang and C.~D. Roberts, ``{Empirically Charting Dynamical Chiral Symmetry
  Breaking},'' {\em AIP Conf. Proc.}, vol.~1261, pp.~25--30, 2010.

\bibitem{Bicudo:2001aw}
P.~Bicudo, ``{Proof of the Weinberg theorem for chiral quark models},''
  nucl-th/0110052.

\bibitem{Bicudo:2001jq}
P.~Bicudo {\em et~al.}, ``{Chirally symmetric quark description of low energy
  {$\pi \pi$} scattering},'' {\em Phys. Rev.}, vol.~D65, p.~076008, 2002.

\bibitem{Brodsky:2010ur}
S.~J. Brodsky, G.~F. de~Teramond, and A.~Deur, ``{Nonperturbative QCD Coupling
  and its $\beta$ function from Light-Front Holography},'' {\em Phys. Rev.},
  vol.~D81, p.~096010, 2010.

\bibitem{Aguilar:2010gm}
A.~C. Aguilar, D.~Binosi, and J.~Papavassiliou, ``{QCD effective charges from
  lattice data},'' {\em JHEP}, vol.~07, p.~002, 2010.

\bibitem{Richards:2010ck}
C.~M. Richards, A.~C. Irving, E.~B. Gregory, and C.~McNeile, ``{Glueball mass
  measurements from improved staggered fermion simulations},'' {\em Phys.
  Rev.}, vol.~D82, p.~034501, 2010.

\bibitem{Eichmann:2011ej}
G.~Eichmann, ``{Baryon form factors from Dyson-Schwinger equations},''
  arXiv:1112.4888 [hep-ph].

\bibitem{Jones:1999rz}
M.~K. Jones {\em et~al.}, ``{G(E(p))/G(M(p)) ratio by polarization transfer in
  e(pol.) p --> e p(pol.)},'' {\em Phys. Rev. Lett.}, vol.~84, pp.~1398--1402,
  2000.

\bibitem{Cahill:1988dx}
R.~T. Cahill, C.~D. Roberts, and J.~Praschifka, ``{BARYON STRUCTURE AND QCD},''
  {\em Austral. J. Phys.}, vol.~42, pp.~129--145, 1989.

\bibitem{Cahill:1987qr}
R.~T. Cahill, C.~D. Roberts, and J.~Praschifka, ``{CALCULATION OF DIQUARK
  MASSES IN QCD},'' {\em Phys. Rev.}, vol.~D36, p.~2804, 1987.

\bibitem{Maris:2002yu}
P.~Maris, ``{Effective masses of diquarks},'' {\em Few Body Syst.}, vol.~32,
  pp.~41--52, 2002.

\bibitem{Maris:2004bp}
P.~Maris, ``{Electromagnetic properties of diquarks},'' {\em Few Body Syst.},
  vol.~35, pp.~117--127, 2004.

\bibitem{Buck:1992wz}
A.~Buck, R.~Alkofer, and H.~Reinhardt, ``{Baryons as bound states of diquarks
  and quarks in the Nambu-Jona-Lasinio model},'' {\em Phys. Lett.}, vol.~B286,
  pp.~29--35, 1992.

\bibitem{Bentz:2007zs}
W.~Bentz, I.~C. Clo{\"e}t, T.~Ito, A.~W. Thomas, and K.~Yazaki, ``{Polarized
  structure functions of nucleons and nuclei},'' {\em Prog. Part. Nucl. Phys.},
  vol.~61, pp.~238--244, 2008.

\bibitem{Suzuki:2009nj}
N.~Suzuki {\em et~al.}, ``{Disentangling the Dynamical Origin of P-11 Nucleon
  Resonances},'' {\em Phys. Rev. Lett.}, vol.~104, p.~042302, 2010.

\bibitem{Gasparyan:2003fp}
A.~M. Gasparyan, J.~Haidenbauer, C.~Hanhart, and J.~Speth, ``{Pion nucleon
  scattering in a meson exchange model},'' {\em Phys. Rev.}, vol.~C68,
  p.~045207, 2003.

\bibitem{Aznauryan:2011td}
I.~G. Aznauryan, V.~D. Burkert, and V.~I. Mokeev, ``{Nucleon Resonance
  Electrocouplings from the CLAS Meson Electroproduction Data},''
  arXiv:1108.1125 [nucl-ex].

\bibitem{Gothe:2011up}
R.~W. Gothe, ``{Experimental Challenges of the N* Program},'' arXiv:1108.4703
  [nucl-ex].

\bibitem{Roberts:2011ym}
C.~D. Roberts, I.~C. Clo{\"e}t, L.~Chang, and H.~L.~L. Roberts,
  ``{Dressed-quarks and the Roper resonance},'' arXiv:1108.1327 [nucl-th].

\bibitem{Pauli:1957}
W.~Pauli, ``{On the conservation of the Lepton charge},'' {\em Nuovo Cim.},
  vol.~6, pp.~204--215, 1957.

\bibitem{Gursey:1958}
F.~G{\"u}rsey, ``{Relation of charge independence and baryon conservation to
  Pauli's transformation},'' {\em Nuovo Cim.}, vol.~7, pp.~411--415, 1958.

\bibitem{Oettel:1999gc}
M.~Oettel, M.~Pichowsky, and L.~von Smekal, ``{Current conservation in the
  covariant quark-diquark model of the nucleon},'' {\em Eur. Phys. J.},
  vol.~A8, pp.~251--281, 2000.

\bibitem{Kelly:2004hm}
J.~J. Kelly, ``{Simple parametrization of nucleon form factors},'' {\em Phys.
  Rev.}, vol.~C70, p.~068202, 2004.

\bibitem{Roberts:1993ks}
C.~D. Roberts, R.~T. Cahill, M.~E. Sevior, and N.~Iannella, ``{pi pi scattering
  in a QCD based model field theory},'' {\em Phys. Rev.}, vol.~D49,
  pp.~125--137, 1994.

\bibitem{Gayou:2001qd}
O.~Gayou {\em et~al.}, ``{Measurement of G(E(p))/G(M(p)) in e(pol.) p {$\to$} e
  p(pol.) to $Q^2 = 5.6\,$GeV$^2$},'' {\em Phys. Rev. Lett.}, vol.~88,
  p.~092301, 2002.

\bibitem{Qattan:2004ht}
I.~A. Qattan {\em et~al.}, ``{Precision Rosenbluth measurement of the proton
  elastic form factors},'' {\em Phys. Rev. Lett.}, vol.~94, p.~142301, 2005.

\bibitem{Punjabi:2005wq}
V.~Punjabi {\em et~al.}, ``{Proton elastic form factor ratios to $Q^2 =
  3.5\,$GeV$^2$ by polarization transfer},'' {\em Phys. Rev.}, vol.~C71,
  p.~055202, 2005.
\newblock [Erratum-ibid.C71:069902,2005].

\bibitem{Puckett:2010ac}
A.~J.~R. Puckett {\em et~al.}, ``{Recoil Polarization Measurements of the
  Proton Electromagnetic Form Factor Ratio to $Q^2 = 8.5\,$GeV$^2$},'' {\em
  Phys. Rev. Lett.}, vol.~104, p.~242301, 2010.

\bibitem{Puckett:2011xg}
A.~J.~R. Puckett {\em et~al.}, ``{Reanalysis of Proton Form Factor Ratio Data
  at $\mathbf{Q^2 =}$ 4.0, 4.8, and 5.6 GeV$\mathbf{^2}$},'' arXiv:1102.5737
  [nucl-ex].

\bibitem{Riordan:2010id}
S.~Riordan {\em et~al.}, ``{Measurements of the Electric Form Factor of the
  Neutron up to $Q^2=3.4\,$GeV$^2$ using the Reaction
  $^3$He(e,e$^\prime$n)pp},'' {\em Phys. Rev. Lett.}, vol.~105, p.~262302,
  2010.

\bibitem{Cates:2011pz}
G.~D. Cates, C.~W. de~Jager, S.~Riordan, and B.~Wojtsekhowski, ``{Flavor
  decomposition of the elastic nucleon electromagnetic form factors},'' {\em
  Phys. Rev. Lett.}, vol.~106, p.~252003, 2011.

\bibitem{Bloch:2003vn}
J.~C.~R. Bloch, A.~Krassnigg, and C.~D. Roberts, ``{Regarding proton form
  factors},'' {\em Few Body Syst.}, vol.~33, pp.~219--232, 2003.

\bibitem{Hen:2011rt}
O.~Hen, A.~Accardi, W.~Melnitchouk, and E.~Piasetzky, ``{Constraints on the
  large-$x$ d/u ratio from electron-nucleus scattering at $x>1$},''
  arXiv:1110.2419 [hep-ph].

\bibitem{Holt:2010zz}
R.~J. Holt and J.~R. Arrington, ``{From the nucleus to the quarks},'' {\em AIP
  Conf. Proc.}, vol.~1261, pp.~79--84, 2010.

\end{thebibliography}

\end{document}